\documentclass[apj]{emulateapj}
\usepackage{apjfonts,amsmath,natbib}
\usepackage[breaklinks,colorlinks,urlcolor=blue,citecolor=blue,linkcolor=blue]{hyperref}
\makeatletter

\newcommand{\Rmnum}[1]{\expandafter\@slowromancap\romannumeral #1@}
\makeatother

\newcommand{\gps}{\ensuremath{g_{\rm P1}}}
\newcommand{\gpo}{\ensuremath{g_{\rm P1,\ 0}}}
\newcommand{\rps}{\ensuremath{r_{\rm P1}}}
\newcommand{\rpo}{\ensuremath{r_{\rm P1,\ 0}}}
\newcommand{\ips}{\ensuremath{i_{\rm P1}}}
\newcommand{\zps}{\ensuremath{z_{\rm P1}}}

\newcommand{\griz}{\gps\rps\ips\zps}

\slugcomment{Accepted by ApJ}

\shorttitle{PS1 Monoceros Ring}
\shortauthors{Morganson et al.}

\begin{document}


\title{Mapping the Monoceros Ring in 3D with Pan-STARRS1}


\author{Eric Morganson\altaffilmark{1, 2,3}} 
\author{Blair Conn\altaffilmark{4, 1}} 
\author{Hans-Walter Rix\altaffilmark{1}} 
\author{Eric F. Bell\altaffilmark{5}} 
\author{William S. Burgett\altaffilmark{6}}
\author{Kenneth Chambers\altaffilmark{7}} 
\author{Andrew Dolphin\altaffilmark{8}} 
\author{Peter W. Draper\altaffilmark{9}}
\author{Heather Flewelling\altaffilmark{7}}
\author{Klaus Hodapp\altaffilmark{7}} 
\author{Nick Kaiser\altaffilmark{7}} 
\author{Eugene A. Magnier\altaffilmark{7}} 
\author{Nicolas F. Martin\altaffilmark{10,1}} 
\author{David Martinez-Delgado\altaffilmark{11}}
\author{Nigel Metcalfe\altaffilmark{9}}
\author{Edward F. Schlafly\altaffilmark{1}}  
\author{Colin T. Slater\altaffilmark{5}} 
\author{Richard J. Wainscoat\altaffilmark{7}} 
\author{Christopher Z. Waters\altaffilmark{7}}
\email{ericm@illinois.edu}

\altaffiltext{1}{Max-Planck-Institut f\"ur Astronomie, K\"onigstuhl 17, 69117 Heidelberg, Germany}
\altaffiltext{2}{National Center for Supercomputing Applications, University of Illinois at Urbana-Champaign, 1205 W. Clark Street, Urbana, IL 61801, USA}
\altaffiltext{3}{Harvard Smithsonian Center for Astrophysics, 60 Garden St, Cambridge, MA 02138, USA}
\altaffiltext{4}{Gemini Observatory, Casilla 603, La Serena, Chile}
\altaffiltext{5}{Department of Astronomy, University of Michigan, 500 Church Street, Ann Arbor, MI 48109, USA}
\altaffiltext{6}{GMTO Corp, Suite 300, 251 S. Lake Ave, Pasadena, CA 91101, USA}
\altaffiltext{7}{Institute for Astronomy, University of Hawaii at Manoa, Honolulu, HI 96822, USA}
\altaffiltext{8}{Raytheon Company, 1151 E Hermans Rd, Tucson, AZ 85756, USA}
\altaffiltext{9}{Department of Physics, University of Durham Science Laboratories, South Road Durham DH1 3LE, UK}
\altaffiltext{10}{Observatoire astronomique de Strasbourg, Universit\'e de Strasbourg, CNRS, UMR 7550, 11 rue de l'Universit\'e, F-67000 Strasbourg, France}
\altaffiltext{11}{Astronomisches Rechen-Institut, Zentrum f\"ur Astronomie der Universit\"at Heidelberg, M\"onchhofstr. 12-14, 69120 Heidelberg, Germany}



\begin{abstract} 

Using the Pan-STARRS1 survey, we derive limiting magnitude, spatial completeness and density maps that we use to probe the three dimensional structure and estimate the stellar mass of the so-called Monoceros Ring. The Monoceros Ring is an enormous and complex stellar sub-structure in the outer Milky Way disk. It is most visible across the large Galactic Anticenter region, $120^\circ < l < 240^\circ$, $-30^\circ < b < +40^\circ$. We estimate its stellar mass density profile along every line of sight in 2$\times$2 degree pixels over the entire 30,000 deg$^2$ Pan-STARRS1 survey using the previously developed \textsc{match} software. By parsing this distribution into a radially smooth component and the Monoceros Ring, we obtain its mass and distance from the Sun along each relevant line of sight. The Monoceros Ring is significantly closer to us in the South (6 kpc) than in the North (9 kpc). We also create 2D cross sections parallel to the Galactic plane that show $135^\circ$ of the Monoceros Ring in the South and $170^\circ$ of the Monoceros Ring in the North. We show that the Northern and Southern structures are also roughly concentric circles, suggesting that they may be a wave rippling from a common origin. Excluding the Galactic plane $\sim \pm 4^\circ$, we observe an excess mass of $4 \times 10^6 M_\odot$ across $120^\circ < l < 240^\circ$. If we interpolate across the Galactic plane, we estimate that this region contains $8 \times 10^6 M_\odot$. If we assume (somewhat boldly) that the Monoceros Ring is a set of two Galactocentric rings, its total mass is $6 \times 10^7 M_\odot$. Finally, if we assume that it is a set of two circles centered at a point 4 kpc from the Galactic center in the anti-central direction, as our data suggests, we estimate its mass to be $4 \times 10^7 M_\odot$. 

\end{abstract}





\section{Introduction}\label{sect:intro}

Our knowledge of the outer disc of the Milky Way has undergone a revolution in the last decade as new deep photometric surveys have probed close to the Galactic plane revealing new structures and new mysteries regarding their formation.  Today we recognize five major stellar substructures in the outer disc, The Monoceros Ring, Triangulum-Andromeda, the Anti-Center Stream, the Eastern Banded Structure and a tidal arm of the Sagittarius Dwarf galaxy (the Sagittarius Stream).

The first to be discovered and studied extensively was a large stellar overdensity located between 14-18 kpc from the Galactic center and across $60^\circ < l < 280^\circ$ at Galactic heights of $|z| <$ 5 kpc. This structure was dubbed the Monoceros Ring\footnote{Other names for this structure include: Galactic Anticenter Stellar Stream; Galactic Anticenter Stellar Structure; Monoceros stream; Monoceros Overdensity}, and many studies to trace and map its full extent have followed its discovery. While initially constrained to the SDSS footprint \citep{NEWB++02,YANN++03}, many smaller surveys probed the Galactic plane for its signature and through this an appreciation of its large scale was developed \citep{IBAT++03,CONN++05a,CONN++05b,CONN++07,CONN++08,CONN++12,SOLL++11}. With the Two Micron All-Sky Survey \citep[2MASS,][]{SKRU++06}, \cite{ROCH++03} isolated the MR M-giants and mapped the structure between 12$^{\circ}< b < +36^{\circ}$ and 100$^{\circ} < l < 270^{\circ}$ The most recent spectroscopic studies have determined a velocity dispersion of 15 km/s and a metallicity of [Fe/H] $\sim -0.8\pm0.01$ \citep{LI++12}. 

Explanations for the MR's size and extent were initially split between those who proposed a Milky Way origin for the MR and those who favored a disrupting satellite scenario. The disrupting satellite proponents were encouraged when the discovery of the Canis Major dwarf galaxy candidate was announced as a potential progenitor for forming the MR \citep{MART++04,MART++06,CONN++07}. Efforts to characterize Canis Major began in earnest, and this area of the Galaxy was heavily studied \citep{FORB++04, BELL++04, BELL++06, MART++04b, DINE++05, MART++05, BUTL++07, DEJO++07}. In parallel, many investigated how extended galactic discs, and more specifically the MR, could be formed as a result of a dwarf galaxy accretion event \citep{HELM++03, MART++04, PENA++05,PENA++06, SOLL++11}. For those interpreting the evidence as belonging to the warp, flare or spiral arms of the Milky Way, their focus was predominantly on Canis Major \citep{MOMA++04,MOMA++06,MOIT++06,LOPE++07,PIAT++08,REYL++09}, with only a few studies specifically related to the MR in the outer disc \citep{LOPE++14, KALB++14}.  These were countered by a series of papers outlining the reasons why standard Galactic structures were insufficient to explain the exact properties of the MR \citep{MART++06, CASE++08, CASE++10, DEJO++10, CONN++12}. A third smaller group discussed the possibility of the MR being formed from a caustic in the dark matter profile of the Milky Way \citep{SIKI03, NATA++07, DUFF++08}. Finally, a fourth group has been investigating how various density waves can propagate through the Galactic disc, and there is mounting evidence that the Galactic disc has both internal and external processes which can influence the density and location of stars in the outer disc. The internal Galactic processes for forming stellar structures in the outer disc mostly involve the radial migration of stars driven by the bar \citep{MINC++12b, MINC++12a} while the external model involves the influence of a dwarf galaxy passing nearby or through the disc. Modeling of a Sagittarius-like dwarf galaxy with a disc has demonstrated that such collisions or fly-bys can cause "ringing" in the disc and drive the formation of rings and streams in the outer disc \citep{KAZA++08, YOUN++08, MICH++11, PURC++12, GOME++12, GOME++13}. The evidence for such modes in the disc is building with density and velocity asymmetries having been reported by \citet{WIDR++12}, \citet{CARL++13}, \citet{YANN++13}, \citet{WILL++13}, \citet{WIDR++14} and \citet{WIDR++15}. Most recently,  \citet{XU++15} report on the presence of a radial wave in the disc as detected by analysis of the SDSS data, supporting a prediction made by \citet{IBAT++03} in the earliest days of this field.

The other large outer disk structures of the Anti-Center Stream \citep{ROCH++03, CRAN++03, FRIN++04} and Triangulum Andromeda \citep[TriAnd,][]{ROCH++04, MAJE++04} were both initially discovered when searching for overdensities of M-giant stars in 2MASS.  However, the initial discovery of the actual ACS was hampered by its confusion with the MR. Due to the poor spatial sampling of the M-giant population along with the MR and ACS having the same distance \citet{ROCH++04} confused the two and reported the ACS as synonymous with the MR. The details of the MR were so loosely constrained at this stage that it was not until \citet{GRIL06} and \citet{GRIL++08} that a clearer picture of the ACS began to develop.  The ACS is now thought to be a tidal stream and is visible above the plane, extending from (l,b) = $(151,+38)^{\circ}$ to $(224.8,+20)^{\circ}$. 

The Triangulum-Andromeda stellar structure is located between $100^\circ < l < 150^\circ$ and $−20^\circ > b > −40^\circ$ at a distance of 15-30 kpc from the Sun. TriAnd is thus more distant and has a larger line of sight depth than the MR. \citet{ROCH++04} concluded it was the remnant of a dwarf galaxy merger due to its very cold velocity dispersion of $\sigma \sim$17 km.s$^{-1}$.  \citet{MART++07} detected TriAnd in the foreground of M31 and resolved it into two structures, one at $\sim$25 kpc (TriAnd1) and $\sim$33 kpc (TriAnd2) covering at least 76 square degrees. \citet{CHOU++11} performed a chemical analysis study of TriAnd stars and confirm that it is indeed a separate structure to the ACS and MR. \citet{SHEF++14} continued the study into TriAnd finding that the nearer TriAnd component (TriAnd1) is younger (6-10 Gyr) than the more distant older component (TriAnd2) at 10-12 Gyrs. They propose that both TriAnd1 and TriAnd2 are material from a dwarf galaxy accretion event which were formed during two distinct pericentric passages. Finally, \citet{PRIC++15} proposes that TriAnd-like substructures can be formed by concentric rings propagating outwards through the Galactic disc. This lends support to \citet{XU++15} who have proposed such a scenario for the formation of the MR.

Finally, the Eastern Banded Structure (EBS) discovered by \citet{GRIL06} is now recognized as another tidal stream located in the Galactic Anticenter and very close to the ACS [$(l,b)^\circ$ = $\sim$(229,+30)$^{\circ}$ to $\sim$(217,30)$^{\circ}$]. It was first thought to be part of the ACS tidal arms \citep{GRIL++08} but it is now recognized as being associated with a potential dwarf galaxy candidate Hydra I \citep{GRIL11, HARG++15}.

The work presented here focuses mostly on the MR and the challenges involved in studying such a large stellar structure in the Galactic disc. This is the first study with contiguous optical data of sufficient depth near the Galactic plane to study the MR where it is most pronounced. Although initially discovered in the Sloan Digital Sky Survey \citep[SDSS,][]{NEWB++02}. SDSS \citep{YORK++00} excluded the Milky Way plane and thus only uncovering the northern edge of the MR. \citet{MOMA++06} used 2MASS to study the outer Milky Way (including the disk and MR), but 2MASS's depth limited this work to luminous red clump and red giant stars which necessarily limited its precision. Conversely, \citet{DEJO++10} used the Sloan Extension for Galactic Understanding and Exploration (SEGUE) stripes which cross the Milky Way plane and have sufficient depth to observe main sequence MR stars. But these stripes only provide relatively localized pictures of the MR. \citet{MART++06,CONN++12} and other works mentioned above have used deep photometry and spectroscopy to measure metallicities, stellar types and velocities in the MR, but a lack of sufficiently deep and wide optical data has prevented us from studying the global structure of the MR with great precision. Without large contiguous datasets near the Galactic plane, even estimating the total mass of the MR has been out of reach. 
The Panoramic Survey Telescope and Rapid Response System 1 Survey \citep[PS1][]{KAIS++10} 3$\pi$ dataset is an ideal dataset to study the MR near the plane and produce the first large scale, contiguous map of the MR. PS1 has sufficient depth in the \gps\ and \rps\ bands to measure the luminosity and color of main sequence stars, particularly the blue edge stars which are plentiful and suffer less from the dwarf/giant degeneracy of redder stars, and it covers approximately 75\% of the Galactic plane. \citet{SLAT++14} used the PS1 dataset to take the first large scale contiguous look at the MR and qualitatively compare the MR results to computational models of the MR as a satellite accretion \citep{PENA++05} and the MR as a disrupted disk \citep{KAZA++09}.

In this paper, we extend the work from \citet{SLAT++14} using the CMD-fitting techniques from \citet{DEJO++10} to produce a more quantitative, three dimensional analysis of the MR. We observe the large scale structure of the MR and particularly find major asymmetries in mass and distance in the North and South. We also estimate the total observed mass and extrapolate the total mass of the MR for a variety of models. In Section \ref{sect:data}, we discuss the PS1 dataset, its advantages and limitations when mapping the MR. In Sections \ref{sect:density} and \ref{sect:los} we will discuss how we bin stars across the sky and use the \textsc{match} software to measure densities along the line of sight of every 2 $\times$ 2 degree (constant area of 4 square degrees) pixel. In Sections \ref{sect:map} and \ref{sect:map3d} we discuss our techniques for making large, contiguous, 2D and 3D maps of the Milky Way and Monoceros Ring. We estimate the total mass of the Monoceros Ring and in Section \ref{sect:mass}.

\section{Data}\label{sect:data}

PS1 has produced an ideal catalog to study stellar densities in the Monoceros Ring. Like SDSS, PS1 has sufficient optical depth in the $g$ and $r$ filters to measure the magnitudes and colors of main sequence stars at MR distances of roughly 10 kpc. Unlike SDSS, PS1 does not avoid the Milky Way disk. Instead it covers the entire 30,000 deg$^2$ above declination $-30^\circ$. This includes roughly 75\% of the plane of the Milky Way. At very low latitudes, PS1's depth is limited by dust and the light from foreground stars. But PS1 can make reliable measurements of MR stars down to Galactic latitudes of $\pm4^\circ$ while SDSS was mainly restricted to be more than $30^\circ$ from the plane.

\begin{table}
\centering
\begin{tabular}{cccc}
	\hline
Filter &  SDSS    &   PS1 Single Exposure & PS1 Average \\
	\hline
$u$   & 21.3 & --   & --   \\
$g$   & 22.3 & 21.3 & 21.6 \\
$r$   & 21.9 & 21.1 & 21.5 \\
$i$   & 21.4 & 20.8 & 21.2 \\
$z$   & 19.9 & 20.1 & 20.7 \\
$y$   & --   & 19.1 & 19.5 \\
	\hline
\end{tabular}
\caption{\rm{10$\sigma$ Limiting AB Magnitudes of point sources in SDSS and PS1 3$\pi$. PS1 Average results are made by averaging all detections together. Similarly-named filters from SDSS and PS1 are not exactly the same.}}\label{tab:limmags}
\end{table}

The PS1 \griz\ filters (described in \citet{TONR++12}) are fairly similar to the analogously named SDSS filters, although the \gps\ filter (including the camera sensitivity) is a bit redder than the SDSS $g$ filter and the \zps filter has a sharper red cutoff than the SDSS $z$ filter. SDSS-PS1 filter transformations are discussed in \citet{MORG++14}.

In Table \ref{tab:limmags}, we show the 10$\sigma$ limiting magnitudes of the SDSS and PS1 surveys. In this paper, we used the average (median) of at least two (typically ten) 5$\sigma$ detections that pass a quality cut (no flag matching the PS1-defined 0X00003f98 and at least 85\% of the PSF flux is on reliable, unmasked pixels) within the single exposure catalogs as opposed to the detections made from stacked images. PS1 is currently making a catalog from the stacked images that will significantly increase its depth, but this catalog was not available at the time of this work. We calculate these limiting magnitudes by tiling the sky in 2$\times$2 degree squares (accounting for latitude so that all pixels are equal area), finding the mean object with photometric uncertainty of approximately 0.1 magnitudes and then taking the median value across all pixels. Our isochrone fitting (Section \ref{sect:density}) works best if we limit ourselves to stars that are brighter (and generally bluer) than mid G-type stars, which have absolute \gps\ magnitudes of $6.4$  and \gps0-\rps\ of $0.5$. The MR is typically less than 10 kpc (distance modulus 15) away from us in the Anticenter direction requiring a limiting magnitude of $\gps = 21.4$ to sample well. So both PS1 and SDSS have enough depth to probe main sequence stars in the MR across a significant area.

\begin{figure*}[ht]
\plottwo{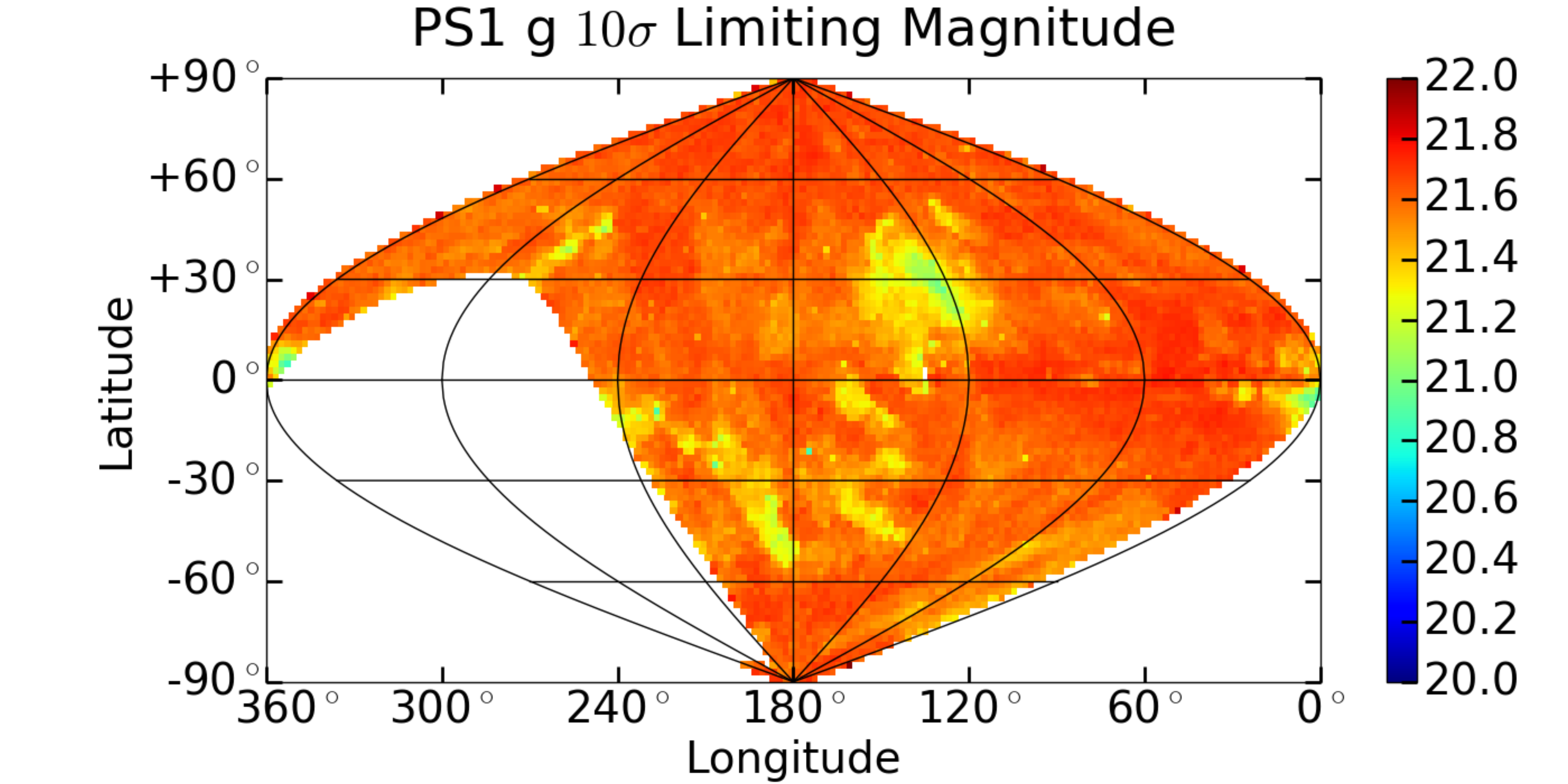}{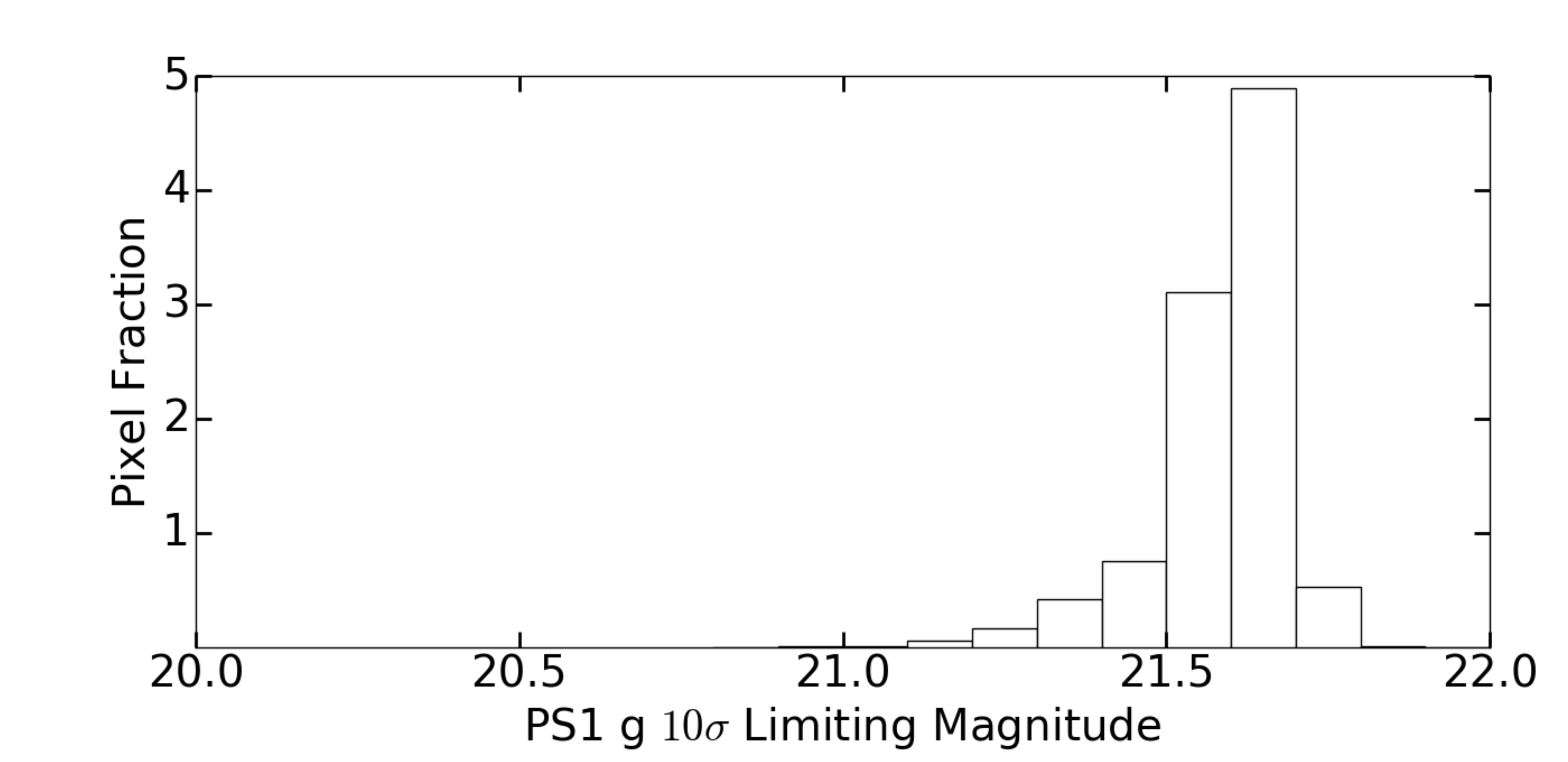}
\plottwo{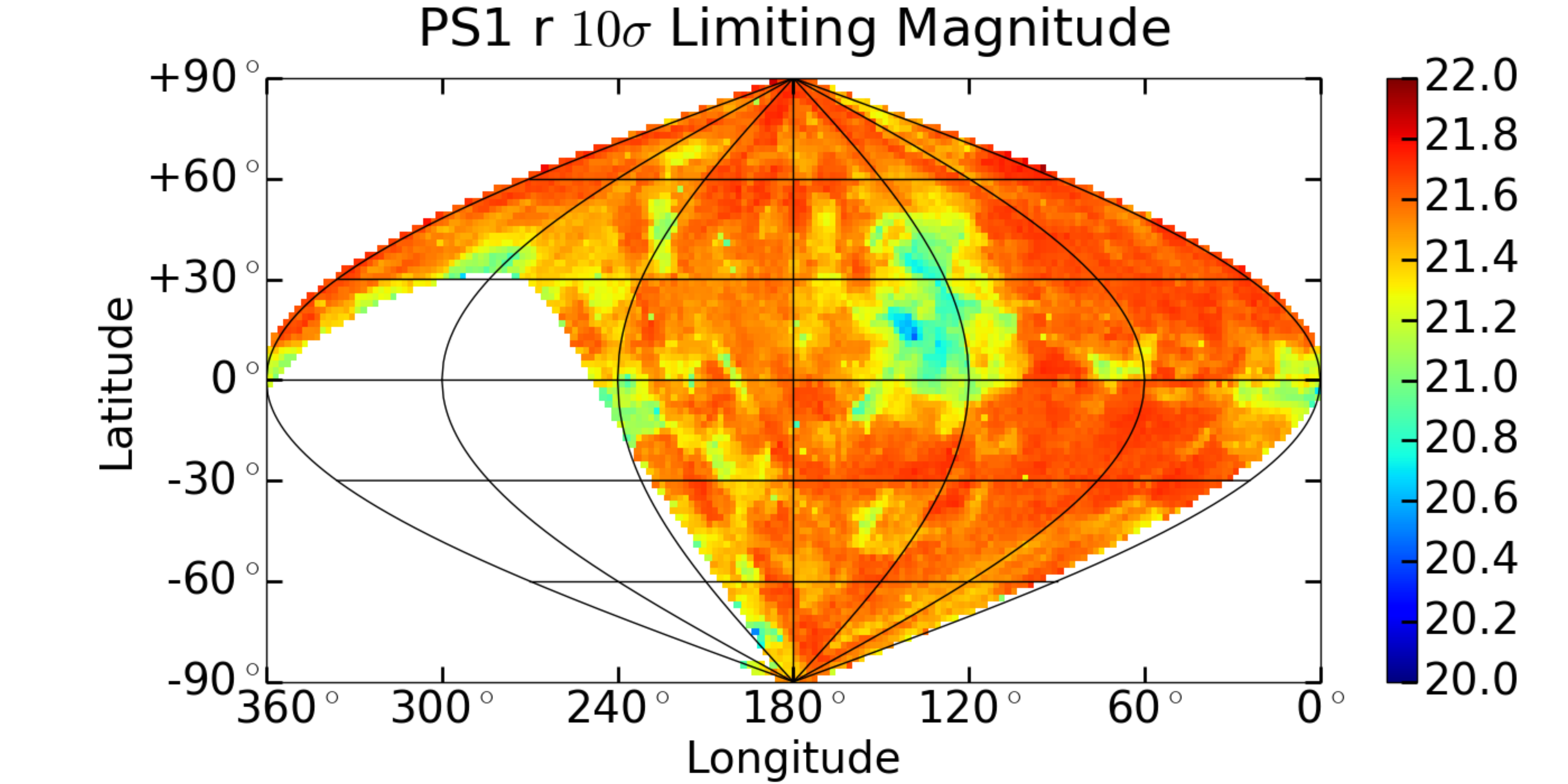}{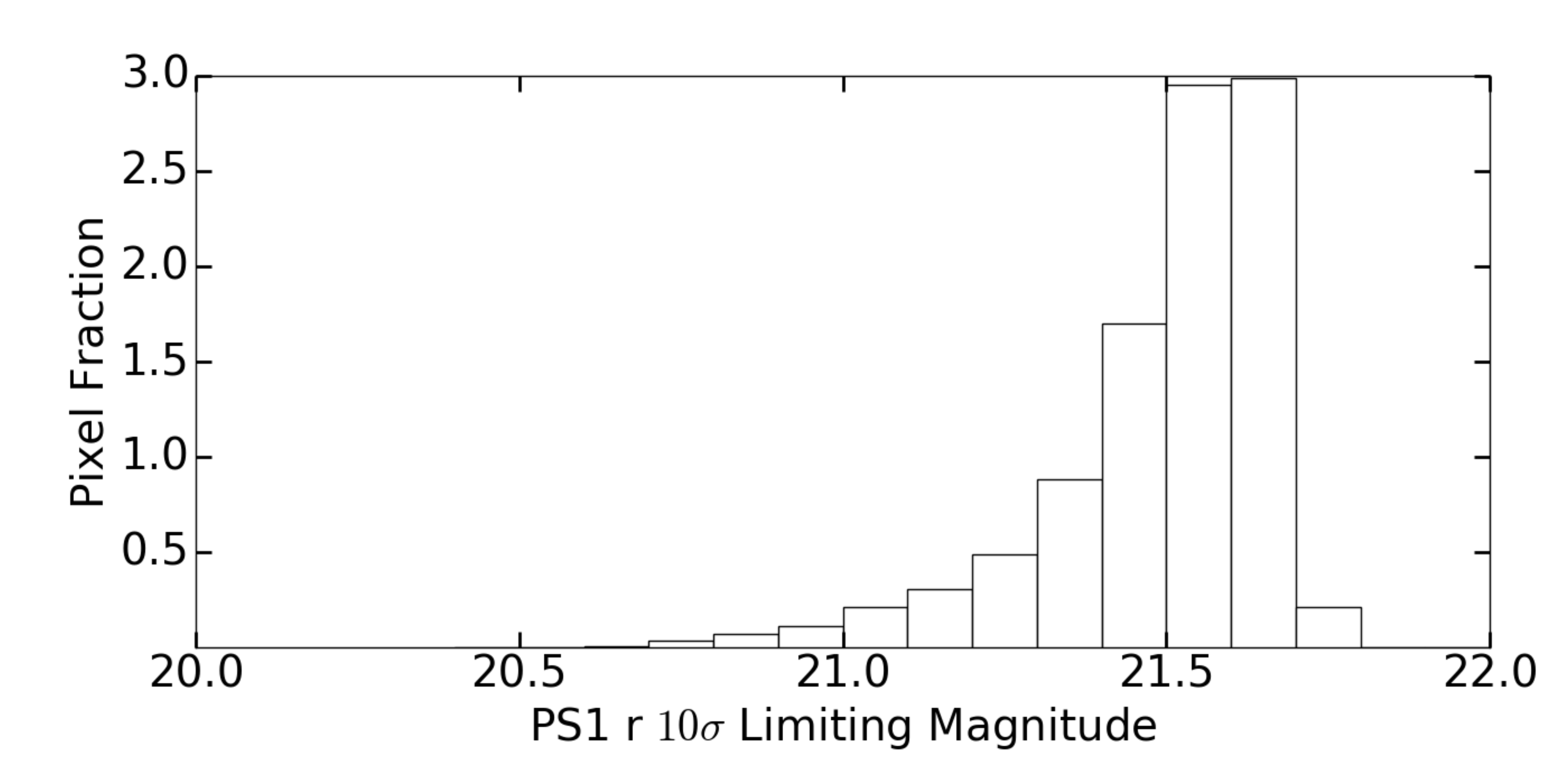}
\caption{\rm{The \gps\ band 10$\sigma$ point source limiting magnitude across the sky in Galactic coordinates (top left) and the distribution (fraction per magnitude so that the integral is unity) of limiting magnitude of the different pixels (top right). We show the analogous quantities for the \rps\ filter in the two bottom panels.}}
\label{fig:depth}\end{figure*}

In Fig.\ \ref{fig:depth}, we show the \gps\ and \rps\ 10$\sigma$ limiting magnitudes in Galactic coordinates. We see that PS1 covers roughly 75\% of the Milky Way plane including the Galactic Anticenter, which is the easiest place to detect the MR. The limiting magnitude is also relatively uniform, only rarely going below $\gps = 21$ where depth becomes a major issue for detecting the MR. 

\begin{figure*}[ht]
\plottwo{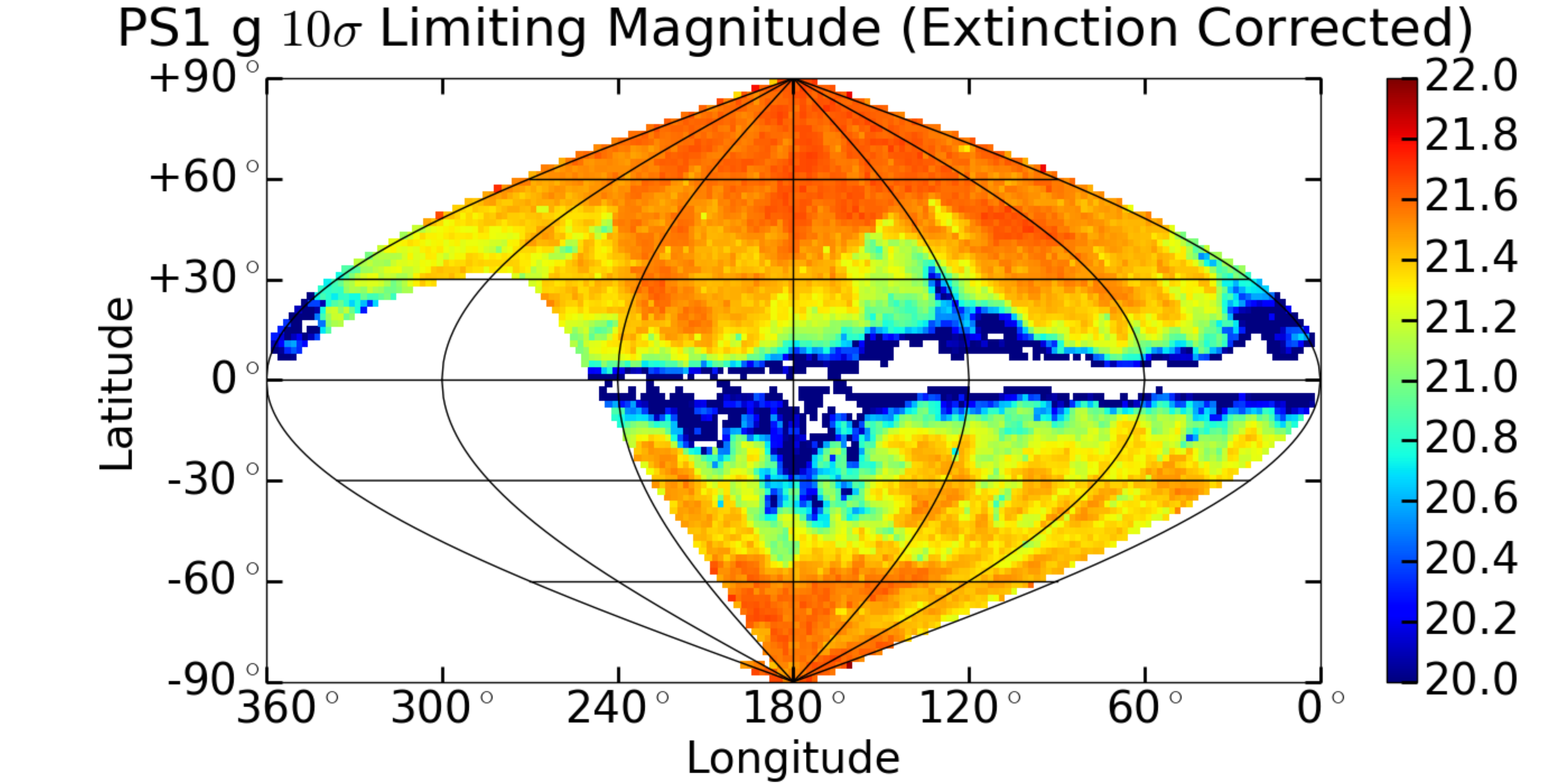}{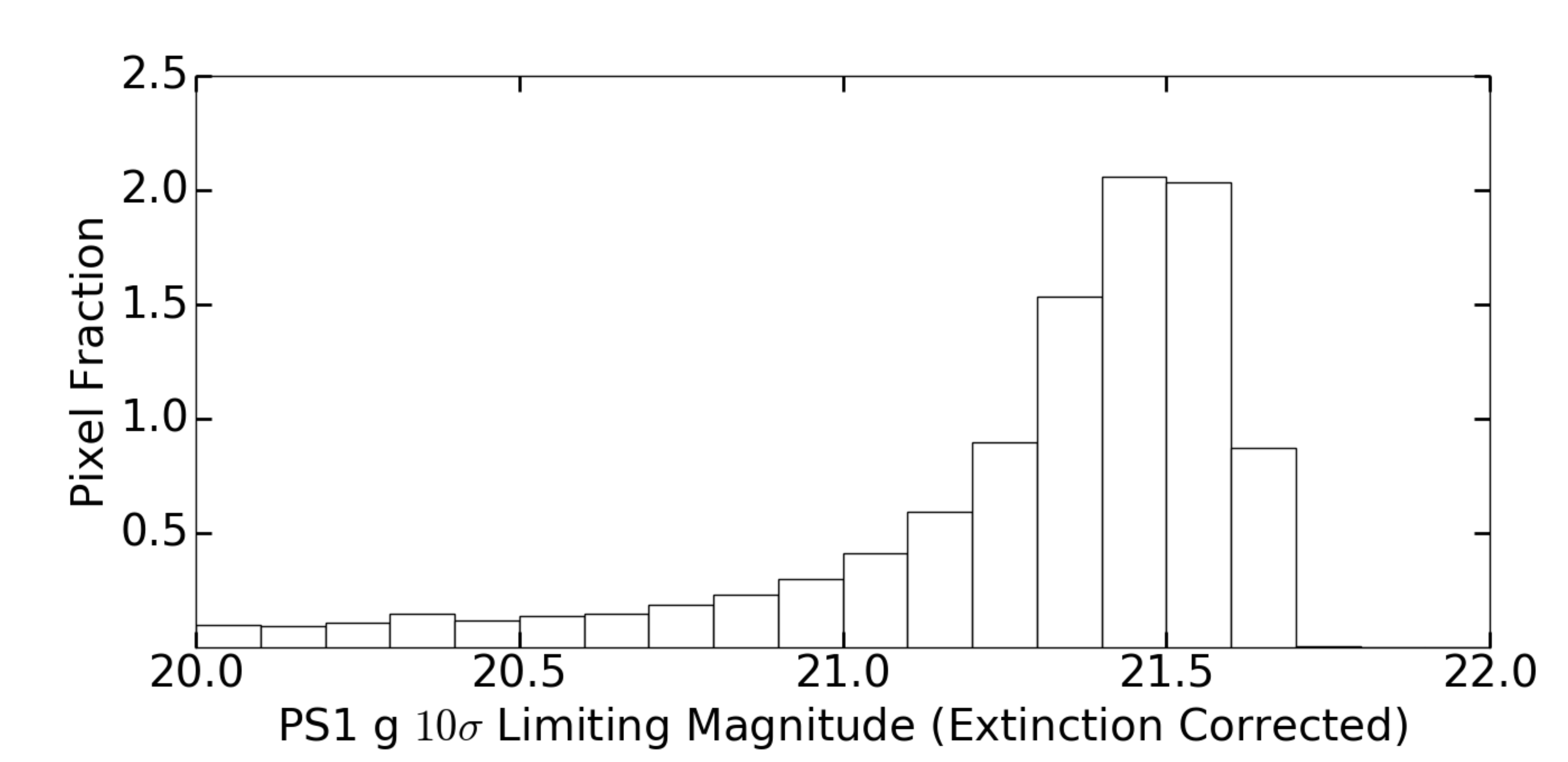}
\plottwo{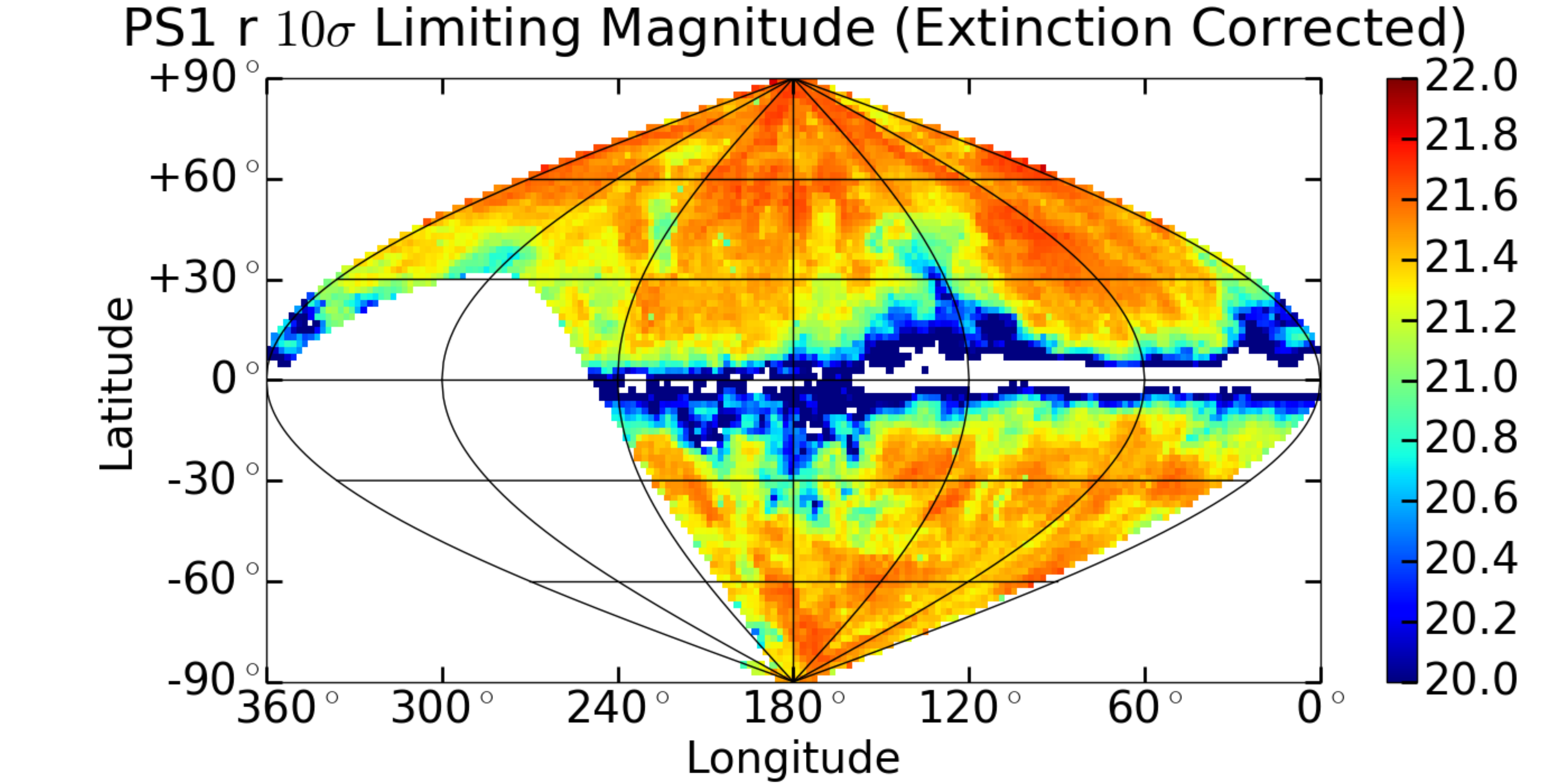}{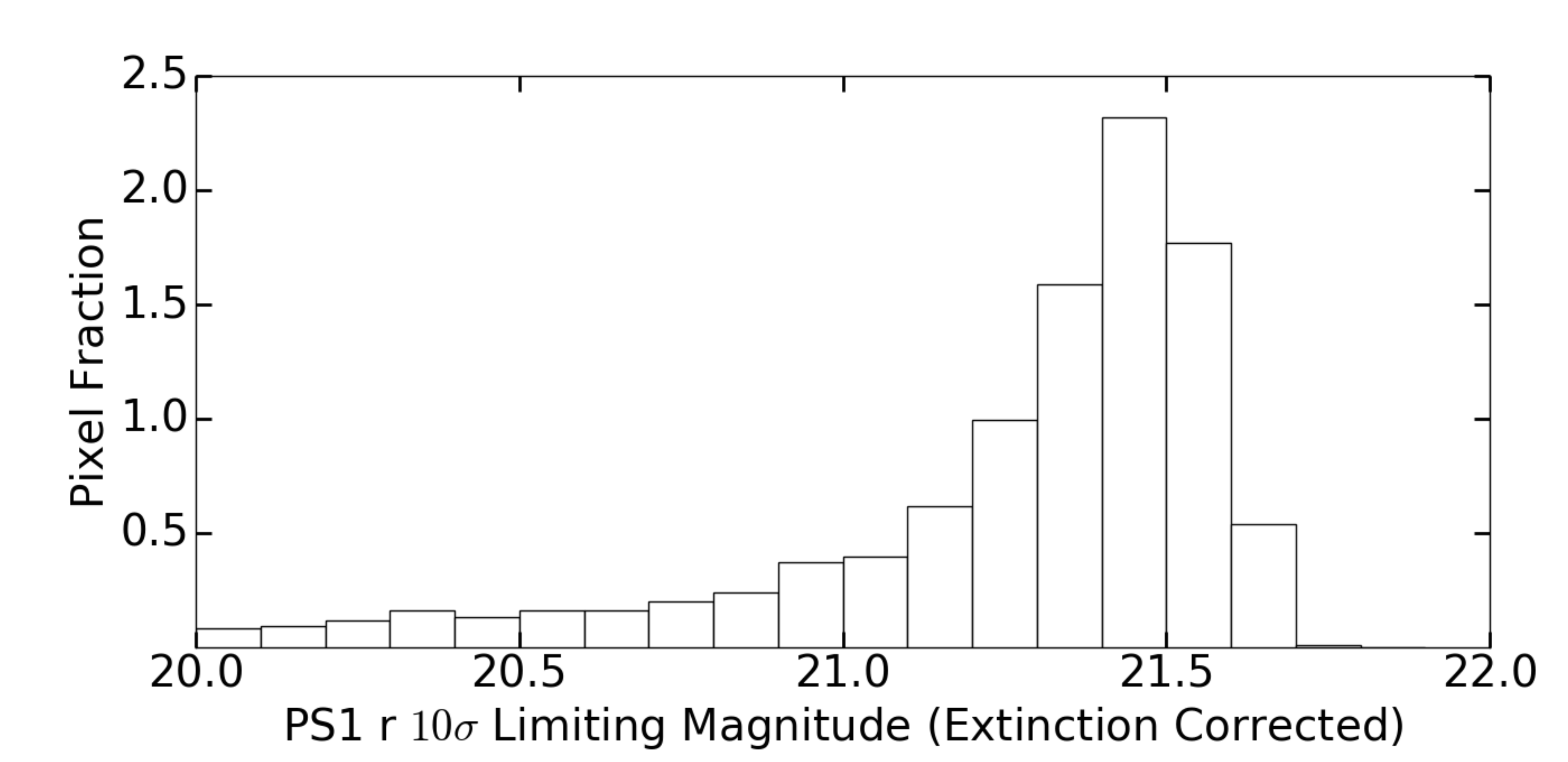}
\caption{\rm{The dust-corrected \citep{SCHL++98} \gpo\ band 10$\sigma$ point source limiting magnitude across the sky in Galactic coordinates (top left) and the distribution (fraction per magnitude so that the integral is unity) of limiting magnitude of the different pixels (top right). We show the analogous quantities for the \rpo\ filter in the two bottom panels. The white areas at low latitude do not have values in the public \citet{SCHL++98} maps.}}
\label{fig:dustdepth}\end{figure*}

Since the MR is roughly aligned with the Galactic plane where dust plays a significant role, we must compensate for dust to optimize our measurement of the MR. Since the MR is typically 15 kpc from the Galactic Center, we can use the two dimensional Schlegel-Finkbeiner-Davis \citep[SFD,][]{SCHL++98} dust map that includes all of the MW dust along each line of sight. We multiply the E(B-V) number reported by SFD by the PS1 extinction coefficients from \citet{SCHL++11} with $R_V = 3.1$ to correct our magnitudes. In the crucial \gps\ and \rps\ bands, these coefficients are 3.172 and 2.271, respectively. In Fig.\ \ref{fig:dustdepth}, we show the dust-corrected PS1 10$\sigma$ depth. Even accounting for dust absorption, our limiting magnitude satisfies $\gpo > 21 (20.5)$ for 78\% (87\%) of the sky more than $4^\circ$ away from the Galactic plane. In our unmasked regions (see Eq. \ref{eq:mask}) our limiting magnitude satisfies $\gpo > 21 (20.5)$ for 85\% (94\%) of the sky.

We always bin the sky in roughly 2 $\times$ 2 degree pixels. Our pixel height ($\Delta b$) is always exactly 2 degrees. Our pixel width ($\Delta l$) is constant for each $b$ and is chosen so that the area of each pixel is as close to 4 square degrees as possible while also having an integer number of pixels for each $b$. In practice, this means that pixel area varies from 4 square degrees by roughly $1\%$ over most of the sky. We correct for pixel area in all density calculations, although in the text we just assume the area is 4 square degrees.

\subsection{PS1 Spatial Completeness}\label{sect:completeness}

In order to accurately estimate the density of stars in the Monoceros Ring, we must accurately account for the spatial completeness of the PS1 survey. PS1 does not cover the sky uniformly. Gaps between chips in the camera and between exposures combine with bad weather to produce small holes in the PS1 survey. To measure the fraction of the sky covered by PS1, we cross-match our PS1 data with the 2MASS point source catalog \citep{SKRU++06} and calculate the fraction of stars in 2MASS that are detected in the \gps\ and \rps\ bands. 

We use the fraction of 2MASS stars we detect in each 2 $\times$ 2 degree pixel as a proxy for the spatial completeness in PS1 in that pixel. In order for the fraction of 2MASS stars detected by PS1 to accurately represent the spatial completeness of PS1, we must ensure that our 2MASS stars are real stars. We thus require that our 2MASS objects be internally classified as stars and
\begin{eqnarray}
\rm{SNR}_{J\ \rm{2MASS}} &>& 5,\\ 
\rm{SNR}_{H\ \rm{2MASS}} &>& 5\ \rm{or}\ \rm{SNR}_{K\ \rm{2MASS}} > 5,\nonumber\\
13 < J_{\rm{2MASS}} &<& 15\nonumber
\end{eqnarray}  
SNR is the signal to noise ratio in each 2MASS filter, and this requirement is simply saying that a source must be a 5$\sigma$ detection in the $J$ and either the $H$ or $K$ filter. The $13 < J_{\rm{2MASS}} < 15$ requirement ensures that the source should be observed in PS1 as $g'-J$ colors are between 0 and 6.5 across the main sequence \citep{DAVE++14}. We must also use our actual PS1 source requirement for MR detections:
\begin{eqnarray}
\rm{Err}_{g\ \rm{P1}},\ \rm{Err}_{r\ \rm{P1}} &<& 0.2,\label{eq:req}\\
\rm{N}_{g\ \rm{P1\ Detect}},\ \rm{N}_{r\ \rm{P1\ Detect}} &>& 1.\nonumber
\end{eqnarray}
Here Err is statistical error and $\rm{N}_{\rm{P1\ Detect}}$ is the number of detections in a given filter. $\rm{N}_{\rm{P1\ Detect}}$  can be as high as 16 but has a typical value of 10.

\begin{figure}[ht]
\includegraphics[width=0.99\columnwidth]{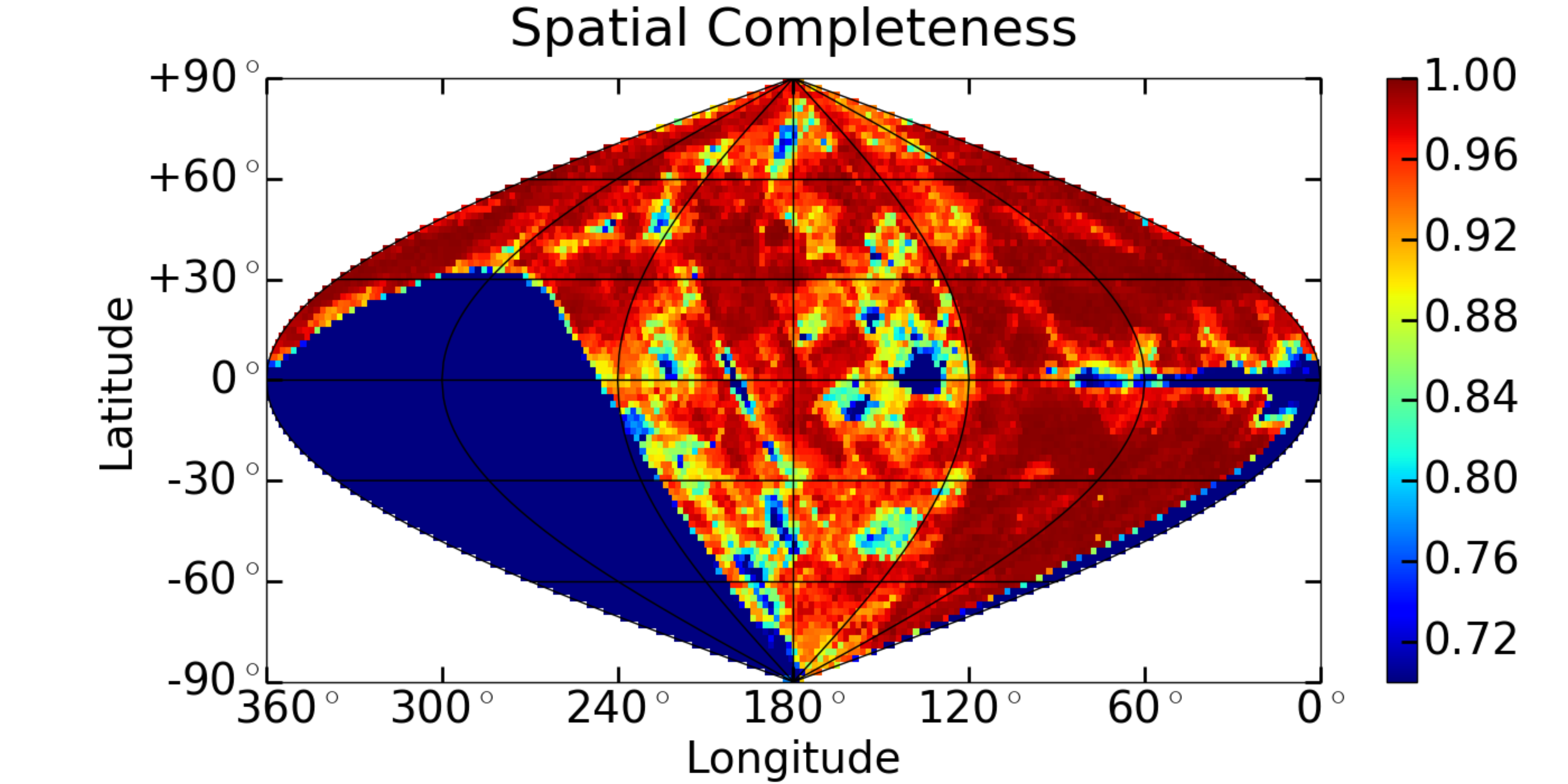}
\includegraphics[width=0.99\columnwidth]{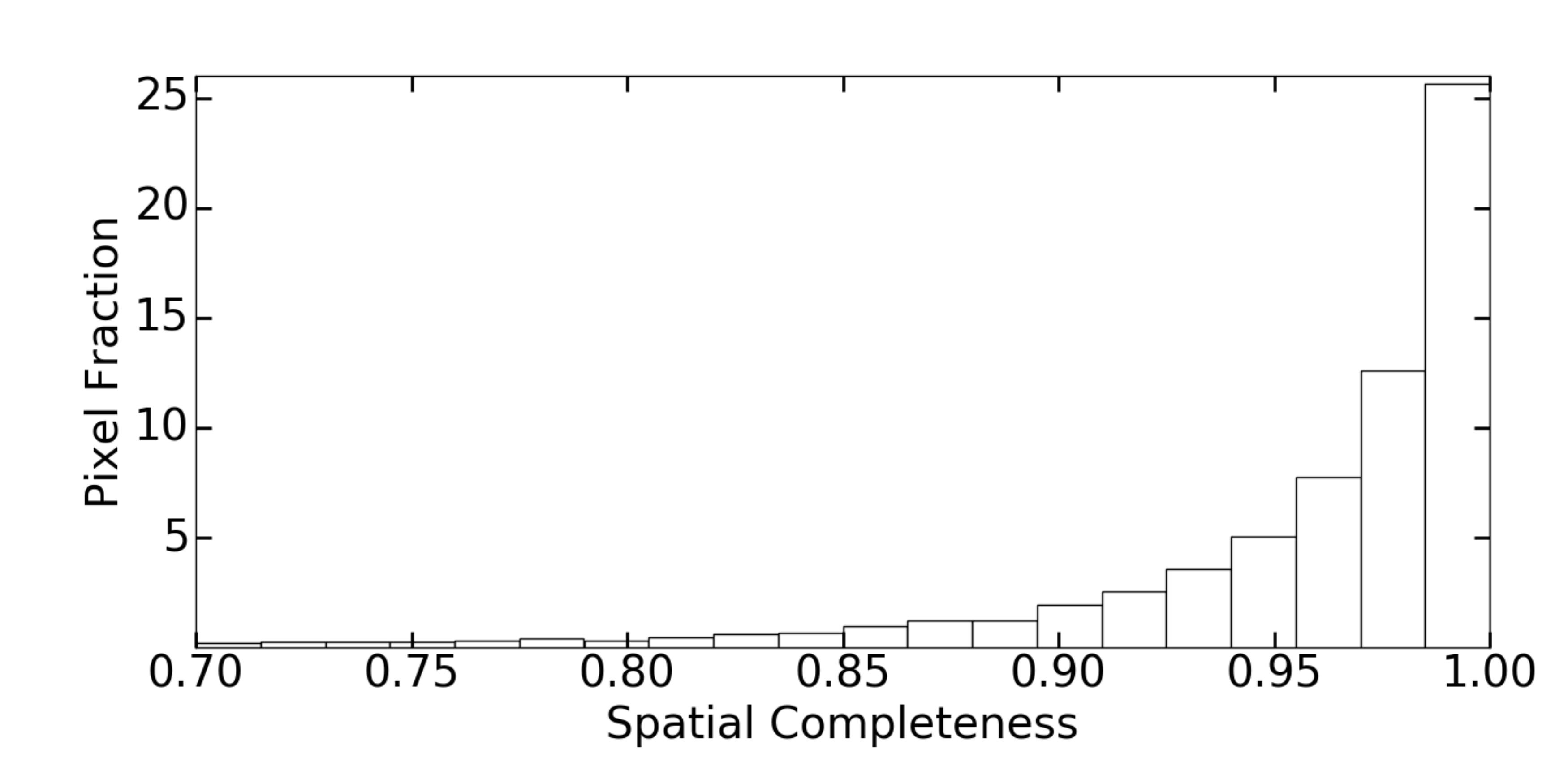}
\caption{\rm{Spatial completeness of the PS1 g-r overlap across the survey as estimated from the fraction of 2MASS stars detected by PS1.}}
\label{fig:completeness}\end{figure}

In Fig.\ \ref{fig:completeness}, we show our estimated spatial completeness across the sky and the distribution of pixel completenesses. We see a few areas of very low completeness along the Galactic plane where crowding and extreme dust will hide the outer Milky Way. Beyond this, the majority of the sky is well-sampled with roughly 75\% of pixels being 90\% or more complete. 

\subsection{Masking the Data}\label{sect:mask}

Despite the extensive coverage and generally high quality of PS1 data, we cannot actually use it to analyze stellar density across the full 30,000 deg$^2$. At low latitudes, Galactic dust hinders our analysis in several distinct ways. First, over a significant fraction of the low Galactic latitude sky, the \gps\ extinction exceeds 1 magnitude and actually limits our ability to detect outer MW stars. At $|b| = 4^\circ$ and 10 kpc (MR distance), our line of sight is only 0.7 kpc off the Plane, well into the disk which has more complicated structure we choose to avoid. At similar latitudes, there is significant dust beyond the 10 kpc MR distance, and we are over correcting with our extinction correction and making stars too bright and blue. Again at similar latitudes, crowding and bright stars when stellar column density $N > 5\times10^4$ stars deg$^{-2}$ (typical stellar separation 18'') can cause significant systematic photometry problems. We therefore mask-out pixels that don't satisfy:
\begin{eqnarray}
A(g_{\rm{P1}}) &=& 3.172\ \rm{E(B-V)} < 1.8,\label{eq:mask}\\
N &<& 5\times10^4\ \rm{deg}^{-2},\nonumber\\
|b| &>& 4^\circ,\nonumber
\end{eqnarray} 
where $A(g_{\rm{P1}})$ is extinction in the \gps\ band, N is stellar mass density and $b$ is Galactic latitude. These thresholds were decided with reference to the metallicity map, Fig.\ \ref{fig:mapum}, in Appendix \ref{sect:addmap}. In dusty regions along the plane, some stars may be in front of some of dust in the SFD extinction and thus `over-corrected' (made too bright and blue) by SFD. In regions with young (blue) stellar populations, our isochrones cannot fit the population correctly. In both cases our metallicity (as calculated by \textsc{match}) is unphysically low and readily apparent in Fig. \ref{fig:mapum}. These thresholds mask out region where this effect is noticeable. 

\begin{figure}[ht]
\includegraphics[width=0.99\columnwidth]{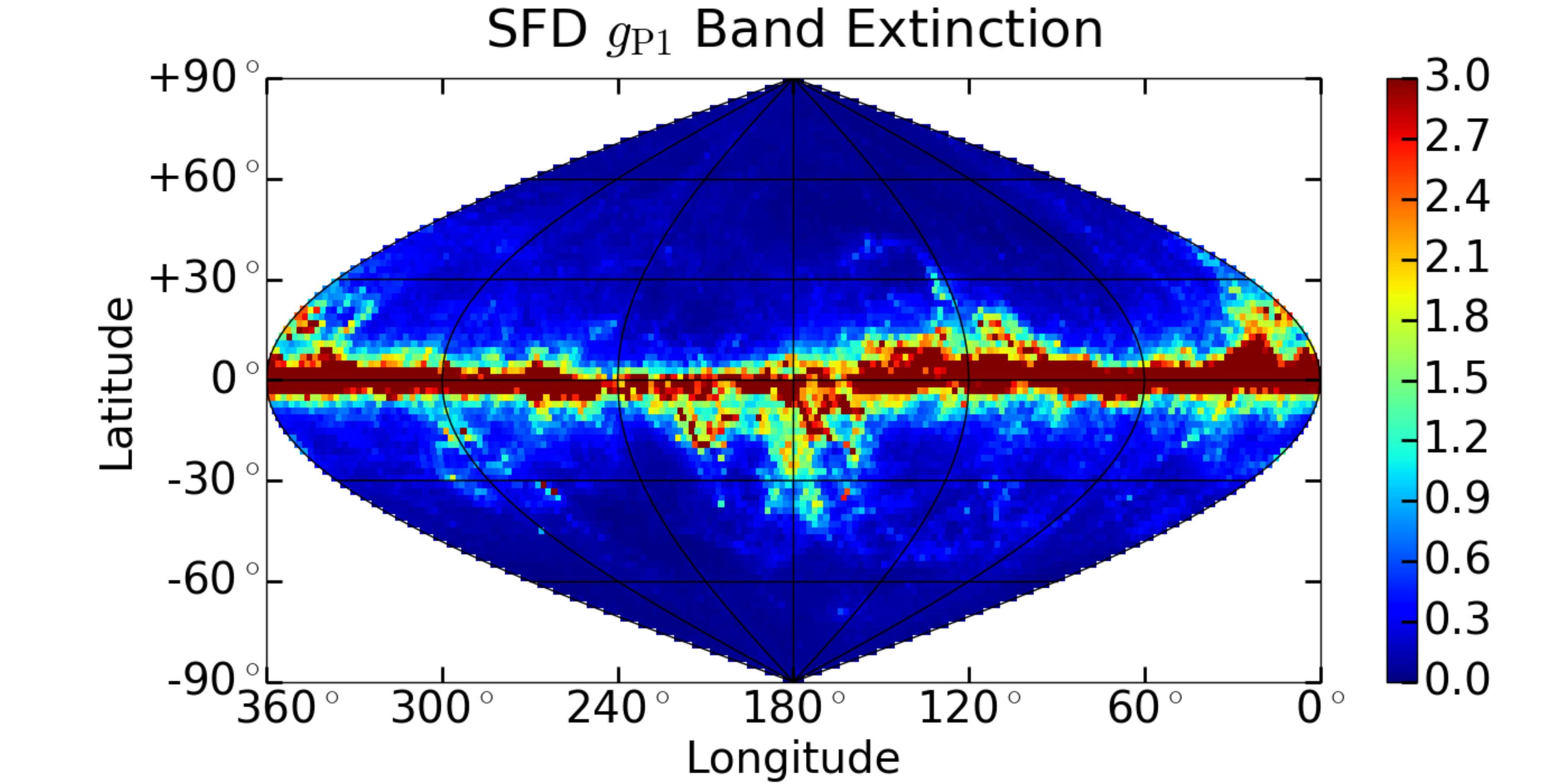}
\includegraphics[width=0.99\columnwidth]{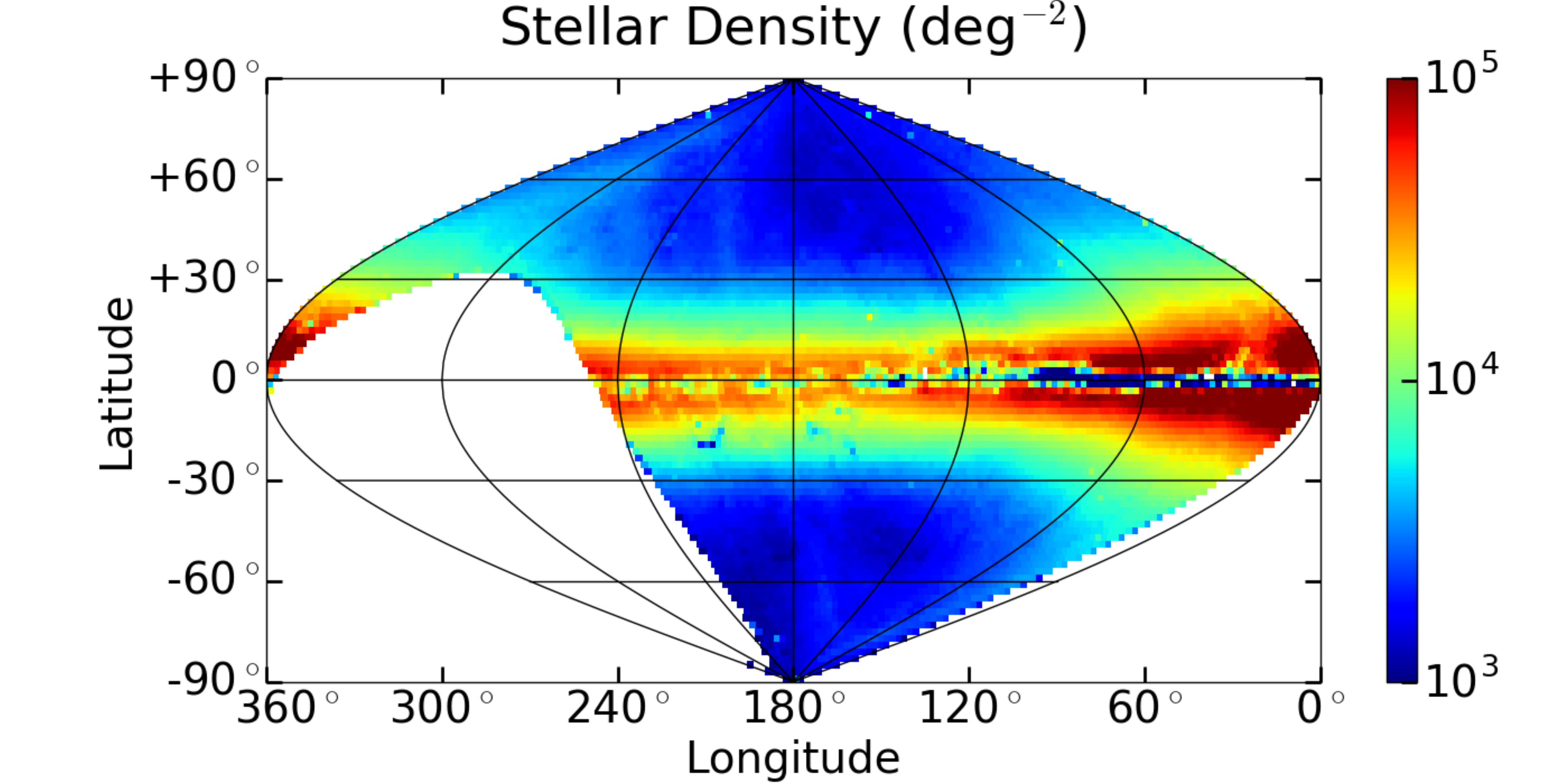}
\includegraphics[width=0.99\columnwidth]{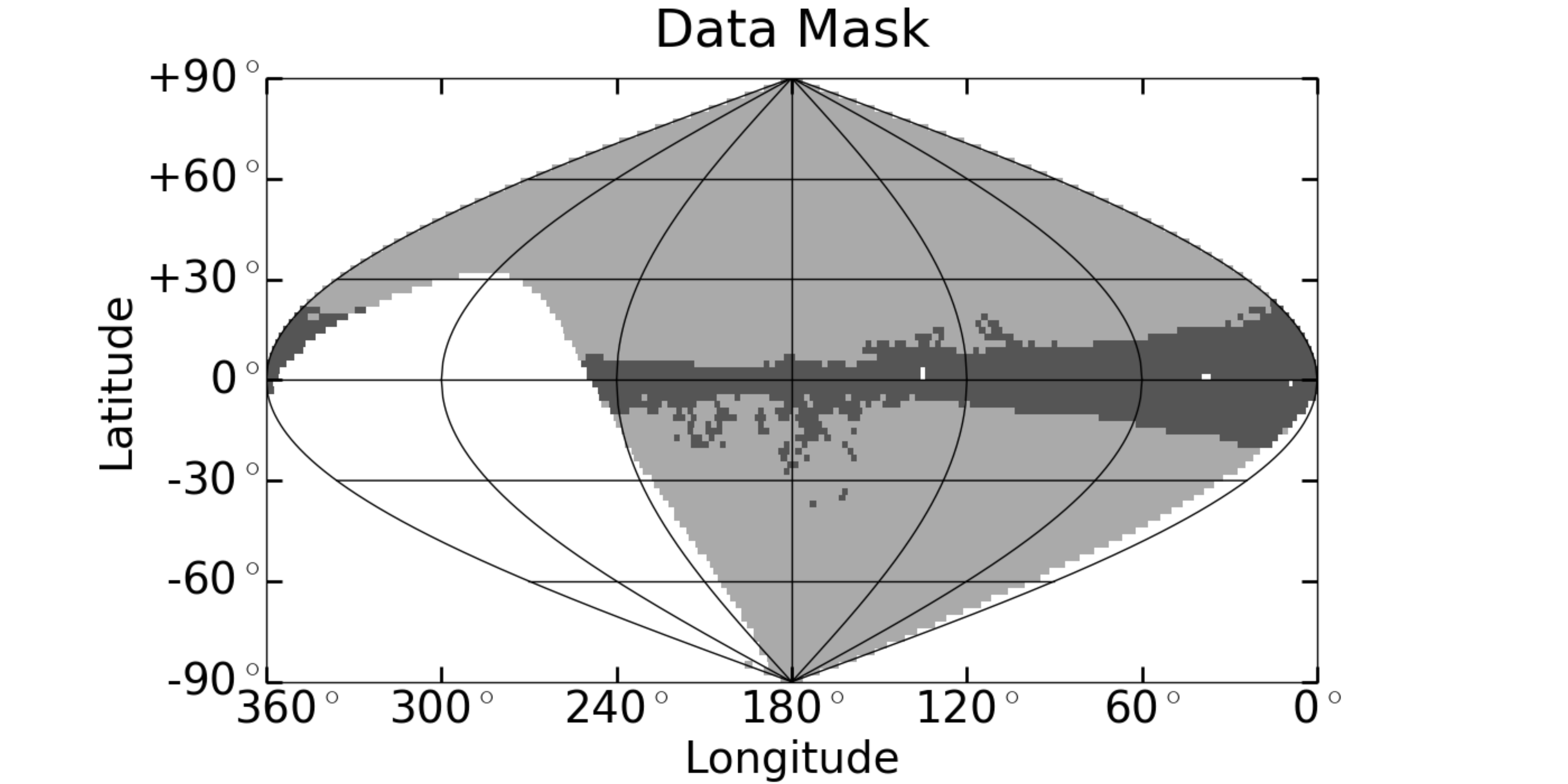}
\caption{\rm{Galactic extinction \citep{SCHL++98} in the \gps\ band (top). Observed number density of PS1 stars deg$^{-2}$ (middle). The mask we make with Eq.\ \ref{eq:mask} to exclude areas in which our analysis would be hindered (bottom). In the mask, light grey indicates that the area contains usable, unmasked data; dark gray indicates masked data; and white indicates areas with no data. There are also a handful or equatorial white pixels for which no extinction coefficient could be calculated.}}
\label{fig:mask}\end{figure}

Fig.\ \ref{fig:mask} shows the \gps\ extinction, the PS1 stellar mass density (where stars are all objects that satisfy Eq.\ \ref{eq:starcut} in the next section, regardless of color) and the mask we make using Eq.\ \ref{eq:mask}. Our mask covers 4{,}848 deg$^2$ including most of the area within $8^\circ$ of the Galactic plane as well as everything within $20^\circ$ of the Galactic center and several significant extensions away from the plane corresponding to dust features. Despite this masking, we retain a significant amount of low latitude area. We retain 5{,}890 deg$^2$ of the 7{,}857 deg$^2$ in the crucial $120^\circ< l< 240^\circ$, $-30^\circ < b < +40^\circ$ region in which the MR is most visible. 

\section{Measuring Stellar Densities}\label{sect:density}

To analyze the roughly 10$^9$ stars observed by PS1, we bin the data into a more condensed form without losing any crucial information about the structure of the Milky Way or the MR. This consists of dividing the sky into manageably-sized pixels, cataloging the (likely) stars and making a color-magnitude diagram (\gpo\ versus \gpo-\rpo) of the stars in each pixel. 
We take our stars from the PS1 average catalog (see Section \ref{sect:data}) and divide the sky into 2$\times$2 degree pixels (equal area pixels in lines of constant Galactic latitude). Our pixels typically contain $10^3-10^5$ total stars depending on the Galactic latitude. In Section \ref{sect:match}, we down-select these stars (with color cuts) by a factor of approximately 2 and bin our stars into 24 distance bins, so the mean bin has only 100 stars at high latitudes. Our current bin and pixel size already introduce significant statistical fluctuation in the analysis of a single pixel, and using smaller pixels makes fitting significantly less robust. In addition, working at the 2 degree scale is sufficient to probe the main Milky Way structure and that of enormous features like the MR. 
Producing a purely stellar catalog is a significant challenge. PS1 star-galaxy separation is still being developed. In lieu of a more advanced star-galaxy separation, we require
\begin{equation}
-0.2 < g_{\rm{P1}}-g_{\rm{P1\ AP}},\ r_{\rm{P1}}-r_{\rm{P1\ AP}} < 0.2\label{eq:starcut},
\end{equation}
where \gps\ and \rps\ are PS1 PSF magnitudes and $g_{\rm{P1\ AP}}$ and $r_{\rm{P1\ AP}}$ are flexible aperture magnitudes designed to measure the total brightness of extended objects. A large, positive PSF and aperture magnitude difference indicates that an object has significant extended source flux while a negative difference generally indicates some kind of image processing problem. Incidentally, this cut also removes stars in crowded fields that also have unreliable photometry. 

Quasars present a more serious contaminant than galaxies as they can be mistaken for blue, high mass stars and disproportionately skew stellar density estimates. \citet{PALA++13} finds 176 quasars deg$^{-2}$ down to $g = 22.5$. In this paper, we probe to $g = 21.6$ and typically find 2500 stars deg$^{-2}$; so quasars could be a contaminant at the level of a few percent. To remove quasars, we cross-match with the data from the Wide-field Infrared Survey Explorer \citep[WISE,][]{WRIG++10} and reject objects that satisfy the WISE-SDSS color cuts set by \citet{WU++12}:
\begin{equation}
z-W_1 > 0.66(g-z)+2.01.
\end{equation}
Here $W_1$ is the WISE 3.1 $\mu m$ filter and $g$ and $z$ are typically extinction-corrected SDSS filters. We take advantage of the similarity of the PS1 filters to the SDSS filters and use the same cut. This cut is less effective at $i > 20.5$ (roughly $g >  21$ for typical quasars), but it removes 60 quasars deg$^{-2}$, roughly what we would expect.

\begin{figure}[ht]
\includegraphics[width=0.32\columnwidth]{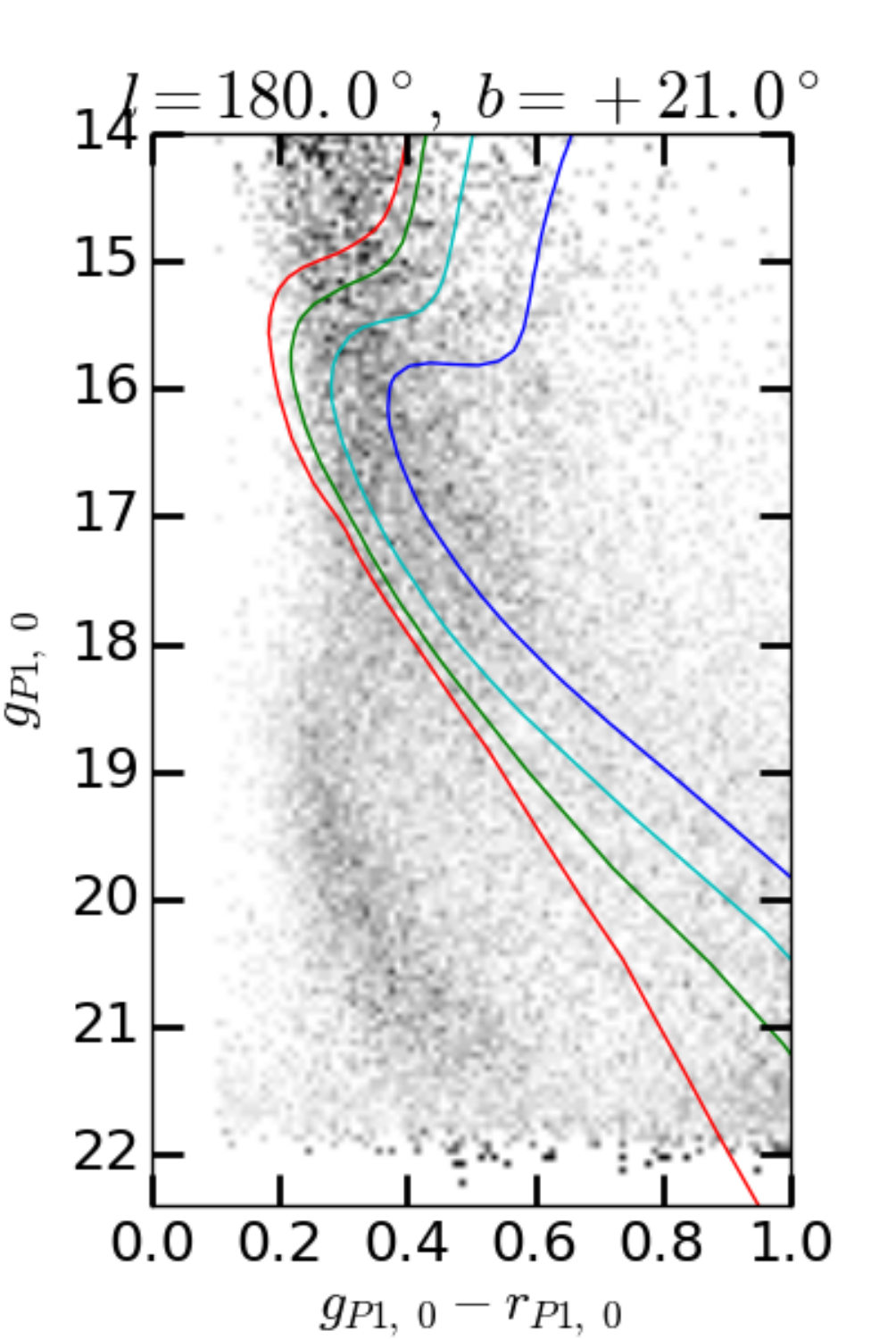}
\includegraphics[width=0.32\columnwidth]{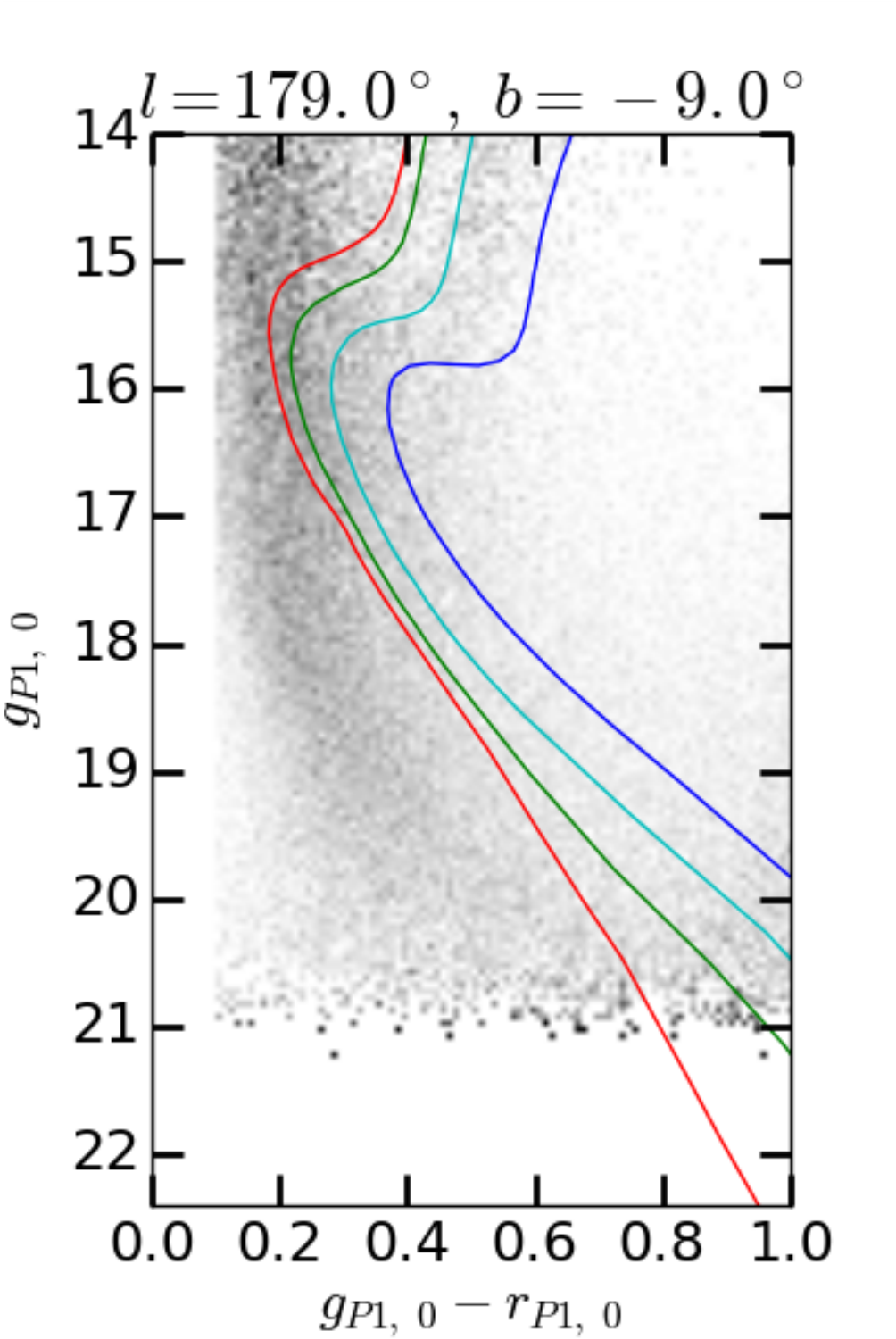}
\includegraphics[width=0.32\columnwidth]{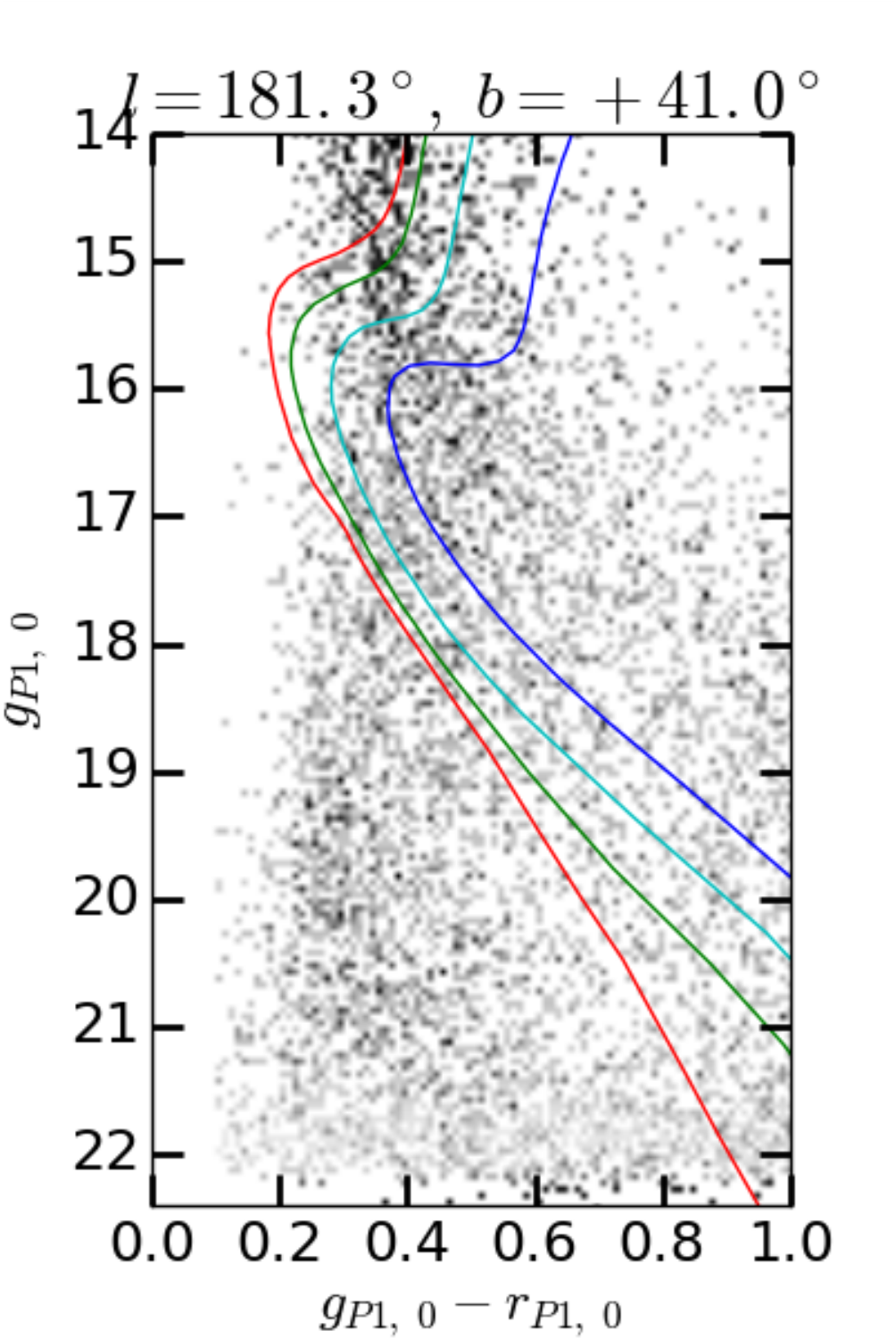}
\caption{\rm{Three \gpo\ versus $\gpo-\rpo$ Color Magnitude Diagrams (CMDs) of the stars in 4 deg$^2$ pixels at $l \approx 180^\circ$. We show the PS1 main sequence isochrones for populations of stars at 2 kpc (distance modulus 11.5) aged 13.32 billion years with metallicity -0.3 (red, leftmost), -0.8 (green), -1.4 (cyan) and -2.1 (blue, rightmost). The left panel shows the stars at $l = 180^\circ,\ b = +21^\circ$. We see both the main Galactic population bounded by the 2 kpc line and another distinct population roughly 2 magnitudes fainter. The middle panel shows the stars at $l = 179^\circ,\ b = -9^\circ$.  Here the main Galactic population extends out well past the 2 kpc line. The right panel shows the stars at $l = 181^\circ,\ b = +41^\circ$. Only the main Galactic population is apparent. The transparency of each point is normalized by the total number of points with similar \gpo\ magnitude in order to make the color distributions visually more comparable across different panels and magnitude ranges.}}
\label{fig:hess}\end{figure}

Fig.\ \ref{fig:hess} shows three \gpo\ versus $\gpo-\rpo$ Color Magnitude Diagrams (CMDs) of the stars from our 4 deg$^2$ pixels. We also plot the four isochrones \citep{BRES++12} that we use in CMD fitting for stars 2 kpc from the Sun (distance modulus 11.5). These isochrones are obtained directly from http://stev.oapd.inaf.it/cgi-bin/cmd). All three diagrams are from longitude $l \approx 180^\circ$, where MR is most discernible. But the first two panels are taken from $b= +21^\circ$ and $b = -9^\circ$ well within the previously detected MR region while the second is taken from $b=+41^\circ$, well above the MR. We see that if we look along the ``blue edge'' at $0.1 <\gpo -\rpo < 0.3$ of the stellar population, the $b=+21^\circ$ (left) panel has one diffuse population covering $\gpo < 17$ and a more discrete population in $19 < \gpo < 20$. This is consistent with a continuous main sequence population (the main MW structure) out to 2 kpc and a second population roughly 8 kpc from the Sun (the MR structure). The middle diagram appears to contain one continuous population out to $\gpo = 19.5$ (d = 5 kpc). The SFD correction along this line of sight appears to make our stars too blue. We allow our fitting code to compensate for this later in our pipeline (see Section \ref{sect:match}). Our rightmost diagram, at higher Galactic latitude, shows only the main MW population. The transparency (boldness) of individual points in this diagrams is weighted by the number of points in a similar \gpo\ range. This prevents the diagram from being saturated at the fainter \gpo\ which probe exponentially more volume than the brighter \gpo. Converting this qualitative description of MR densities along the line of sight into a quantitative one is one of our main challenges in this paper.

\section{Fitting Line of Sight Densities}\label{sect:los}

To convert our CMDs (Fig.\ \ref{fig:hess}) from densities in \gpo, $\gpo-\rpo$ space into line of sight stellar mass density estimates, we use a program called \textsc{match} \citep{DOLP02}. With significant fine-tuning, \textsc{match} can fit not only the line of sight densities, but also metallicities for large volumes of the Milky Way. After modeling the density uncertainty in each bin, we can fit these line of sight densities and simultaneously estimate the Milky Way density and the MR overdensity. 

\subsection{The MATCH Software}\label{sect:match}

The \textsc{match} software matches color magnitude diagrams to fit stellar age, metallicity and density as a linear combination of model isochrones (we use those from \citet{CION++06a,CION++06b}) with fixed age, metallicity and distance. While it was initially used to probe the metallicity and age of stars in nearby dwarf galaxies or other localized structures with essentially fixed distances, it was adapted to model stellar mass densities with a smaller set of fixed age and metallicity combinations. Since we are concerned with the outer Milky Way, which has a relatively old stellar populations, we use 4 isochrones of stars of age 13.3 billion years ($10^{10.1}$ -- $10^{10.15}$ years in the settings file) with median metallicities ([Fe/H]) of -0.3, -0.8, -1.4 and -2.1 (in $\log_{10}$ units relative to $Z_\odot$). Each population has total metallicity width of 0.2. The difficulty of distinguishing between nearby red dwarfs and distant red giants and the change in the ratio of early to late type stars across the MW make it difficult for \textsc{match} to use the same isochrones everywhere. To minimize the impact of these problems, we restrict ourselves to the $0.1 < \gpo - \rpo < 0.5$ area of color space that excludes main sequence stars later than mid-G type. This excludes 40\% of total stars at our magnitude limit, but the vast majority of the redder stars are fairly local red dwarfs closer than 8 kpc and are not useful for probing the outer MW. 

\begin{table}
\centering
\begin{tabular}{cccc}
	\hline
\textsc{match} Setting &  Minimum & Maximum & Precision \\
	\hline
Metallicity ($\log_{10}$) & -2.1 & -0.3 & $\approx$ 0.6 \\
Stellar Age ($\log_{10}$) & 10.1 & 10.15 & -- \\
Distance Modulus & 10 & 17.6 & 0.05 \\
\gpo & 14.0 & 21.6 & 0.1 \\
\rpo & 13.5 & 21.5 & -- \\
$\gpo-\rpo$ & 0.1 & 0.5 & 0.05 \\
Extinction (\gps\ mags)  & -0.2  & 0.2 & 0.05 \\
	\hline
\end{tabular}
\caption{\rm{A summary of the \textsc{match} settings we use. The exact metallicity binning is described in the text. We automatically correct for dust extinction with SFD, but also allow \textsc{match}'s de-reddening tool to correct errors in SFD where the SFD \gps\ extinction is greater than 0.9 mags.}}\label{tab:match}
\end{table}

\textsc{match} sometimes produces nonphysical line of sight densities in which a bin with zero density will be surrounded by two bins with large densities. To avoid this, we use fine distance bins, $\Delta \mu = 0.05$, and smooth our results along the line of sight with a $\sigma_\mu = 0.4$ magnitude Gaussian. We then bin our results into $\Delta \mu = 0.4$ bins. With this final bin size, each distance bin is 20\% farther than its predecessor. This is appropriate for MR analysis and the statistical precision of our data. The actual \textsc{match} analysis uses CMD bin sizes of $\Delta \gpo = 0.1$ and $\Delta (\gpo-\rpo) = 0.05$. As previously mentioned, we use SFD dust extinction maps. But in cases where the expected \gps\ extinction is greater than 0.9 we allow \textsc{match} to apply additional extinction correction between -0.2 and 0.2 \gps\ magnitudes. We summarize these settings in Table \ref{tab:match}.

\begin{figure*}[ht]
\includegraphics[width=0.32\columnwidth]{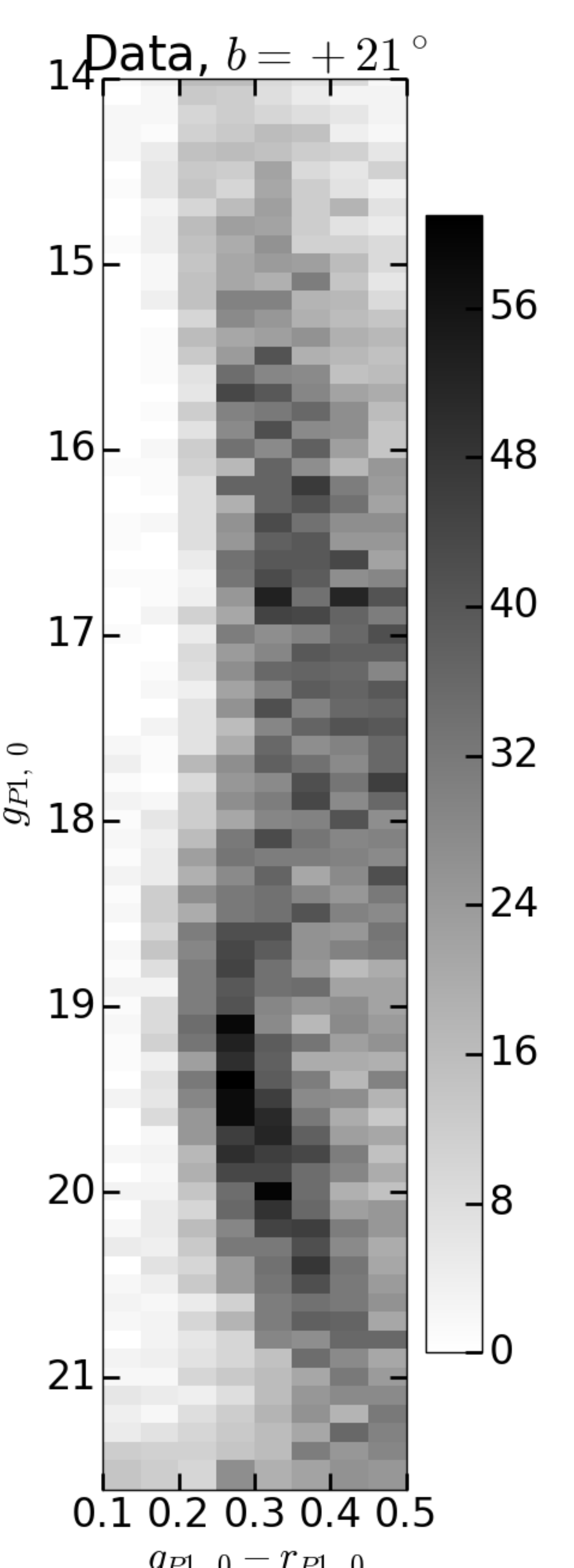}
\includegraphics[width=0.32\columnwidth]{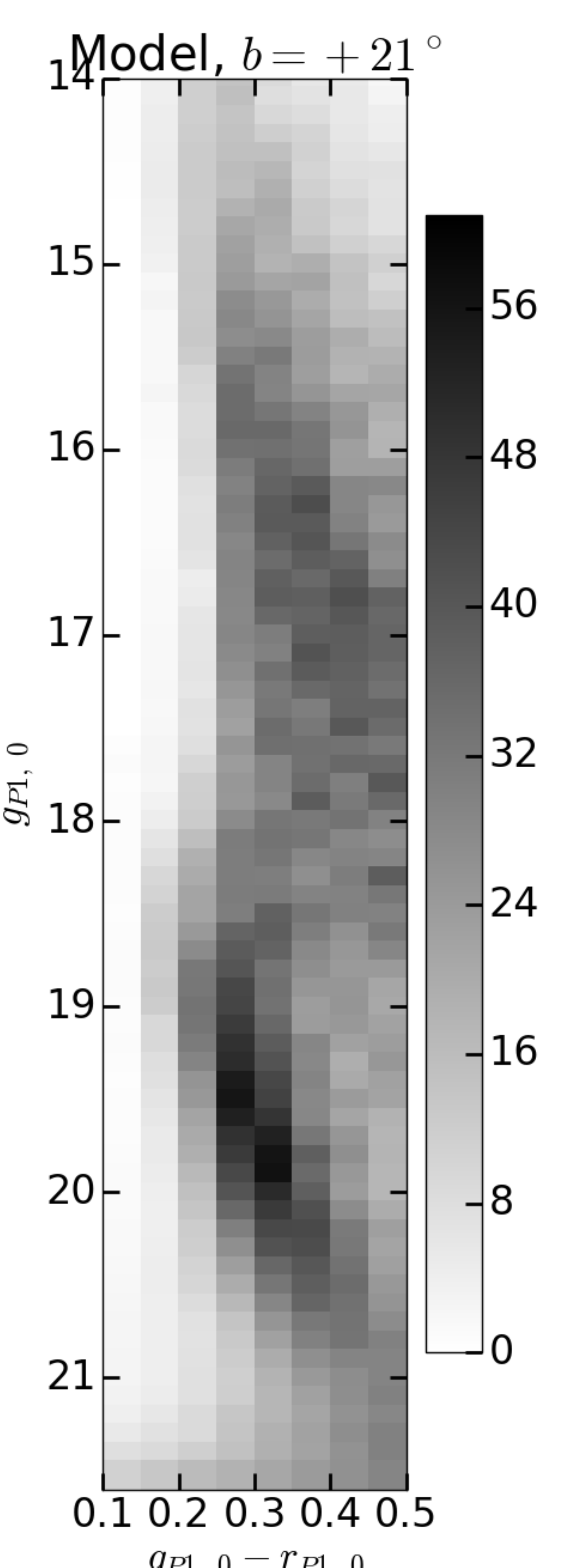}
\includegraphics[width=0.32\columnwidth]{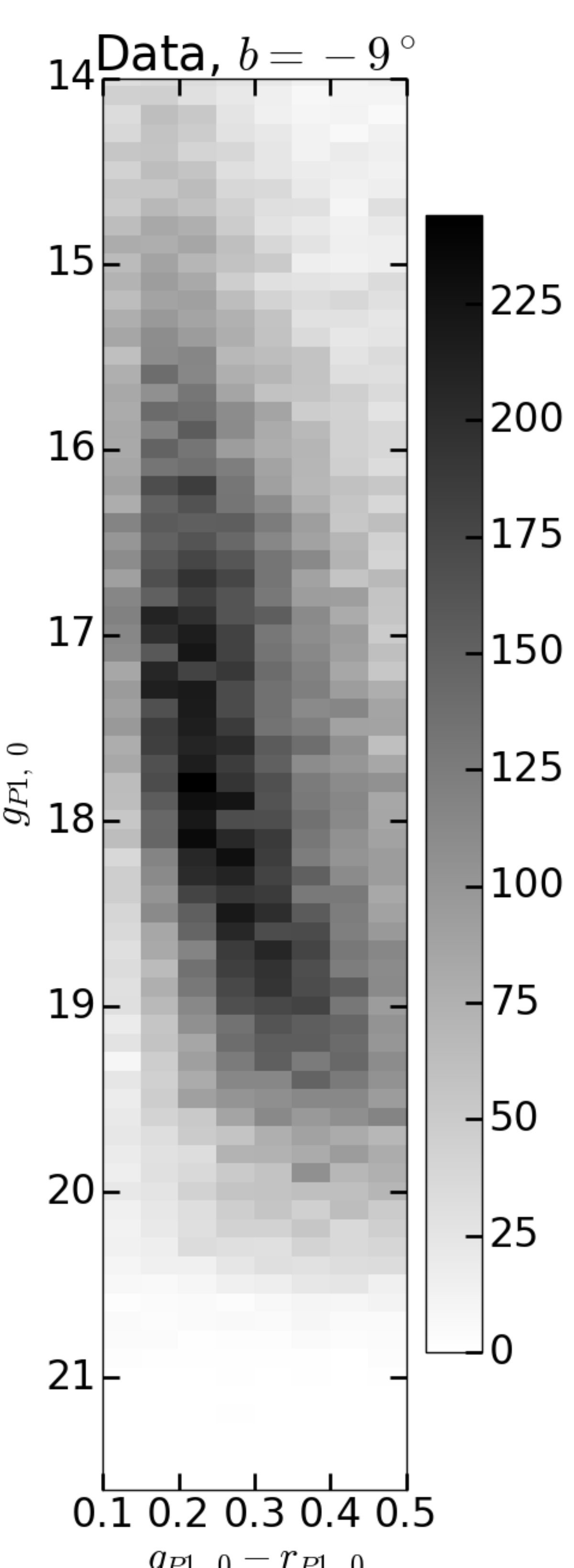}
\includegraphics[width=0.32\columnwidth]{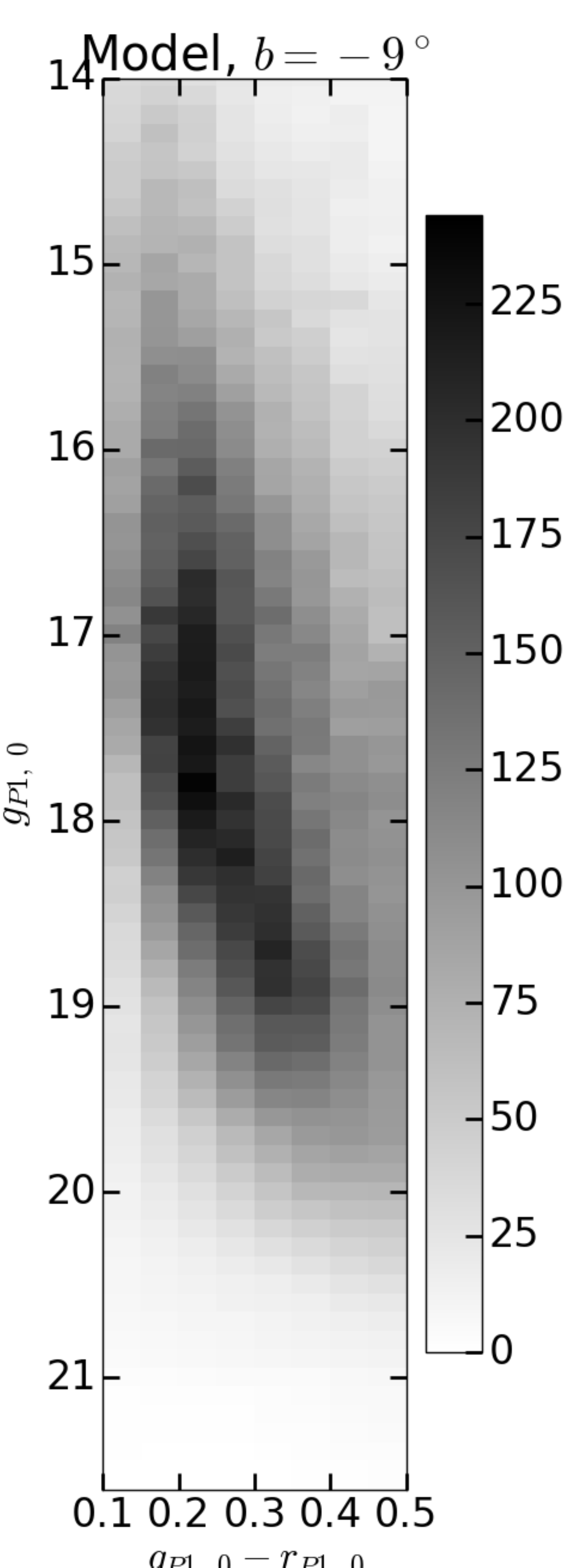}
\includegraphics[width=0.32\columnwidth]{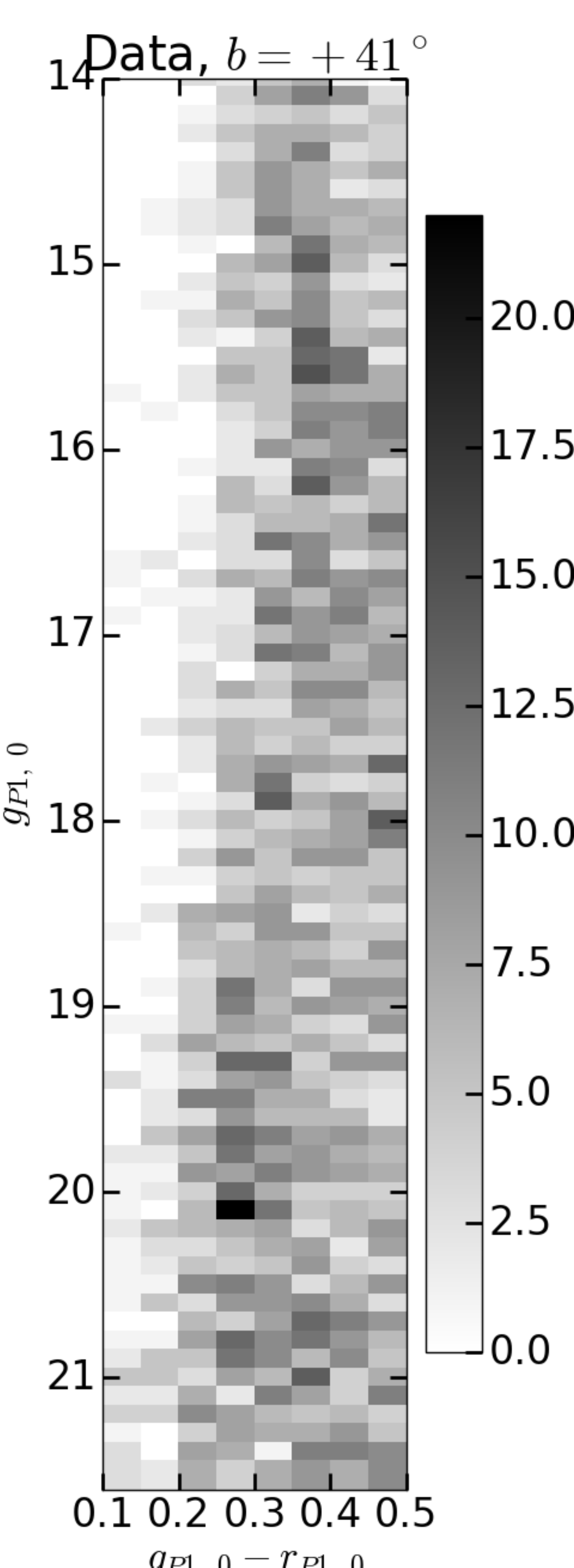}
\includegraphics[width=0.32\columnwidth]{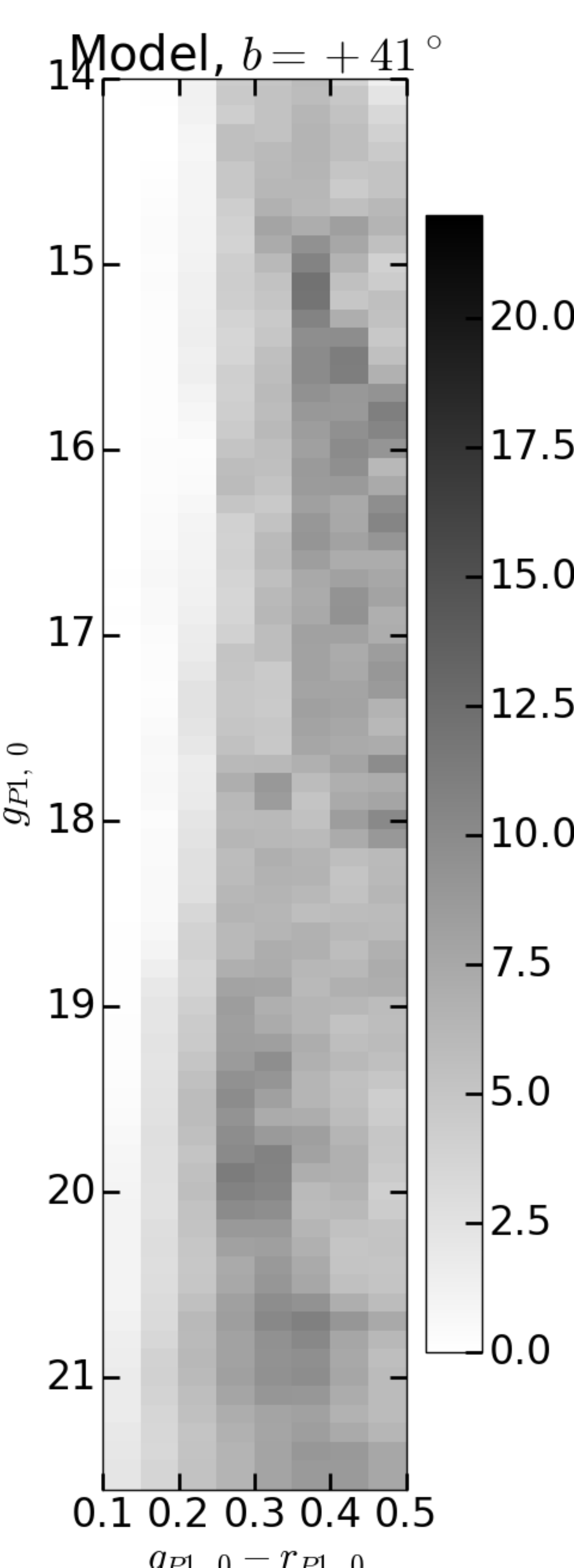}
\caption{\rm{Three \gpo\ versus $\gpo-\rpo$ color magnitude diagrams (CMDs) of the stars in different 4 deg$^2$ pixels at $l \approx 180^\circ$ as binned and modeled by the \textsc{match} program. The first two panels show the data and model stars at $l = 180^\circ,\ b=+21^\circ$. As in Fig.\ \ref{fig:hess} we see a distinct population roughly 2 magnitudes fainter than the main MW population. The middle two panels show the stars at $l = 179^\circ,\ b = -9^\circ$. We see that the Galactic population extends out much farther. The final two panels show the stars at $l = 181^\circ,\ b=+41^\circ$. Only the main Galactic population is apparent. We see that in all three cases the ``Model'' diagram is similar to the ``Data'' diagram, with the differences being fairly consistent with random noise. Unlike Fig. \ref{fig:hess}, the data here is not normalized in any way. Each bin is contain the total number of stars in that bin.}}
\label{fig:match}\end{figure*}

In addition to a catalog of $\gpo$ versus $\gpo - \rpo$ for each object in a pixel and a settings file from Table \ref{tab:match}, \textsc{match} also requires a simulated input and output catalog. This catalog accounts for incompleteness at fainter magnitudes and estimates the measurement uncertainty at each position in color magnitude space. Ideally this would be produced by placing synthetic point sources of known magnitudes in each of our images and recording the input and output magnitudes. This was computationally prohibitive to do across 30{,}000 deg$^2$. Instead, we produced purely simulated catalogs in which the input sources covered $14 < \gpo < 21.6$ versus $-1.0 < \gpo - \rpo < 2.0$ uniformly and randomly. In the output catalog, we added a Gaussian noise term with standard deviation 
\begin{equation}
\sigma = \left(0.01^2+0.1^2\ 10^{0.8(\rm{mag}-\rm{mag}_{10\sigma})}\right)^{1/2}.
\end{equation} 
Here, we are simply adding 0.01 magnitudes of calibration error in quadrature with background noise. This model uses the fact that background noise is dominant for sources with uncertainty above 0.01 mags in PS1. The $\rm{mag}$ term could be either \gps or \rps and $\rm{mag}_{10\sigma}$ is the 10$\sigma$ limiting magnitude of the appropriate filter in each pixel. In each of our two significant filters, we assign a probability that the source is detected: 
\begin{equation}
P_{\rm{detect}} = \frac{1}{2}\left(1-\tanh\left(0.4*\left(\rm{mag}-\rm{mag}_{10\sigma}+0.9\right)\right)\right).
\end{equation} 
This probability sensibly decreases as input object magnitudes go fainter than the 10$\sigma$ limit. The values 0.4 and 0.9 were found to minimize \textsc{match}'s reported goodness of fit to the data across a representative area. Qualitatively, minimizing this goodness of fit ensured that \textsc{match}'s model had the same faint magnitude cutoff as the data (e.g. in Fig. \ref{fig:match}). Sources that are not detected in either filter would not be in our sample, and are assigned magnitude 0 which \textsc{match} recognizes as being undetected. 

Fig.\ \ref{fig:match} shows two of the main \textsc{match} data products: the binned CMD and the best fit \textsc{match} model of the CMD. We show the \textsc{match} products for the same pixels as an Fig. \ref{fig:hess} so that the leftmost pair ($b=+21^\circ$) has a a prominent MR population. The center pair ($b=-9^\circ$) has a less distinct extension out to 6 kpc that we will later identify as MR. Here, \textsc{match} has also corrected the extinction by $\Delta \gps = 0.15$ magnitudes as described in the last section, and the extinction-corrected limiting magnitudes limit us to \gps < 20. The rightmost pair ($b=+41^\circ$) is consistent with a drop-off in density coupled with an increasing volume element. The input CMDs match the model CMDs quite well with the data and model CMDs being roughly consistent with Poisson noise. Fig.\ \ref{fig:match} is not row-normalized like Fig. \ref{fig:hess}, so one of the main trends we see is that there are more stars in each CMD as we go to fainter \gpo's. The MR population is less visually distinct without this normalization. We require more thorough analysis of the line of sight density to prove, quantitatively, that the MR overdensity is real.

\begin{figure}[ht]
\includegraphics[width=0.99\columnwidth]{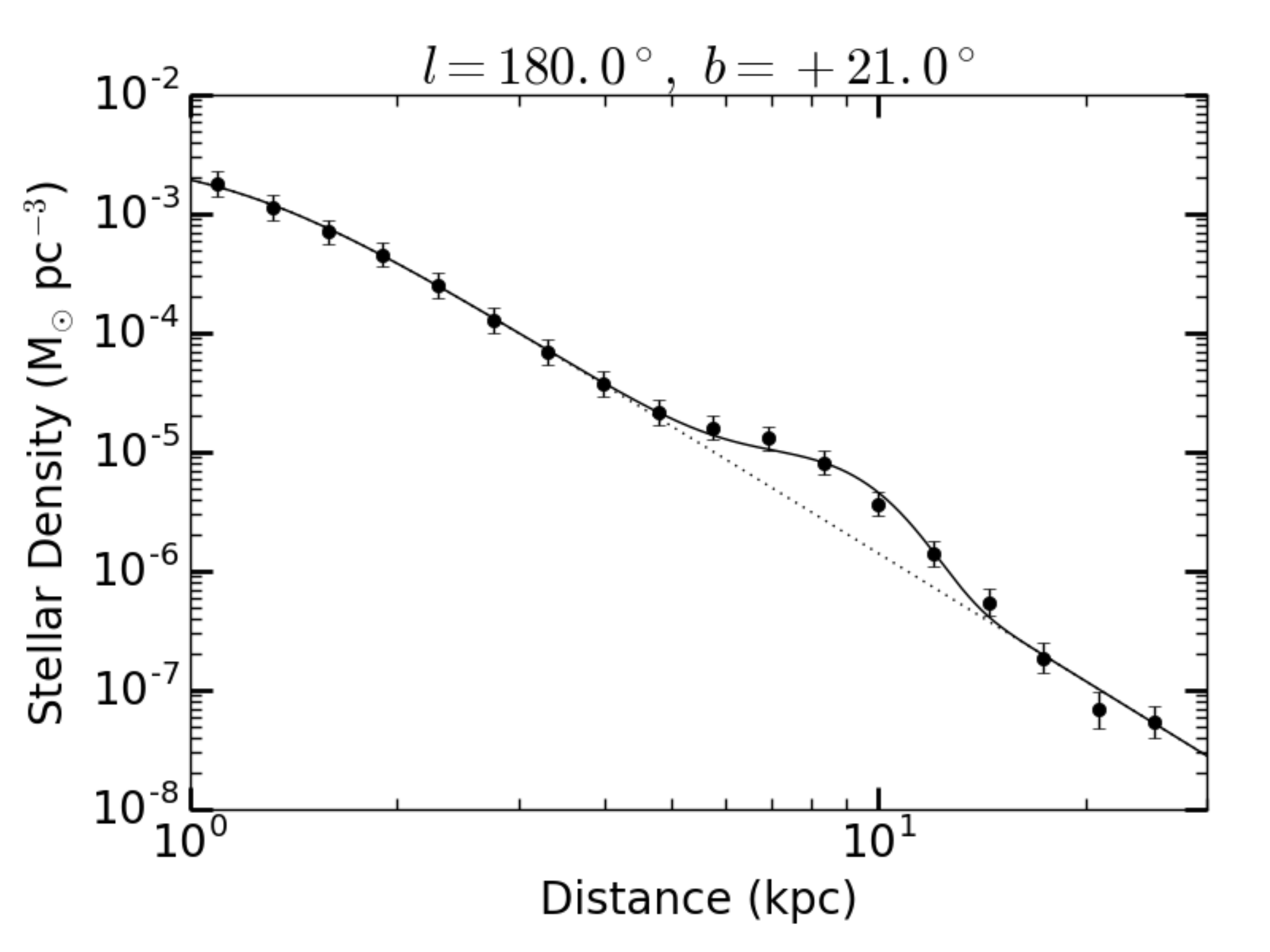}
\includegraphics[width=0.99\columnwidth]{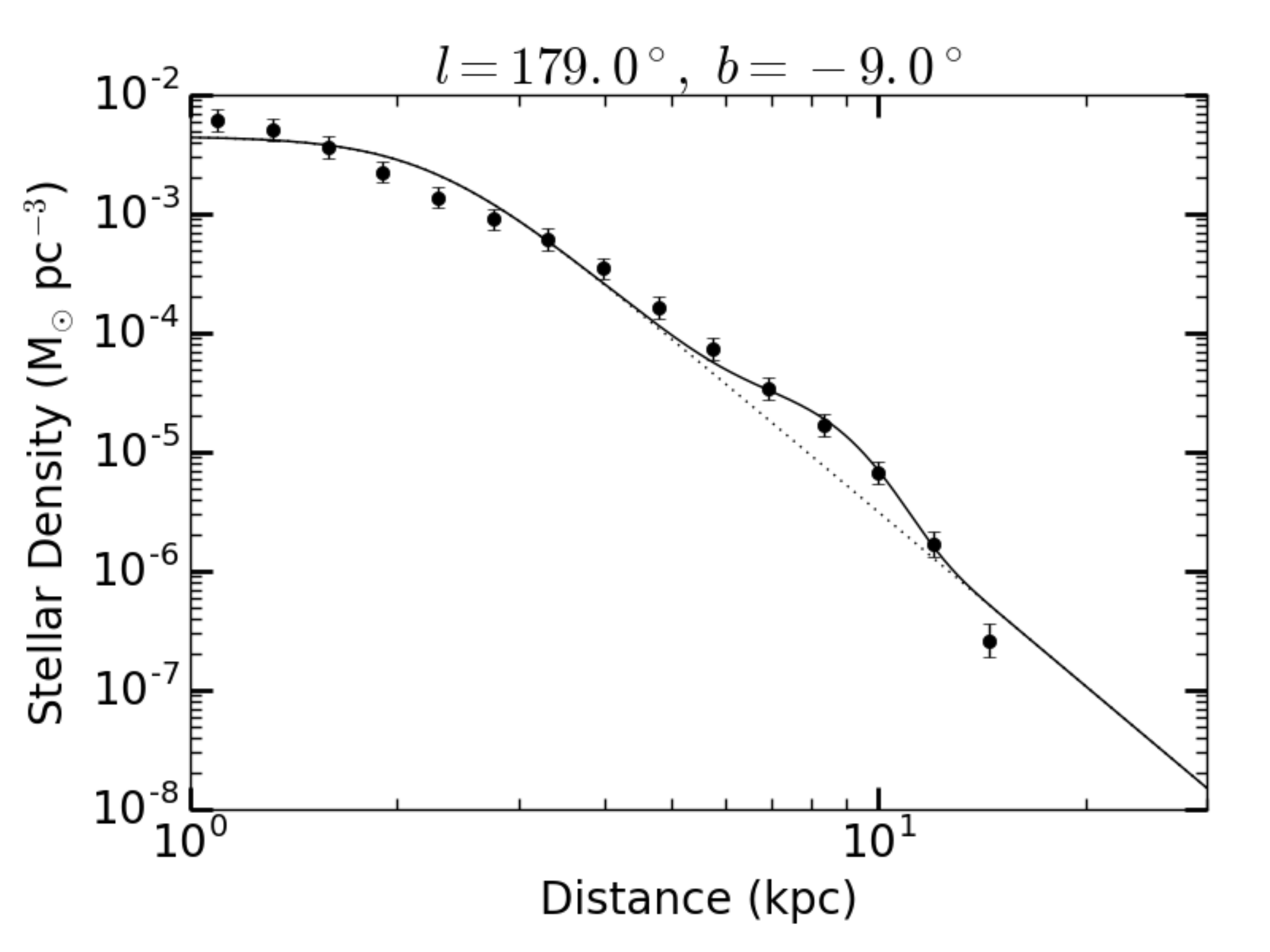}
\includegraphics[width=0.99\columnwidth]{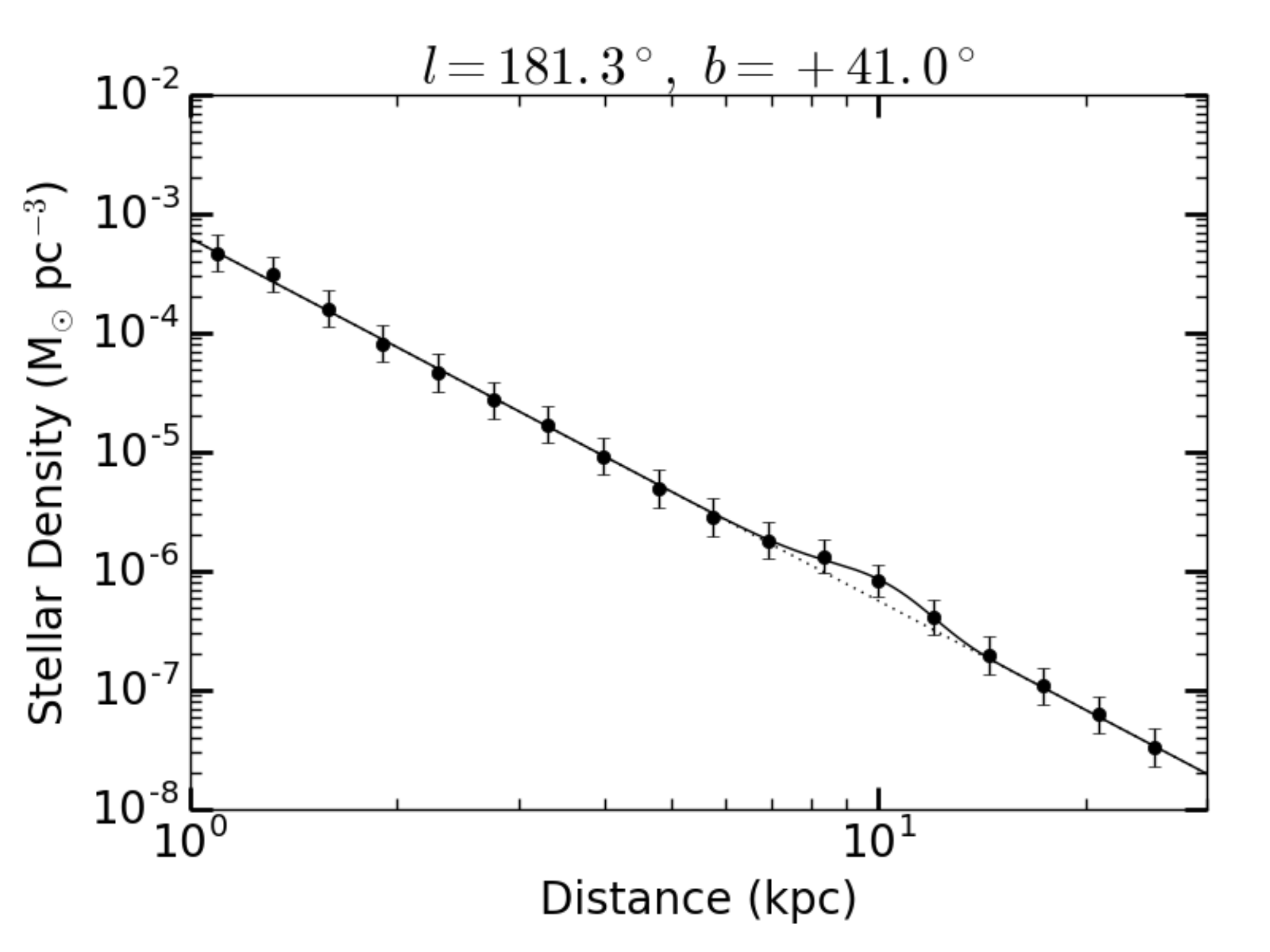}
\caption{\rm{Three stellar mass density diagrams of the stars in 4 deg$^2$ pixels at $l \approx 180^\circ$. We fit each population as a modified power law with an additional Gaussian lump of stars. We show this fit as a solid line and the power law profile alone as a dotted line. The top panel shows the stars at $l = 180^\circ,\ b = +21^\circ$. We see that there are a significant number of `extra' stars at d = 8 kpc. The middle panel shows the stars at $l = 179^\circ,\ b = -9^\circ$. The `extra stars' lump is less distinct, but the data prefer a significant overdensity at $d = 7$ kpc. The bottom panel shows the stars at  $l = 181^\circ, b = +41^\circ$. The `extra' stars are insignificant. }}
\label{fig:distance}\end{figure}

Fig.\ \ref{fig:distance} shows stellar mass density as a function of distance in the \textsc{match} model CMDs for our $b=+21^\circ$, $b= -9^\circ$ and $b = +41^\circ$ pixels. Consistent with Fig. \ref{fig:match}, we see a pronounced overdensity in the North at $b=+21^\circ$ a less distinct overdensity at $b= -9^\circ$ and almost no overdensity at $b = +41^\circ$. 

\subsection{Estimating Uncertainties}\label{sect:errors}

In order to fit and quantify our stellar overdensity in Fig. \ref{fig:distance}, we must estimate the density uncertainty along each line of sight. At the time of this analysis, \textsc{match} only returned the projected mass of stars in each bin. We attempted to model our uncertainties using a Bootstrap process \citep{EFRO79}, producing 25 alternate CMDs for each pixel and taking the standard deviation in each distance bin as the uncertainty. This method failed, often showing no variance (and correspondingly no uncertainty) in a given bin. \citet{DOLP13} showed a Monte Carlo method for estimating error uncertainties, but this method is too computationally intensive for us to use over 8{,}000 relevant lines of sight. We opted to model our uncertainties analytically. 

We initially modeled the uncertainty, $\sigma_{M_{\rm{bin}}}$, in the mass of each bin, $M_{\rm{bin}}$, along the line of sight as Poisson noise. The typical mass of stars in our color range is $1 M_{\odot}$ and $M_{\rm{bin}} \approx N_{\rm{bin}}$, the estimated number of stars in a bin. So in solar units, we would expect our statistical variance in a given mass bin to be equal to the mass in that mass bin. By comparing equivalent distance bins along neighboring high latitude lines of sight, we find that this term actually underestimates our data variance by a factor of 3. Scatter around high latitude fits with no MR or other known structures also was also a factor of 3 higher than estimated from simple Poisson noise. This may be because \textsc{match} models entire isochrones instead of individual stars. Additionally, in very dense regions where Poisson noise is not significant, we find that stellar mass in neighboring pixels varies by 1\%. This is likely real degree-scale astrophysical variation, but within our precision, we can treat this as noise. Finally, we add one ``quantum'' of noise in quadrature to prevent us from having bins with 0 uncertainty and obtain our semi-analytic uncertainty estimator for the mass in a given bin:
\begin{equation}
\sigma_{M_{\rm{bin}}} = \left(3 + 3 M_{\rm{bin}} +\left(0.01M_{\rm{bin}}\right)^2 \right)^{1/2}.\label{eq:errorbars}
\end{equation}

To avoid systematic errors at the faint end, we mask out data points whose distance modulus, $\mu$ does not satisfy:
\begin{equation}\label{eq:mulim}
\mu < g_{\rm{P1,\ 0}}(10\sigma)-4.
\end{equation}
Here, $g_{\rm{P1,\ 0}}(10\sigma)$ is the extinction-corrected 10$\sigma$ limiting magnitude of the pixel. This statement masks out datapoints for which we cannot see a star of absolute magnitude 4 (the magnitude of the blue edge for our isochrones) with 10$\sigma$ precision. This corresponds to roughly $15$ kpc for some of our dustier pixels (as in the middle panel of Fig. \ref{fig:distance}).

\subsection{Fitting Line of Sight Densities}\label{sect:fitting}

Having produced a stellar mass and uncertainties along every line of sight, we are finally able to model (fit) the line of sight stellar mass density. First, we must convert from stellar mass to stellar mass density, by dividing the mass and uncertainty in each bin by a volume element:
\begin{equation}
V = \frac{1}{3}\left(r_n^3-r_{n-1}^3\right) \left(\frac{2\ \rm{deg}\pi}{180\ \rm{deg}}\right)^2 C\label{eq:volume}.
\end{equation}
Here, $r_n$ is the maximum radius of the $n$th distance bin (and $r_{n-1}$ is the minimum distance of that bin. The terms in the second set of parentheses are merely converting degrees to radians, and the final $C$ is geometric completeness in the pixel as determined from cross-matching with 2MASS (Subsection \ref{sect:completeness}) and is applied uniformly along the line of sight.  

We fit the data with a combined Milky Way and Monoceros Ring radial density described by:
\begin{eqnarray}
\rho(d) &=& MW(d)+MR(d)\label{eq:linefit}\\
MW(d) &=&\frac{\rho_0}{1+\left(d/d_0\right)^\gamma}\nonumber\\
MR(d) &=& \frac{\delta \rho (1 kpc)}{\left(2\pi W_{MR}^2\right)^{1/2}} \exp\left(\frac{(d-d_{MR})^2}{2 W_{MR}^2}\right).\nonumber
\end{eqnarray}
Here, $d$ is distance along the line of sight from the Sun. $MW(d)$, the main Milky Way density, as a function of distance, is fit as a modified power law with distance scale, $d_0$, and exponent, $\gamma$, as free parameters. This profile performed slightly better (produced lower $\chi^2$'s) than the more traditional S{\'e}rsic profile \citep{SERS63} for our data. $MR(d)$ is our model Monoceros perturbation, a Gaussian that peaks at distance $d_{MR}$ from the Sun with width $W_{MR}$ and amplitude $\delta \rho$. The ``1 kpc'' term accounts for units. In Section \ref{sect:clusters}, we show that while $d_{MR}$ is a mathematically convenient term, it is a biased estimator of the distance to the MR center of mass, $d_{mass}$, a more physical quantity. To prevent our $\chi^2$ algorithm from settling on unphysical solutions, we actually fix $W_{MR}$ to 1.4 kpc (a typical value across our MR area), fit $d_{MR}$ and then run our fit again with constant $d_{MR}$ and all other parameters (including $W_{MR}$) free. Our extra MR overdensity from Figs. \ref{fig:hess} and \ref{fig:match} do reassuringly correspond to a significant overdensity at $d_{MR} = 8$ kpc and $7$ kpc in our low latitude ($b = +21^\circ,\ -9^\circ$) pixels, but only a spurious detection in our higher latitude ($b = +41^\circ$) pixel.  

Eq. \ref{eq:linefit} fails in the direction of the Galactic Center, where a more sophisticated, global MW model is needed. Specifically, it uses the ``MR'' component to fit the bulge. But we are ultimately only interested in probing the Monoceros Ring over roughly $120^\circ < l < 240^\circ$ and ignore our fits near the Galactic Center. 

\subsection{Using Clusters to Estimate Distance Errors}\label{sect:clusters}

While scanning the sky to detect the Monoceros Ring, we also detect many smaller overdensities with \textsc{match} (similar to \citet{DEJO++08}) including many globular clusters and open clusters listed in  \citet{HARR96} and \citet{KHAR++13}, respectively. Table \ref{tab:cluster} shows previously measured distances to a series of known clusters which we also detect (see Fig. \ref{fig:hemi} in Appendix \ref{sect:addmap}). We also show the distance to peak overdensity, $d_{\rm{MR}}$ from Eq. \ref{eq:linefit}, and well as the distance to the center of mass our overdensity:
\begin{eqnarray}
d_{mass} &=& \frac{\int MR(x) x^3 dx }{\int MR(x) x^2 dx}, \label{eq:com}\\
d_{mass} &=& d_{MR}\frac{d_{MR}^2 + 3 W_{MR}^2 }{ d_{MR}^2 + W_{MR}^2 },\nonumber
\end{eqnarray} 
where $MR(x)$, $D_{MR}$ and $W_{MR}$ are from Eq. \ref{eq:linefit} and our result here is just the quotient of two Gaussian integrals. As previously noted, $d_{MR}$ is a biased distance estimator that tends to slightly underestimate the distance to known clusters, with the average $d_{MR}/d = 0.93$ and a standard deviation of 0.05. The center of mass distance performs significantly better with average $d_{mass}/d = 1.02$ and a standard deviation of 0.07. We thus use $d_{mass}$ as our canonical distance measurement in this paper and estimate that our distance error are 7\% in a single pixel and 2\% over larger areas.

\begin{table}
\centering
\begin{tabular}{cccccc}
	\hline
Name & $l$ (deg) & $b$ (deg)  & $d$ (kpc) & $d_{MR}$ (kpc) & $d_{mass}$ (kpc) \\
	\hline
NGC 288  & 152.30 & -89.38 & 8.9  & 8.6  & 9.3 \\
NGC 1904 & 227.23 & -29.35 & 12.9 & 11.3 & 12.1 \\
NGC 4590 & 299.63 & 36.05  & 10.3 & 9.7  & 11.2 \\
NGC 5053 & 335.70 & 78.95  & 17.4 & 15.4 & 16.7 \\
NGC 5272 & 42.22  & 78.71  & 10.2 & 9.3  & 10.2 \\
NGC 5466 & 42.15  & 73.59  & 16.0 & 12.9 & 14.4 \\
NGC 5904 & 3.86   & 46.80  & 7.5  & 6.8  & 7.5 \\
NGC 6205 & 59.01  & 40.91  & 7.1  & 7.2  & 7.9 \\
NGC 6341 & 68.34  & 34.86  & 8.3  & 7.6  & 8.4 \\
NGC 7078 & 65.01  & -27.31 & 10.4 & 9.9  & 10.8 \\
NGC 7089 & 53.37  & -35.77 & 11.5 & 11.1 & 13.0 \\
NGC 7099 & 27.18  & -46.84 & 8.1  & 7.6  & 8.0 \\
	\hline
\end{tabular}
\caption{\rm{The locations of and distances to known clusters. The final two columns are the distances as measured by \textsc{match}. The first is the distance measured in Eq. \ref{eq:linefit}. The second is the center of mass distance from Eq. \ref{eq:com}. }}\label{tab:cluster}
\end{table}

\begin{figure}[ht]
\includegraphics[width=0.99\columnwidth]{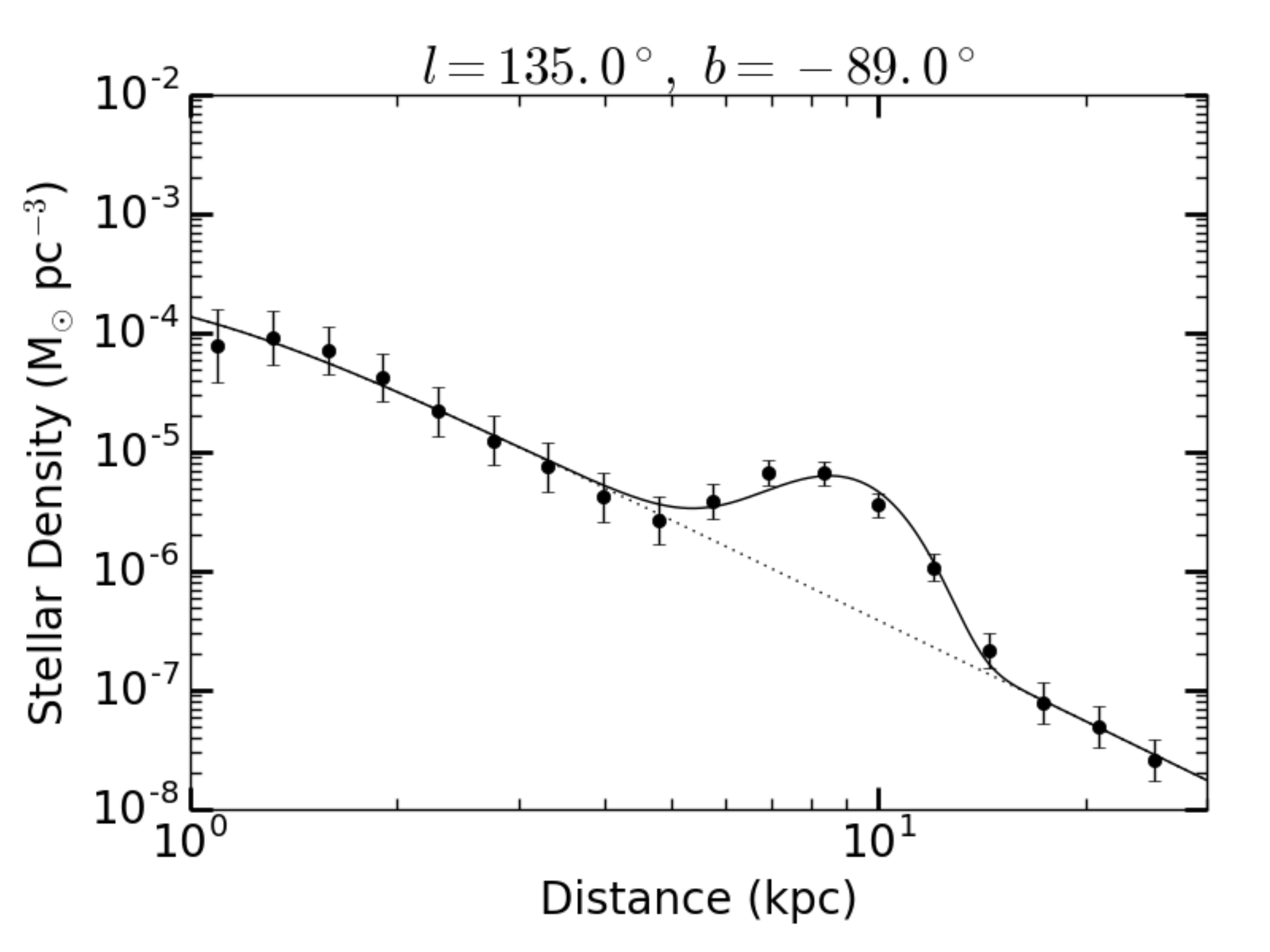}
\caption{\rm{The line of sight density in the direction of NGC 288 ($l = 152.3^\circ,\ b = -89.4^\circ$). This figure actually includes all stars within $90^\circ < l < 180^\circ,\ -90^\circ < b < -88^\circ$. We see a pronounced overdensity at roughly the expected distance of $d =$ 8.9 kpc.}}
\label{fig:ngc}\end{figure}

As a basic confirmation of our cluster fitting technique, we show the line of sight density in the $90^\circ < l < 180^\circ,\ -90^\circ < b < -88^\circ$ direction in Fig \ref{fig:ngc}. Because of the extremely negative value of $b$, this pixel is still roughly 4 square degrees despite the large $l$ range. This region includes NGC 288 ($l = 152.3^\circ,\ b = -89.4^\circ$) which is clearly visible at $d =$ 8.9 kpc. Our results here also show that we can produce reasonable distance estimates even in the moderately crowded fields in Table \ref{tab:cluster}. 

\subsection{Measuring Metallicity}\label{sect:metal}

In addition to producing a total density as a function of distance, $\rho(d)$, our pipeline produces a density for each of the four stellar populations noted in the previous section. We label these densities by their metallicities $\rho_{-0.3}$, $\rho_{-0.8}$, $\rho_{-1.4}$ and $\rho_{-2.1}$, and we can estimate $Z$ as the weighted average of these metallicities:
\begin{equation}
Z(d) = \frac{-0.3 \rho_{-0.3} -0.8 \rho_{-0.8} -1.4 \rho_{-1.4} -2.1 \rho_{-2.1}}{\rho_{-0.3}+\rho_{-0.8}+\rho_{-1.4}+\rho_{-2.1}}.\label{eq:metal}
\end{equation} 
This metallicity estimate is far from perfect. Isochrone systematics, unaccounted variation in reddening, non-stellar sources sneaking into our CMD and Galactic variations in stellar age will all be aliased as metallicity. We discuss these metallicity measurements more in Appendix \ref{sect:addmetal}. 

\section{Mapping the Monoceros Ring in 2D}\label{sect:map}

While the focus of this paper is quantitative analysis of the MR, global imaging of the MR region is important to qualitatively inform our analysis, and we present our large scale MR (and more general Milky Way) maps here. We display our maps as Heliocentric spherical shells of thickness 0.4 magnitudes of distance modulus and focus on a radius of 8.3 kpc, were the MR features appear most distinctly. This is essentially equivalent to taking a single data point from Fig.\ \ref{fig:distance} along every line of sight. In this section we present the total density map (with limited 3D information) and the stellar metallicity as defined by Eq.\ \ref{eq:metal}. These maps all use the mask defined in Section \ref{sect:metal}. The unmasked versions of these maps may be of significant use in other applications and are presented in Appendix \ref{sect:addmap}. These images are similar to those presented by \citet{SLAT++14} with an earlier PS1 release.

\begin{figure}[ht]
\includegraphics[width=0.99\columnwidth]{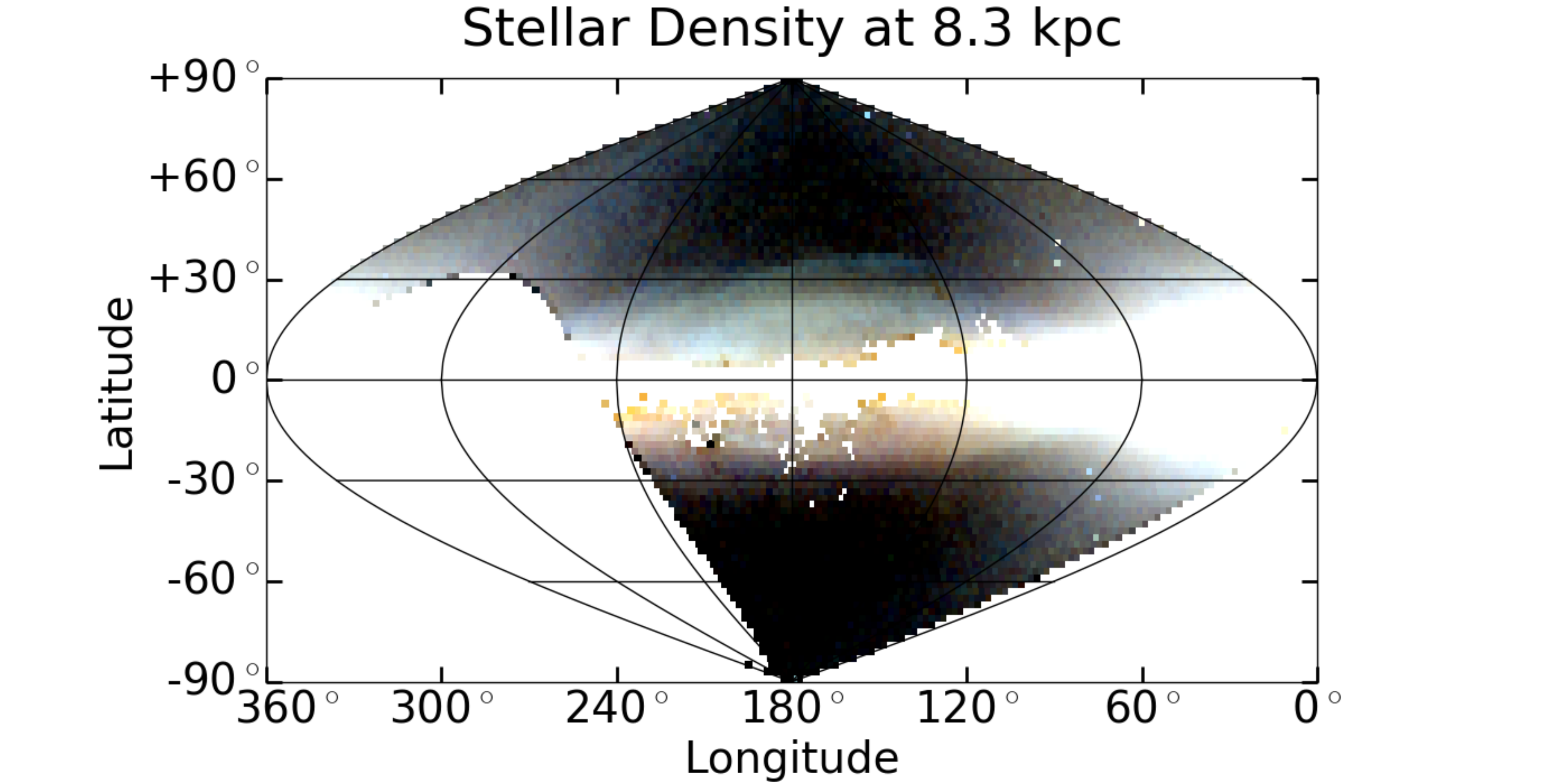}
\includegraphics[width=0.99\columnwidth]{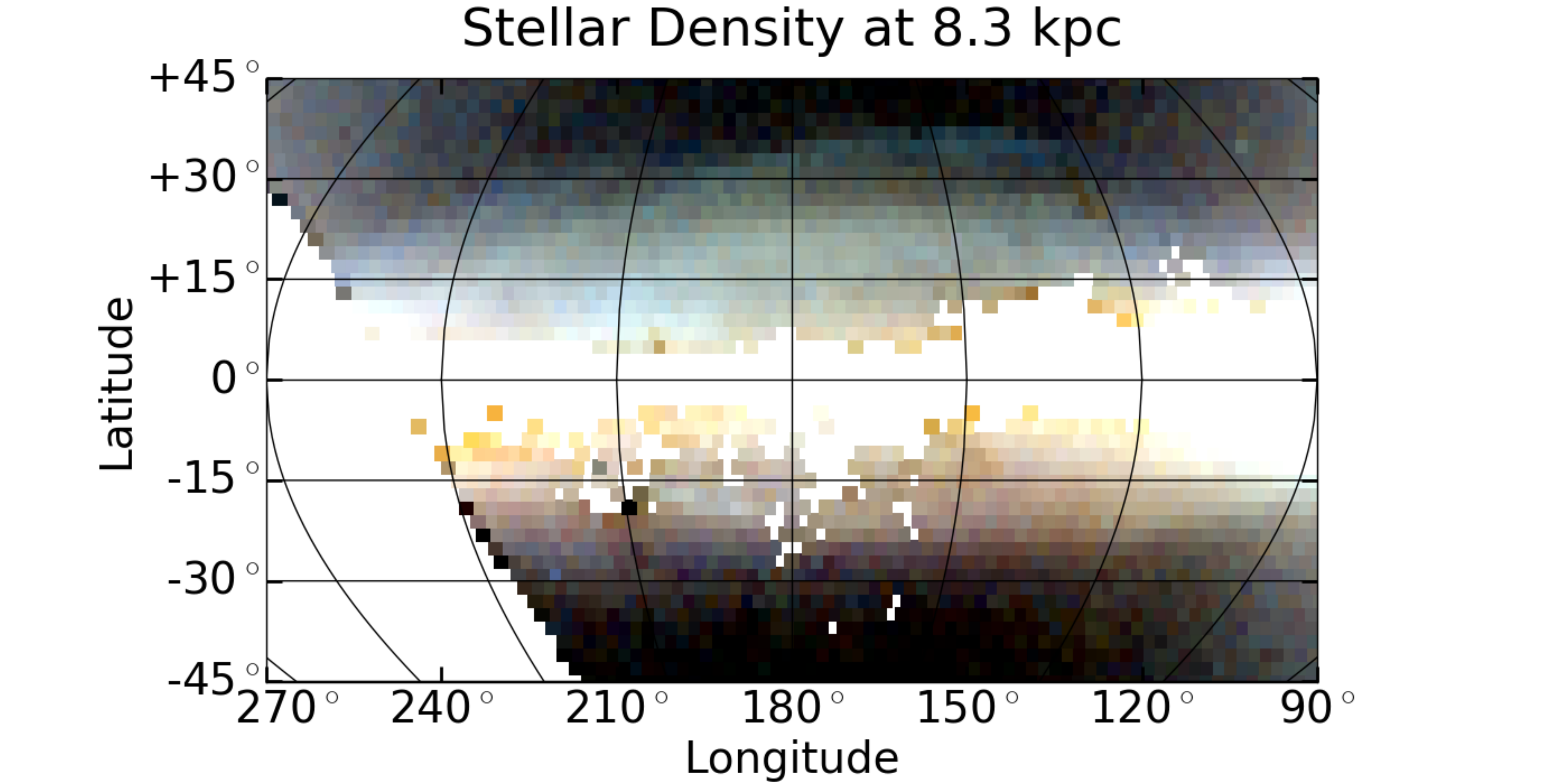}
\caption{\rm{The (masked) stellar mass density across the sky. The Red/Green/Blue (RGB) channels represent density at 6.9, 8.3 and 10 kpc, respectively (Heliocentric). Each channel is scaled logarithmically with the minimum and maximum set at the level of the 10th and 95th percentile (masked). In units of $10^{-6} M_\odot \rm{pc}^{-3}$, this corresponds to 1.9 and 46 in the R channel, 0.97 and 22 in the G channel and 0.55 and 11 in the B channel.}}
\label{fig:densitymap}\end{figure}

Fig.\ref{fig:densitymap} shows the total stellar mass density at 6.9-10 kpc. It is a logarithmic RGB plot where the three red, green and blue channels are stellar mass density at 6.9, 8.3 and 10 kpc, respectively. The channels are scaled for maximum contrast. The two most distinct features are the white cloud of stars that covers everything (that is not masked) within roughly $60^\circ$ of $l=0^\circ,\ b=0^\circ$ and the smaller cloud of stars that covers the $120^\circ < l< 240^\circ$, $-30^\circ < b < +40^\circ$ region. The first is, of course, the Galactic Bulge. Its stars essentially saturate our image, and it is not the subject of this paper. The second is the Monoceros Ring. As in \citet{SLAT++14}, we note the structure is sharply bound at $b = +40^\circ$ and $b = -30^\circ$. {In the North, there are three distinct stream-like features dominating the anticenter region. Between $15^{\circ} < b < 30^{\circ}$ and $130^{\circ} < l < 220^{\circ}$ there is a large broad feature which is associated with the Monoceros Ring. Above this feature is a clear, thin, stream-like overdensity arcing from $(l,b)$ = $\sim(240,15)^{\circ}$ to $\sim(90,15)^{\circ}$. This is the Anti-Center Stream (ACS) from \citet{GRIL06} and \citet{GRIL++08} and it clearly extends far beyond SDSS survey edge at $(l,b)$ $\sim(224.8,+20)^{\circ}$, as per the initial discovery. We see that the ACS becomes more distant as it approaches the Galactic Center. The Eastern Banded Structure (EBS), now associated with the Hydra I dwarf galaxy candidate \citet{GRIL11} was previously only visible between $(l,b)$ = $\sim$$(229,+30)$$^{\circ}$ to $\sim$$(217,30)$$^{\circ}$. The EBS is visible from $(l,b) \sim(217,30)^{\circ}$ to $(l,b) \sim(260,15)^{\circ}$ and is approximately 10 kpc distant in this direction. The Southern MR appears more uniform, with no visible discrete structures. There is also a notable North-South asymmetry, with the MR being both more extensive and slightly farther away in the North.

\begin{figure}[ht]
\includegraphics[width=0.99\columnwidth]{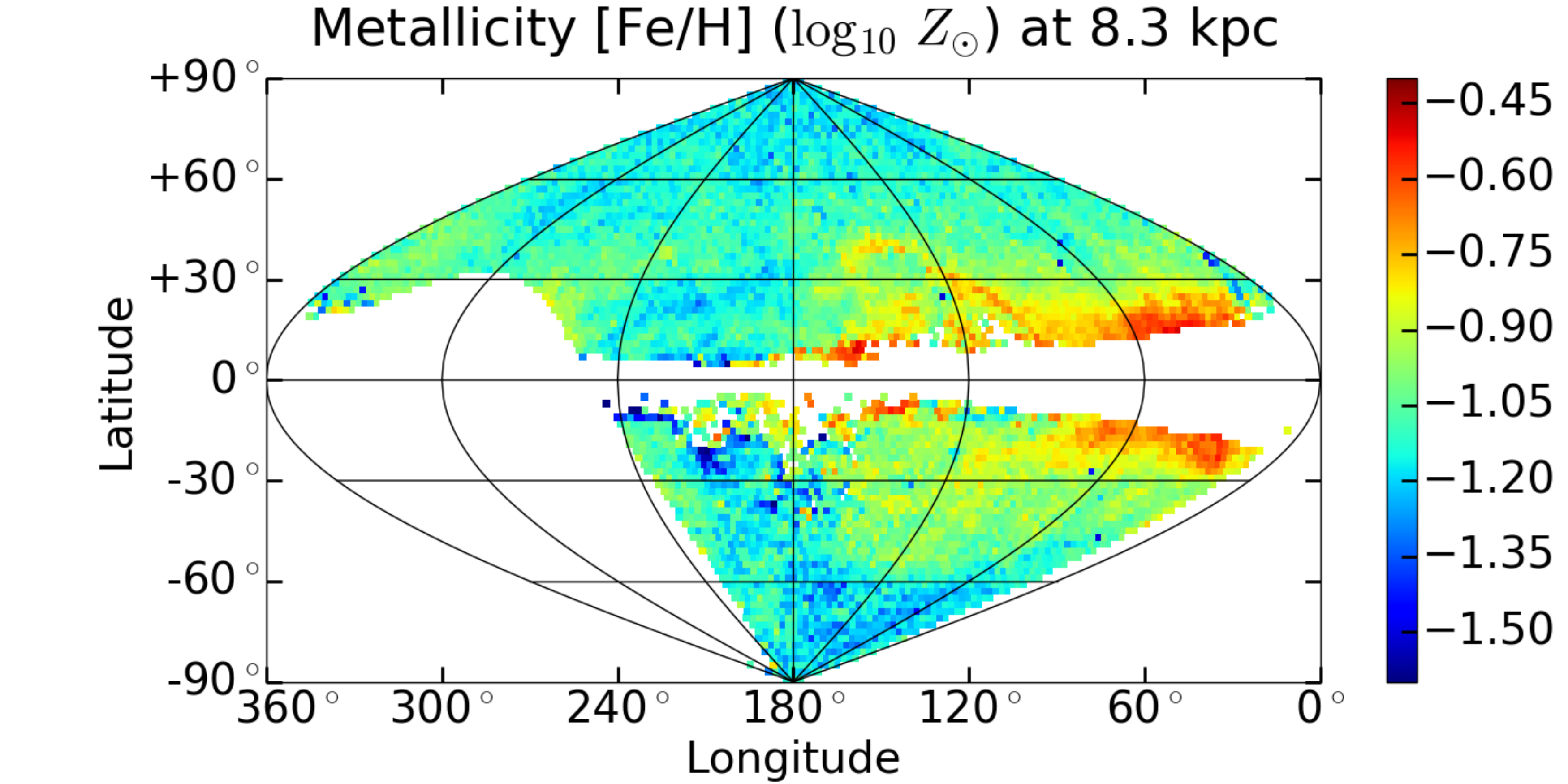}
\includegraphics[width=0.99\columnwidth]{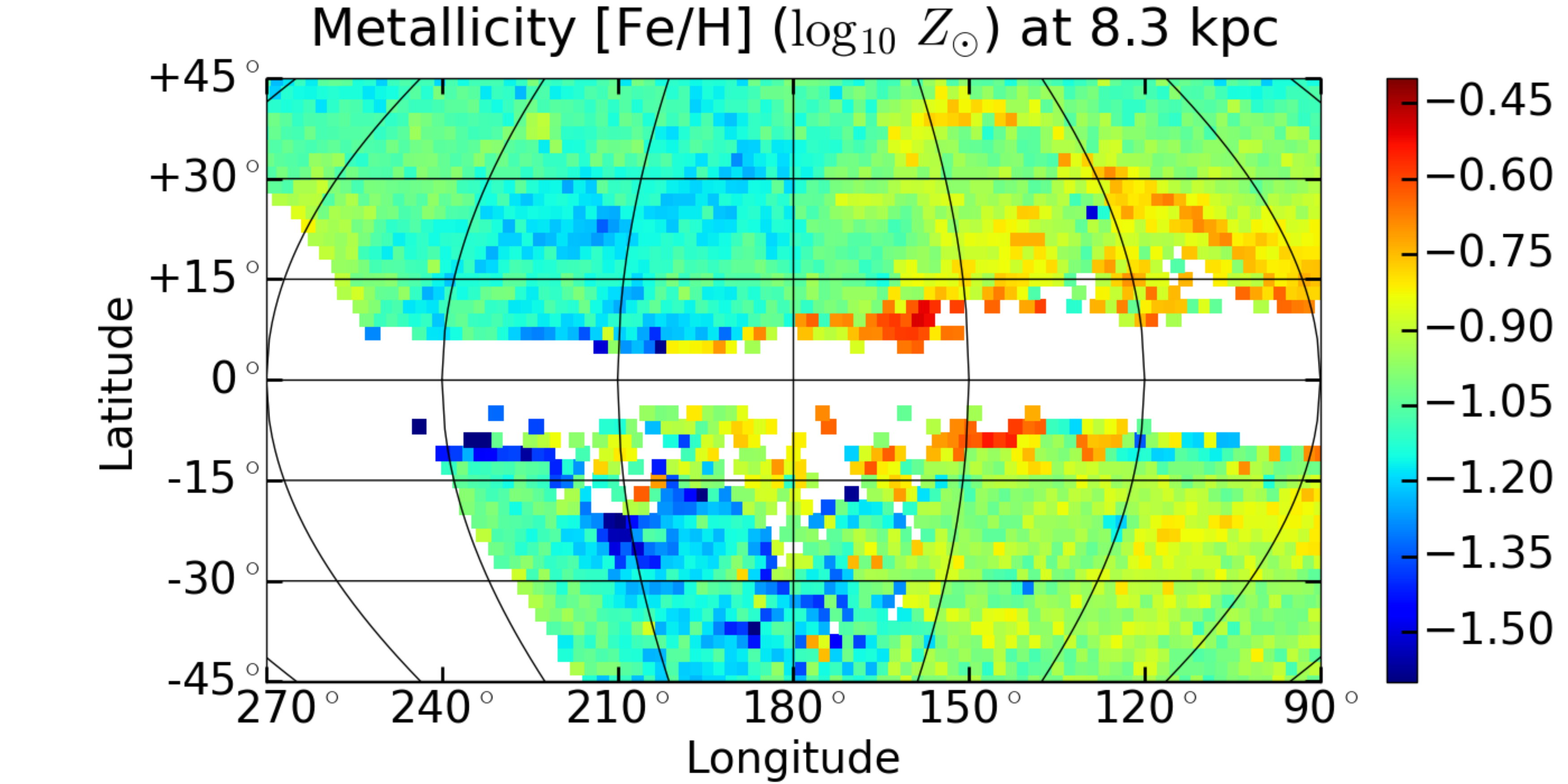}
\includegraphics[width=0.99\columnwidth]{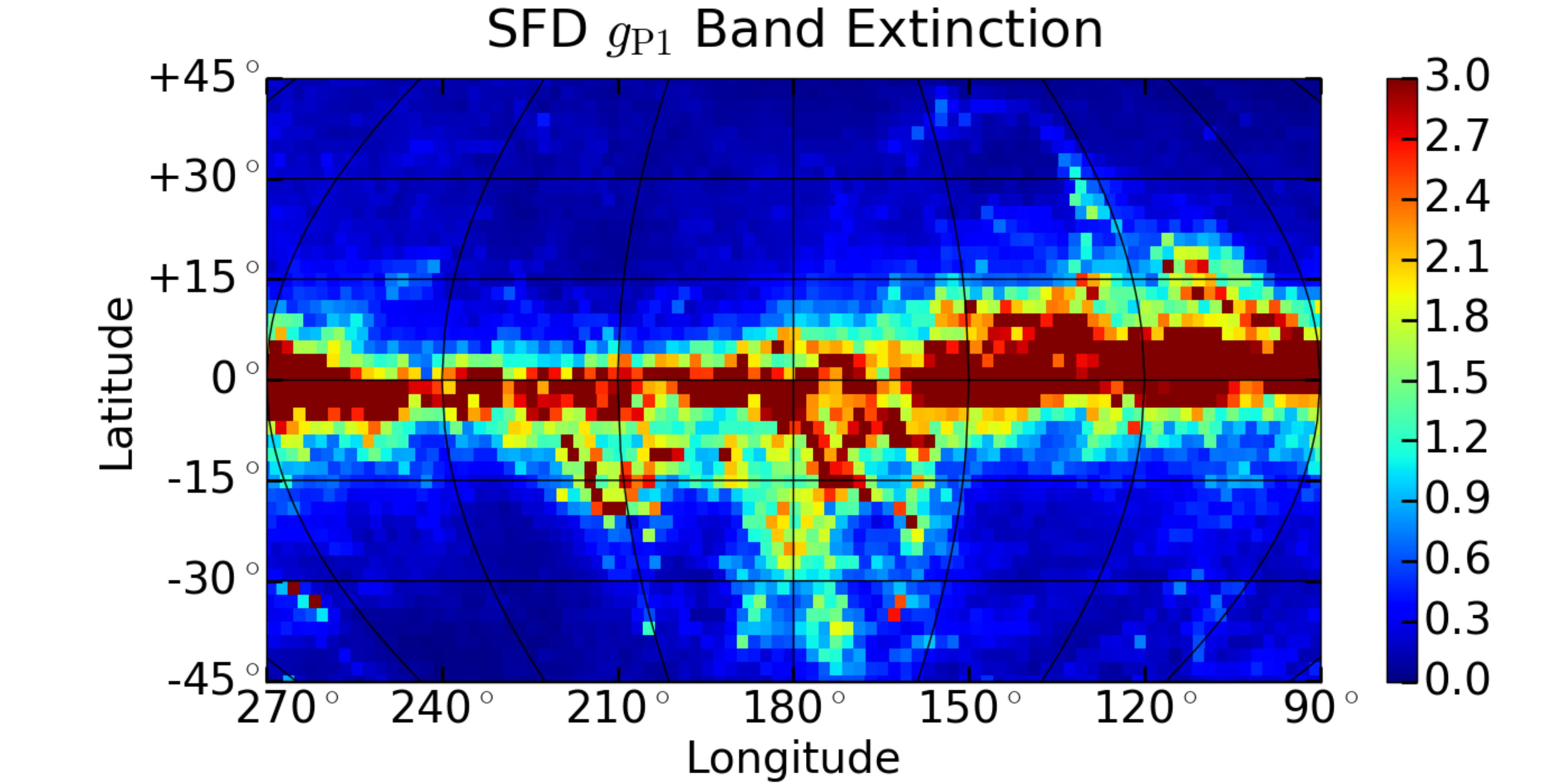}
\caption{\rm{The (masked) stellar metallicity at 8.3 kpc (Heliocentric) as calculated with Eq.\ \ref{eq:metal}. Note that with our settings, \textsc{match} aliases errors in stellar age or Galactic extinction as (typically lower) metallicity. Our masking removes the worst of these aliasing problems. In the zoomed-in metallicity plot (middle) we see a clear spur of apparent high metallicity that along $l = 120^\circ$ in the North that corresponds to an SFD dust feature (bottom).}}
\label{fig:metalmap}\end{figure}

Fig.\ \ref{fig:metalmap} shows our calculated metallicity across the sky. $Z$ goes from -0.4 to -0.8 near the Galactic center and then fall to -1.2 over most of the outer MW, halo region. The MR region has $-1.3 < Z < -0.9$ consistent, with the high latitude halo population in this figure. Along the edge of the mask, a small number of pixels have anomolously high or low $Z$. This is likely due to a younger stellar populations or Galactic extinction over-correction being aliased as low metallicity. Particularly, there is a looping high metallicity structure stretching from the Galactic plane through $(l,b) = (120^\circ, +30^\circ)$ that corresponds to a dust structure that is apparently being under-corrected by the SFD extinction map. There is a notable high metallicity feature stretching from $(l, b) = (240^\circ, 15^\circ)$ to $(l, b) = (160^\circ, 35^\circ)$. This corresponds to the ACS and suggests (as previously noted by \citet{LI++12}) that the ACS may be of distinct origin from the main MR. 

Figs.\ \ref{fig:densitymap} and \ref{fig:metalmap} show just a slice of our 3D stellar mass density maps which stretches from 1 kpc to 30 kpc with similar quality to what we see here. In the future, we may use this 3D map to parameterize the structure of the MW globally and to examine the outer MW Sagittarius stream \citep{YANN++00,SLAT++13} in detail.

\subsection{Deeper 2D Milky Way Mapping}\label{sect:mapdeep}

In addition to probing probing MW structure at the key Heliocentric distance of roughly 8 kpc, we can examine it (with considerably less precision) at larger distances. Fig. \ref{fig:deepmap} shows the analogous stellar mass density map as Fig. \ref{fig:densitymap} at distances between 14.5, 17.4 and 20.9 kpc in its red, green and blue channels, respectively. Since these data are mapping fainter stars than those in Fig. \ref{fig:densitymap}, our effective masking extends farther from the plane where dust prevents us from detecting sufficiently faint stars (see Eq. \ref{eq:mulim} ). The Sagittarius Stream is obvious as a blue vertical stream near the center of the image which wraps around the lower right and upper left edges of the plot. We show another view of the Sagittarius Stream in Appendix Appendix \ref{sect:addmap}. Interestingly, we do not see any evidence of the TriAnd structure which we would expect to see in our deepest distance bin. This may be because its stars are effectively blurred across a large distance range at there faint magnitudes. We will discuss deeper MW structures as observed by PS1 in more depth the upcoming paper Conn et al. 2016 (in preparation). 

\begin{figure}[ht]
\includegraphics[width=0.99\columnwidth]{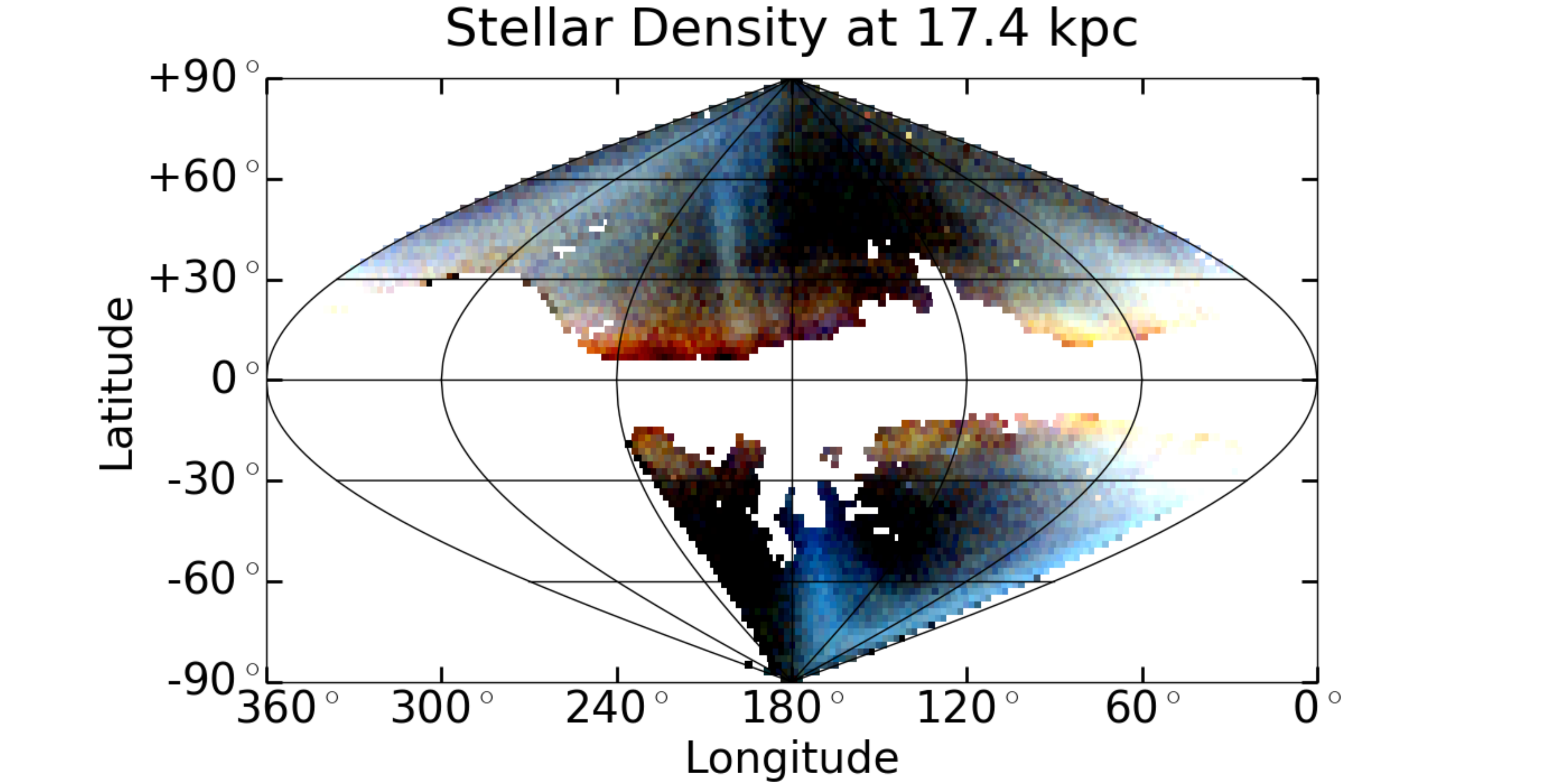}
\includegraphics[width=0.99\columnwidth]{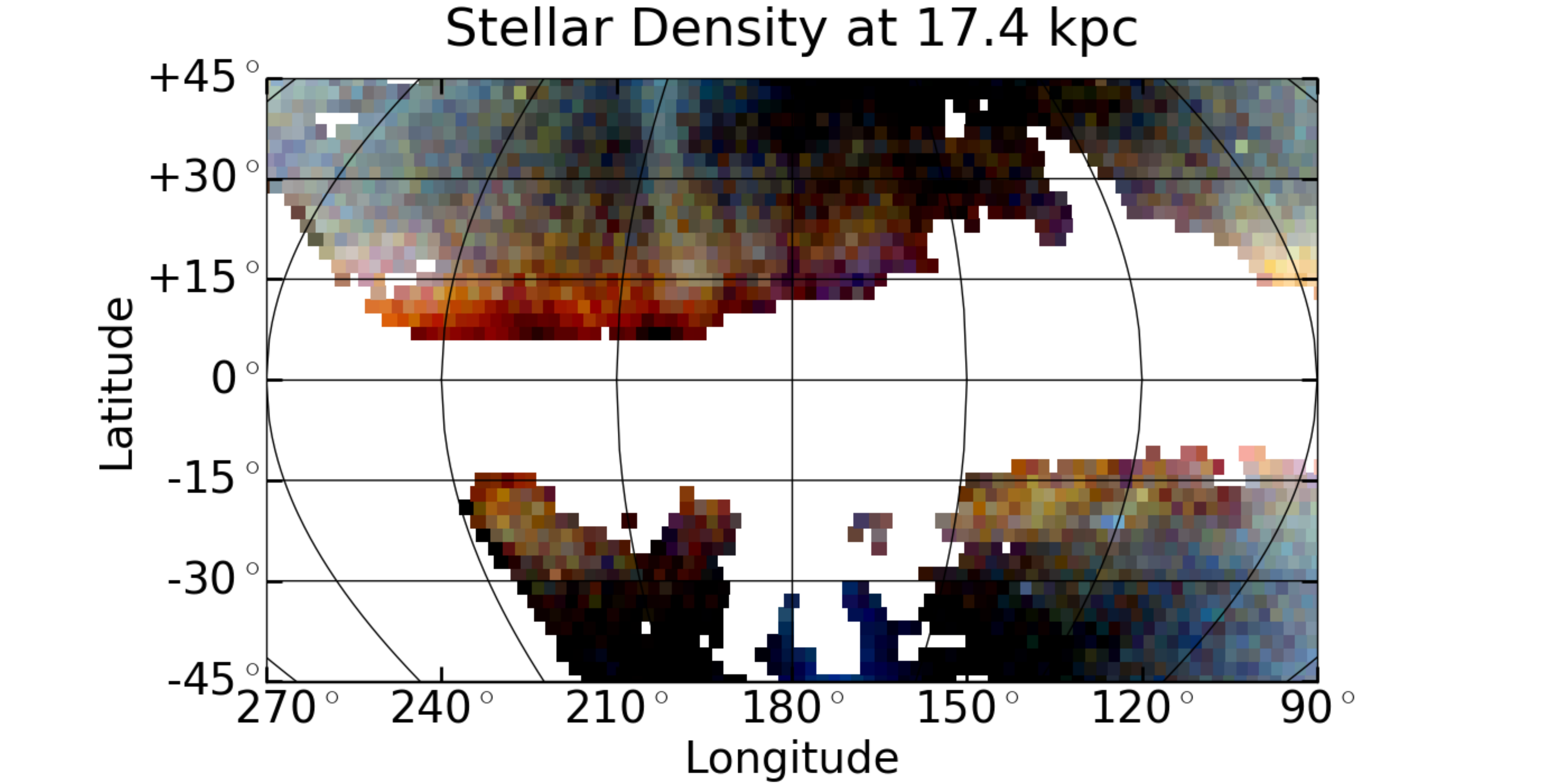}
\caption{\rm{The (masked) stellar mass density across the sky. The Red/Green/Blue (RGB) channels represent density at 14.5, 17.4 and 20.9 kpc, respectively (Heliocentric). Each channel is scaled logarithmically with the minimum and maximum set at the level of the 10th and 95th percentile (masked). In units of $10^{-6} M_\odot \rm{pc}^{-3}$, this corresponds to 0.23 and 1.9 in the R channel, 0.13 and 0.84 in the G channel and 0.068 and 0.40 in the B channel.}}
\label{fig:deepmap}\end{figure}

\section{Mapping the Monoceros Ring in 3D}\label{sect:map3d}

\citet{NEWB++02} and \citet{SLAT++14} have already produced valuable 2D maps and qualitative 3D maps of the Monoceros Ring similar to Fig. \ref{fig:densitymap}. Our handling of dust and different metallicity populations with \textsc{match} has allowed us to probe the new PS1 area more precisely. This analysis allows us to study the MR quantitatively and in three dimensions. Below we present quantitative density and MR distance maps, meridional cross sections of the MW which reveal how the MR structure changes along our lines of sight and planar cross-sections which fully reveal the roughly arcing structure of the MR. These different views of the MR are all consistent with its structure being two concentric planar circles, one in the South and the other in the North. The Southern MR is significantly closer and denser than the Northern MR. 

\subsection{Quantitative Projection Maps}\label{sect:quant}

Our line of sight analysis with \textsc{match} allows us to make more quantitative measurements of the Monoceros Ring than previous analyses. Having fit the main Milky Way and MR populations as separate functions along each line of sight with Eq. \ref{eq:linefit}, we have numerical estimates of both the total MR mass and distance to the MR along every (unmasked) line of sight in PS1. Both the MR mass and distance show quantifiable North-South asymmetry. 

The fitting form (Eq. \ref{eq:linefit}) provides an MR density scale, $\delta \rho$, a distance with peak excess density, $d_{MR}$ and a width, $W_{MR}$. To obtain an excess mass estimate for each pixel, we integrate the Gaussian-shaped overdensity along the line of sight volume, accounting for the integrated line of sight volume element (including the $d^2$)  to obtain a total excess mass per pixel of
\begin{equation}
M_{pix} = \frac{\delta\rho_{pix}}{\pi^2}\left(d_{MR}^2+W_{MR}^2\right).\label{eq:pixmass}
\end{equation}
Here the factor of $(d_{MR}^2+W_{MR}^2)$ is derived from a Gaussian integral and the factor of $(2/2\pi)^2 = \pi^{-2}$ accounts for our $2 \times 2$ degree pixel. We use $d_{mass}$, the MR center of mass distance from Eq. \ref{eq:com}, as our distance quantity in these maps.

\begin{figure}[ht]
\includegraphics[width=0.99\columnwidth]{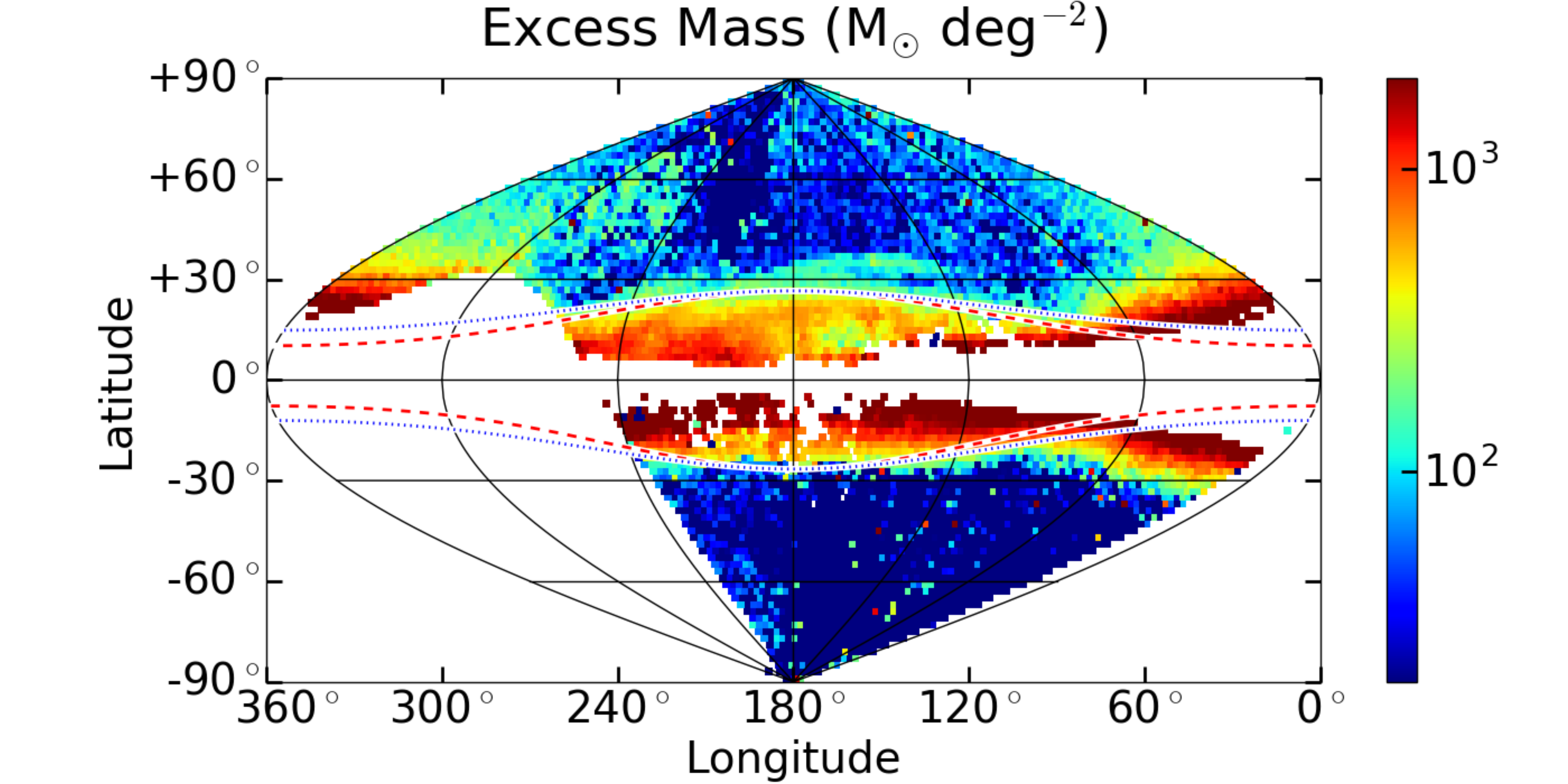}
\includegraphics[width=0.99\columnwidth]{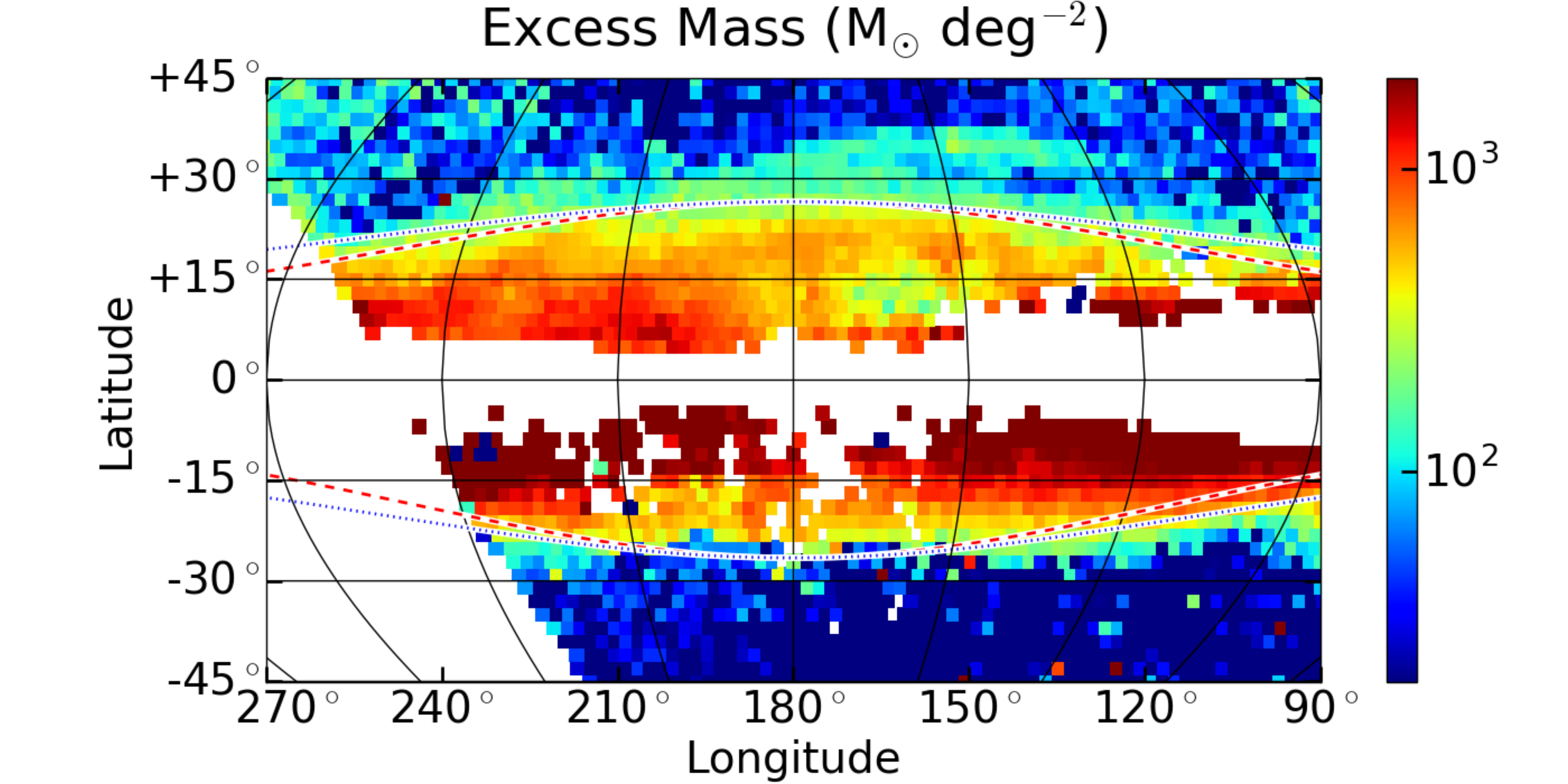}
\includegraphics[width=0.99\columnwidth]{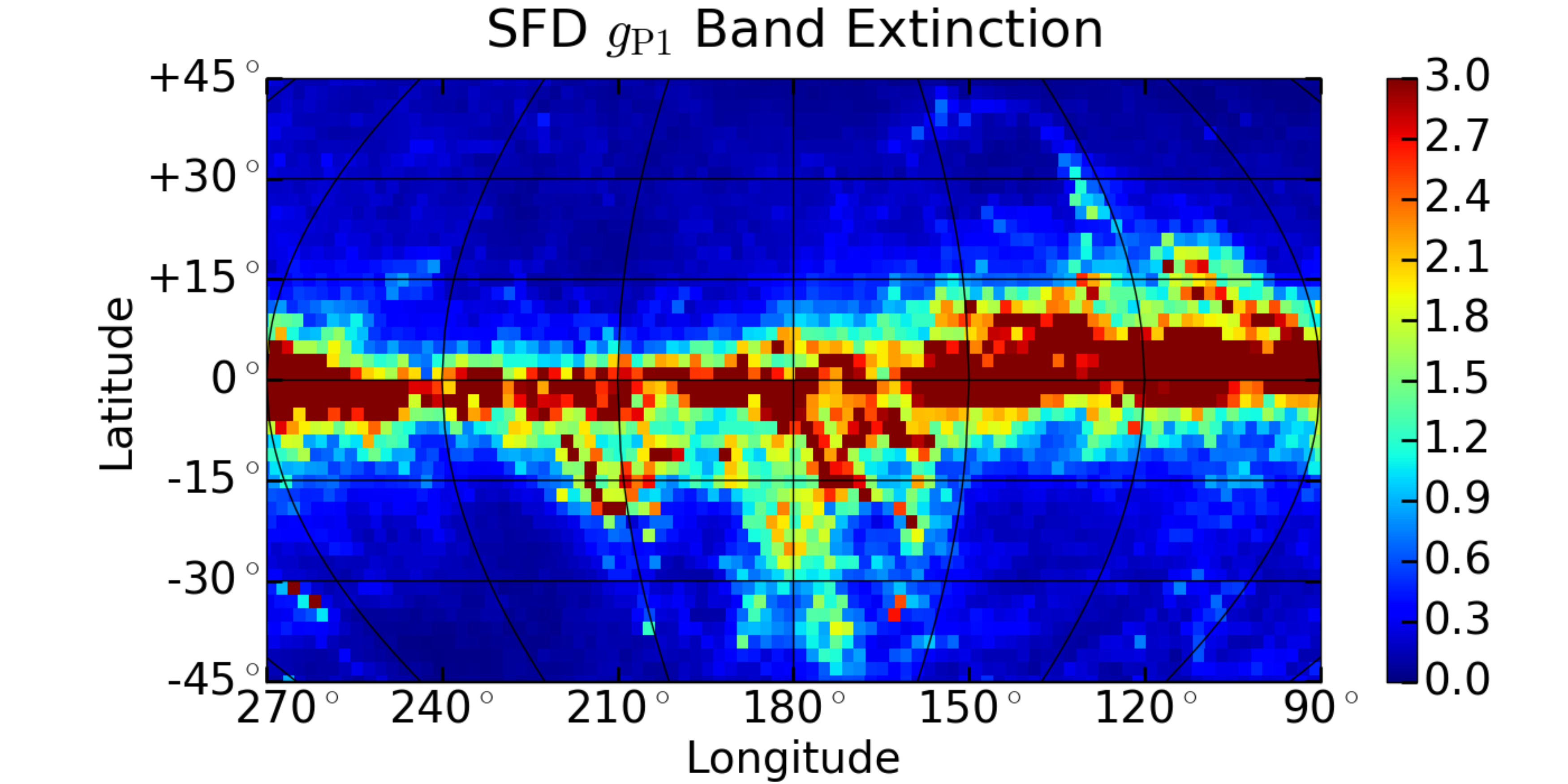}
\caption{\rm{The integrated total excess mass (in M$_\odot$ deg$^{-2}$) associated with the Monoceros Ring as fit by Eq.\ \ref{eq:linefit}. Each pixel represents the integrated excess mass per square degree in that 4 degree area, so that the total excess mass is just the sum of the pixel values times 4. The middle figure is a zoom-in of the top figure The bottom figure shows the SFD extinction in the \gps\ filter and is a zoom-in of Fig. \ref{fig:mask}. We do not see a strong correlation between MR features and the dust map. The red dashed line shows the edge of a Galactocentric cylinders (centered 8 kpc from the Sun at $l = 0^\circ$) with radius 17 kpc (14 kpc) and height 4.5 kpc (-3 kpc) in the North (South). The blue dotted line shows the edge of a cylinder centered 4 kpc from Sun at $l = 0^\circ$ with radius 13 kpc (10 kpc) and height 4.5 kpc (-3 kpc) in the North (South).}}
\label{fig:massmap}\end{figure}

Fig. \ref{fig:massmap} shows the excess mass, $M_{pix}$, from Eq. \ref{eq:linefit} and Eq. \ref{eq:pixmass}. Near $l = 0^\circ$ our fitting routine clearly uses the Monoceros Ring bump to fit the excess stars at the Galactic Center. We ignore this region in our MR analysis and zoom in to the Anticenter region in the middle panel. First, the Monoceros Ring is simply more massive in the South, having densities of more than 3{,}000 $M_\odot \rm{deg}^{-2}$ over much of its area while the Northern density rarely exceeds 1{,}000\ $M_\odot \rm{deg}^{-2}$. Several Northern features from \citet{SLAT++14} stand out. The ACS is apparent, stretching across $120^\circ < l < 180^\circ$ at $b = +32^\circ$. In addition the main stream at $b=+20^\circ$ is obvious, although it could also be interpreted as being due to a void around ($160^\circ,\ +10^\circ$). The EBS is also visible as a sharp edge stretching from ($250^\circ,\ +15^\circ$) to ($220^\circ,\ +35^\circ $). There are no distinct features in the South, only a smooth gradient. In the bottom panel, we show a zoomed-in version of our dust map. None of the distinct Northern features are traced by dust features, suggesting that they are not due to problems with our dust correction or the foreground features that tend to trace dust. 

To better understand the geometry of the MR in physical space, Fig. \ref{fig:massmap} shows the projected angular height of different cylinders with constant physical extents above and below the plane, $z_{MR}$. Using cylindrical coordinates, (cylindrical radius, $r$; Galactic longitude, $l$; and the height above the Galactic plane, $z$), we can derive the observed angular height of such a cylindrical MR model,  
\begin{equation}
b_{MR}(l) = \arctan\left(z_{MR}/r_H(l)\right),
\end{equation}
where $r_H(l)$ is the Heliocentric cylindrical distance to MR (the distance to MR along the Galactic plane), and $z_{MR}$ is the (constant) height of a cylindrical MR model. We can derive this quantity with respect the $r_{MR}$, the cylindrical radius of the Monoceros Ring, and $x_\odot$ the distance from the Sun to the center of the Monoceros Ring cylinder (which we assume to be along the $l = 0^\circ$ line):
\begin{eqnarray}
r_{MR}^2 &=& \left(r_H(l) \cos(l)-x_\odot \right)^2 + r_H(l)^2 \sin^2(l),\\
0 &=& r_H^2(l) + 2\cos(l) x_\odot\ r_H(l) - \left( r_{MR}^2-x_\odot^2\right)\nonumber,\\
r_H(l) &=& \left( r_{MR}^2-\sin^2(l) x_\odot^2 \right)^{1/2}+\cos(l) x_\odot.\nonumber
\end{eqnarray}
In Fig. \ref{fig:massmap}, we show Galactocentric cylinders (assuming $x_{MR} = 8$ kpc) as red dashed lines. These lines have $z_{MR}$ = 4.5 (-3) and $r_{MR}$ = 17 (14) in the North (South). These particular radii are measured in Fig. \ref{fig:distancemap} and discussed in Section \ref{sect:cross}. Qualitatively, the Northern Galactocentric cylinder matches the Northern MR edge fairly well. But the Southern MR edge is much too flat (in our spherical projection) for the cylindrical fit. Physically, this means that the Southern MR may actually come up towards the plane near $l = 180^\circ$ so that it appears flat in our projection. In Section \ref{sect:cross} we will see that the MR is better fit as a cylinder centered roughly 4 kpc from the Sun, $x_{MR} = 4$. We show these cylinders with radii $r_{MR}$ = 13 (10) in the North (South) as blue dashed lines. They are indeed slightly better fits to the MR edge in both the North and the South. 

\begin{figure}[ht]
\includegraphics[width=0.99\columnwidth]{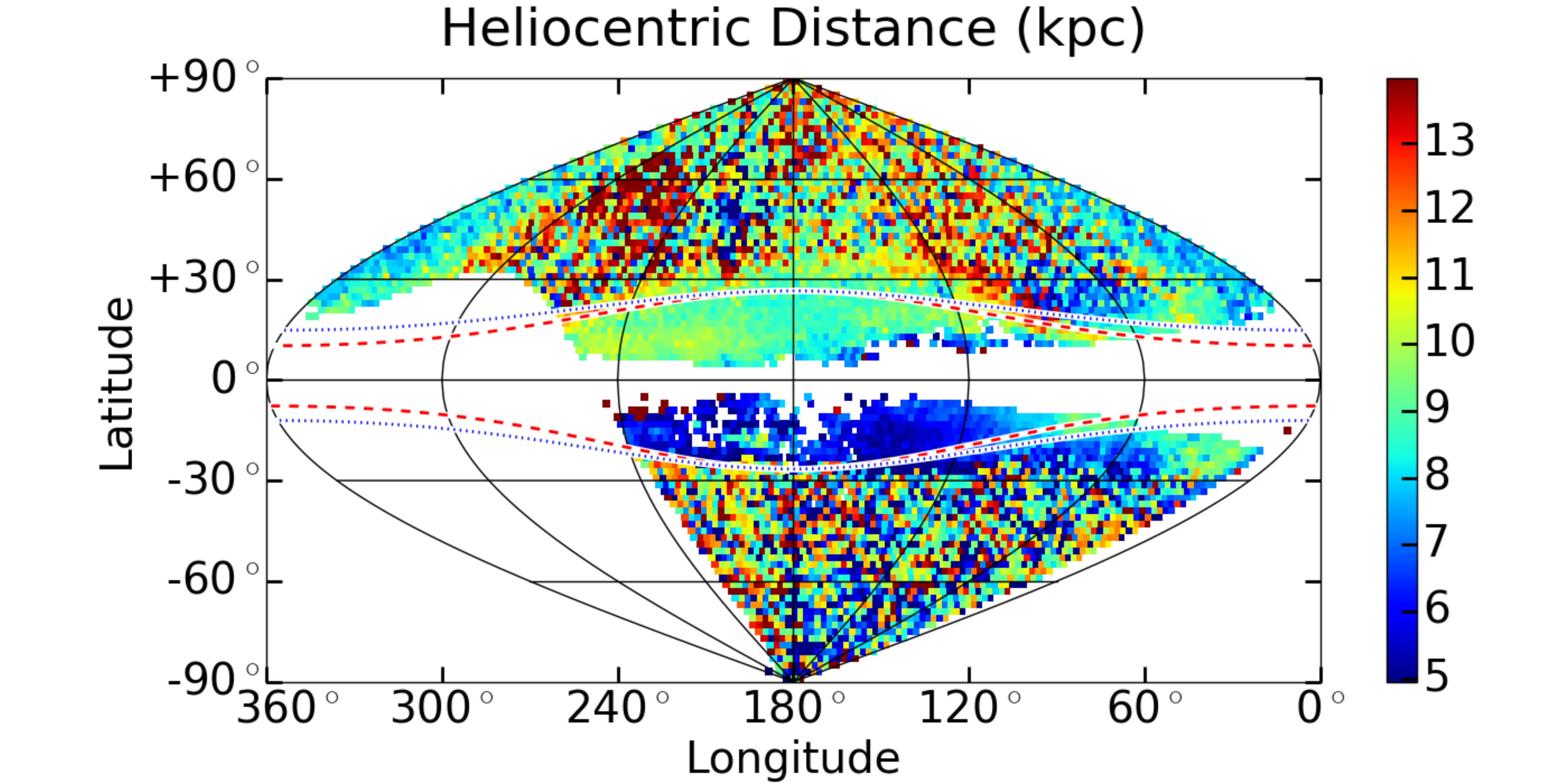}
\includegraphics[width=0.99\columnwidth]{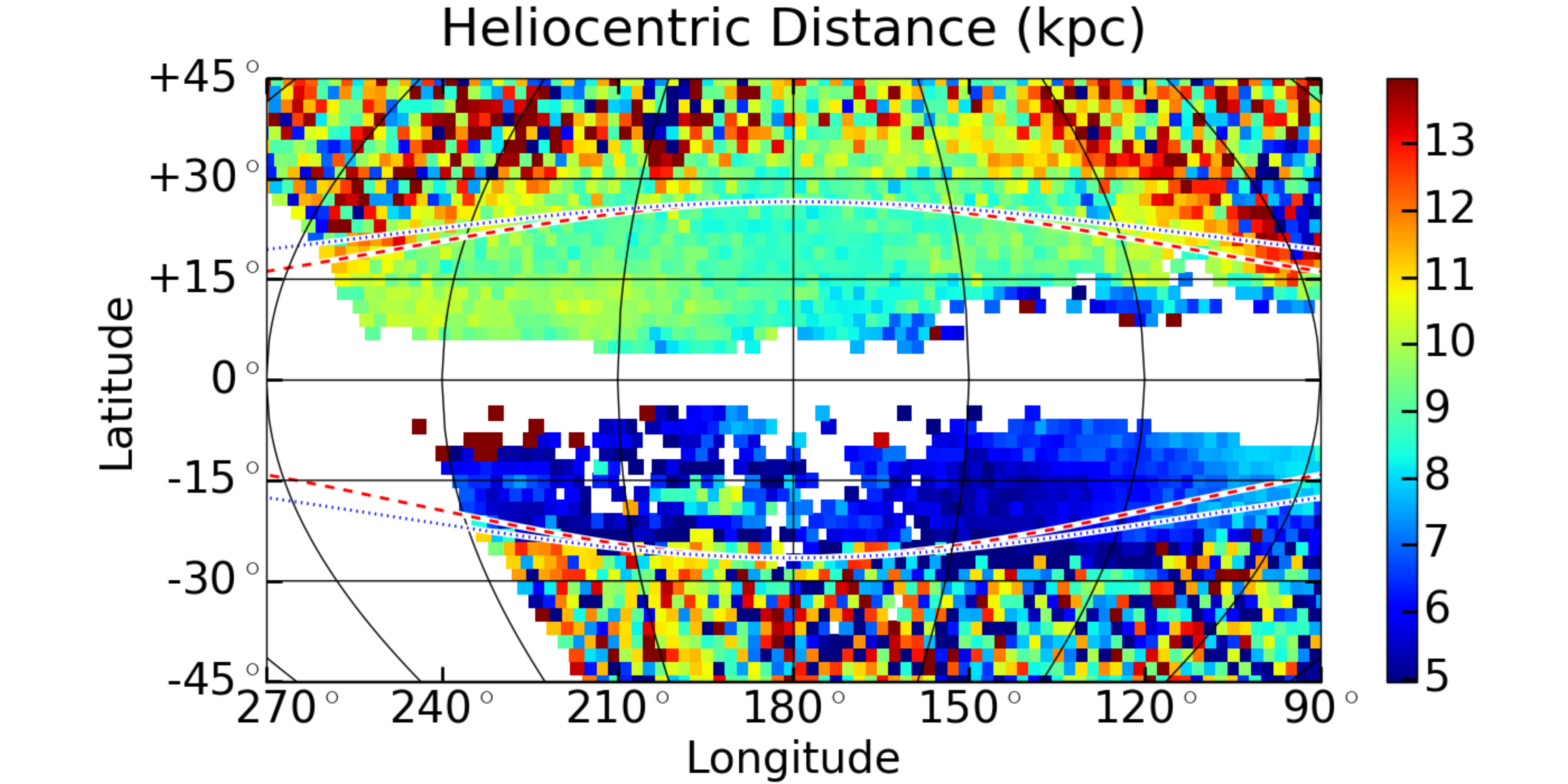}
\includegraphics[width=0.99\columnwidth]{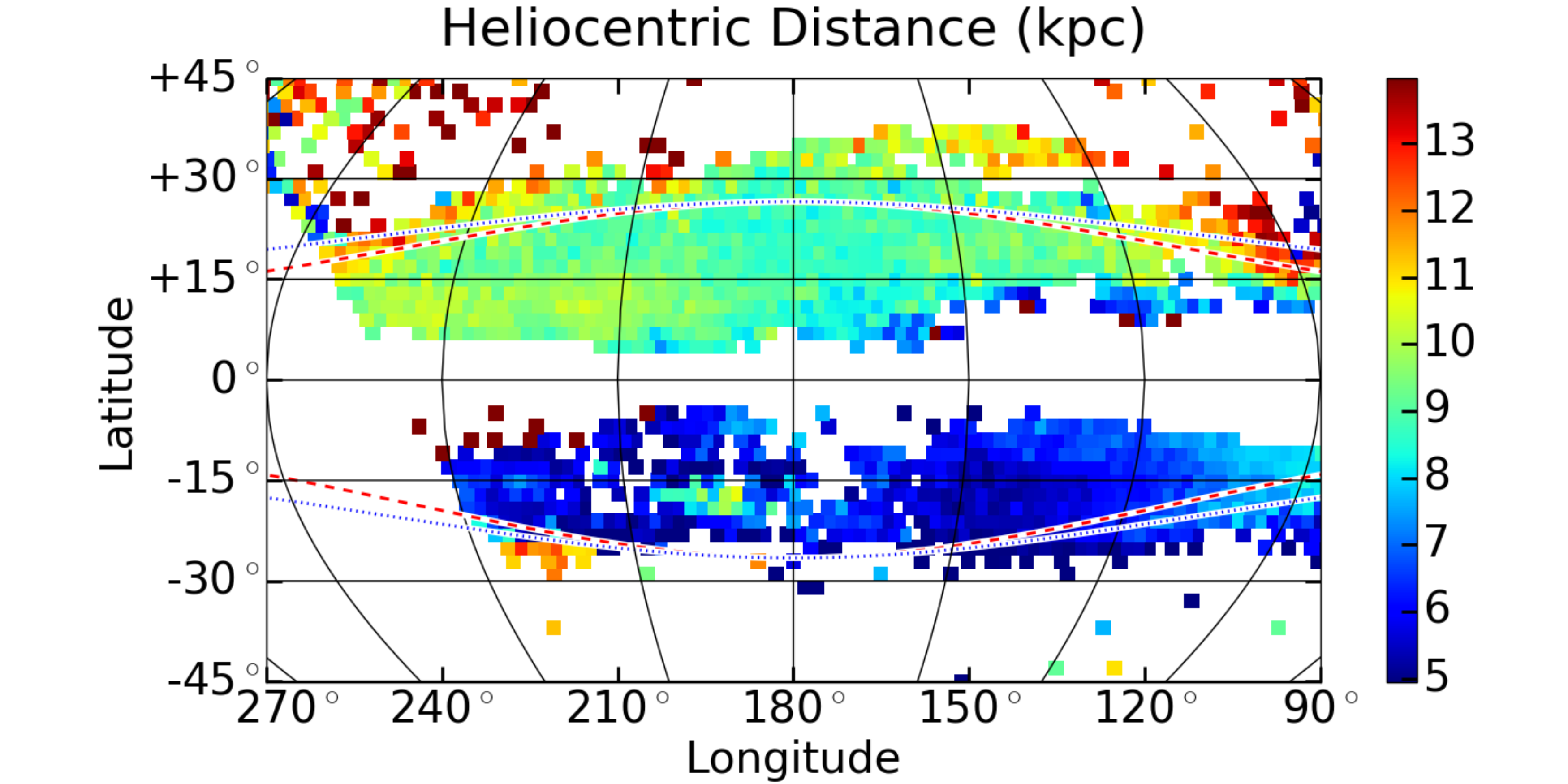}
\caption{\rm{The Heliocentric distance (in kpc) to the Monoceros Ring along each line of sight as fit by Eq.\ \ref{eq:linefit}. Areas outside the MR or GC region with only small overdensities have essentially random distances. The middle figure is a zoom-in of the top figure, and in the bottom figure we show only pixels with more than 100 $M_\odot$ excess mass.}}
\label{fig:distancemap}\end{figure}

Fig. \ref{fig:distancemap} shows the Heliocentric distance to the MR center of mass along each line of sight. Again, it is most useful to zoom in on the main MR area in the bottom panels. We immediately see a clear split between the North and South with Northern MR being being roughly 9 kpc away while the Southern MR is 6 kpc. While the transition from the Northern MR to Southern MR is masked by the plane, there is no obvious gradient in distance as we travel from the North to South, suggesting a sudden transition in distance. Across $100^\circ < l < 200^\circ$, we see a small amount of $d = 6$ area just North of the masked region. This area corresponds to anomalously high density regions in the North, suggesting that some of the more local, denser Southern population may cross the Galactic plane. 

\subsection{Meridional (Vertical Slice) Cross-Sections}\label{sect:vert}

While the Heliocentric radial projection in Section \ref{sect:quant} allows us to measure the mass and distance of the Monoceros Ring, it does not tell us much about the three dimensional structure of the MR. To examine this structure, we produce meridional cross-sections, slices of constant $l$ (and $l+180^\circ$) that cut through the Sun. These cross-sections, shown in Fig. \ref{fig:sliceplot}, are stellar mass densities as a function of Galactic latitude, $b$, and Heliocentric distance, $d$. They are similar to those made in \citet{DEJO++10}. We calculate them as the median density of all points at that $b$ within $10^\circ$ along that line of latitude:
\begin{equation}
\rho(b, d, l_0) = {\rm median}\left( \rho(b,d,l),\ {\rm for}\ |l-l_0| < 10^\circ/\cos b\right).
\end{equation}
Here, the $\cos b$ term ensures that we are averaging across the same area (number of pixels) at all latitudes. When making these bins, we wrap around the $l = 0^\circ,\ 360^\circ$ line appropriately. To make our cross section a full circle, we actually plot $\rho(b, d, l_0)$ and $\rho(b, d, l_0+180^\circ)$, again accounting for the wrap around $360^\circ$. These circular disks correspond to actual circular slices in physical space. Lines of Galactic latitude, $b$, with fewer than 3 unmasked pixels in their $20^\circ$ $l$ range are masked out. 

\begin{figure*}[ht]
\includegraphics[width=0.66\columnwidth]{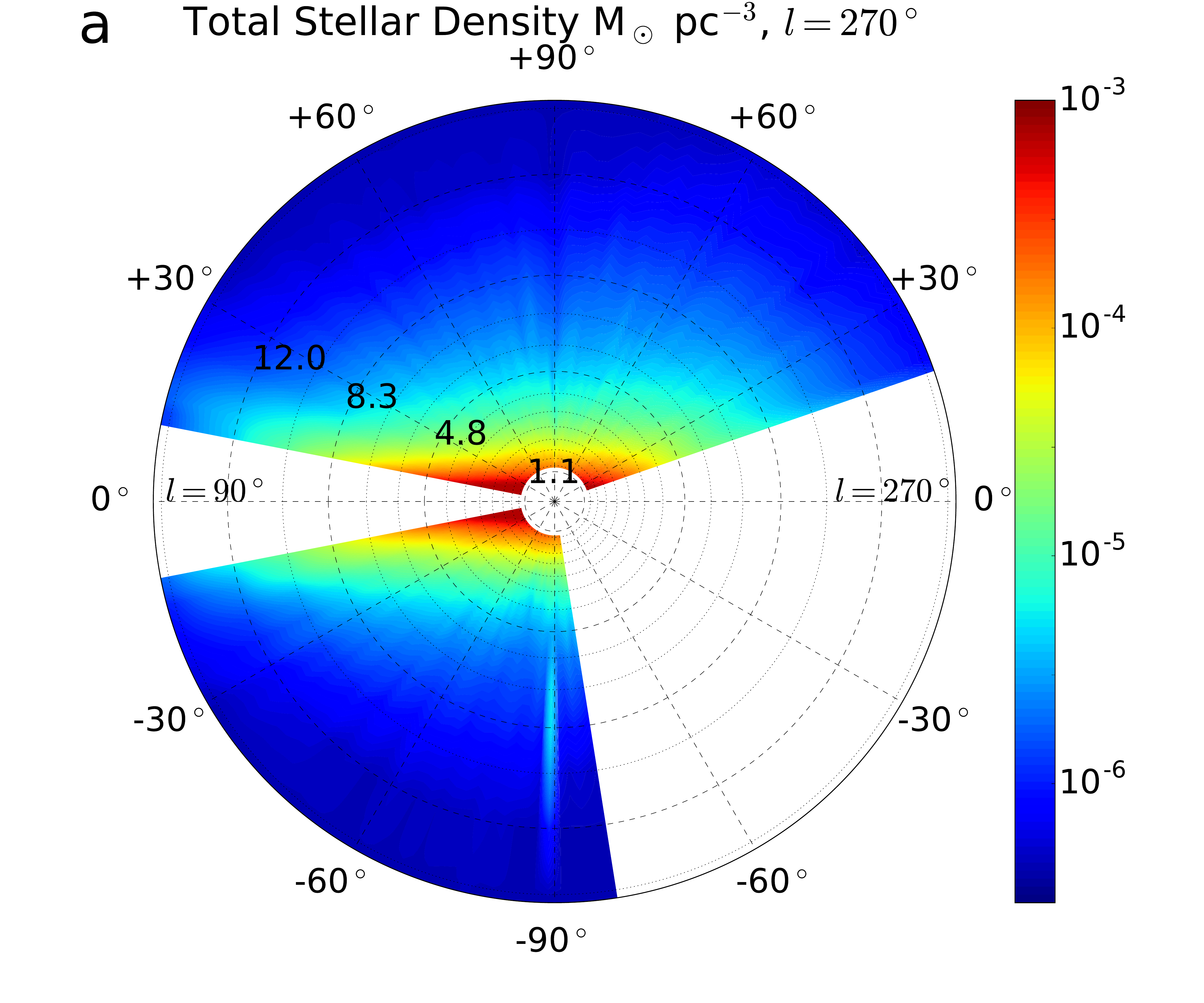}
\includegraphics[width=0.66\columnwidth]{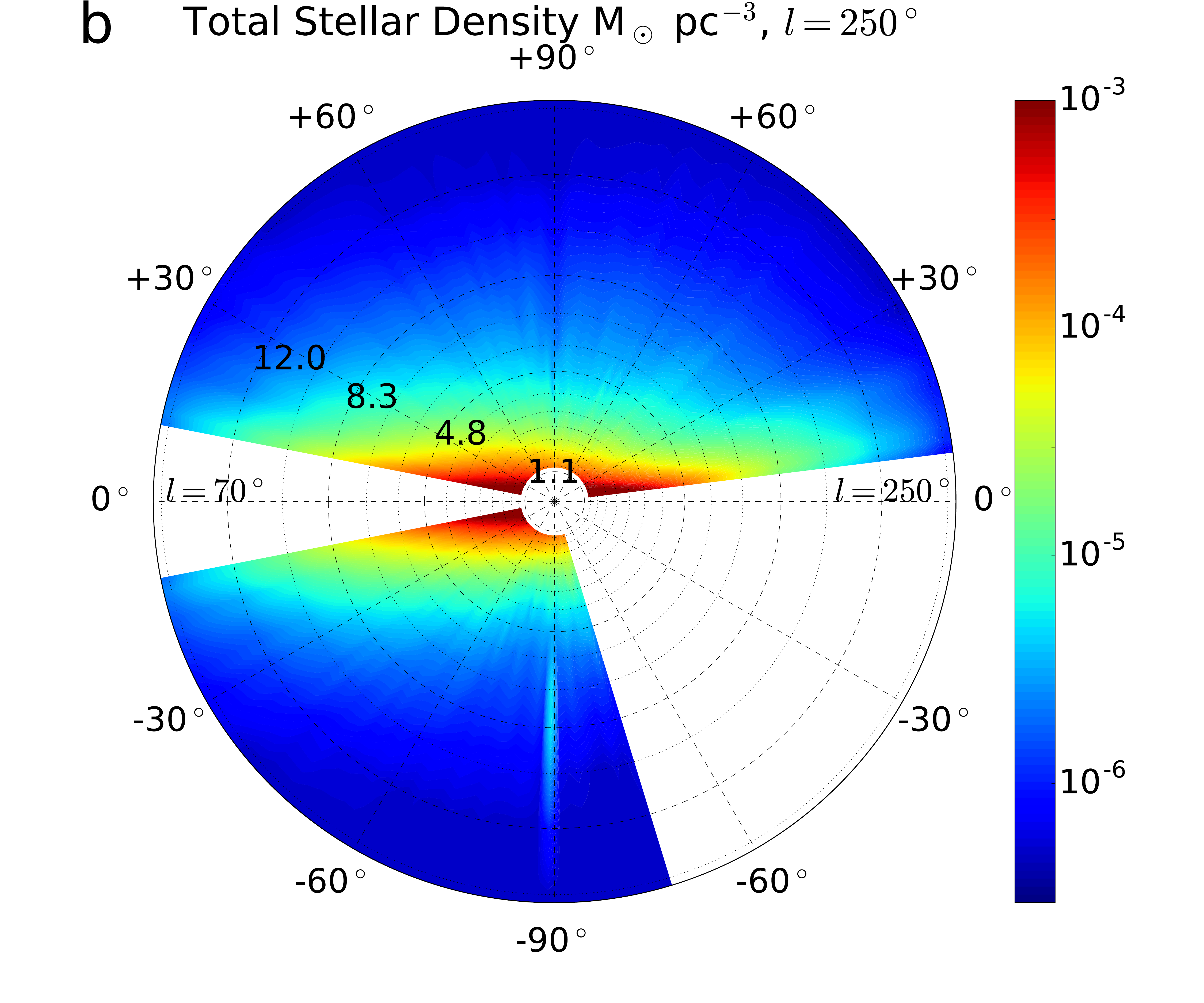}
\includegraphics[width=0.66\columnwidth]{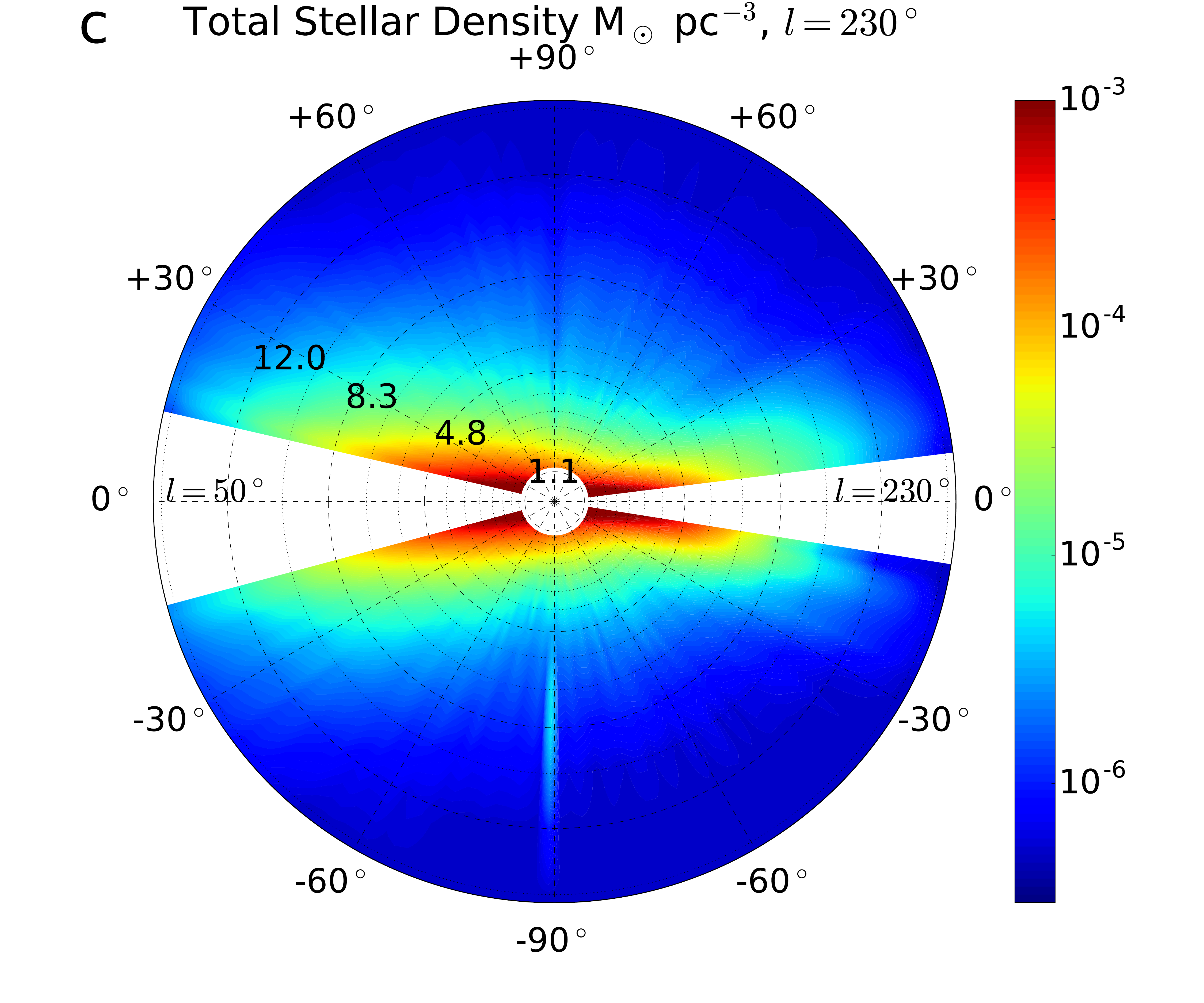}
\includegraphics[width=0.66\columnwidth]{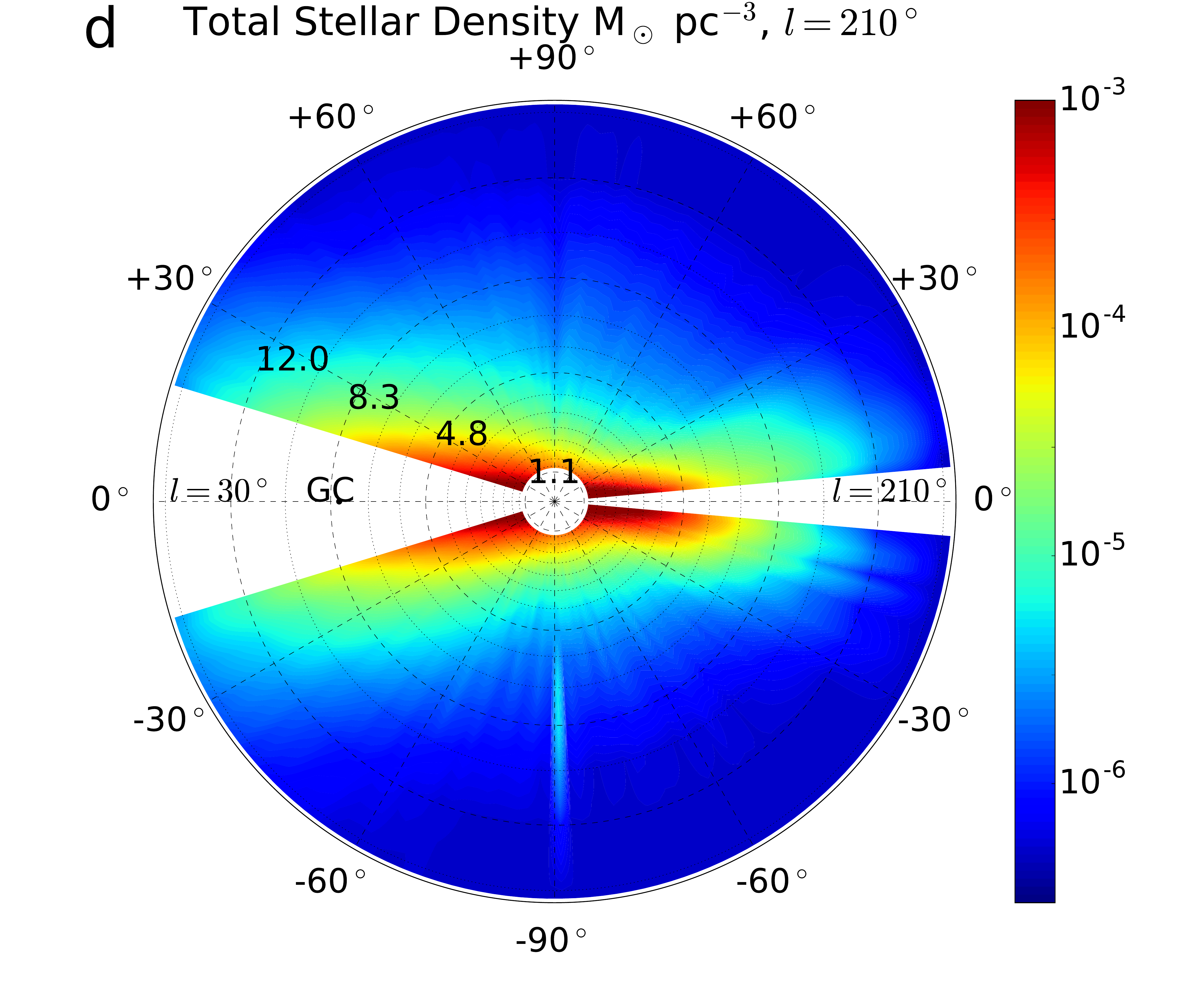}
\includegraphics[width=0.66\columnwidth]{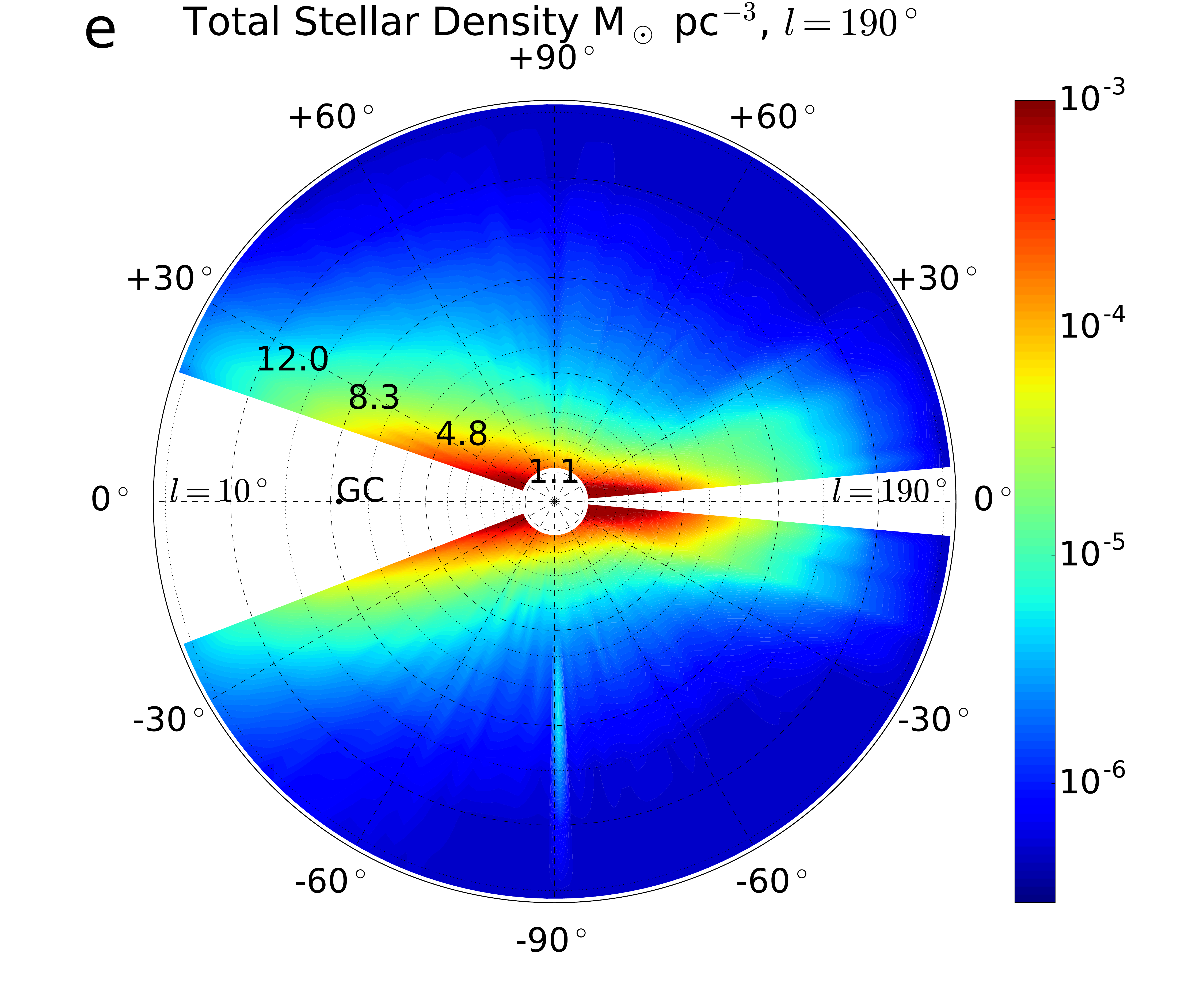}
\includegraphics[width=0.66\columnwidth]{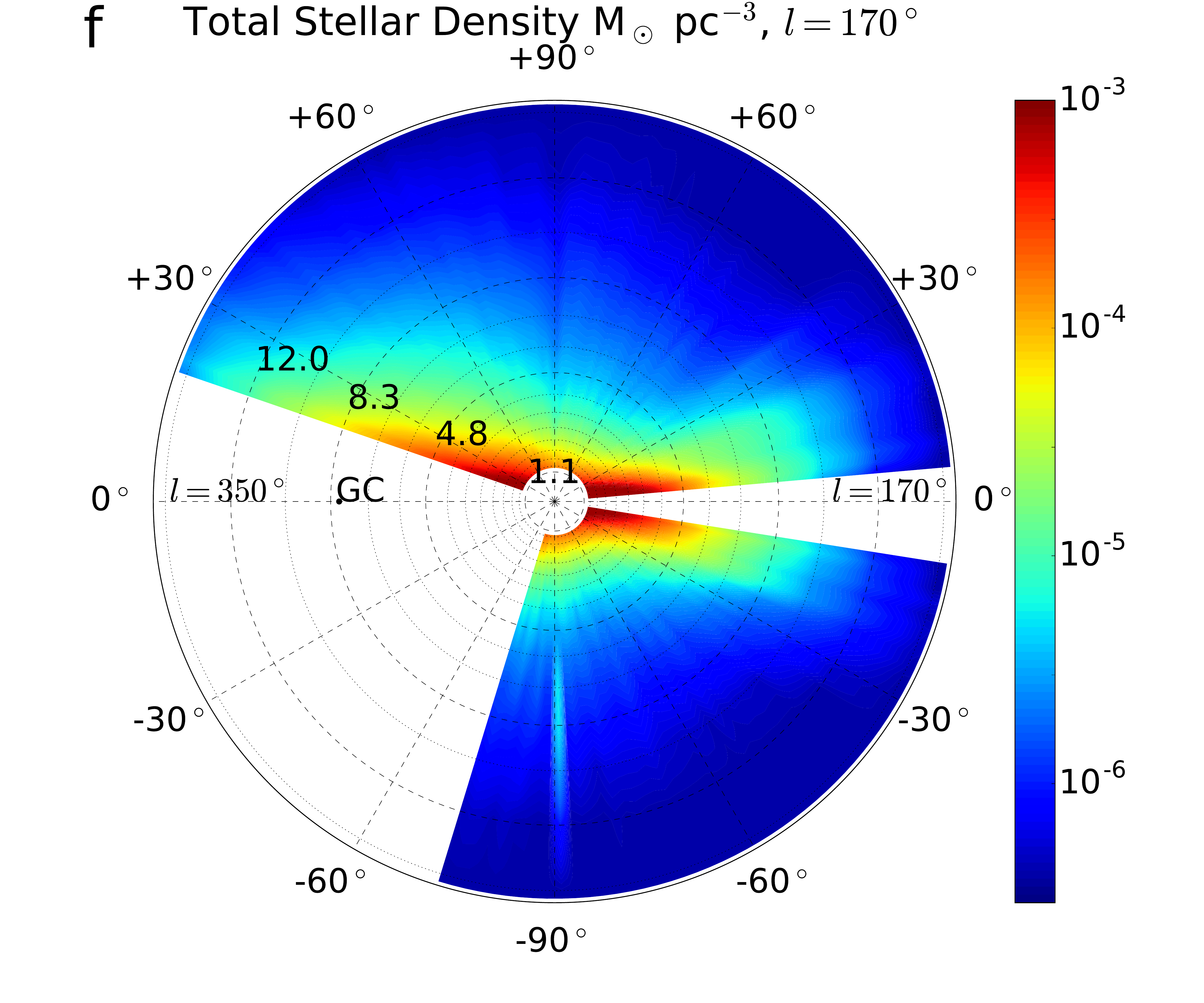}
\includegraphics[width=0.66\columnwidth]{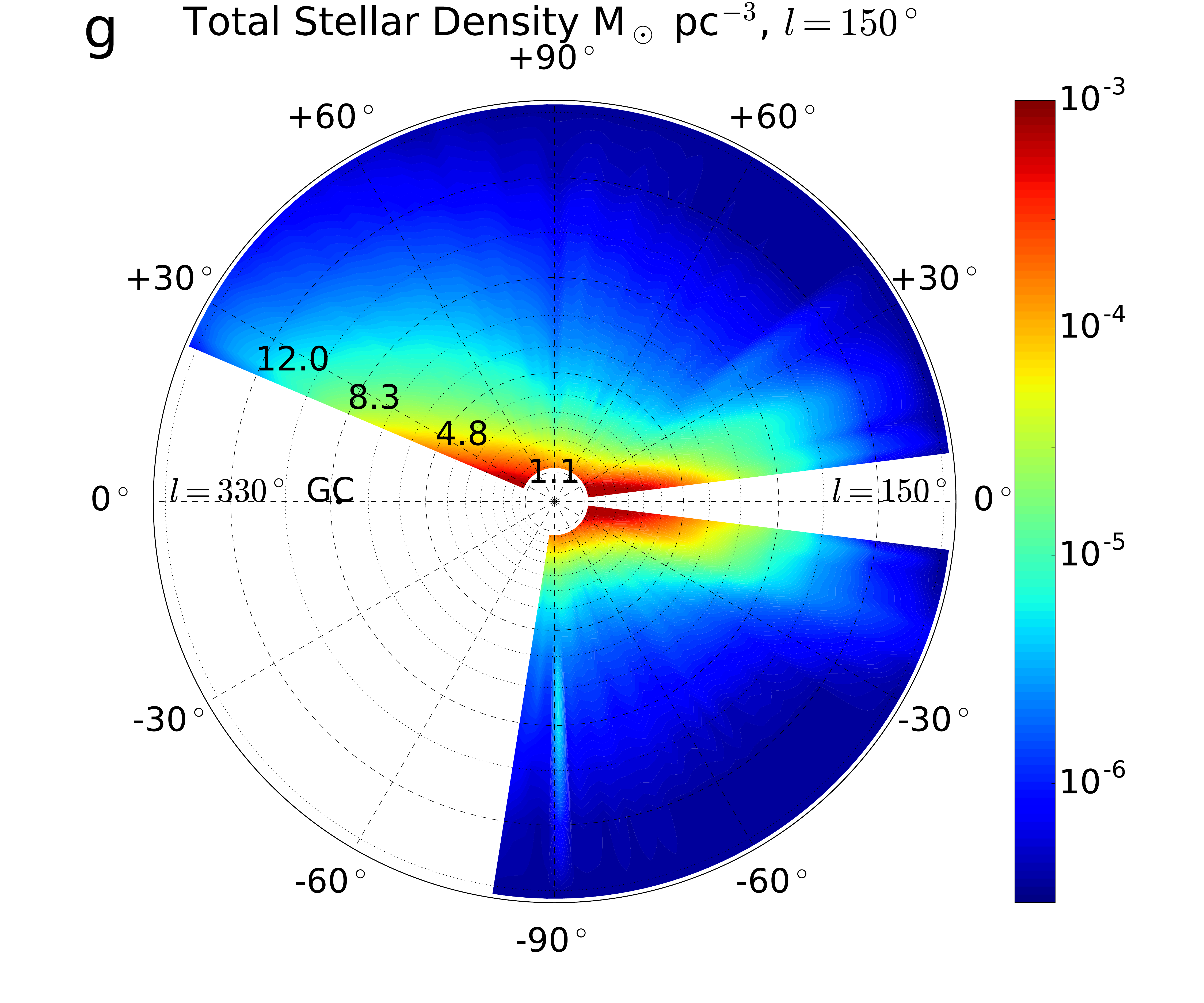}
\includegraphics[width=0.66\columnwidth]{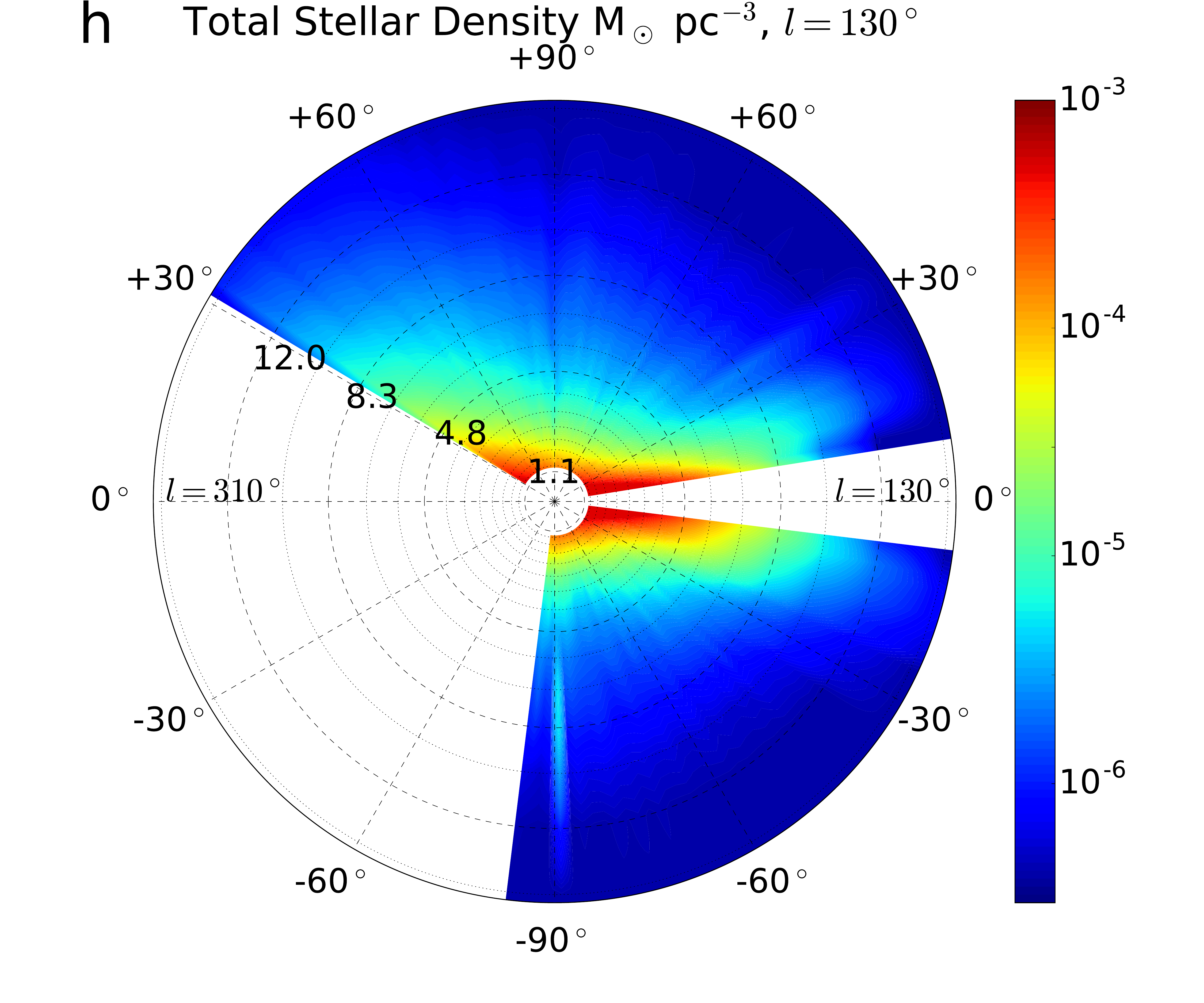}
\includegraphics[width=0.66\columnwidth]{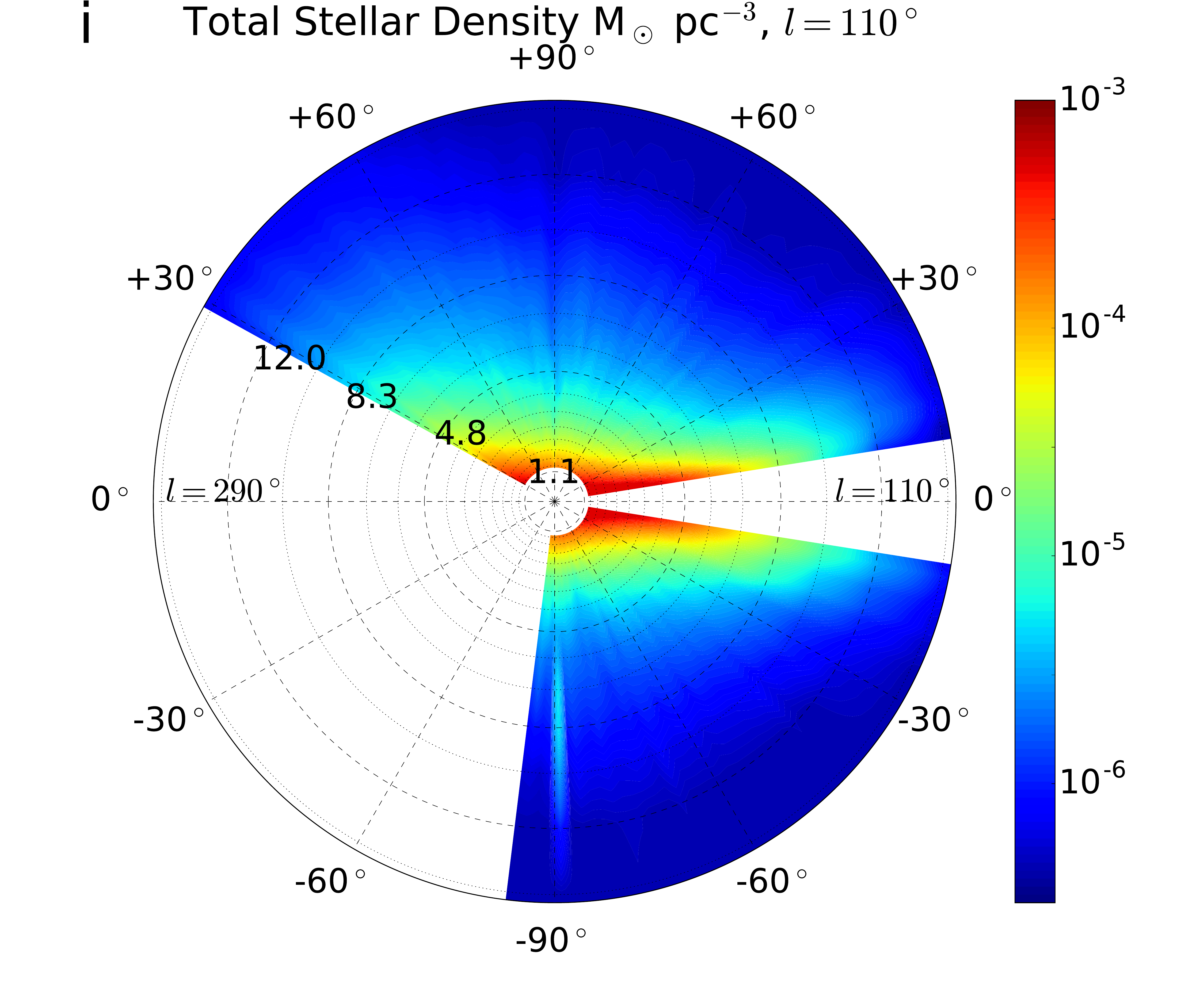}
\caption{\rm{Meridional cross-sections of the total stellar mass density in the Milky way The radial coordinate is Heliocentric radius in kpc. In row 1, we scan (left to right) from $l_0 = 270^\circ$ to $230^\circ$. Given that our bins are $20^\circ$ wide, this actually covers the data from $280^\circ$ to $l = 220^\circ$. In row 2, we scan (left to right) from $l_0 = 210^\circ$ to $170^\circ$. This covers the area from $l = 220^\circ$ to $160^\circ$. In row 3, we scan (left to right) from $l_0 = 150^\circ$ to $110^\circ$. This covers the area from $l = 160^\circ$ to $100^\circ$. Since this projection shows both halves of the galaxy, this $180^\circ$ span covers the whole sky. A projection of the Galactic Center is marked ``GC'' on the left and the MR is on the right in each figure.}}
\label{fig:sliceplot}\end{figure*}

Fig. \ref{fig:sliceplot} shows the meridional cross-sections for our total stellar mass density. When interpreting Fig. \ref{fig:sliceplot}, it is worth noting that the point spread Function in Fig. \ref{fig:sliceplot} has significant extent along the line of sight. For reference, NGC 288 \citep{HARR96}, conveniently located at the South Galactic Pole ($l = 151.3^\circ,\ b = -89.4^\circ,\ d = 8.8$ kpc), appears as a vertical spike at $b = -90^\circ$ in every image.  In each panel of Fig. \ref{fig:sliceplot}, the Galactic Center is on the left and the Anticenter and observed Monoceros Ring is on the right. This distinction is most pronounced in Fig. \ref{fig:sliceplot}c-h. As we might expect, the total density produces a smooth distribution with a denser and more extensive population on the Galactic Center side. On the right, we see a sharp, vertical cutoff. The height of this cutoff is probably due in part to low, relatively local ($b = +20^\circ,\ d = 6$ kpc) structure being projected out to unphysical heights by the line of sight convolution.

\begin{figure*}[ht]
\includegraphics[width=0.66\columnwidth]{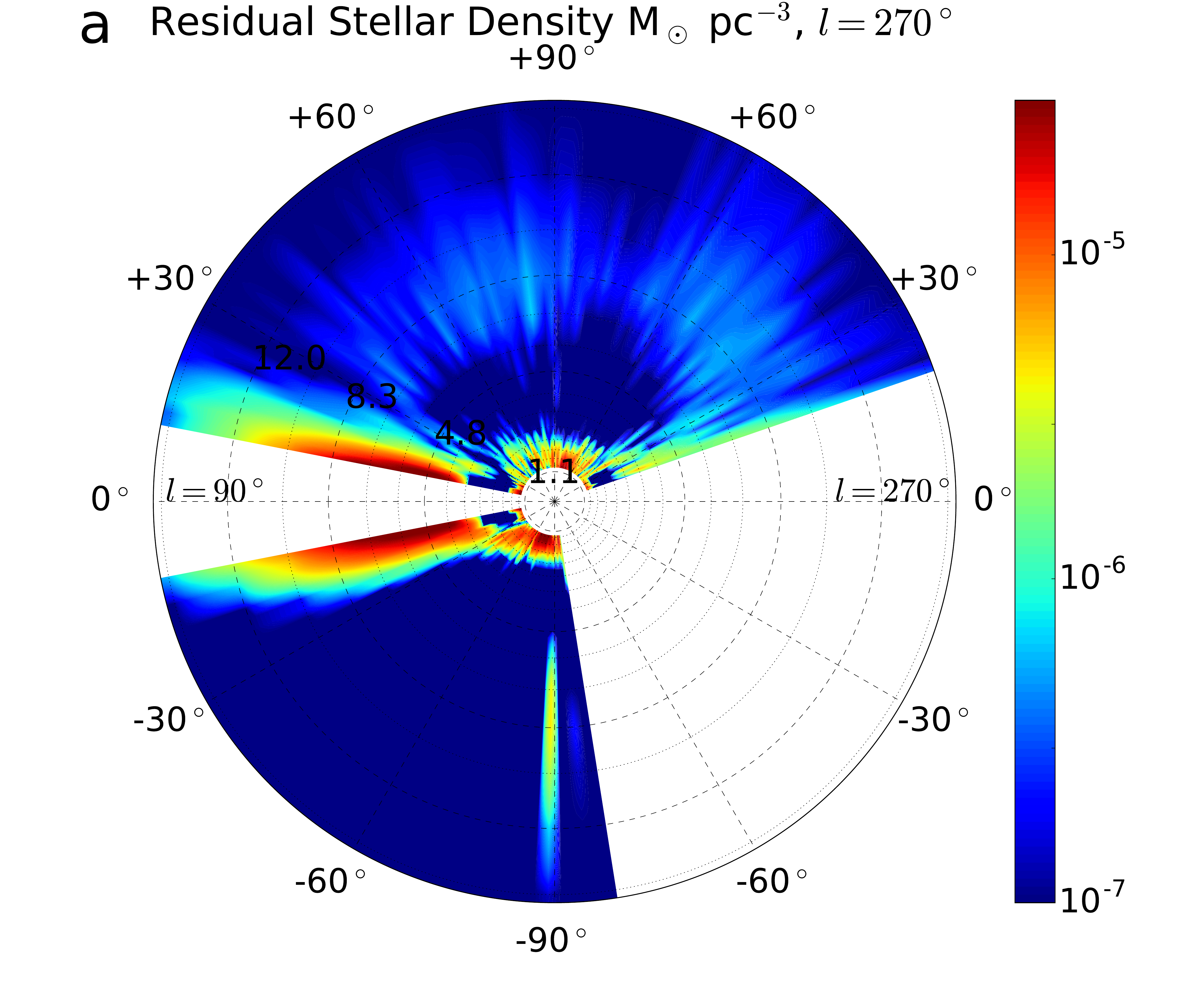}
\includegraphics[width=0.66\columnwidth]{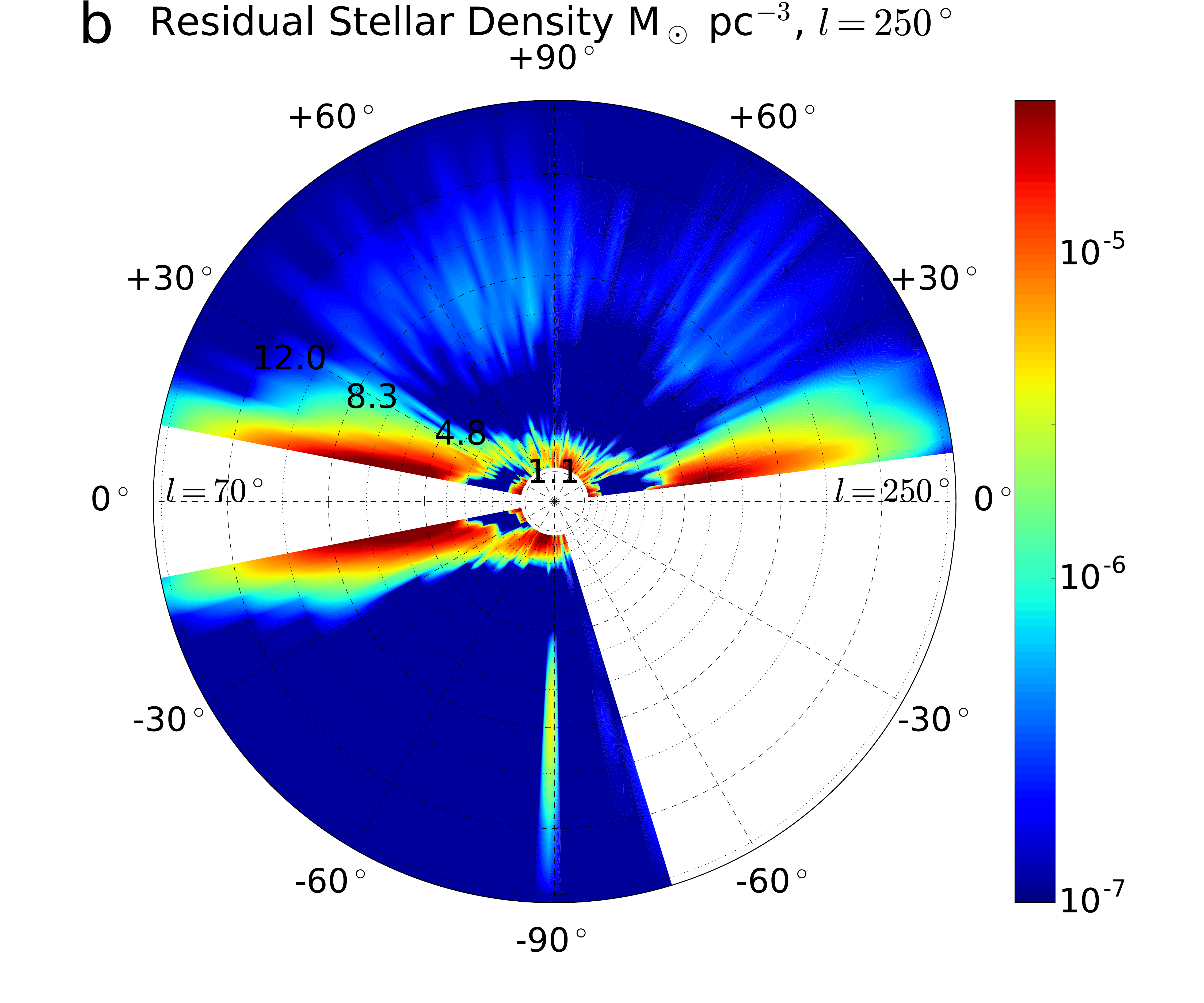}
\includegraphics[width=0.66\columnwidth]{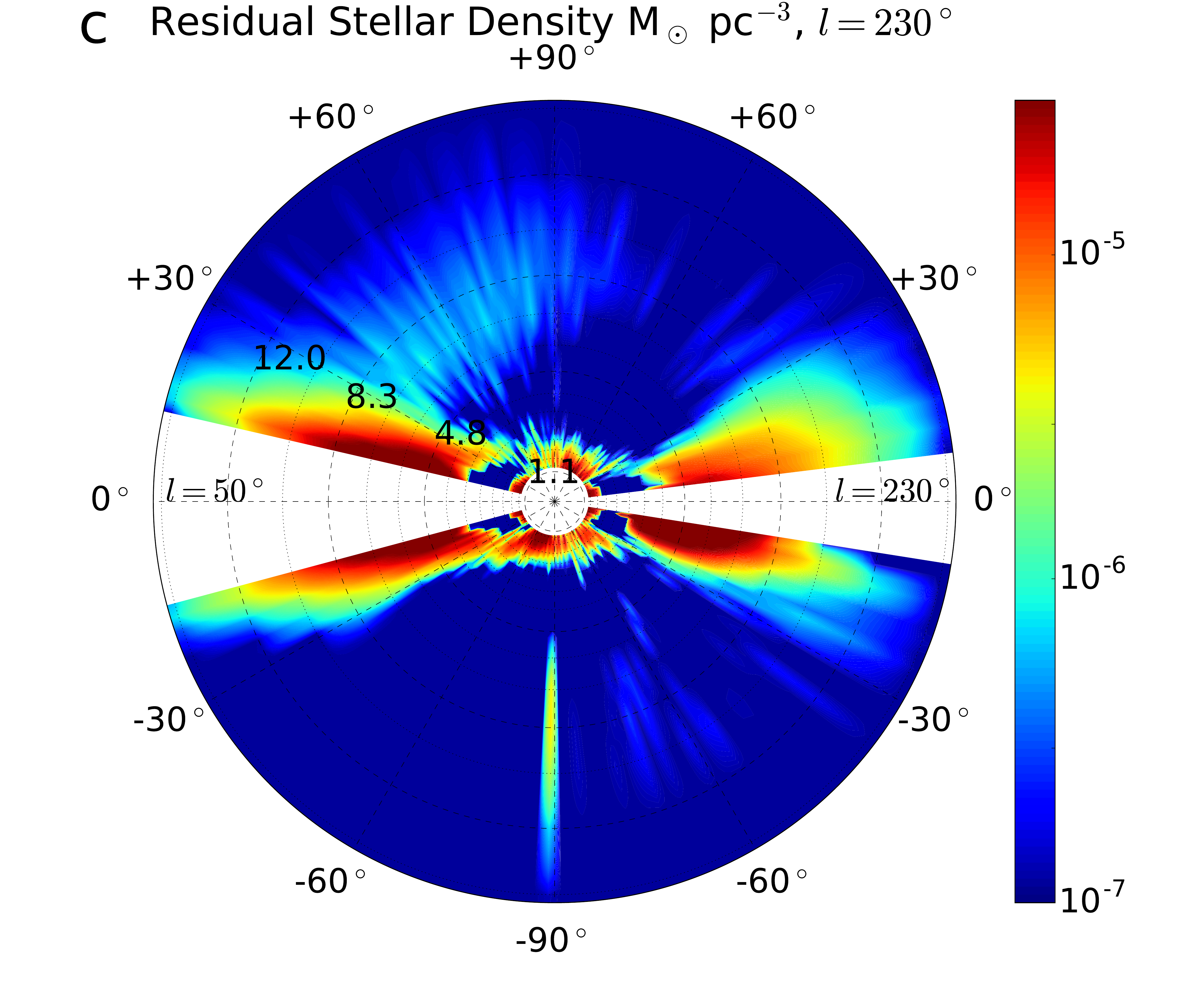}
\includegraphics[width=0.66\columnwidth]{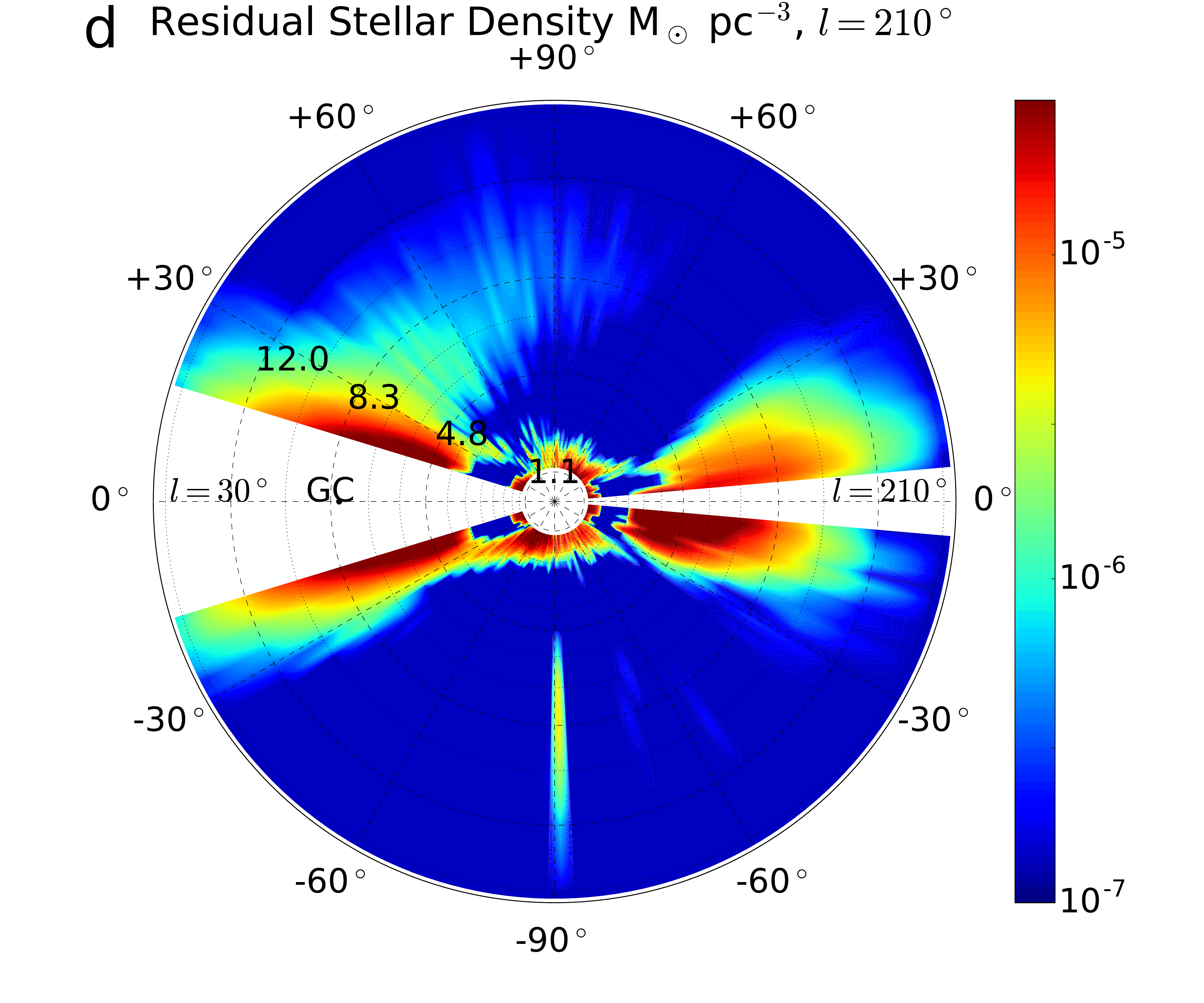}
\includegraphics[width=0.66\columnwidth]{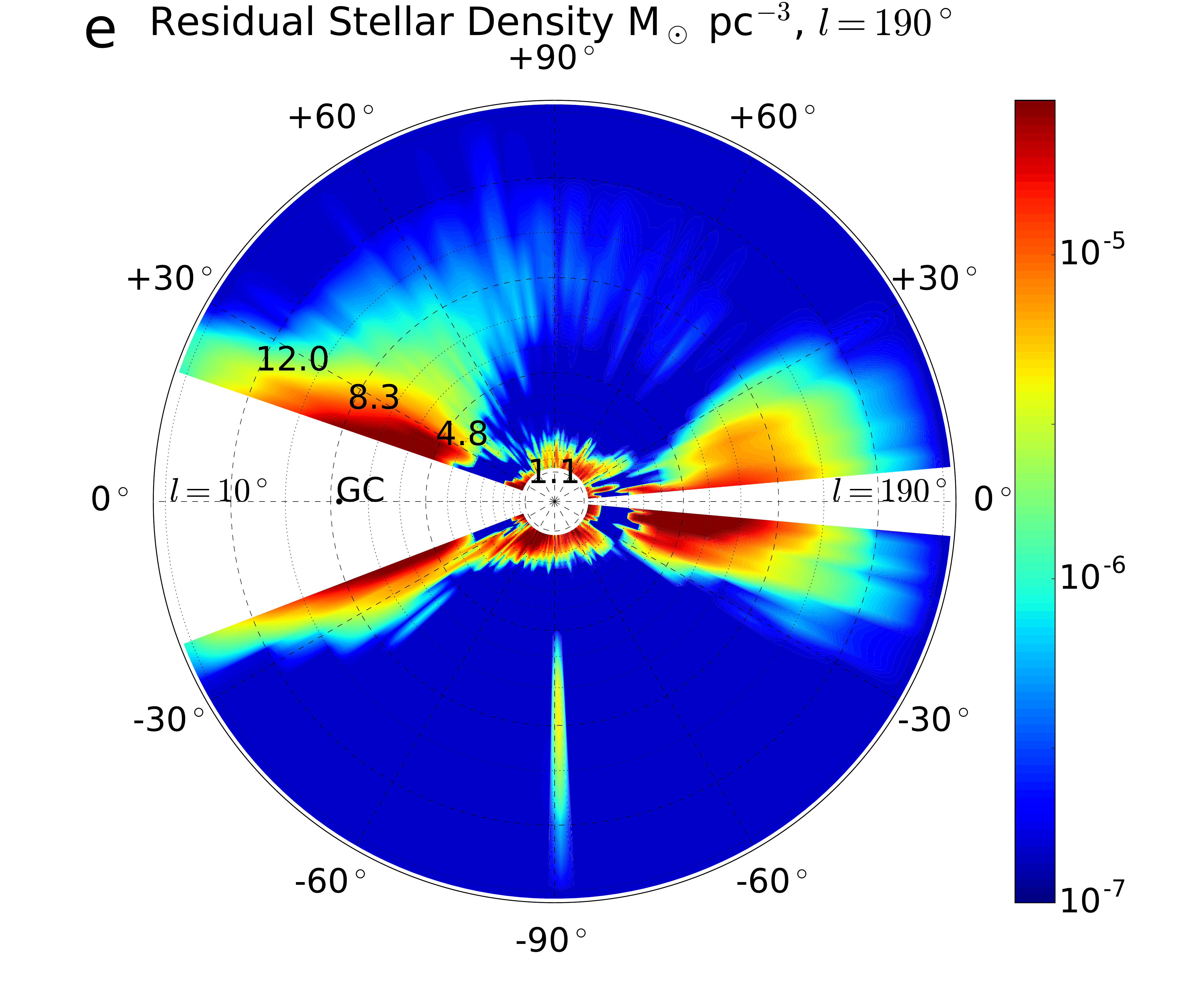}
\includegraphics[width=0.66\columnwidth]{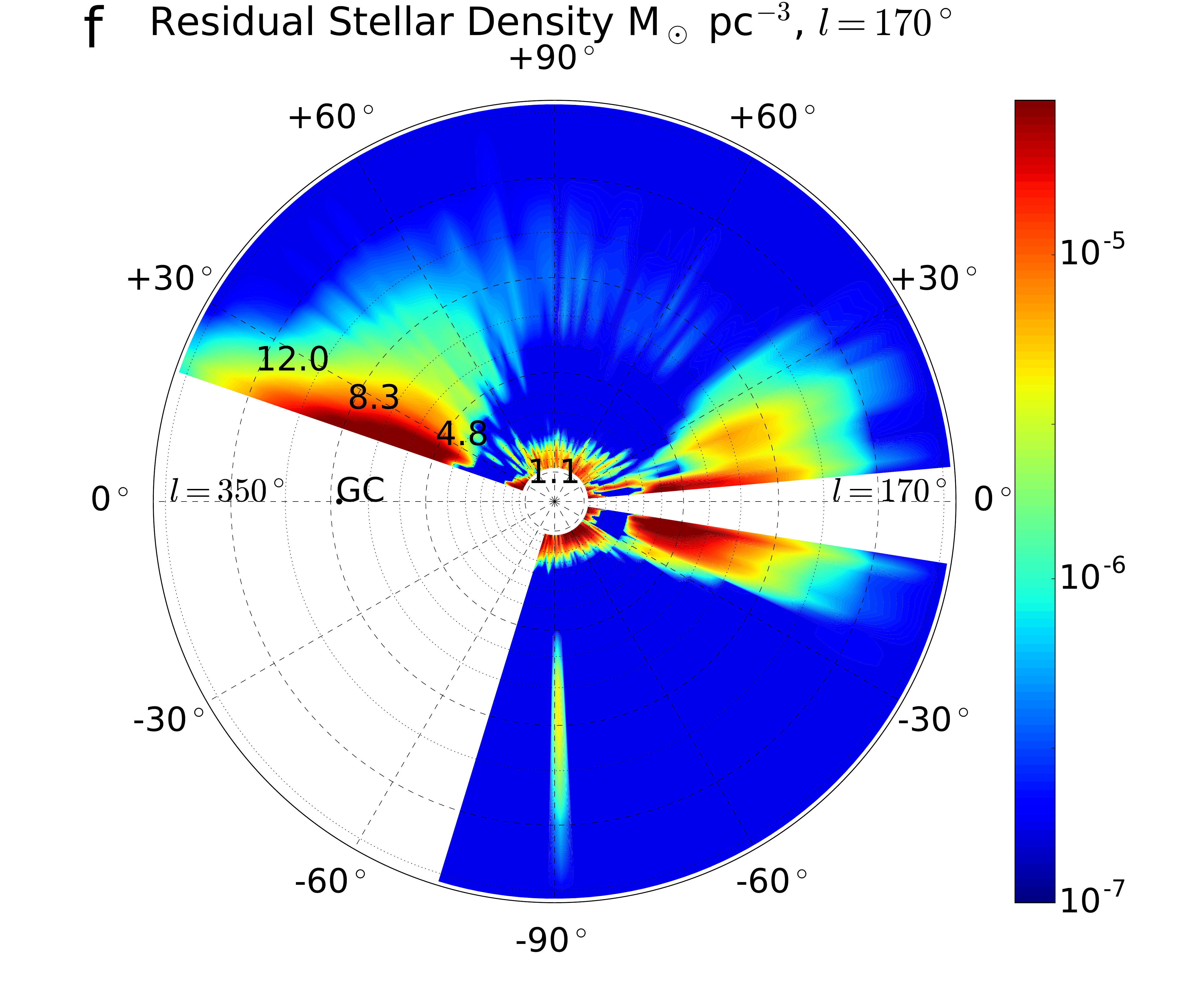}
\includegraphics[width=0.66\columnwidth]{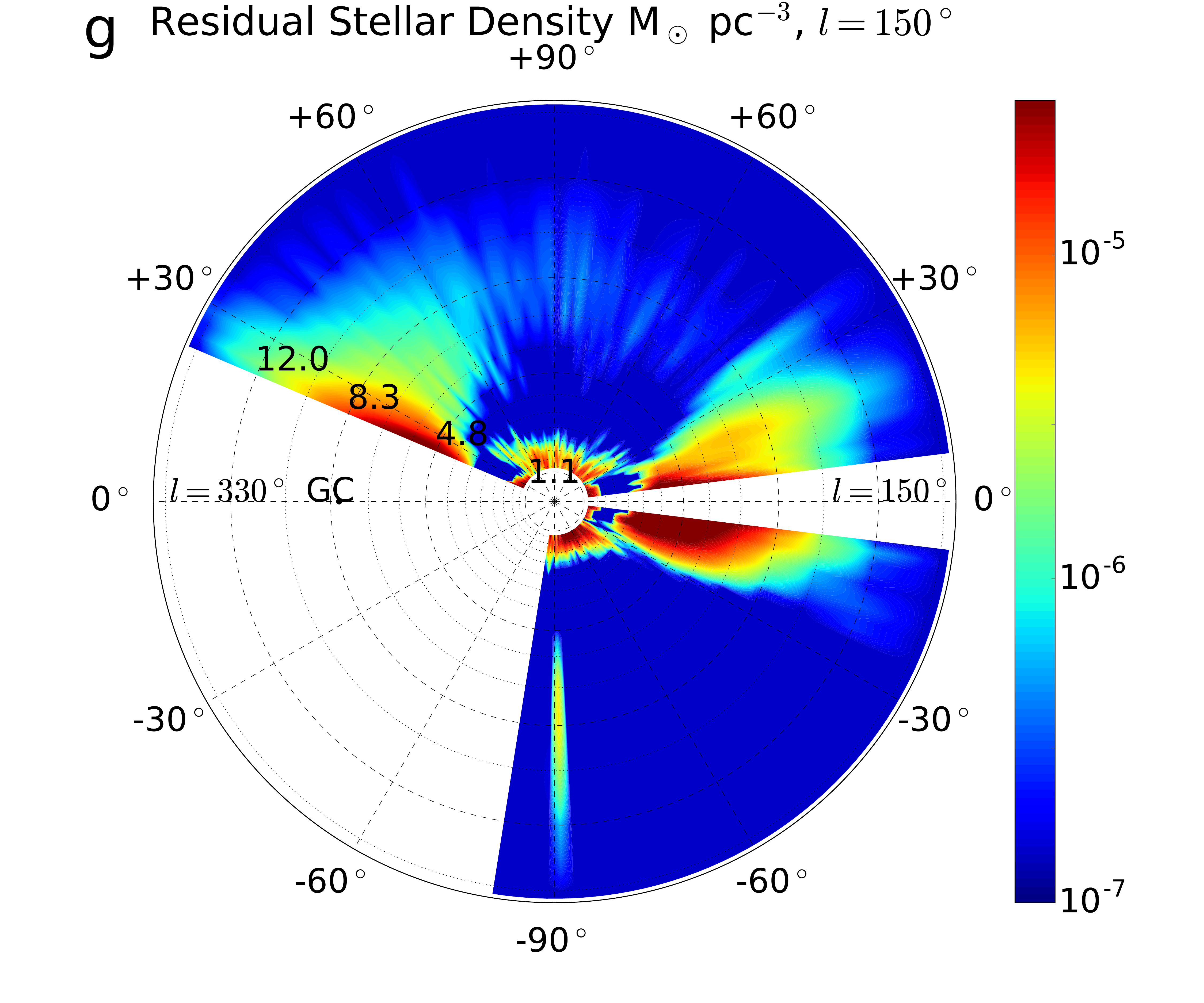}
\includegraphics[width=0.66\columnwidth]{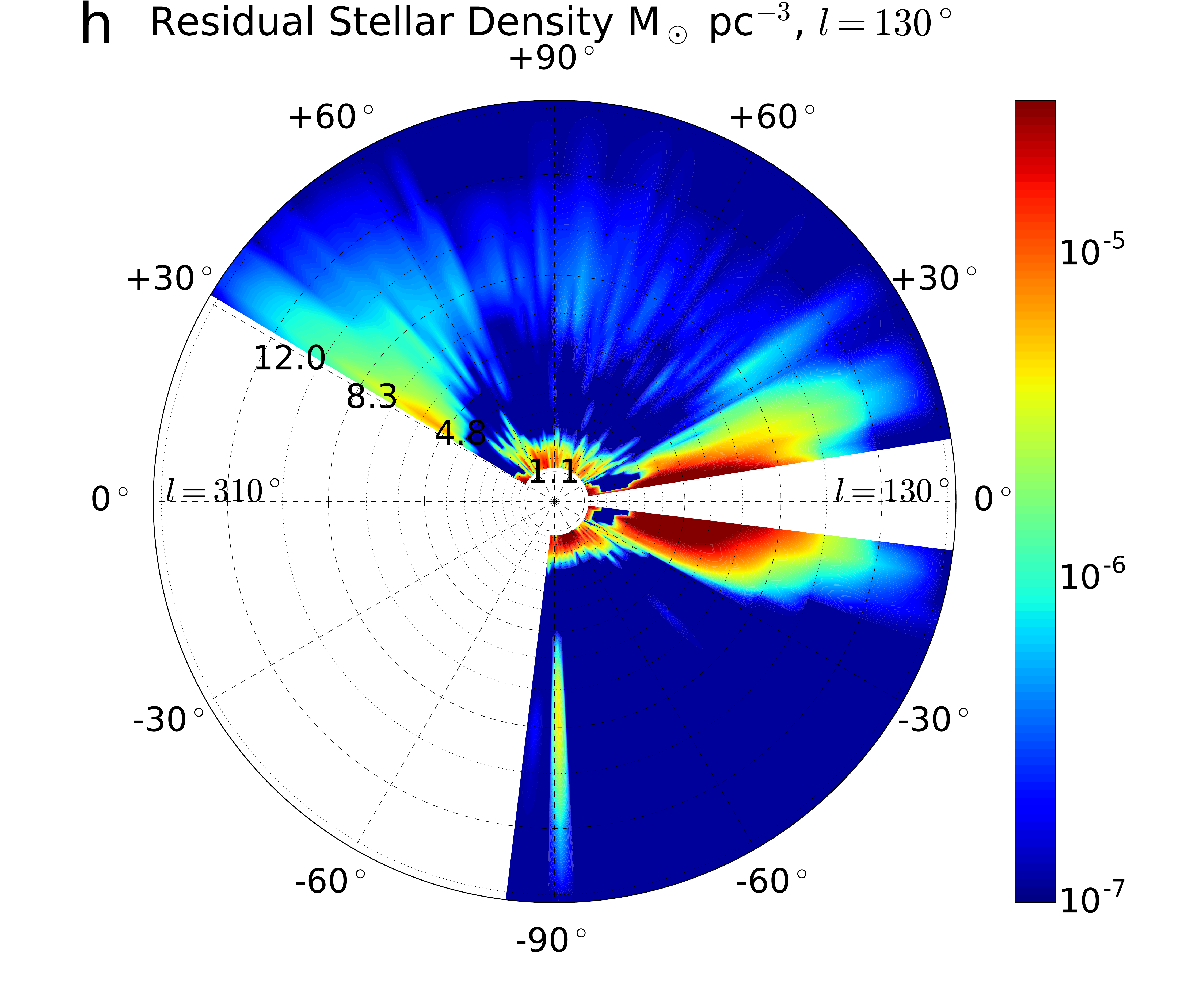}
\includegraphics[width=0.66\columnwidth]{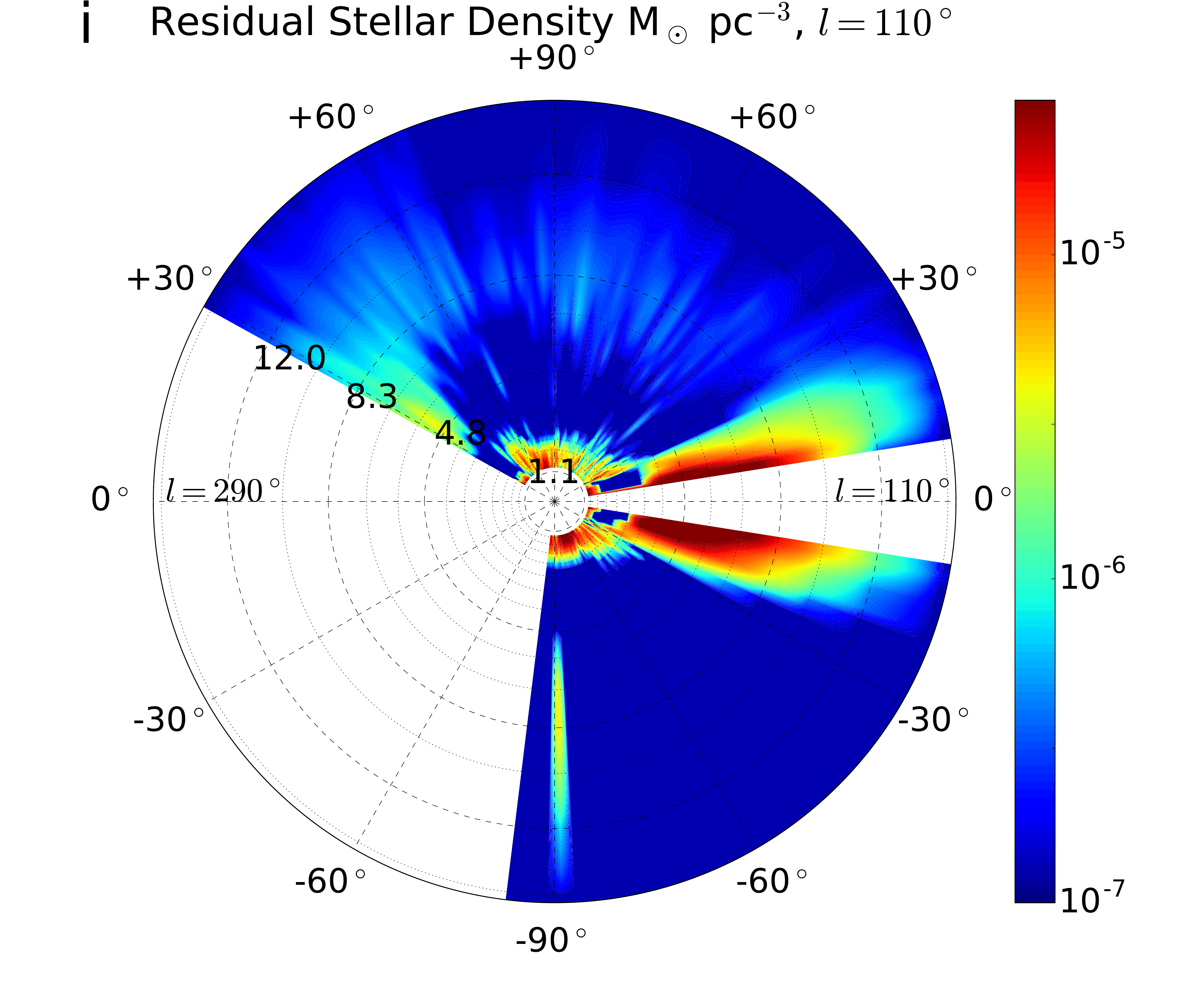}
\caption{\rm{Meridional cross-sections of the residual stellar mass density after removing a power law Milky Way fit. The radial coordinate is Heliocentric radius in kpc. In row 1, we scan (left to right) from $l_0 = 270^\circ$ to $230^\circ$. Given that our bins are $20^\circ$ wide, this actually covers the data from $280^\circ$ to $l = 220^\circ$. In row 2, we scan (left to right) from $l_0 = 210^\circ$ to $170^\circ$. This covers the area from $l = 220^\circ$ to $160^\circ$. In row 3, we scan (left to right) from $l_0 = 150^\circ$ to $110^\circ$. This covers the area from $l = 160^\circ$ to $100^\circ$. Since this projection shows both halves of the galaxy, this $180^\circ$ span covers the whole sky.  A projection of the Galactic Center is marked ``GC'' on the left and the MR is on the right in each figure.}}
\label{fig:sliceplot_resid}\end{figure*}

Fig. \ref{fig:sliceplot_resid} shows our residual density after subtracting our power law Milky Way fit from Eq. \ref{eq:linefit}. This residual fit is essentially the sum of our MR fit and any actual residuals from our MW+MR fit. It is dominated by the MR fit, but its shape is not artificially constrained to be a Gaussian along the line of sight. Examining the MR in different figures, moving from left to right, corresponds to scanning left to right in Figs. \ref{fig:massmap} and \ref{fig:distancemap}. 

The power of our MW subtraction can be seen by comparing the North and South. We immediately see again that the Southern overdensity, while more compact and closer to the plane, contains significantly more mass than the Northern over overdensity. Again, the Southern MR peaks at around 6 kpc. These images also show an extension of the Southern structure, also at roughly 6 kpc, that appears on the Northern side of the Galactic plane and masked region up to $b = +10^\circ$. The Northern MR appears to be an essentially separate stream, starting as a Northern lobe at ($b = +15^\circ,\ d = 8$ kpc) in Fig. \ref{fig:sliceplot_resid}c and becoming totally distinct from the main Southern structure in Figs. \ref{fig:sliceplot_resid}f, g and h. The ACS is also clearly visible in Figs. \ref{fig:sliceplot_resid}f, g and h as a cyan spike at $b =+35^\circ$. These images give the strong impression that the Monoceros Ring is composed of a large, mostly Southern structure that is contiguous with the Galactic plane and several smaller, more distant Northern streams. The Sagittarius Stream \citep{YANN++00,SLAT++13} and the Virgo Overdensity \citep{JURI++08} can also be seen arcing across the northern hemisphere in Figs. \ref{fig:sliceplot_resid}.

\subsection{Planar (Top Down) Cross-Sections}\label{sect:cross}

We have produced a Heliocentric projection of the Monoceros Ring that is fairly close to vertical slice through the $l = 180^\circ,\ d = 8.3$ kpc line for the crucial Galactic Anticenter region where the Ring is most visible in Sections \ref{sect:map} and \ref{sect:map3d}. We have also produced genuine meridional (vertical) cross sections through lines of constant $l$ in Section \ref{sect:vert}. It is instructive to produce planar cross sections of the Milky Way and Monoceros Ring and get a ``bird's eye view'' of the Galaxy. 

It is difficult to make planar cross-sections from Heliocentric (non-planar) lines of sight. We thus abandon our line of sight fitting and \textsc{match} and use a simpler model to estimate the Heliocentric distance to each star. We use the single isochrone (from \citet{BRES++12}) and the accompanying website to fit distance modulus, $\mu$, as:
\begin{eqnarray}
\mu &=& \left(g_{\rm{P1,\ 0}}-4.34\right) + \left(\left(g_{\rm{P1,\ 0}} -r_{\rm{P1,\ 0}} \right) -0.278\right)^{0.603},\\
\rm{for}&\ &0.278 < g_{\rm{P1,\ 0}} -r_{\rm{P1,\ 0}} < 0.5,\label{eq:isofit}\nonumber\\
\mu &=& \left(g_{\rm{P1,\ 0}}-4.34\right),\nonumber\\ 
\rm{for}&\ &g_{\rm{P1,\ 0}} -r_{\rm{P1,\ 0}} < 0.278.\nonumber
\end{eqnarray}
Using a single isochrone means that we are not modeling different metallicity or age populations or correcting for dust beyond the SFD extinction correction. For stars with metallicity -1.4 with typical age of 13.3 billion years, we are only including stars with $0.278 < g_{\rm{P1,\ 0}} -r_{\rm{P1,\ 0}} < 0.5$ and $4.34 < g_{\rm{P1,\ 0}} <  6.5$ (essentially F and G stars). This is the region where our isochrone is monotonic so that every $g_{\rm{P1,\ 0}} -r_{\rm{P1,\ 0}}$ maps to a specific $g_{\rm{P1,\ 0}}$ and corresponding $\mu$. We assign stars bluer than $g_{\rm{P1,\ 0}} -r_{\rm{P1,\ 0}} = 0.278$ a distance modulus of $g_{\rm{P1,\ 0}}-4.34$. All stars with absolute magnitudes brighter than $4.33628$ are misidentified as their fainter equivalent with identical $0.278 < g_{\rm{P1,\ 0}} -r_{\rm{P1,\ 0}}$. In practice, these are blue OBA and red giant branch stars and should be rare to non-existent in the old stellar populations at the large distances ($> 20$ kpc) at which they would be confused for more local dwarfs in the MR. But given all these limitations, we restrict ourselves to large scale qualitative analysis. 

Unlike the rest of the MR imaging and analysis in this paper, this simple isochrone fitting does not use \textsc{match} or our line of sight fitting. It is thus a mostly independent analysis of the PS1 data that shows the same basic qualitative results as the rest of our analysis (albeit from a different angle). However, this less sophisticated analysis has several disadvantages. Uncertainties in color cause distance measurements to an individual star to be very imprecise, so structures are projected along the line of sight. This means that at higher (lower) Galactic heights, structures will appear farther (closer) to the Sun. In addition, since there are more faint red stars than brighter blue stars, stellar populations are asymmetrically scattered to appear farther (see \citet{NEWB++11}). Because of these limitations, we use these maps qualitatively. 

\begin{figure*}[ht]
\plottwo{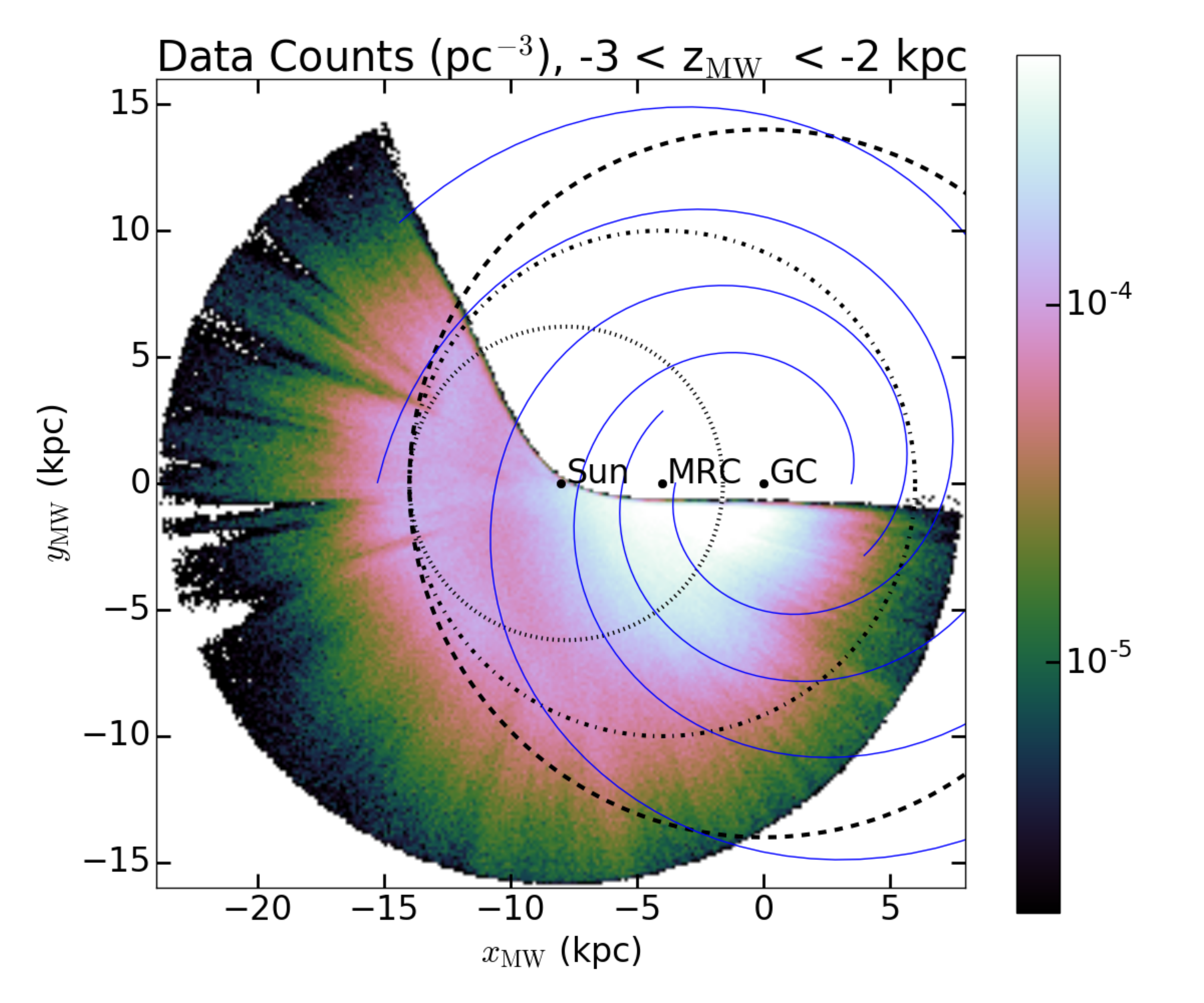}{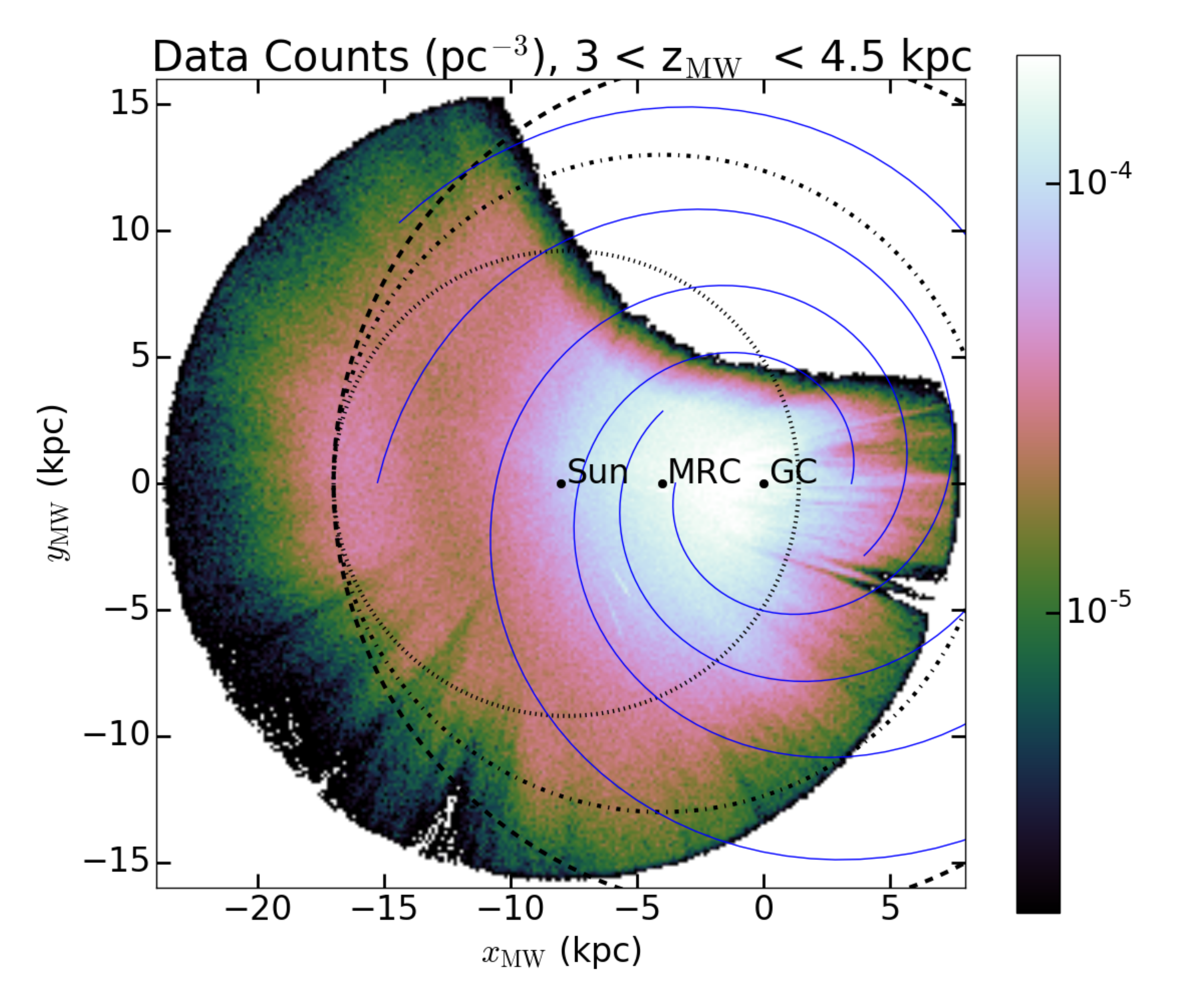}
\plottwo{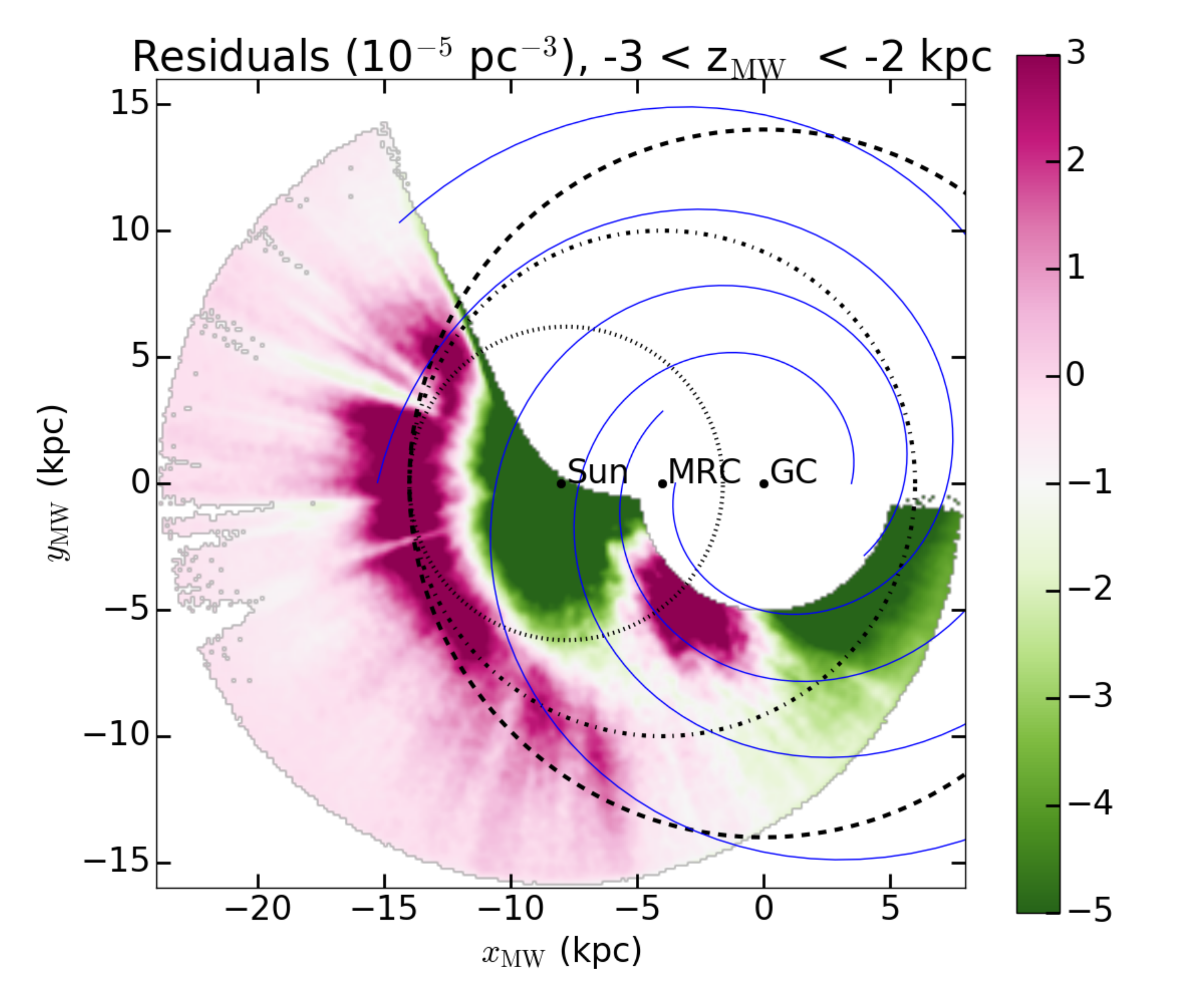}{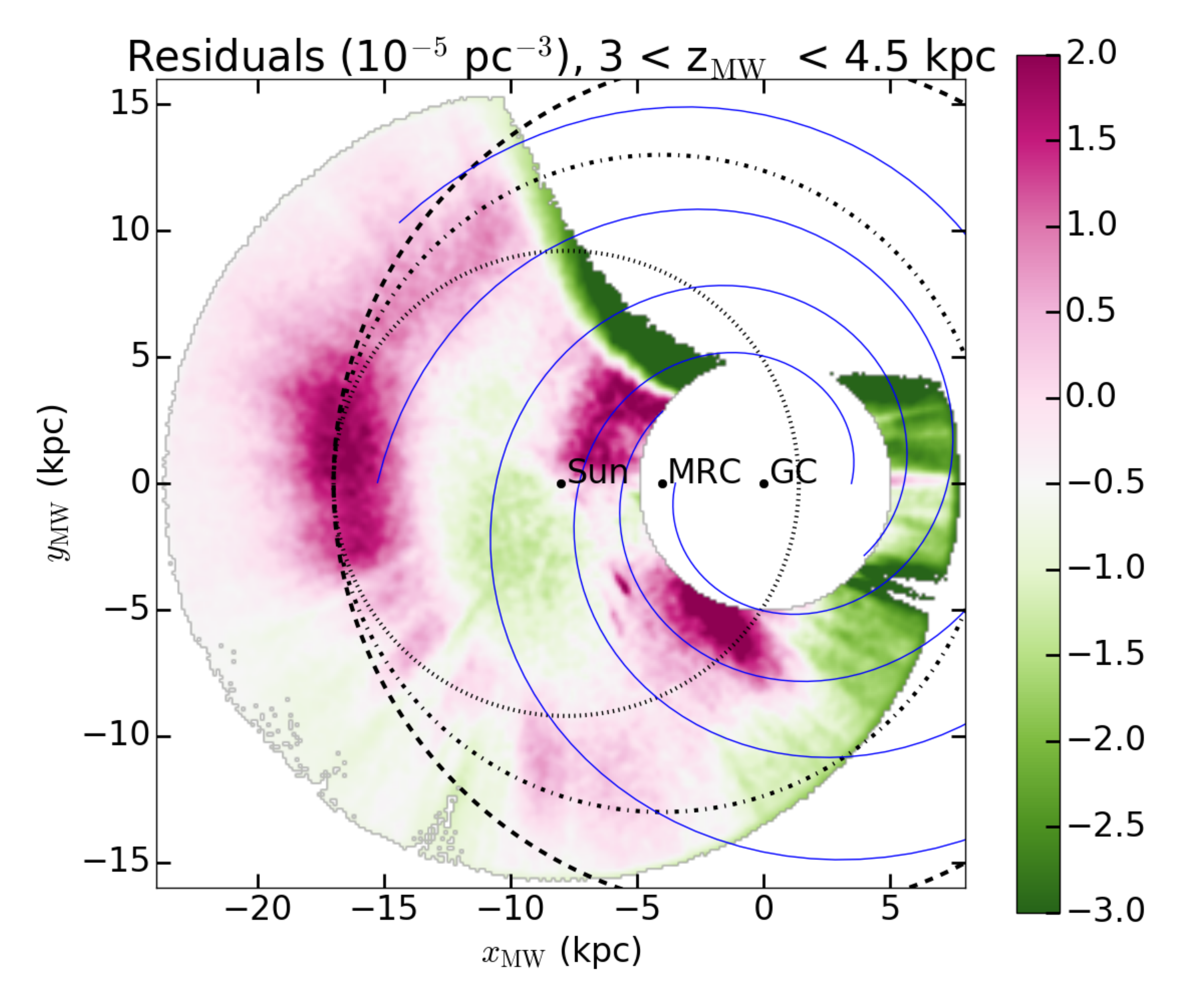}
\caption{\rm{In the top row, we have a map of all $0.1 < g_{\rm{P1}} -r_{\rm{P1}} < 0.5$ stars at Galactic height $-3.0 < z_{\rm{MW}} < -2$ kpc (left) with a Galactocentric cylinder with radius 14 kpc (dashed line), an alternate cylinder 4 kpc from the Galactic center with radius 10 kpc (dot-dashed line) and a Heliocentric circle with radius 6 kpc (dotted line) on the left. We also have a map of all $0.1 < g_{\rm{P1}} -r_{\rm{P1}} < 0.5$ stars at Galactic height $3 < z_{\rm{MW}} < 4.5$ kpc (right) with a Galactocentric cylinder with radius 17 kpc (dashed line), an alternate cylinder 4 kpc from the Galactic center with radius 13 kpc (dot-dashed line) and a Heliocentric circle with radius 9 kpc (dotted line) on the right. We also show model MW arms from \citet{FAUC++06} in blue. In the bottom row, we subtract out a S\'ersic profile to isolate the MR, which appears as a pink circle at the assigned radii (which are the same as in the top row). We do not remove the MR before fitting the S\'ersic profile, so the residuals also include a relatively underdensity just inside the MR. The deviations from our S\'ersic profile appear large (on our linear scale) near the Galactic Center, and we mask this region out as it is not significant to our analysis.} 
}
\label{fig:crosssection}\end{figure*}

The top row of Fig. \ref{fig:crosssection} shows two planar cross-sections of the Galaxy made using this analytical isochrone fit. The left cross-section shows the number density of stars at Galactic height -3 kpc $< z <$ -2 kpc (equivalent to $-30^\circ < b < -21^\circ$ in the Southern MR region), and the right shows the number density of stars at Galactic height 3 kpc $< z <$ 4.5 kpc (equivalent to $+21^\circ < b < +30^\circ$ in the Northern MR region). These height ranges correspond to the detected overdensity regions from Section \ref{sect:quant}. We show a series of cross-sectional slices in the Appendix \ref{sect:addmap}. We assume the Sun is at -8 kpc on the $x$ axis. We see the Galactic center dominates the area around the origin, but that there is a faint arc that runs through the Anticenter at (-14 kpc, 0 kpc) in the South and (-17 kpc, 0 kpc) in the North. 

To improve the contrast of this arc, we subtract off a S\'ersic profile with a Galactic center position $(x_{GC}, y_{GC})$:
\begin{eqnarray}
\rho(x,y) &=& \rho_0 e^{-A (d/1 kpc)^{1/n}}\label{eq:2dfit}\\
d &=&\left(\left(x-x_{GC}\right)^2+\left(y-y_{GC}\right)\right)^{1/2}.\nonumber
\end{eqnarray}
Our fit values are in Table \ref{tab:2dfit}, and the residual densities after subtracting this fit are the lower panels of Fig. \ref{fig:crosssection}. The MR is a distinct purple arc in the North and South. It is 6 kpc and 9 kpc away from the Sun in the South and North, respectively. This is consistent with the distances found in Section \ref{sect:quant}. If the MR were a circular Galactocentric ring, it would be a 14 kpc ring in the South and a 17 kpc ring in the North. Neither this ring nor a Heliocentric ring that meets it at the Galactic Anticenter (i.e. has a radius of 6 kpc in the South and 9 kpc in the North) fit the observed ring well. Qualitatively, the MR is better fit by a circle centered 4 kpc away from the Galactic Center (at -4 kpc, 0 kpc) with a radius of 10 (13) kpc in the South (North). The extent of the arc shown here, roughly $120^\circ$ in the South and $170^\circ$ in the North, has never been seen before and is enabled by our data and particular analysis. Despite the extent of the observed ring, we have not proven that the MR is truly circular, and we could have fit it equally well with an ellipse or parabola (indicating a stream). But even qualitatively, a Galactocentric Ring is obviously not the best fit. Intriguingly, the Galactic bar also extends roughly 4 kpc from the Galactic Center \citep{NIDE++12} suggesting that the MR may be related the the Galactic bar like the MW spiral arms which we show in blue \citep{FAUC++06}. Our background-subtracted MR does not appear to align with the MW arms although there appears to be some coincidental alignment between the Southern MR and the (unlabeled) Norma arm of the Milky Way. 

\begin{table*}
\centering
\begin{tabular}{cccccc}
	\hline
Region & $\rho_0$ ($10^{-6}$ Stars pc$^{-3}$) & $A$ & $n$ & $x_{GC}$ (kpc) & $y_{GC}$ (kpc)  \\
	\hline
Southern & $906 \pm 9$ & $0.072 \pm 0.002$ & $1.071 \pm 0.006$ & $-3.78 \pm 0.01$ & $-0.66 \pm 0.02$ \\
Northern & $348 \pm 3$ & $0.209 \pm 0.005$ & $0.730 \pm 0.004$ & $-2.66 \pm 0.01$ & $-0.18 \pm 0.01$ \\
	\hline
\end{tabular}
\caption{\rm{Our 1D fit value from Eq. \ref{eq:2dfit}. We allow the center of this S\'ersic profile to float to $x_{GC}$ and  $y_{GC}$. In practice, we are not modeling the (dusty, crowded) Galactic bulge, and this floating may account for our inevitable incompleteness in this region. Density, $\rho_0$, is in stars pc$^{-3}$.}}\label{tab:2dfit}
\end{table*}

The alternating concentric circles that comprise the MR are consistent with a ripple emanating from a common center as suggested by \citet{IBAT++03} and \citet{XU++15}. While those papers focused on line of sight modeling, Fig. \ref{fig:crosssection} uses the superior PS1 coverage to show these possible ripples as full 2D structures.

\begin{figure*}[ht]
\plottwo{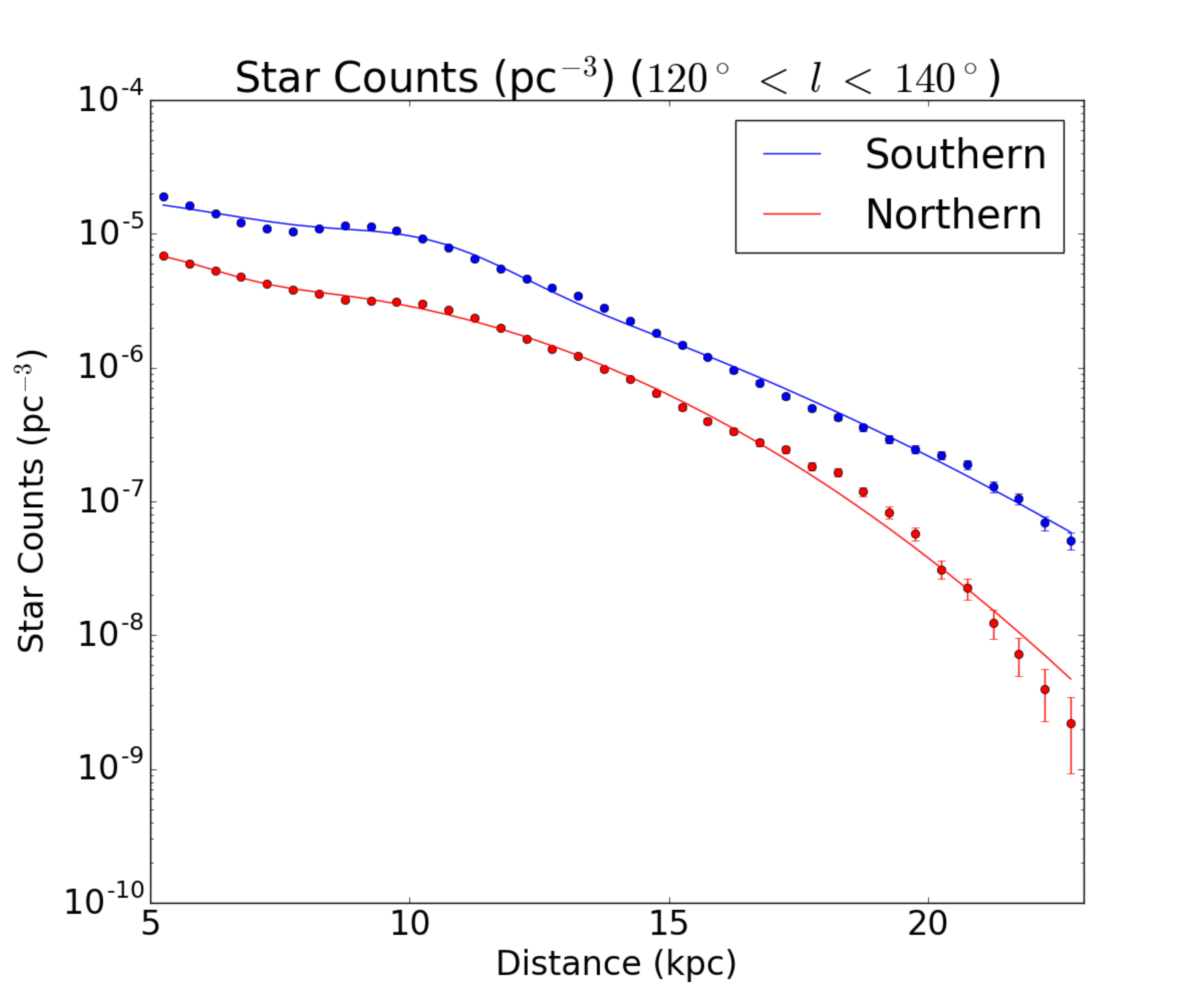}{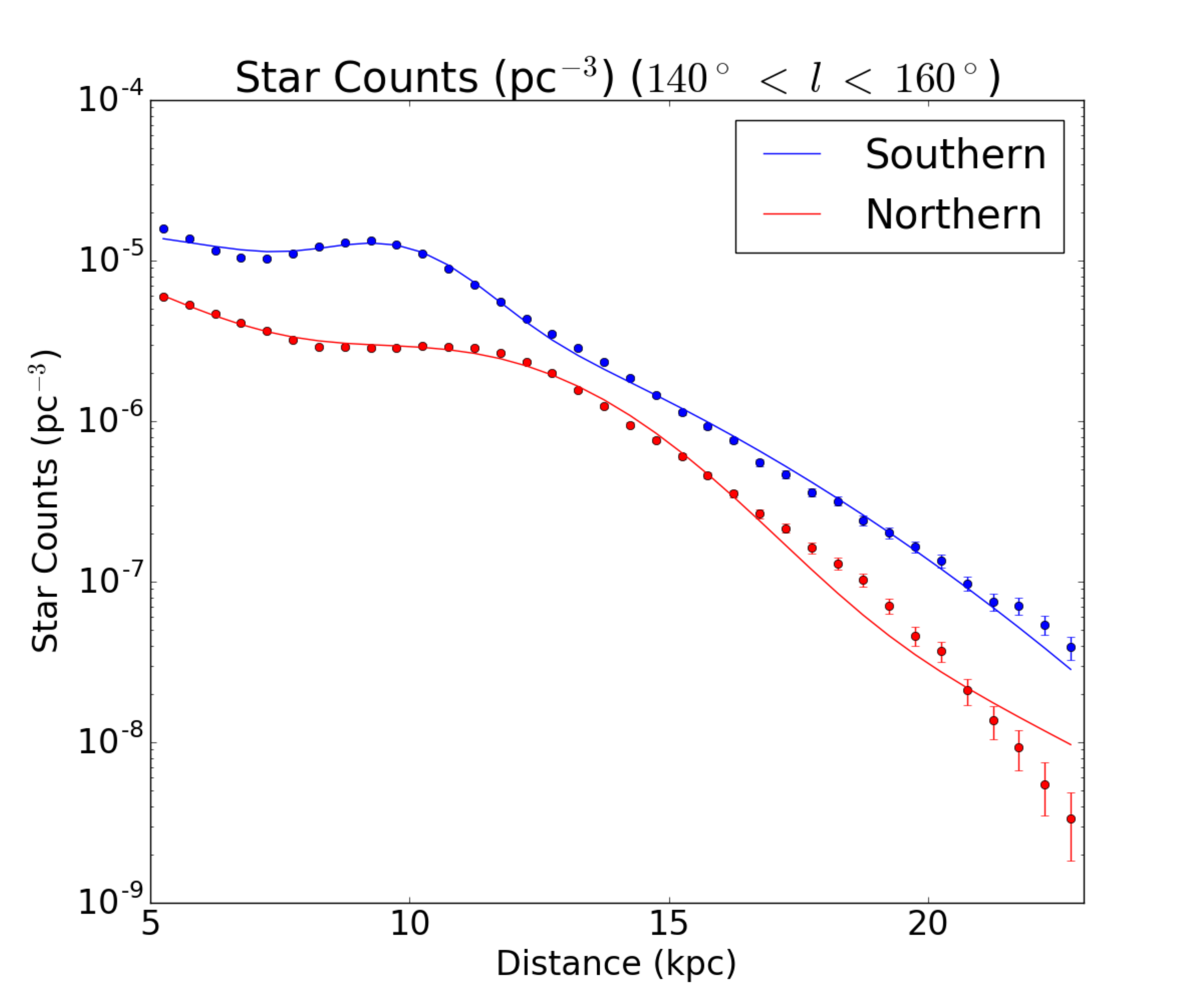}
\plottwo{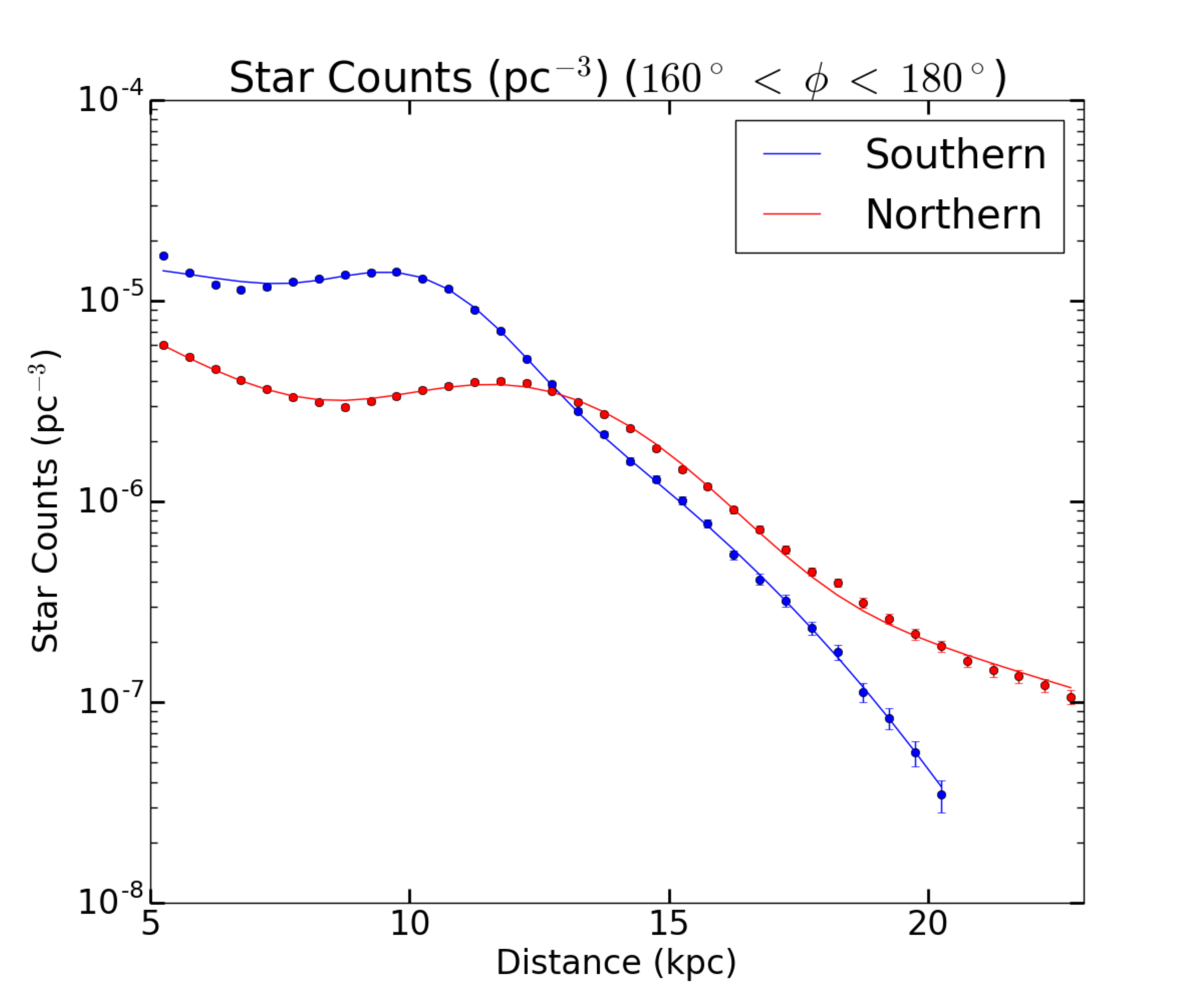}{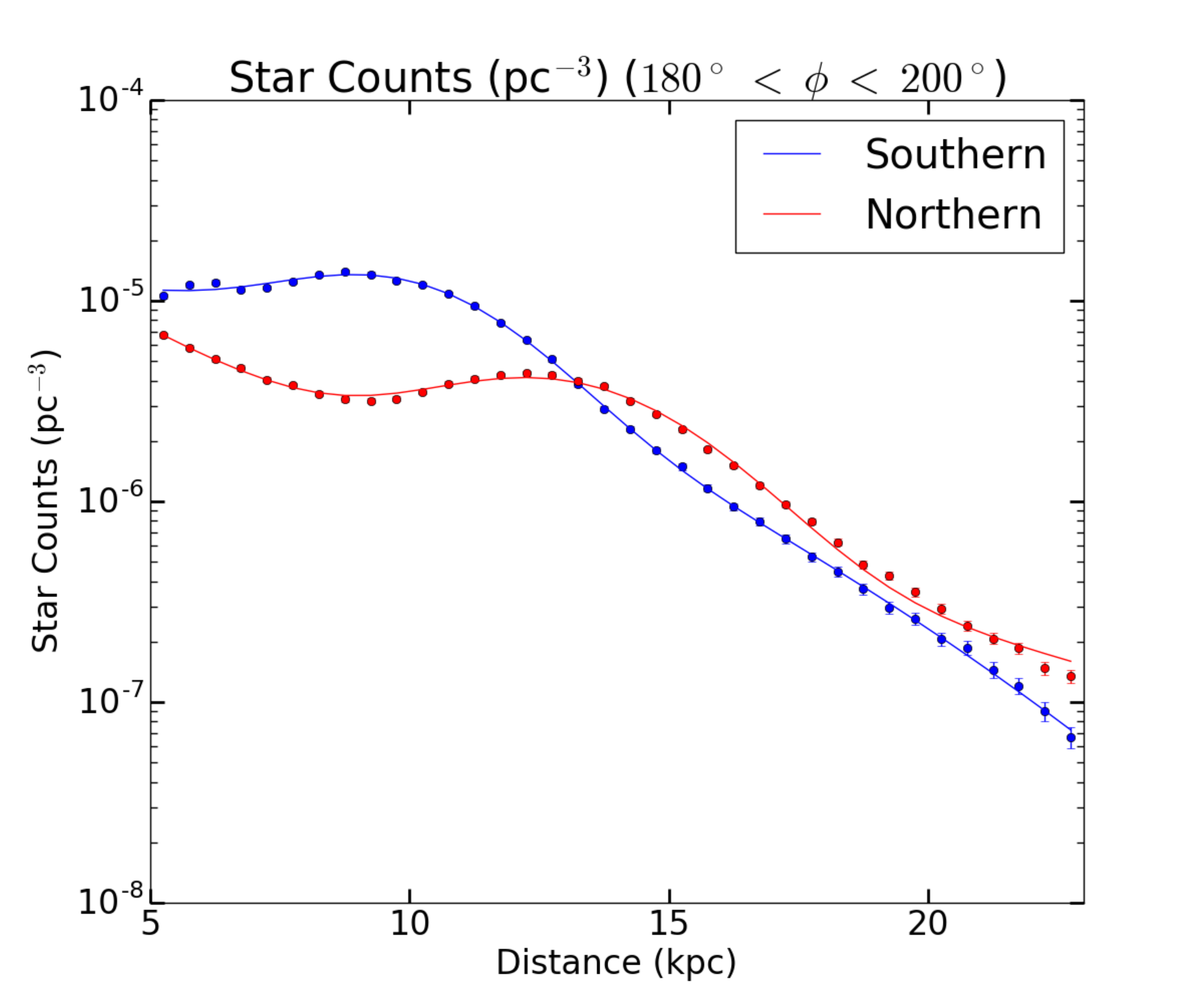}
\caption{\rm{The average stellar mass density (in counts pc$^{-3}$). in a $20^\circ$ wedge taken from the putative MR Center (see Fig. \ref{fig:crosssection}) in the South ($-3 < z_{MW} < -2$ kpc, blue) and the North ($3 < z_{MW} < 4.5$ kpc, red). The x axis is the distance away from the MR center along the $l = 180^\circ$ line. Each curve is fit as a S\'ersic profile plus a Gaussian overdensity. }}
\label{fig:crosssection1d}\end{figure*}

We can examine these ripples more precisely by binning them azimuthally. Fig. \ref{fig:crosssection1d} shows the average density from both the Southern and Northern regions from Fig. \ref{fig:crosssection} in $20^\circ$ wedges stretching from our MR center toward the Galactic Anticenter. Using only the region $d > 5$  kpc away from the MR center and $120^\circ < l < 200^\circ$ excludes areas that are not covered by our survey. We fit each curve with a method similar to our line of sight densities, but we find that a S\'ersic profile works better than a power law as the main MW population:
\begin{eqnarray}
\rho(d) &=& MW(d)+MR(d)\label{eq:1dfit}\\
MW(d) &=&\rho_0 e^{-A (d/1 kpc)^{1/n}}\nonumber\\
MR(d) &=& \frac{\delta \rho (1 kpc)}{\left(2\pi W_{MR}^2\right)^{1/2}} \exp\left(\frac{(d-d_{MR})^2}{2 W_{MR}^2}\right).\nonumber
\end{eqnarray}
Table \ref{tab:1dfit} shows the results of these fits. The main result is that the MR is found to be 9.56-10.15 kpc away from the MR Center in the South and 11.11-12.72 kpc away from the MR Center in the North. At $l = 180^\circ$, this corresponds to 14 and 16 kpc from the Galactic Center and is is roughly consistent with the values of 14 and 17 kpc from Section \ref{sect:quant}, although the different projections make comparisons imprecise. Intriguingly at $d = 18$ kpc, the Northern population is concave (suggesting an underdensity) while the Southern population is convex (suggesting an overdensity). This suggests additional ripples and supports the idea that the MR is the result of a propagating, circular wave \citep{IBAT++03,XU++15}. This effect, if present, is being convolved by our large distance uncertainties, making it difficult to see. Deeper data are needed to observe it more precisely.

\begin{table*}
\centering
\begin{tabular}{cccccccc}
	\hline
median $\phi$ & Hemisphere & $\rho_0$ ($10^{-6}$ Stars pc$^{-3}$) & $A$ & $n$ & $\delta\rho$ ($10^{-6}$ Stars pc$^{-3}$) & $d_{MR}$ (kpc) &  $W_{MR}$ (kpc) \\
	\hline
$130^\circ$ & Southern & $25.11 \pm 0.16$ & $0.021585 \pm 0.000066$ & $0.5555 \pm 0.0003$ & $12.54 \pm 0.36$ & $10.15 \pm 0.06$  & $1.76 \pm 0.05$ \\
            &  Northern & $7.80 \pm 0.06$ & $0.007686 \pm 0.000033$ & $0.4632 \pm 0.0003$ & $1.63 \pm 0.11$ & $11.11 \pm 0.11$  & $1.36 \pm 0.10$ \\
$150^\circ$ & Southern & $18.17 \pm 0.13$ & $0.008435 \pm 0.000030$ & $0.4711 \pm 0.0003$ & $20.04 \pm 0.41$ & $9.73 \pm 0.03$  & $1.33 \pm 0.03$ \\
            & Northern & $25.25 \pm 0.26$ & $0.236088 \pm 0.000886$ & $0.9141 \pm 0.0013$ & $10.20 \pm 0.14$ & $11.17 \pm 0.03$  & $2.29 \pm 0.03$ \\
$170^\circ$ & Southern & $16.53 \pm 0.12$ & $0.001845 \pm 0.000008$ & $0.3714 \pm 0.0002$ & $23.50 \pm 0.44$ & $10.00 \pm 0.03$  & $1.36 \pm 0.02$ \\
            & Northern & $87.15 \pm 0.81$ & $0.968493 \pm 0.002135$ & $1.6275 \pm 0.0023$ & $15.87 \pm 0.17$ & $12.05 \pm 0.03$  & $2.28 \pm 0.02$ \\
$190^\circ$ & Southern & $14.11 \pm 0.12$ & $0.012629 \pm 0.000050$ & $0.5179 \pm 0.0004$ & $41.49 \pm 0.52$ & $9.56 \pm 0.03$  & $2.09 \pm 0.02$ \\
            & Northern & $107.17 \pm 0.94$ & $1.053668 \pm 0.002197$ & $1.7155 \pm 0.0024$ & $18.59 \pm 0.19$ & $12.72 \pm 0.03$  & $2.43 \pm 0.02$ \\
	\hline
\end{tabular}
\caption{\rm{Our 1D fit values for different slices of MR-centered azimuth, $\phi$, from Eq. \ref{eq:1dfit}. Our MW S\'ersic profile is described by $\rho_0$, $A$ and $n$. Our Gaussian MR overdensity is described by $\delta\rho$, $d_{MR}$ (kpc) and $W_{MR}$. We did not model stellar mass distributions so $\rho_0$ and $\delta\rho$ are in counts. The distance errors here are purely statistical. We can also add the 2\% systematic errors from Section \ref{sect:clusters} to the distance measurements.}}\label{tab:1dfit}
\end{table*}

The Heliocentric radial view, the meridional cross-section and the planar cross-section all tell a consistent story: the observed Monoceros Ring is composed of two roughly concentric circles (or arcs which mimic circles across large angles) with the Southern (inner) circle being significantly more massive.

\section{Weighing the Monoceros Ring}\label{sect:mass}

We can use our MR map and fits from this paper to estimate the observed and total stellar mass of the Monoceros Ring. We estimate the total observed mass by adding the excess mass from observed pixels in the region where the overdensity is most strongly detected. To estimate the total stellar mass of the Monoceros Ring, we must extrapolate both through the Galactic plane and around the Milky Way. 

We can add the pixel masses as calculated by Eq. \ref{eq:pixmass} over the $120^\circ < l <240^\circ$, $-30^\circ < b < +40^\circ$ area in which the MR is most cleanly detected. We also add the statistical errors on $M_{pix}$ in quadrature to obtain a total observed excess stellar mass of: 
\begin{equation}
M_{obs} = 4.0 \times 10^6 M_\odot
\end{equation}
with a formal statistical error of 0.5\% which is of course much smaller than the actual uncertainty. This mass estimate excludes the masked areas from Fig. \ref{fig:mask} (the Galactic plane) and areas not in the relatively small Anticenter region. For reference, the  fitted Milky Way population along the same lines of sight contains a stellar mass of $2.2 \times 10^6 M_\odot$ between 6 kpc and 8 kpc heliocentric (a roughly MR width shell). The MR is the dominant source of stellar mass in this region. However, there is $2.4 \times 10^7 M_\odot$ of stellar mass if we extend the projection of our MW population from 0 kpc to 10 kpc in this region.

\begin{figure}[ht]
\includegraphics[width=0.99\columnwidth]{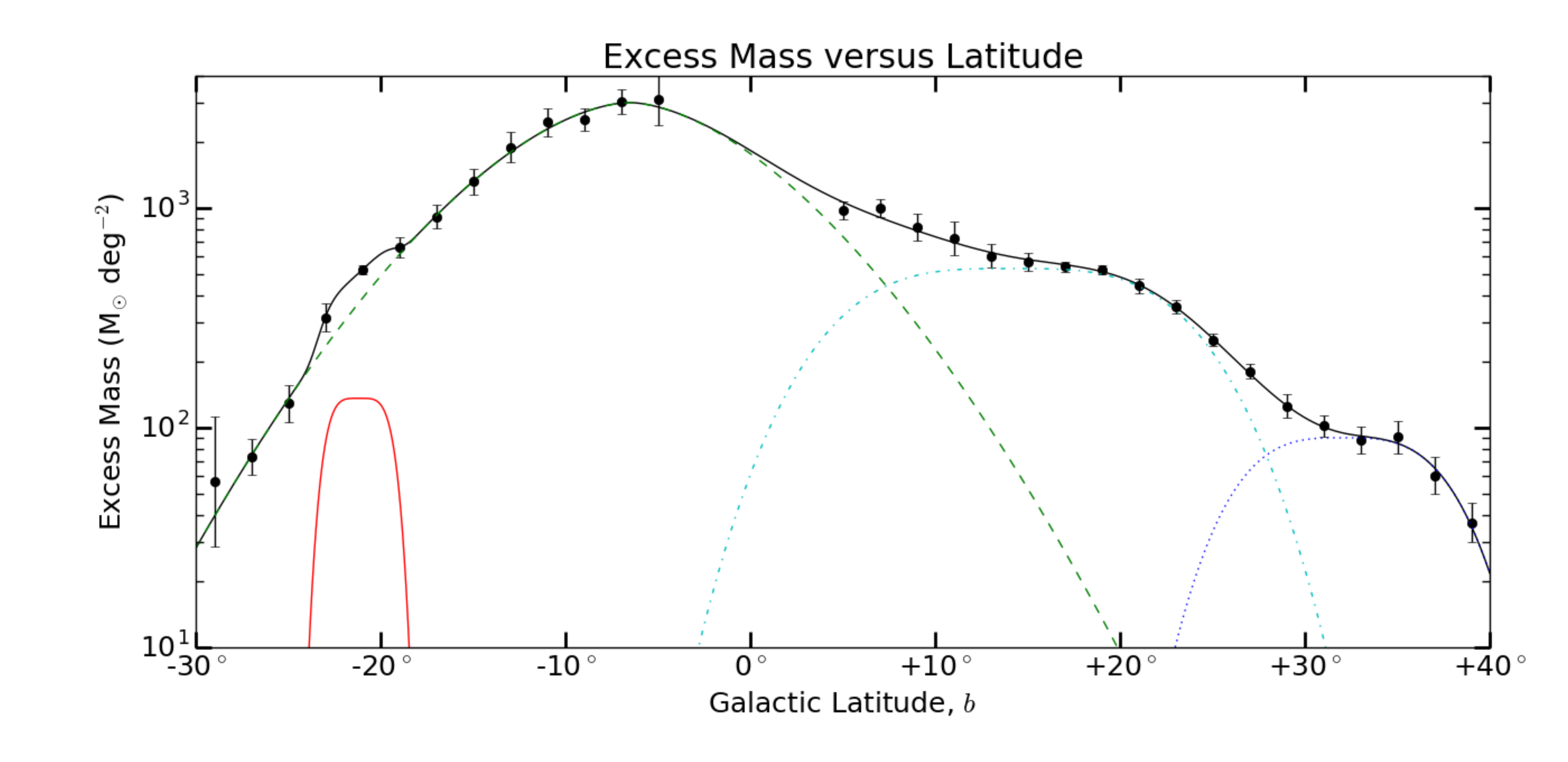}
\includegraphics[width=0.99\columnwidth]{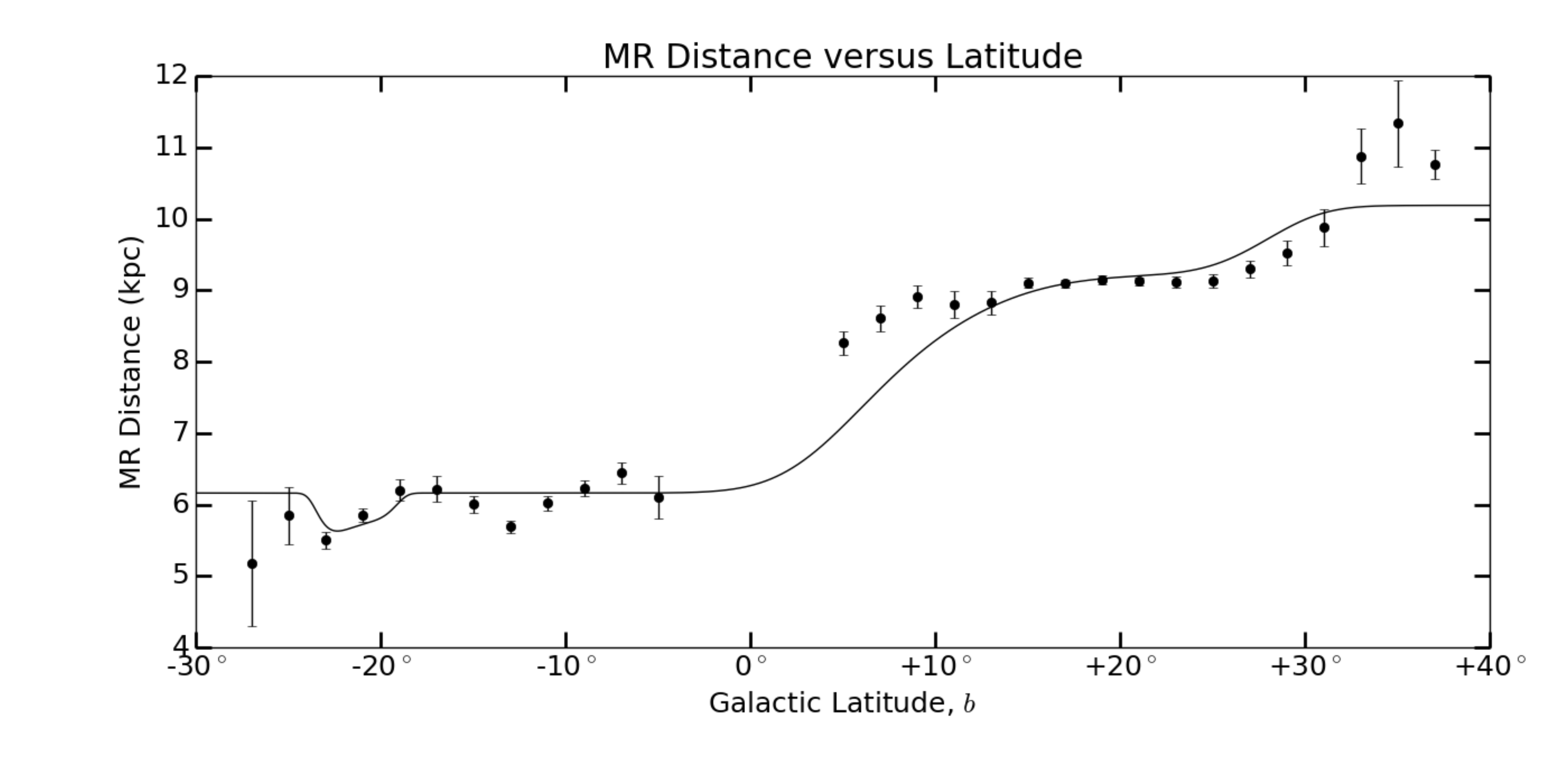}
\caption{\rm{The average excess mass associated with the Monoceros Ring (top) and the average distance to that mass (bottom) as a function of Galactic latitude for the $120^\circ < l < 240^\circ$ region. In the top plot, the solid black line is the total fit which is the sum of the Southern Stream (green dashed line), Northern Stream (cyan dash-doted line), the ACS (blue dotted line) and the Southern Ridge (red solid line). In the bottom plot, the solid black line is an mass average distance assuming each distinct population is at a particular distance (see Eq. \ref{eq:d_b})
}}
\label{fig:profile}\end{figure}

To estimate the total MR mass, we fit the total excess mass observed in our Anticenter region as a function of $b$ and extrapolate across the Milky Way plane and over the rest of the (assumed) circular structure of the Ring. Fig. \ref{fig:profile} shows the mean excess mass (as defined by the Gaussian in Eq. \ref{eq:linefit} and the mean distance (the peak of the Gaussian in Eq. \ref{eq:linefit}) as a function of $b$ in our $120^\circ < l < 240^\circ$. For each value $b$ we assign a log-mean mass and standard deviation-based uncertainties:
\begin{eqnarray}
M(b) &=& e^{<\log M_{pix}>},\\
\sigma(b) &=& \frac{<\log M_{pix}^2>-<\log  M_{pix}>^2}{\left(N_b-1\right)^{1/2}}M(b)\nonumber
\end{eqnarray}
and analogously defined log-mean distance and uncertainties. Here $N_b$ is the number of unmasked pixels at latitude $b$. For distance measurements, we only include pixels with more than 125 $M_\odot$ deg$^{-2}$ to avoid including low significance overdensities in our distance estimate.

\begin{table*}
\centering
\begin{tabular}{ccccccc}
	\hline
Name & Mass ($M_\odot$ deg$^{-2}$ & $b_0$ (deg)  &  $w$ (deg) & $\alpha$ & $d_{mass}$ (kpc) & $d_{MR}$ (kpc) \\
	\hline

Southern Stream & $51020 \pm 850$ & $-6.54 \pm 0.08$ &  $9.40 \pm 0.07$ & $1.70 \pm 0.02$ & $6.16 \pm 0.04$ & $ 4.85 \pm 0.03$\\
Northern Stream & $10740 \pm 140$ & $14.16 \pm 0.08$ &  $11.22 \pm 0.10$ & $3.37 \pm 0.14$ & $9.24 \pm 0.03$ & $8.36 \pm 0.03$ \\
ACS   & $1144 \pm 36$ & $32.08 \pm 0.18$ &  $7.03 \pm 0.22$ & $3.18 \pm 0.62$ & $10.19 \pm 0.12$ & $8.88 \pm 0.10$\\
Southern Ridge  & $538 \pm 44$ & $-21.16 \pm 0.20$ &  $1.93 \pm 0.19$ & $4.33 \pm 3.47$ & $4.46 \pm 0.29$ & $3.74 \pm 0.19$\\
	\hline
\end{tabular}
\caption{\rm{Our fit to the profile of the Monoceros Ring overdensity (excess mass and distance versus Galactic latitude, $b$). We fit Mass deg$^{-2}$ as a function of $b$ with four generalized Gaussians (Eq. \ref{eq:m_b}) and then, assuming each population is at a different distance, fit center of mass distance, $d_{mass}(b)$ as a weighted average of those distance for each b (Eq. \ref{eq:d_b}). We also fit the $d_{MR}$, the distance to peak overdensity with the same technique. The distance errors we present here are purely statistical, and one should also add our 2\% systematic error from Section \ref{sect:clusters} if quoting these results.}}\label{tab:profile}
\end{table*}

We fit the mass distribution in Fig. \ref{fig:profile} as a series of 4 generalized normal distributions (in which the exponent is an independent parameter):
\begin{eqnarray}
M(b) &=& \sum_{i=1}^4 G_i(b)\label{eq:m_b}\\
G(b) &=& \frac{M_0 \beta}{2w\Gamma(1/\beta) }  e^{\left((b-b_0)/w\right)^\beta }.\nonumber
\end{eqnarray} 
Here, $M_0$ is the total mass in each subprofile, $w$ is its width, $b_0$ is its center and $\beta$ controls how fast the subprofile drops off. We show the fit values in Table \ref{tab:profile}. We name each distribution to describe the visually identifiable population it represents. The Southern stream is the main Southern and equatorial population (identified in both Fig. \ref{fig:massmap} and Fig. \ref{fig:sliceplot}). In our fit, it dominates across $-30^\circ < b < +10^\circ$, crossing the disk. The Northern Stream is the main Northern population, dominating $+10^\circ < b < +30^\circ$. It is a more tightly bounded structure (has a sharper exponent) and in other projections appears more distinct from the disk. There is a second smaller Northern stream, the previously identified ACS that stretches across $120^\circ < l < 180^\circ$ at $b > +30^\circ$ in Fig. \ref{fig:massmap}. Finally, these fits work much better with an additional small Southern population at $b = -20.86^\circ$ that creates a small ``Southern Ridge'' in the profile. This last population is co-local with an unusual patch of data at $l=195^\circ,\ b = -20^\circ$ that may correspond to a small independent structure. Adding a fifth Gaussian either produces a curve with an insignificant amount of mass, or a very wide curve that does not represent a distinct population.

To see if our 4 generalized Gaussians correspond to genuinely distinct structures, we fit our center of mass distance profile as
\begin{equation}
d_{mass}(b) = \frac{\sum_{i=1}^4 d_{mass\ i} G_i(b)}{\sum_{i=1}^4 G_i(b)}\label{eq:d_b}.
\end{equation} 
Here the $G_i$'s are taken as a given from the mass profile (Eq. \ref{eq:m_b}) and the 4 $d_{mass\ i}$'s are being fit independently. Essentially, this fit asserts that each of our four populations in Table \ref{tab:profile} is at a distinct (Heliocentric) distance. Fig. \ref{fig:profile} shows that this model is roughly consistent with our distance profile. Specifically, we find that the Southern population is at $d = 6.2$ kpc while the Northern population is $9.2$  kpc, and the transition between the two main populations corresponds to the observed transition in distance. These distances are consistent with our Northern and Southern radii from Section \ref{sect:cross}. The large jump in both distance and density between the Southern population and the Northern population strongly suggests that these two structures are in some real sense distinct. 

With a complete mass and density model in hand, we can estimate the total excess stellar mass of the Monoceros Ring. Just using our fitting formula to interpolate through the plane, we estimate that there is $7.6 \times 10^6 M_\odot$ in the $120^\circ < l < 240^\circ$ region. If we assume that the Monoceros Ring is indeed a uniform circle and extrapolate our results across the entire Milky Way, assuming that the Southern and Northern MR are 13 kpc and 17 kpc Galactocentric circles, respectively, we estimate that the total MR mass is $5.7 \times 10^7 M_\odot$ ($4.8 \times 10^7 M_\odot$ in the South and $8.6 \times 10^6 M_\odot$ in the North). If instead we assume that the MR is centered around a point 4 kpc away from the Galactic Center with radii 9 kpc and 13 kpc (as Fig. \ref{fig:crosssection} seems to prefer), the total MR mass is $4.0 \times 10^7 M_\odot$ ($3.3 \times 10^7 M_\odot$ in the South and $6.4 \times 10^6 M_\odot$ in the North). In addition to this 30\% discrepancy due to the exact radius of the Monoceros Ring (assuming it is a circle), we have not accounted for systematic errors from our interpolation across the Galactic plane, density and thickness variations around the Ring or estimating the excess Milky Way mass along the line of sight. 

\section{Discussion}\label{sect:conc}

Using the PS1 dataset that covers the MW plane, new modeling techniques and novel maps, we have produced a three dimensional and quantitative analysis of the Monoceros Ring. At the root of this work is the Pan-STARRS1 catalog. PS1 is the first optical dataset to examine the large areas of the Milky Way plane with the photometric precision necessary to perform isochrone fitting. Particularly, PS1 observed almost the entire $120^\circ < l < 240^\circ,\ -30^\circ < b < +40^\circ$ region where the MR is most visible. Our construction of depth and completeness maps was essential for precise measurements of stellar mass density. We fit line of sight densities from the \textsc{match} program with a combined Milky Way and Monoceros Ring model to obtain quantitative stellar mass and distance estimates to the Monoceros Ring across PS1.  

In addition to utilizing new data products and applying \textsc{match} across three quarters of the sky, we have developed several new maps of the Milky Way and Monoceros Ring. In Section \ref{sect:map}, we show Heliocentric maps of total stellar mass density and rediscover familiar features from \citet{NEWB++02}, \citet{GRIL06} and \citet{SLAT++14}. In Section \ref{sect:quant}, we map our estimated MR mass and distance in this same projection. We see that the MR is at Heliocentric $d = 6$ kpc in the South but $d = 9$ kpc in the North. In Section \ref{sect:vert}, we produce meridional cross-sections of the Milky that suggest that the Southern MR is contiguous with the Galactic plane while the Northern MR is elevated above the plane. Finally in Section \ref{sect:cross}, we produce planar cross sections that show that the Northern MR is most consistent with a $13$ kpc circle centered $4$ kpc from the Galactic Center in the Anticenter (solar) direction. The Southern MR is consistent with a $10$ kpc circle with the same center. Both MR features could be fit (less well) with Galactocentric circles with larger radii. This is the first time the MR has been shown as a 2D circle covering at least $120^\circ$ in the South and $170^\circ$ in the North . 

Within our observed, unmasked area, we detect $4 \times 10^6 M_\odot$ excess stellar mass associated with the MR. By fitting its mass distribution, and interpolating across the unobserved regions, we can estimate the total stellar mass of the structure, assuming it is a complete circle. This is a strong assumption given that we can only image roughly 40\% of a circle with our current depth. The Northern MR appears to be at least two stellar streams that are $9$ kpc from the Sun and roughly cover the $10^\circ < b < 40^\circ$ region near the Anticenter. Assuming these streams form a complete ring, their total stellar mass would be roughly $8.6 \times 10^6 M_\odot$  if this circle is a Galactocentric ring and $6.4 \times 10^6 M_\odot$ if it is centered around our preferred point 4 kpc closer to the Sun. The Southern overdensity appears to be contiguous with the Galactic plane and even seems to cross the plane and appear in a small number of Northern pixels. This indicates that the Southern MR might be some combination of a Galactic flare and Southern warp (if we are modeling the MR as two structures). It only extends down to $b = -30^\circ$ and is considerably closer to us than the Northern MR, $6$ kpc in the Anticenter direction. But it is also significantly more massive than the Northern stream and would contain $4.8 \times 10^7 M_\odot$ if extrapolated around the Galactic center or $4.0 \times 10^7 M_\odot$ if extrapolated around our alternate circle.

Any astrophysical MR model must account for its profound North-South asymmetry and the roughly circular geometry we observe here. Fig. \ref{fig:crosssection} shows essentially concentric circles alternating in the South and North suggesting a Galactic rippling as though a large mass had ``splashed'' through the Milky Way at a point $4$ kpc away from the Galactic center. This is the 2D extension of the ripple theory put forth by \citet{IBAT++03} and \citet{XU++15}. Alternatively, the Northern and Southern MR may be separate structures that only appear as one contiguous feature when we cannot image the Galactic plane and have limited spatial resolution along the line of sight. One may lose the feeling that this is an unlikely coincidence when one realizes that the two features are separated by roughly $3$ kpc in physical space. \citet{PENA++05} simulates the MR as a tidal dwarf stream that includes several wraps around the Milky Way. It is conceivable that this dwarf could have passed by the Galactic Anticenter at $6$ kpc in the South on one pass and $9$ kpc in the North on the second pass. It would be difficult to model both the Northern and Southern MR as a traditional warp or flare, but a disrupted flare (as per \citet{KAZA++09}) provides a mechanism for creating the observed asymmetry. Deeper data and an improved understanding of Milky Way dust extinction will allow us to constrain the MR further. But even with no improvement in data or analysis, we have shown that any successful model of the Monoceros Ring must include a Northern feature at 9 kpc (Heliocentric), a Southern feature at 6 kpc and a total stellar mass of at least $10^7 M_\odot$.

\section{Acknowledgments}

The PS1 Surveys have been made possible through contributions of the Institute for Astronomy, the University of Hawaii, the Pan-STARRS Project Office, the Max-Planck Society, and its participating institutes, the Max Planck Institute for Astronomy, Heidelberg, and the Max Planck Institute for Extraterrestrial Physics, Garching, The Johns Hopkins University, Durham University, the University of Edinburgh, Queen's University Belfast, the Harvard-Smithsonian Center for Astrophysics, and the Las Cumbres Observatory Global Telescope Network, Incorporated, the National Central University of Taiwan, the National Aeronautics and Space Administration under Grant No. NNX08AR22G issued through the planetary Science Division of the NASA Science Mission Directorate, the National Science Foundation under Grant No. AST-1238877, the University of Maryland, and Eotvos Lorand University F(ELTE).

EM acknowledges funding by Sonderforschungsbereich SFB 881 ``The Milky Way System'' (subproject A3) of the German Research Foundation (DFG).

\bibliography{ms}

\appendix

\section{Additional Maps}\label{sect:addmap}

In this paper, we have optimized all of our imaging for the study of the Monoceros Ring. This included focusing on the $5-10$ kpc Heliocentric distance range, masking out the Galactic plane and centering our maps on the Galactic Anticenter. But we ran \textsc{match} on the whole sky without these restrictions and thus produced imaging across a wider range of distances and latitudes. Here, we present some of these images.

Figs.\ \ref{fig:densitymap}, \ref{fig:metalmap} and our MR mass estimate use density maps with the mask from Section \ref{sect:mask}. It is impossible for us at this time to properly account for the systematic effects Galactic extinction, younger stellar populations and high stellar density in these regions. Examining these unmasked regions shows how necessary our masks were and that our masks efficiently cover problem regions without sacrificing more reliable data.
 
\begin{figure}[ht]
\includegraphics[width=0.49\columnwidth]{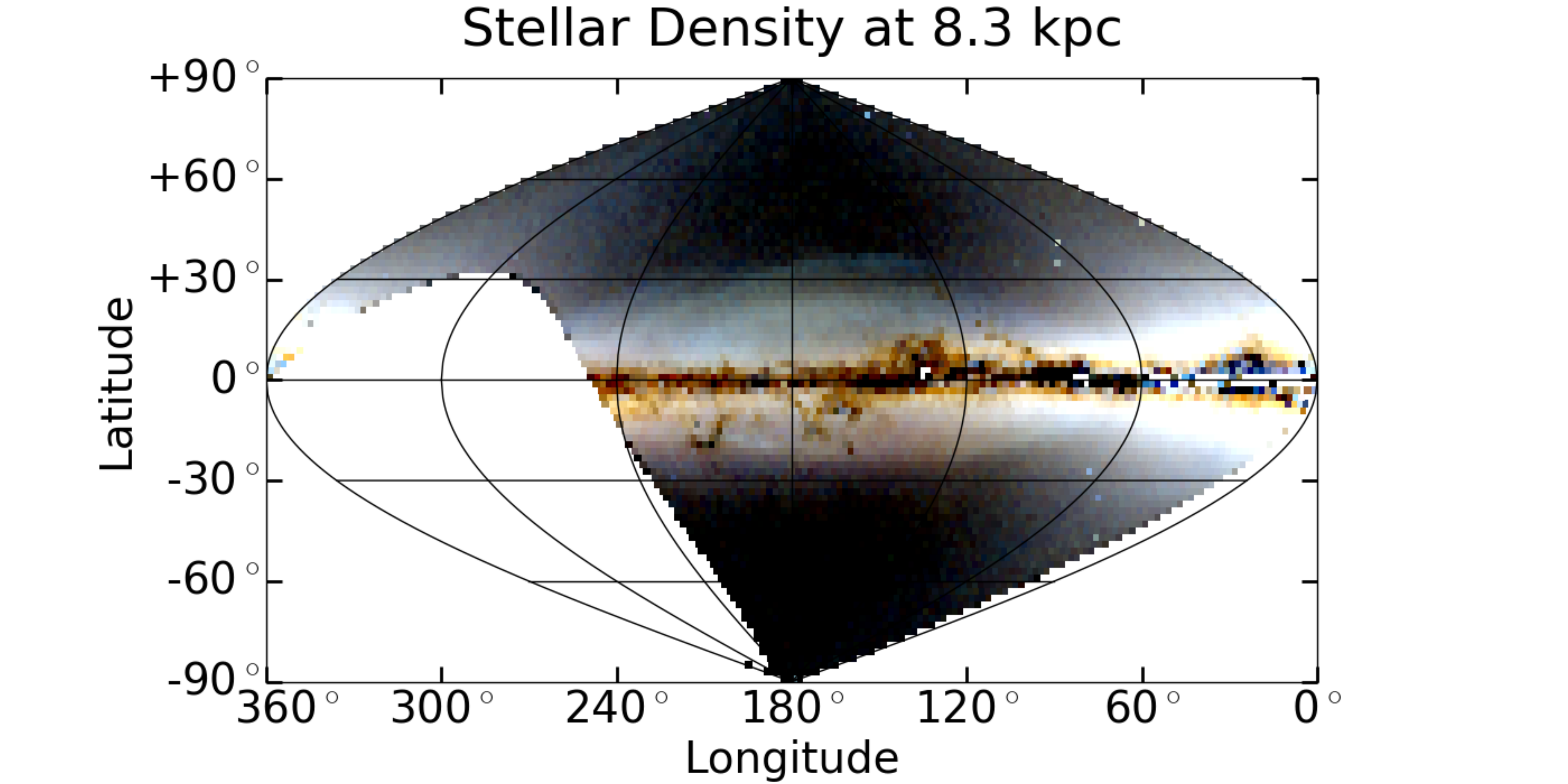}
\includegraphics[width=0.49\columnwidth]{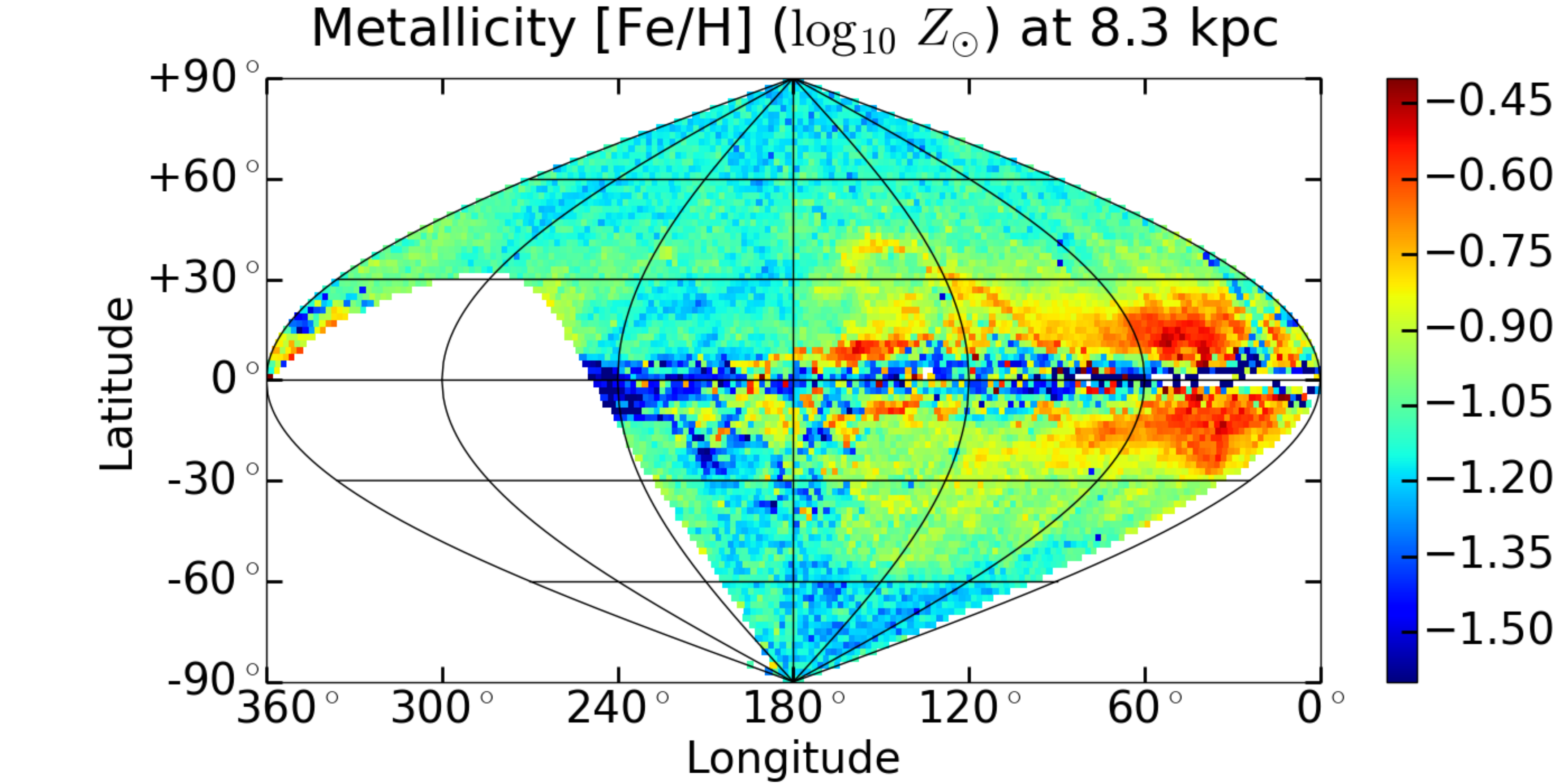}
\caption{\rm{On the left, the (unmasked) stellar mass density across the sky. The RGB channels represent density at 6.9, 8.3 and 10 kpc, respectively (Heliocentric). Each channel is scaled logarithmically with the minimum and maximum set at the level of the 10th and 95th percentile (unmasked). In units of $10^{-6} M_\odot \rm{pc}^{-3}$, this corresponds to 1.7 and 113 in the R channel, 0.98 and 47 in the G channel and 0.53 and 18 in the B channel. On the right, the (unmasked) stellar metallicity at 8.3 kpc (Heliocentric) as calculated with Eq.\ \ref{eq:metal}. Note that with our settings, \textsc{match} aliases errors in stellar age or Galactic extinction as (typically lower) metallicity. This effect dominates the Galactic plane.}}
\label{fig:mapum}\end{figure}

Fig.\ \ref{fig:mapum} (left) shows the complete, unmasked total density map. Over most of the  MR area, the map is smooth and qualitatively ``reasonable looking'' down to $|b| = 4^\circ$. There are significant wisps of apparently low density which correspond to areas of high Galactic extinction. Fig.\ \ref{fig:mapum} (right) shows the unmasked metallicity where dust features are more distinct. The apparent $Z = 1.6$ stars at low latitude are clearly systematic problems, and removing these areas guided our masked thresholds.

\begin{figure*}[ht]
\plottwo{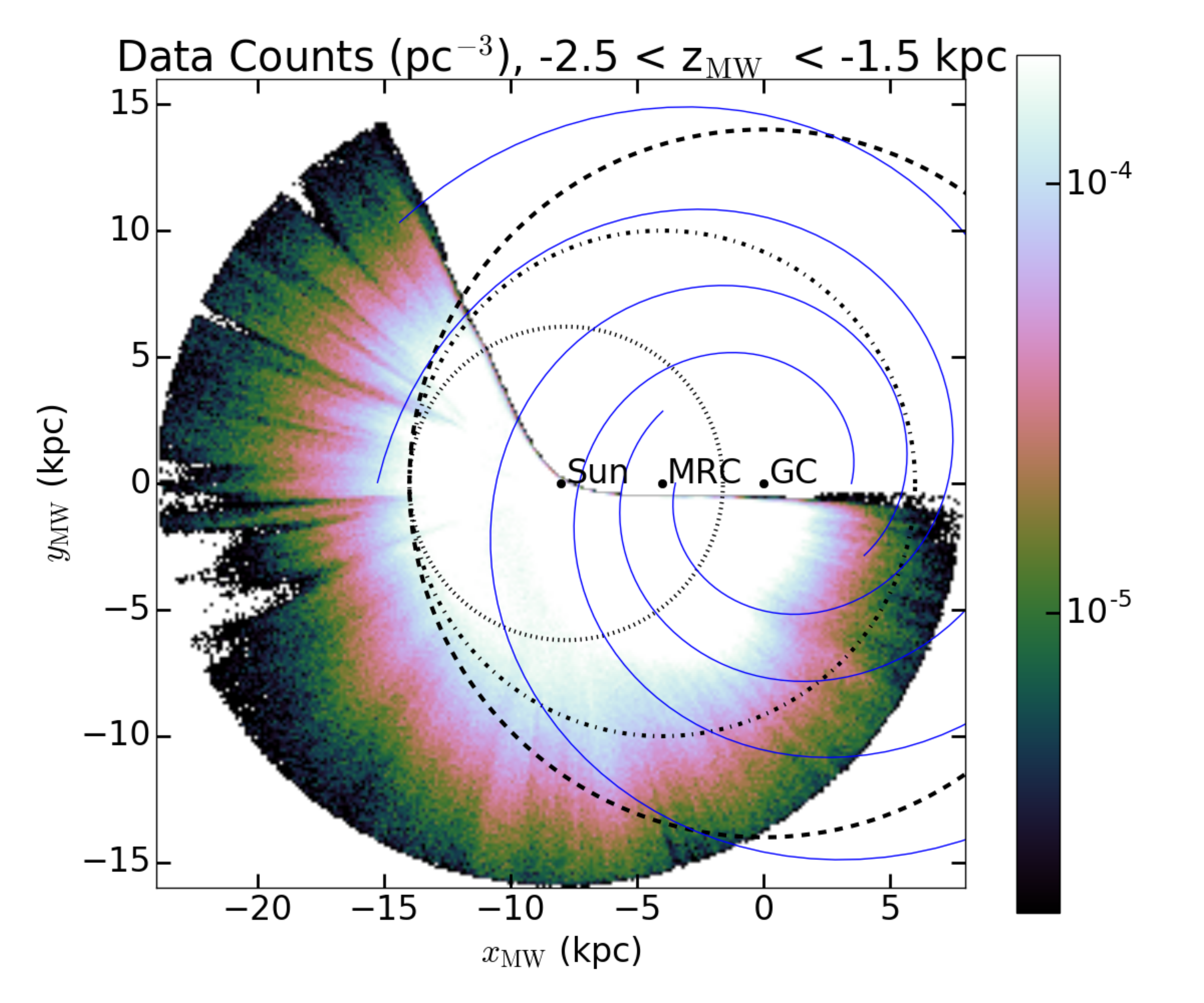}{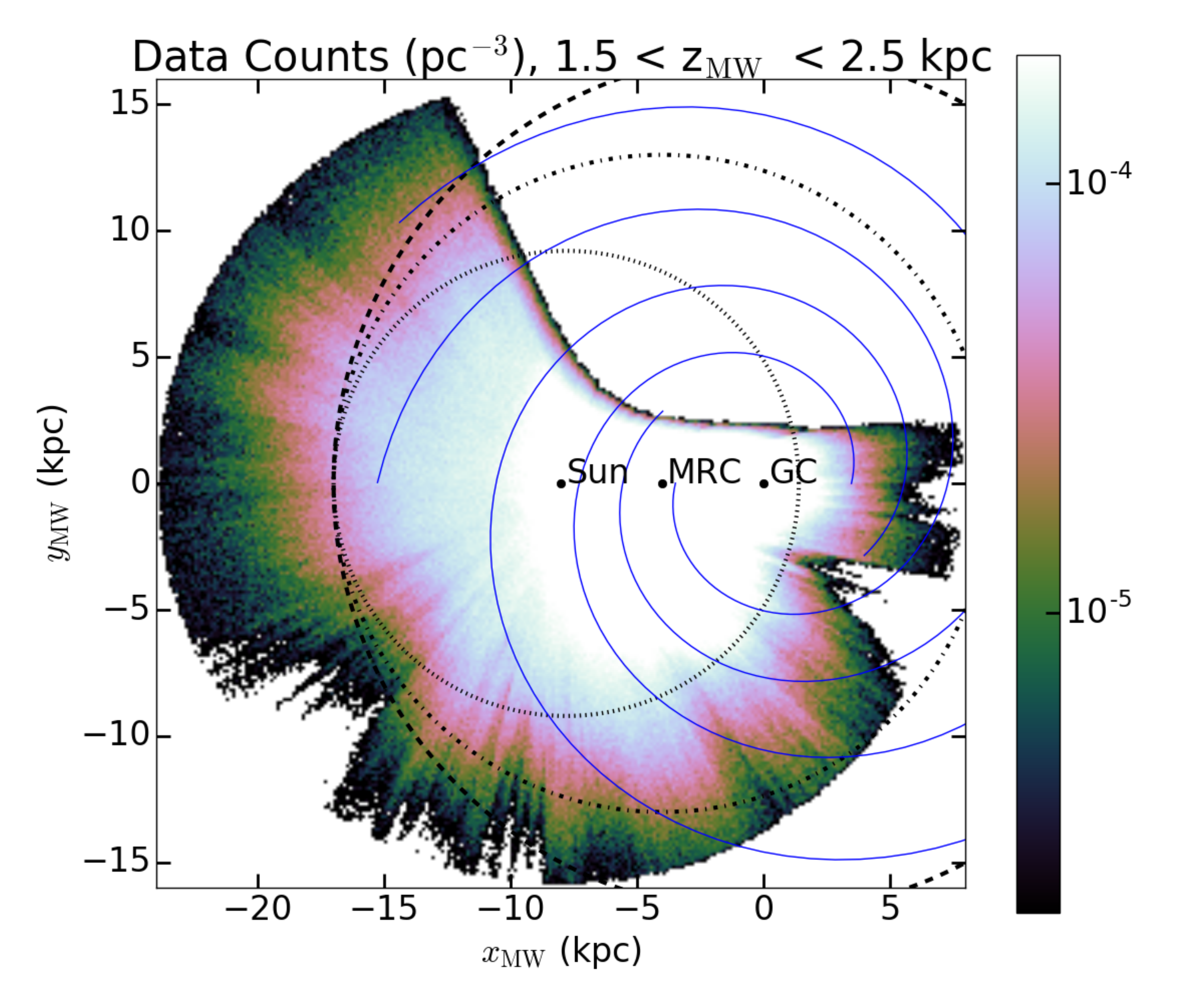}
\plottwo{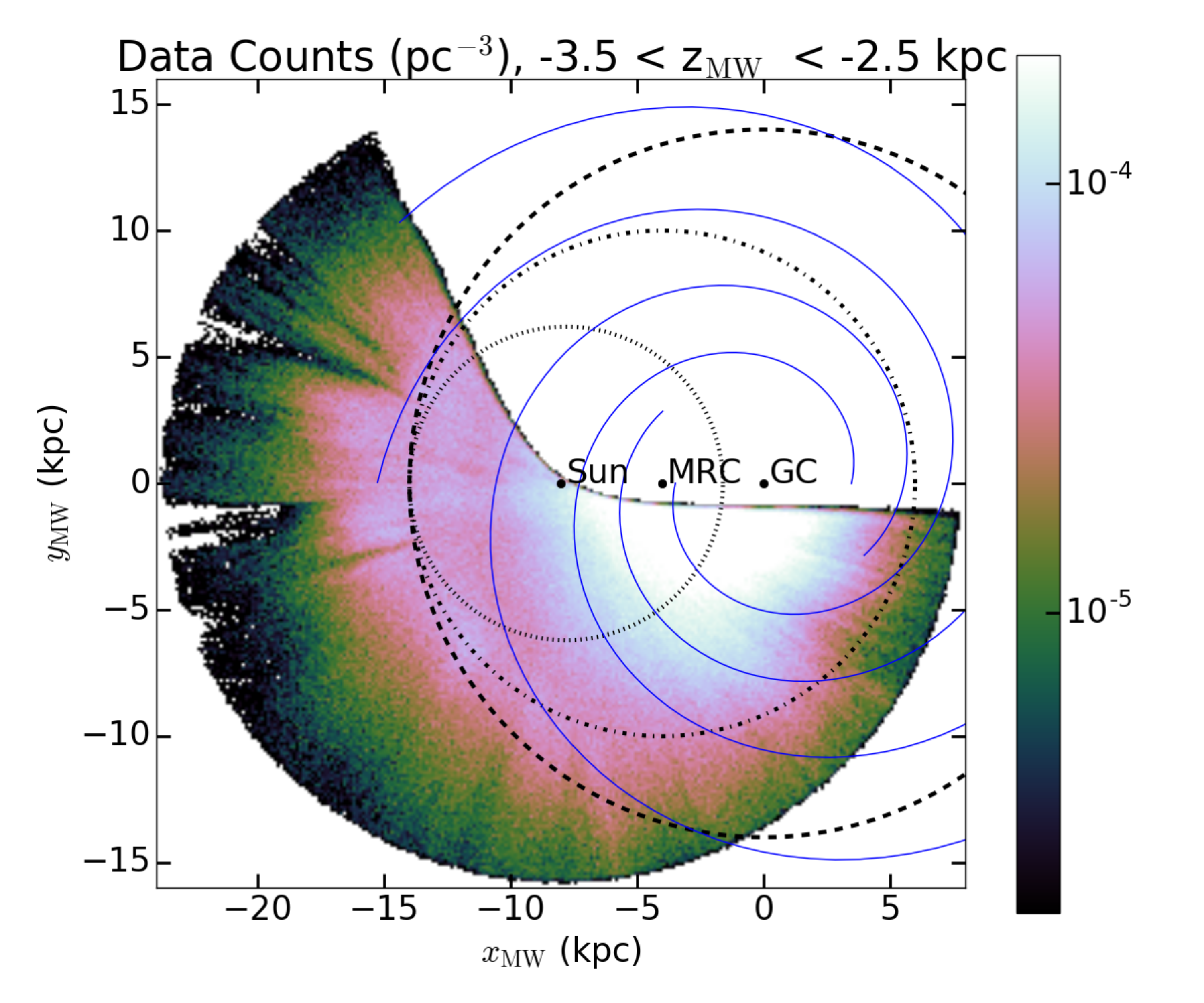}{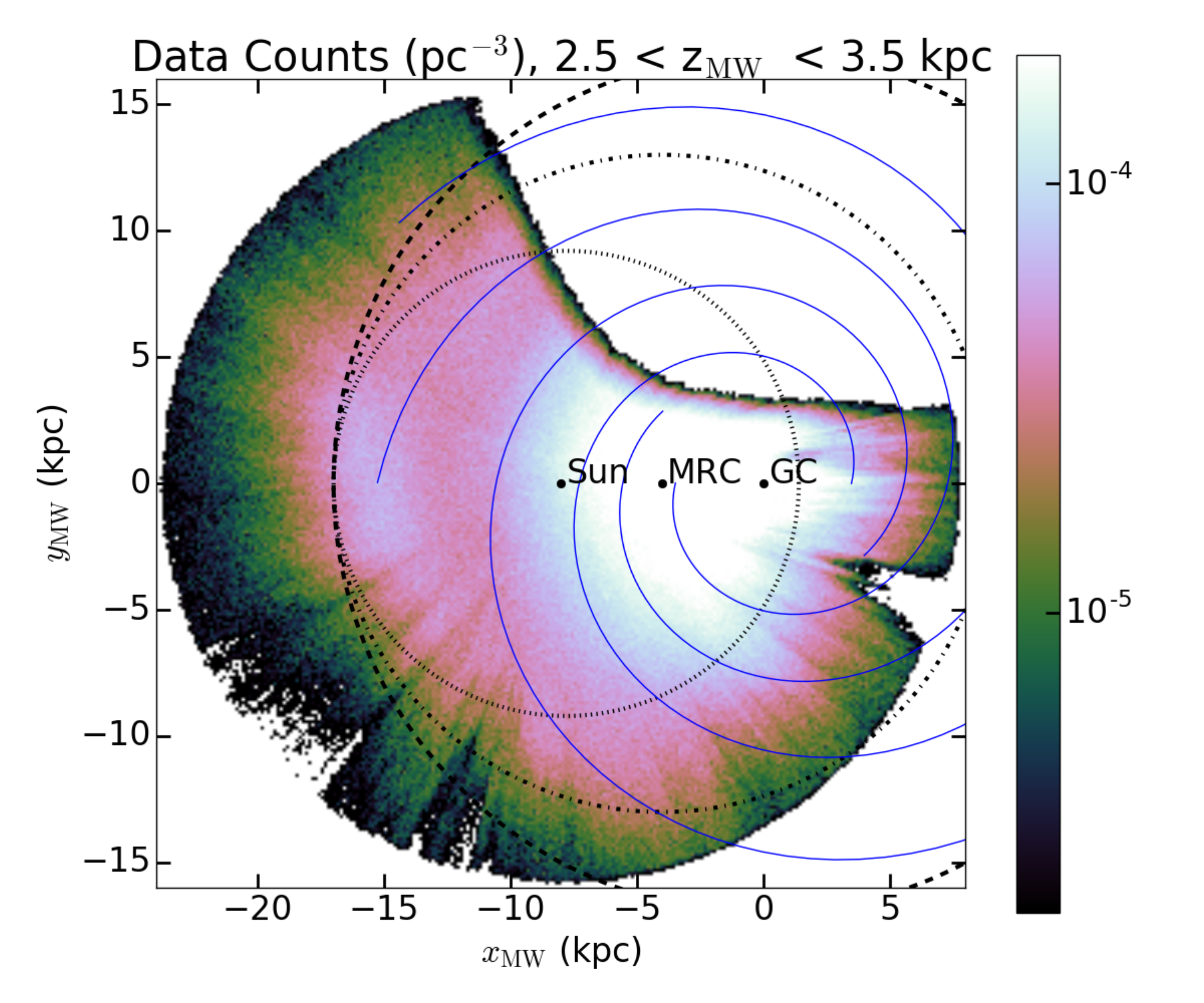}
\plottwo{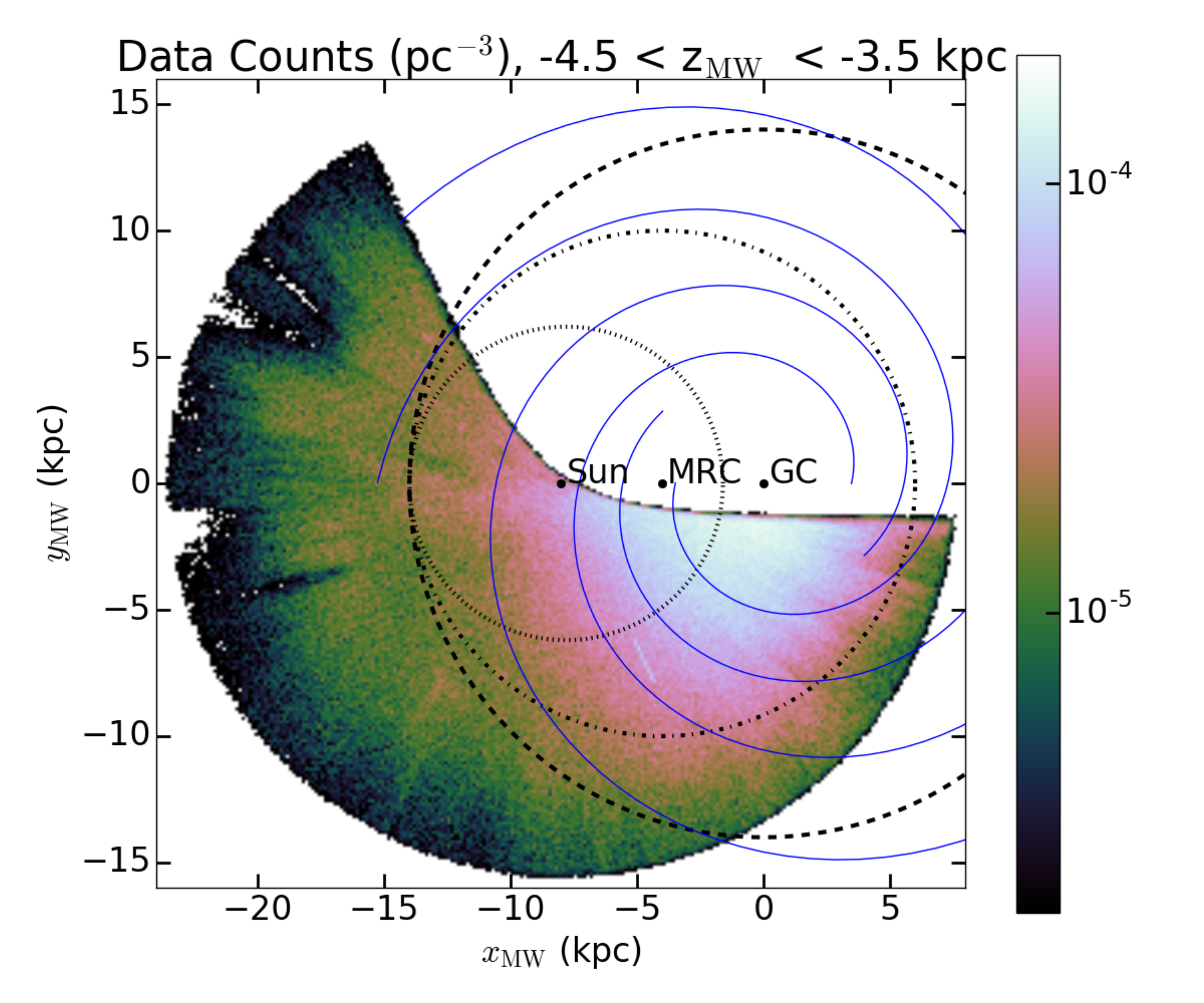}{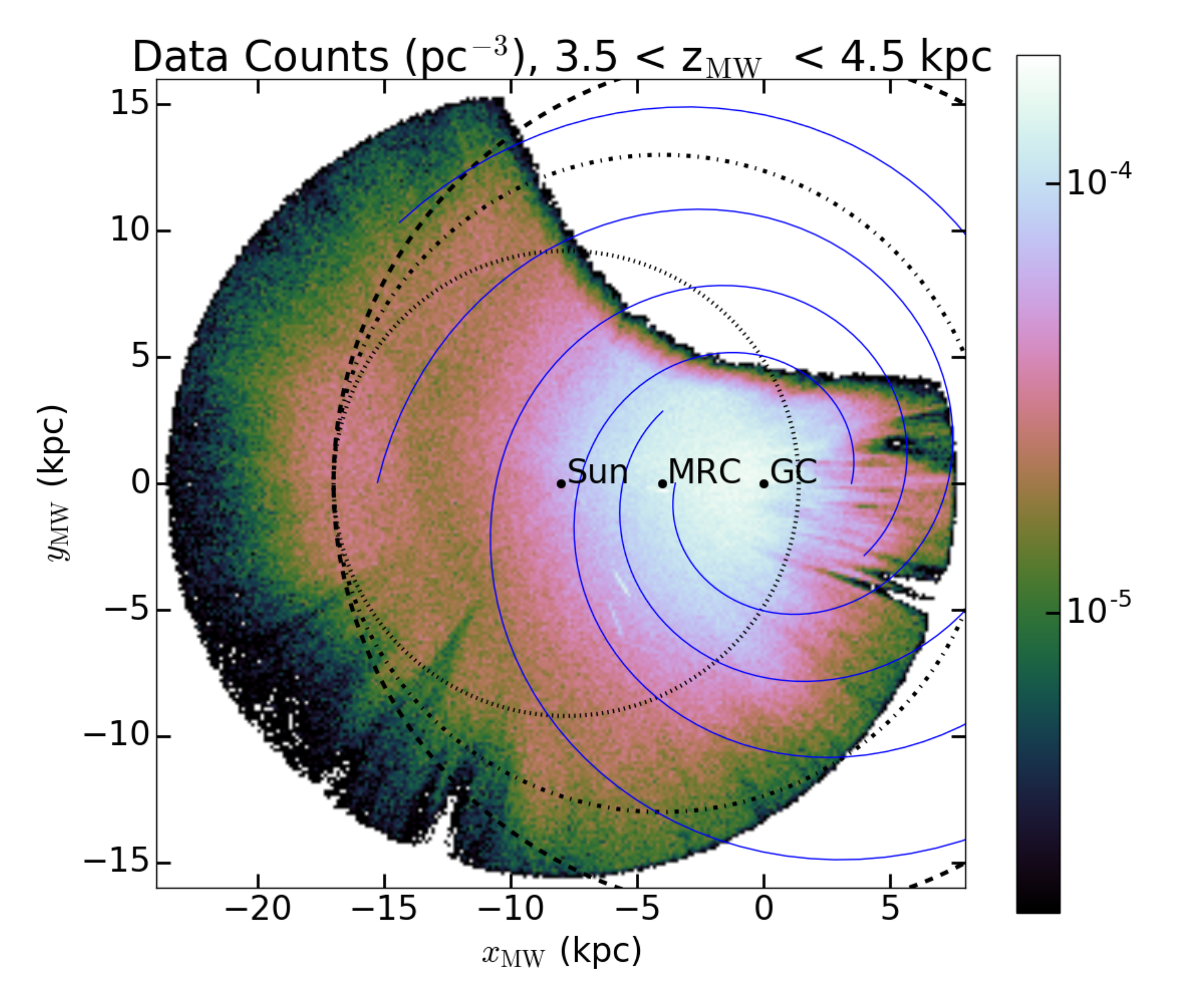}
\plottwo{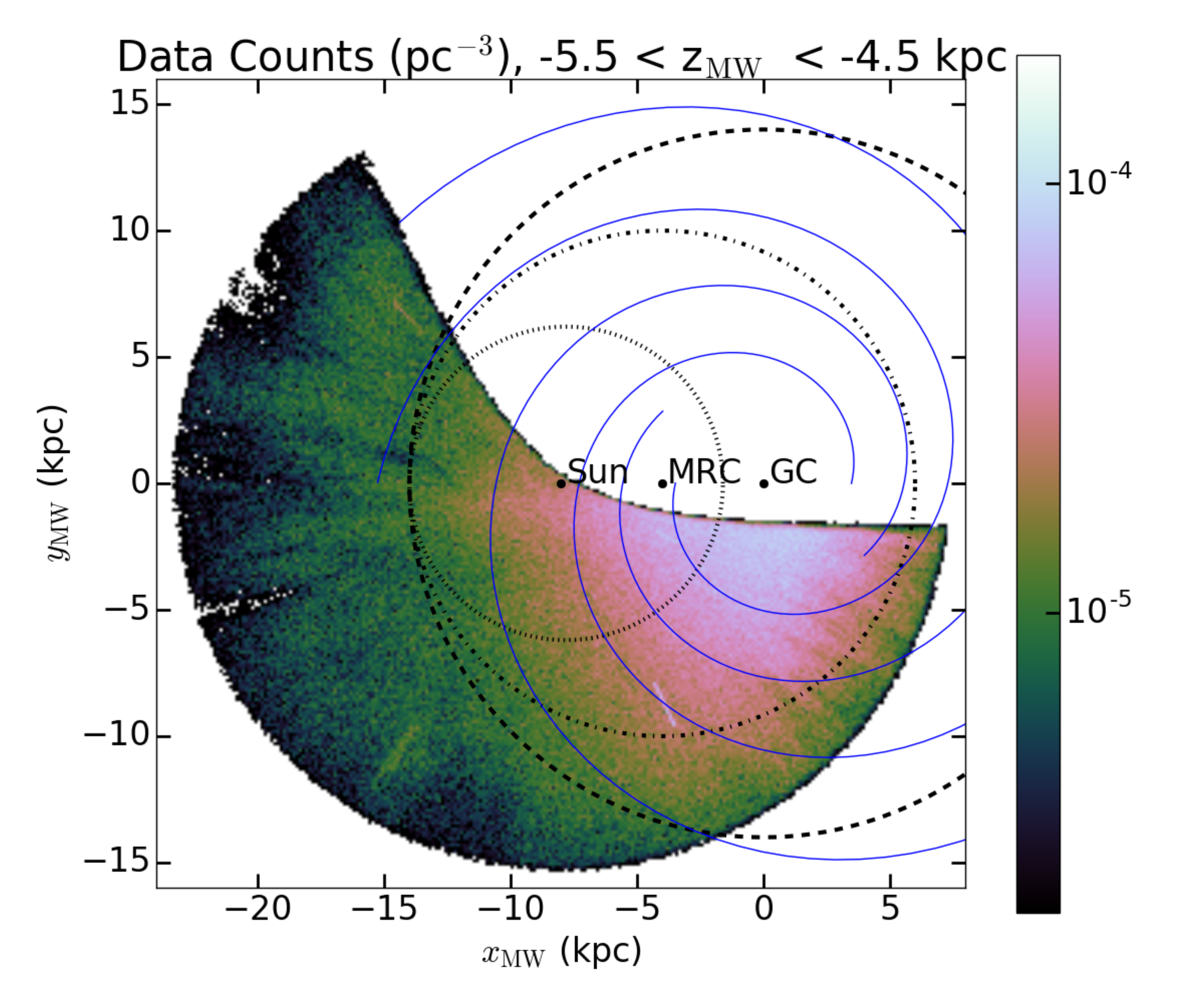}{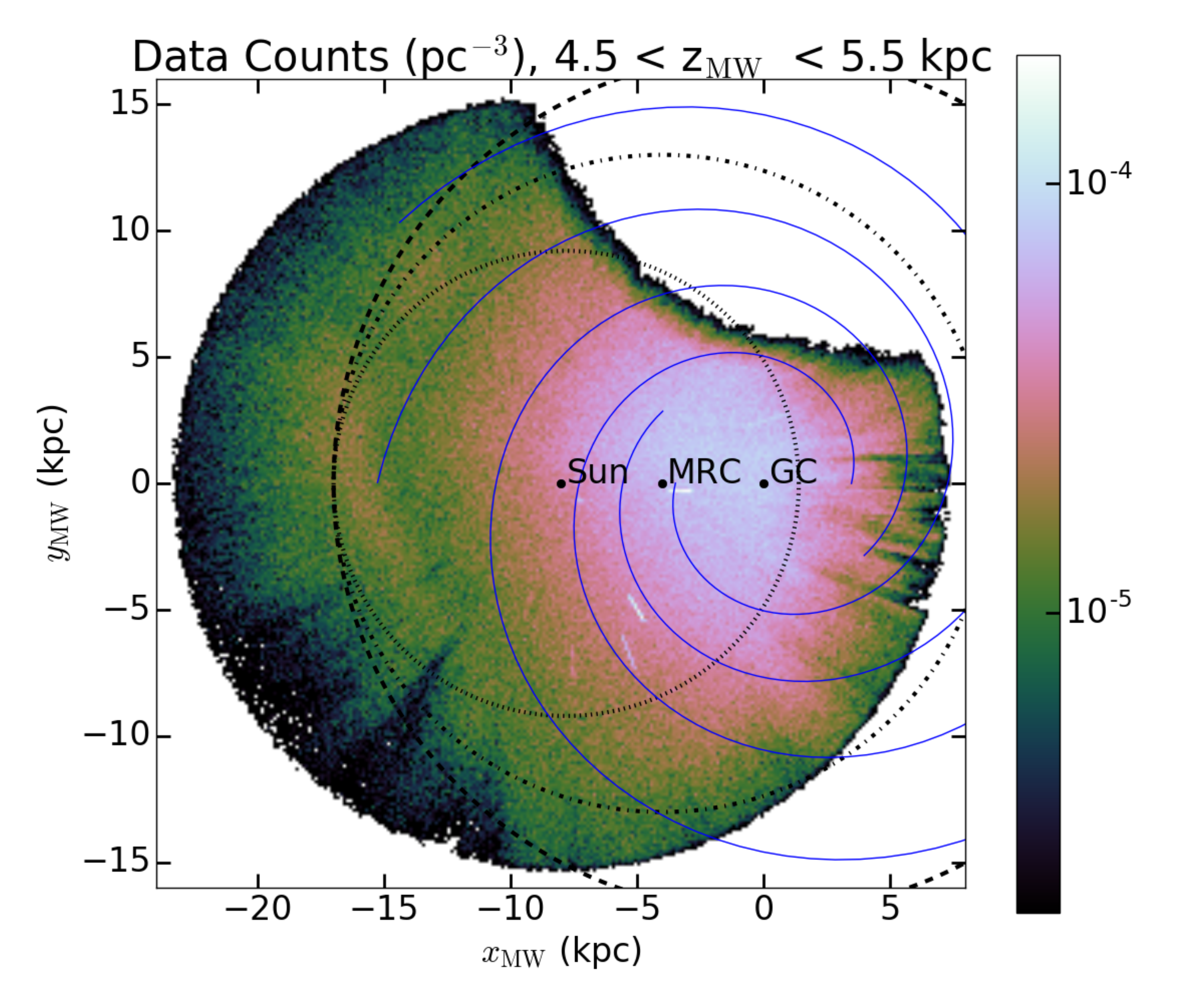}

\caption{\rm{In the top row, we have a map of all $0.1 < g_{\rm{P1}} -r_{\rm{P1}} < 0.5$ stars at Galactic height $-2.5 < z_{\rm{MW}} < -1.5$ kpc with a Galactocentric cylinder with radius 14 kpc (dashed line), an alternate cylinder 4 kpc from the Galactic center with radius 10 kpc (dot-dashed line) and a Heliocentric cylinder with radius 6 kpc (dotted line) on the left. We also have a map of all $0.1 < g_{\rm{P1}} -r_{\rm{P1}} < 0.5$ stars at Galactic height $1.5 < z_{\rm{MW}} < 2.5$ kpc  with a Galactocentric cylinder with radius 17 kpc (dashed line), an alternate cylinder 4 kpc from the Galactic center with radius 13 kpc (dot-dashed line) and a Heliocentric cylinder with radius 9 kpc (dotted line) on the right. The second row contains analogous plots for Galactic height $-3.5 < z_{\rm{MW}} < -2.5$ kpc (left) and $2.5 < z_{\rm{MW}} < 3.5$ kpc (right). The third row contains analogous plots for Galactic height $-4.5 < z_{\rm{MW}} < -3.5$ kpc (left) and $3.5 < z_{\rm{MW}} < 4.5$ kpc (right). The fourth row contains analogous plots for Galactic height $-4.5 < z_{\rm{MW}} < -3.5$ kpc (left) and $3.5 < z_{\rm{MW}} < 4.5$ kpc (right). 
}}
\label{fig:multicross1}\end{figure*}

Figs. \ref{fig:multicross1} and \ref{fig:multicross2} show cross sections of total stellar mass density and residual stellar mass density after a MW model is subtracted. These figures are analogous to Fig. \ref{fig:crosssection}. We see that the ringlike over density appears to shift to larger radii farther from the Galactic plane. This is due to stars being projected along the line of site to larger heights and radii (or alternately smaller heights and radii). Our MATCH analysis accounts for this projection effect by modeling distance uncertainties of individual stars. This is why we primarily rely on MATCH distance measurements (i.e. in Section \ref{sect:quant}) rather than our less reliable individual star method shown here. Despite this, it is clear that the Southern MR structure is at smaller radius for equivalent Galactic heights.

\begin{figure*}[ht]
\plottwo{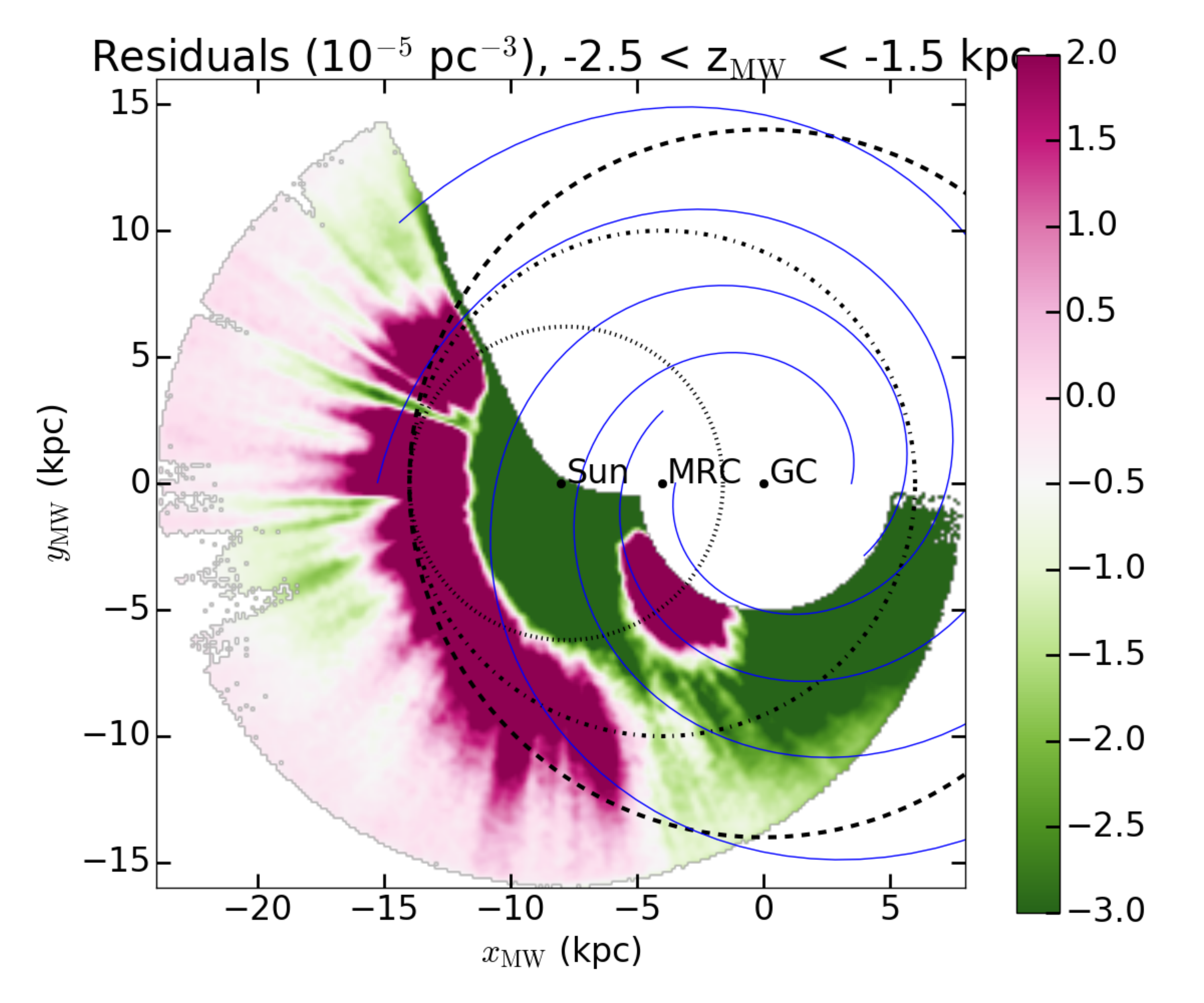}{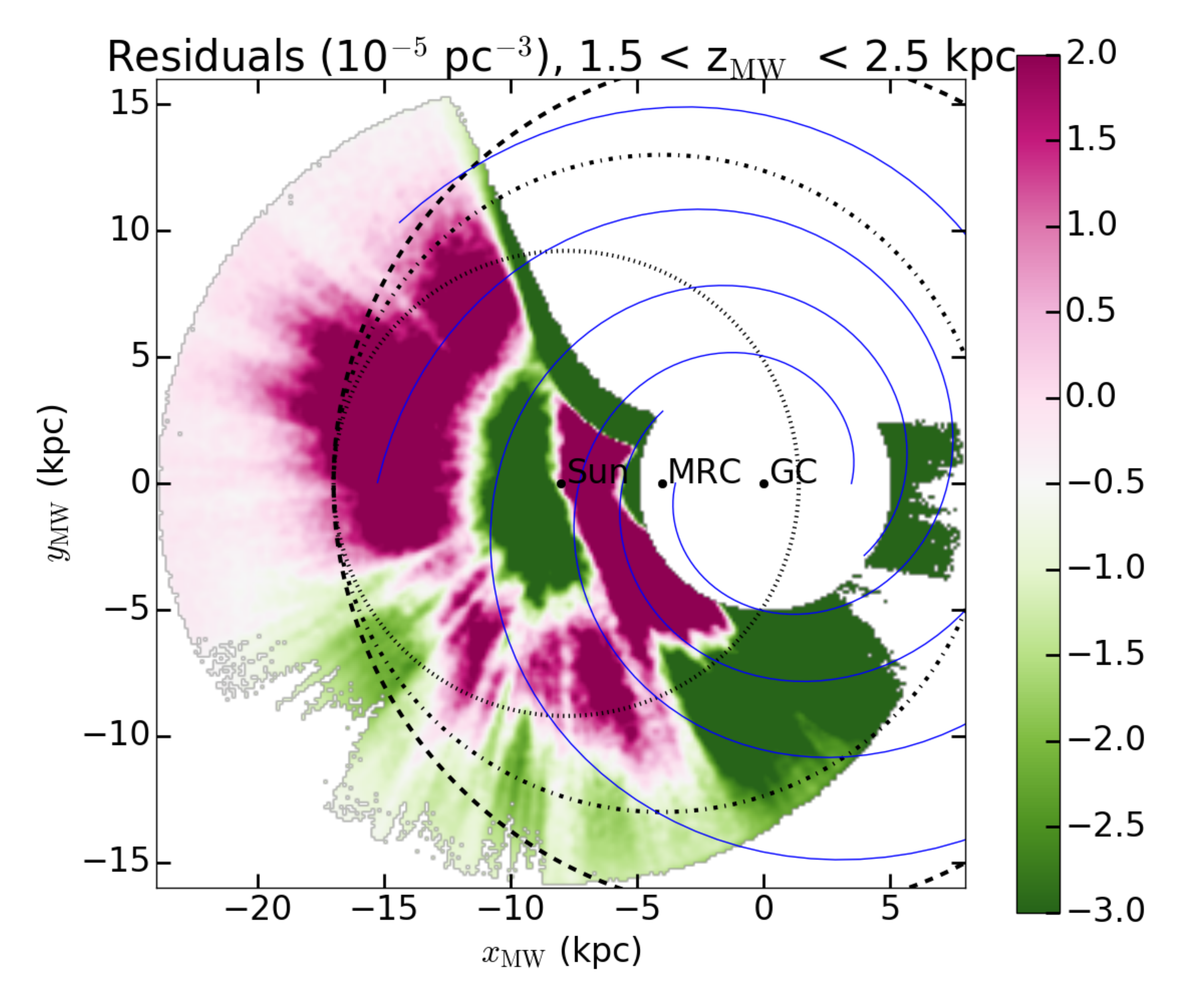}
\plottwo{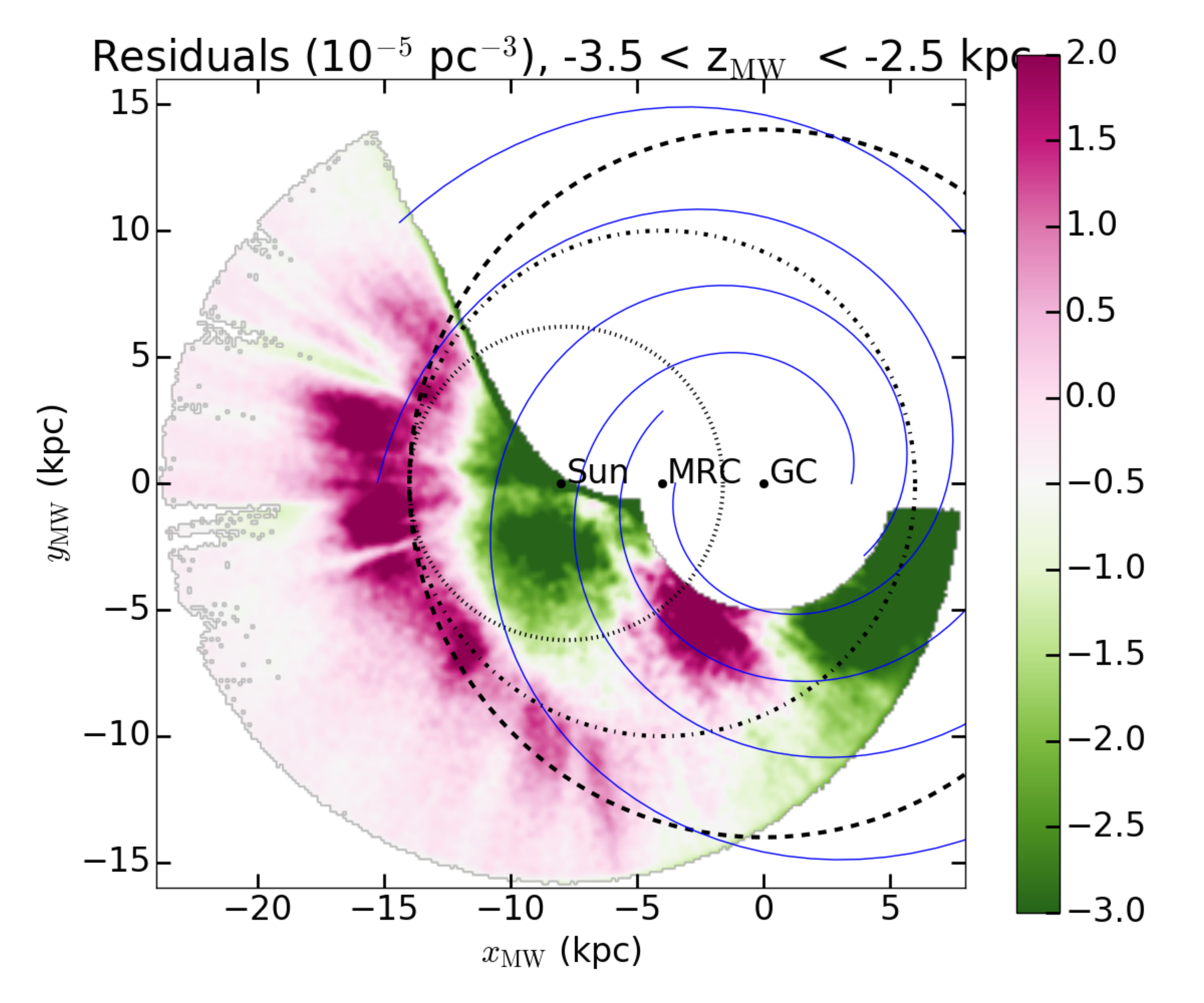}{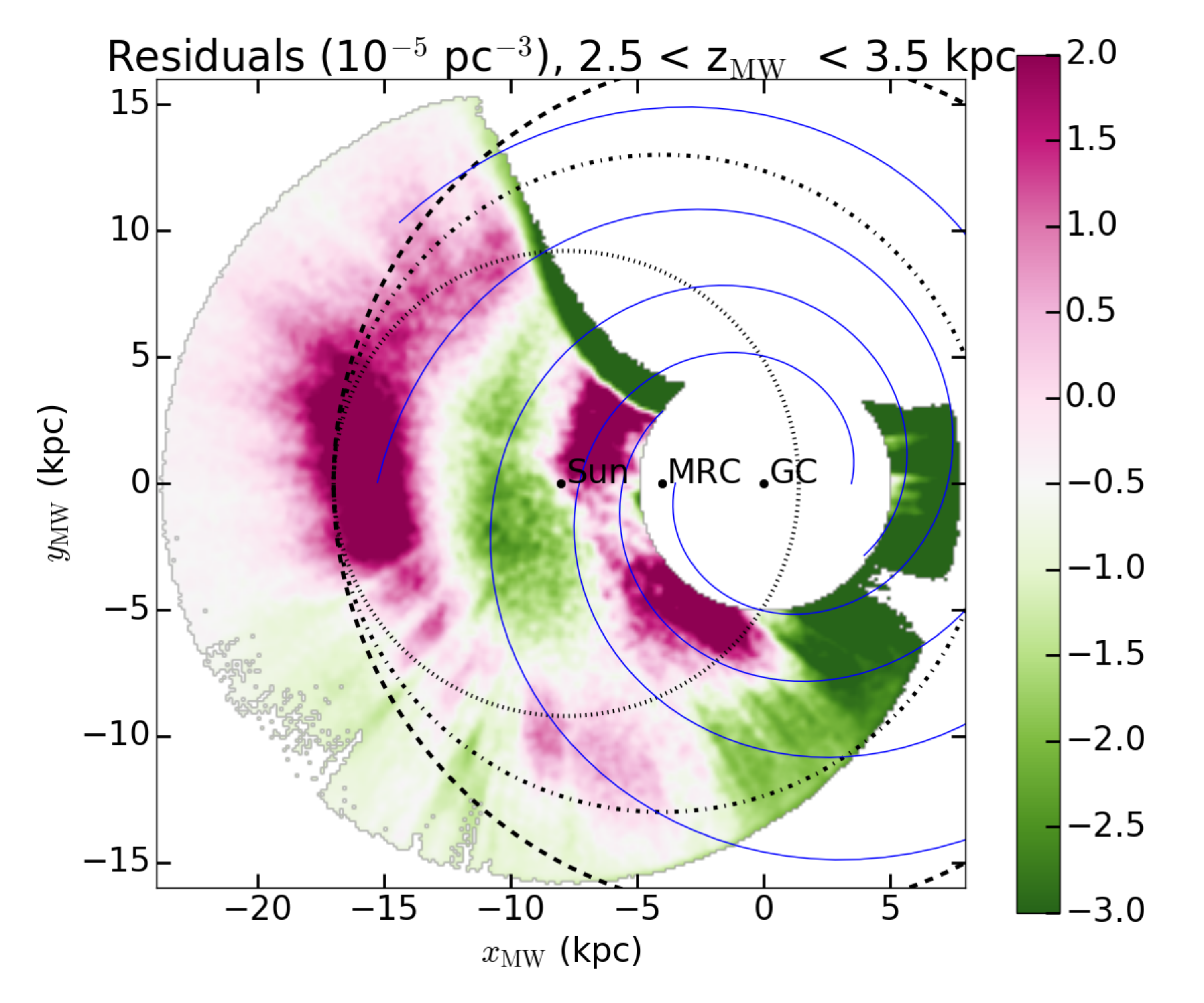}
\plottwo{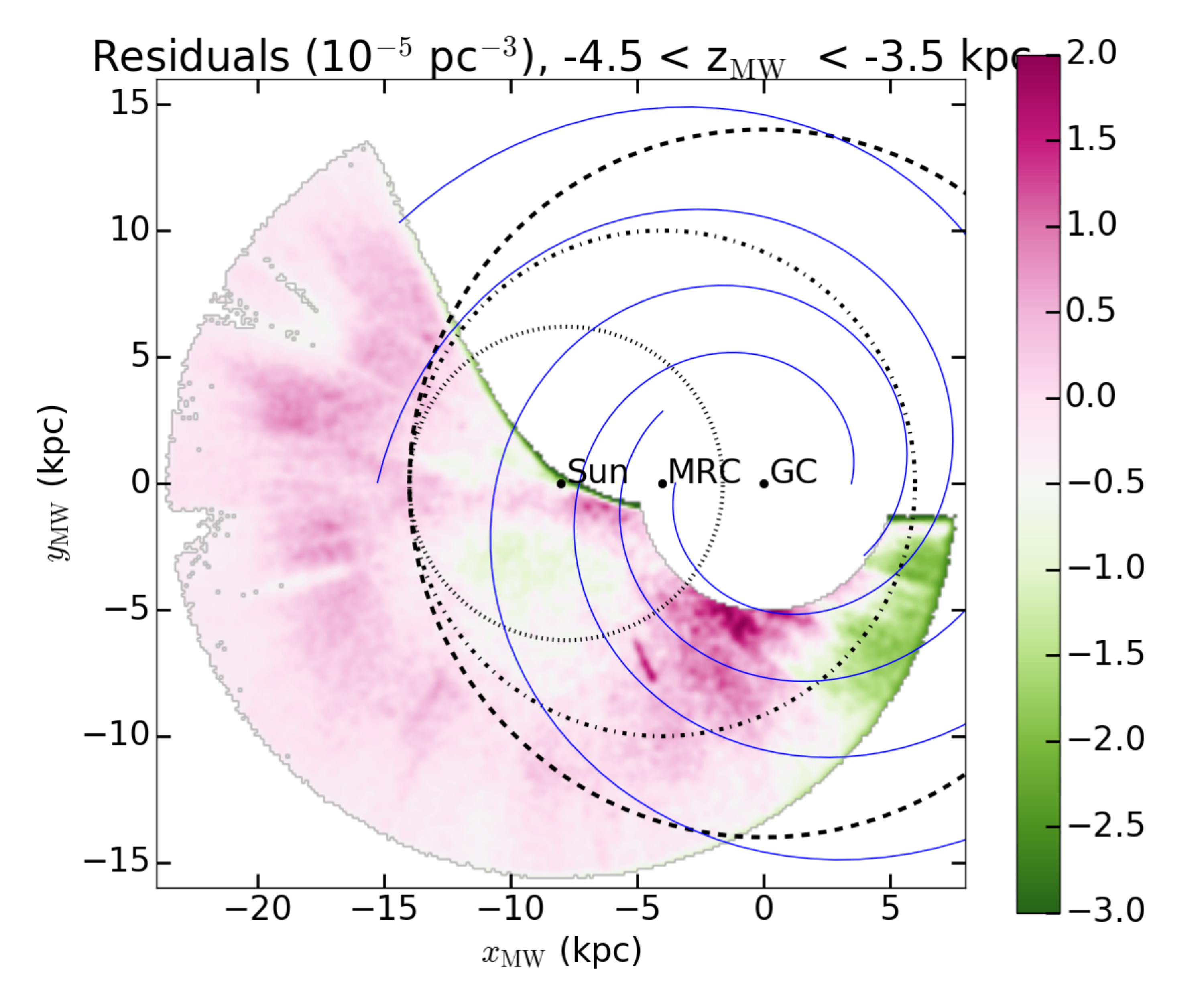}{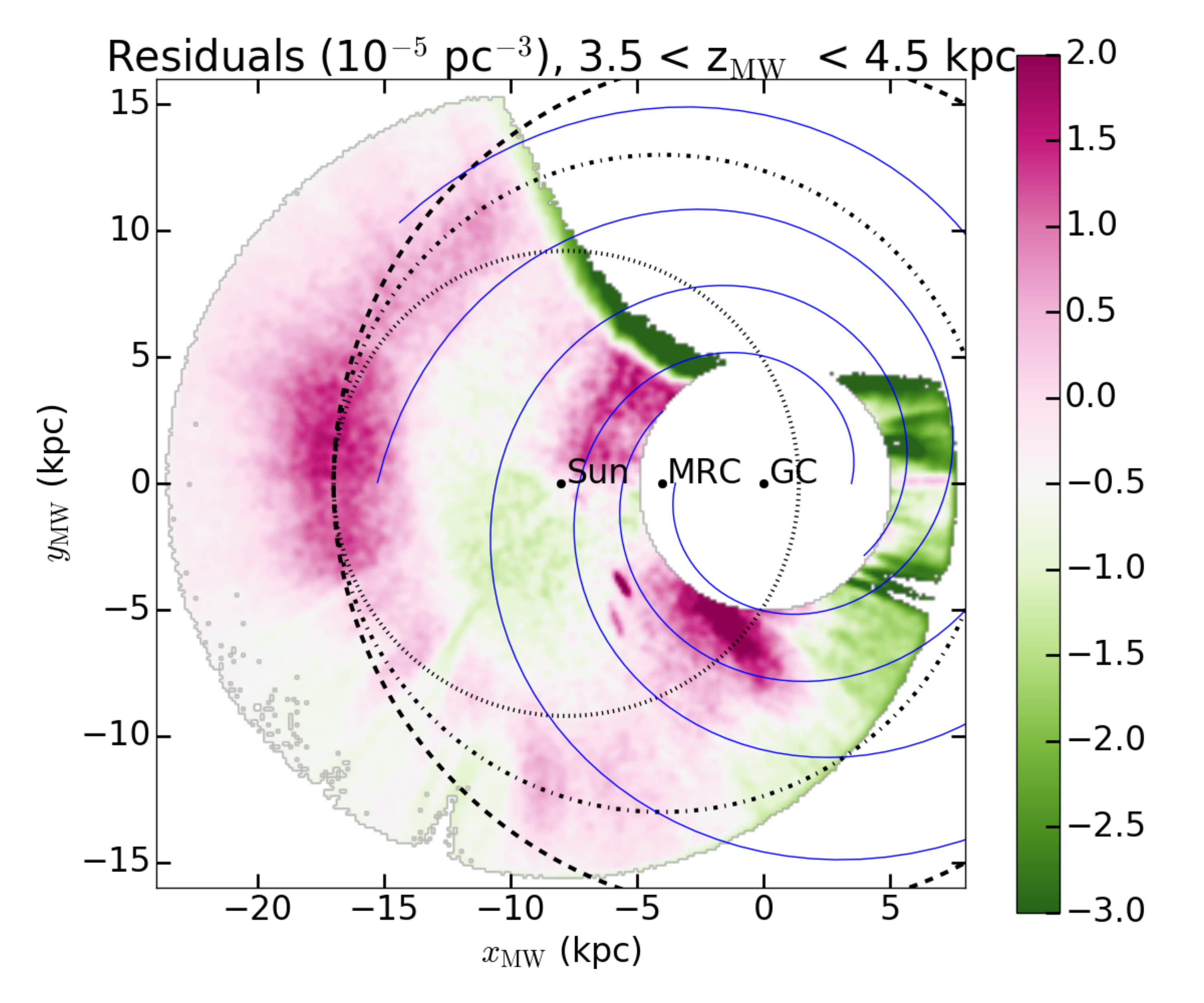}
\plottwo{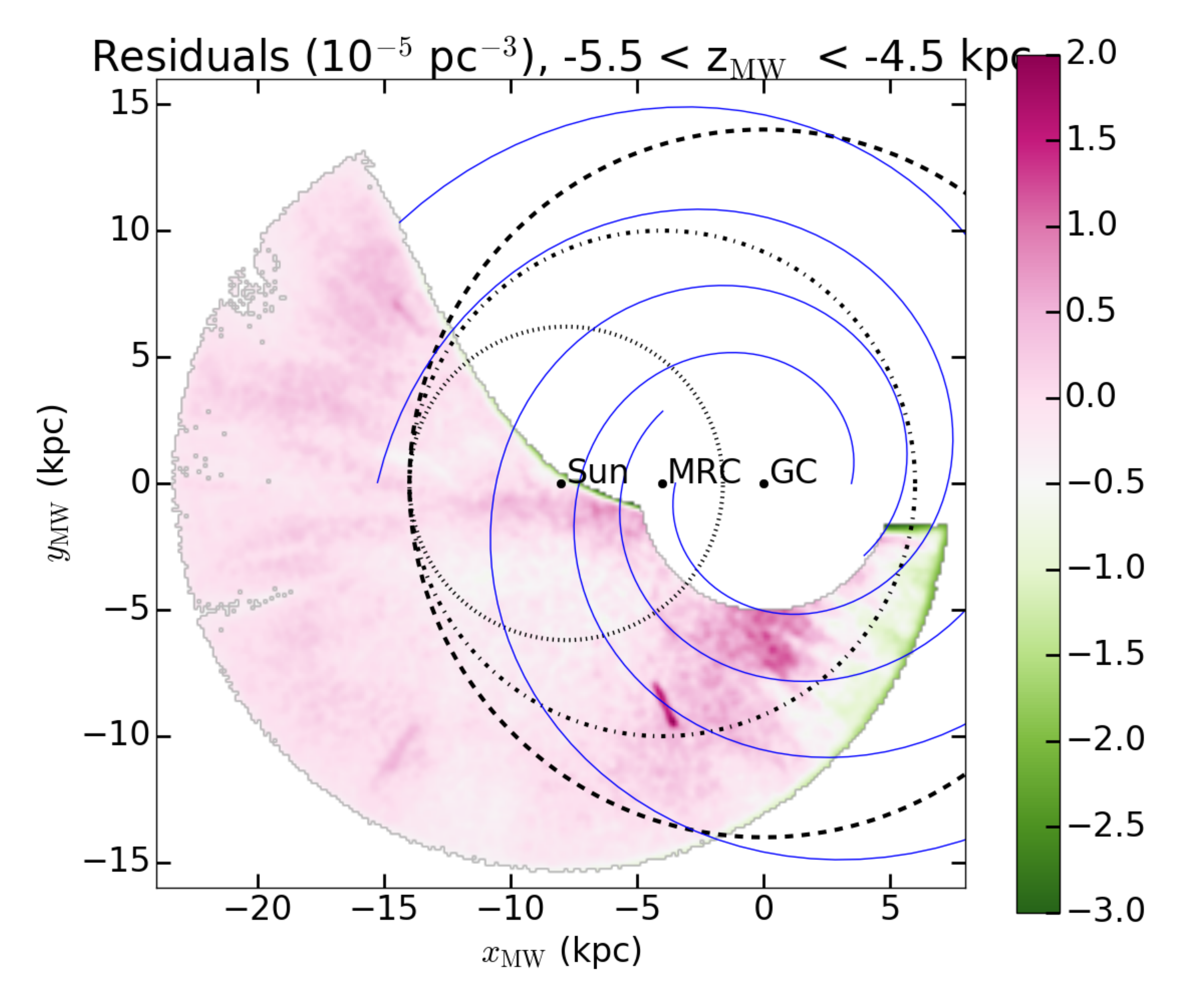}{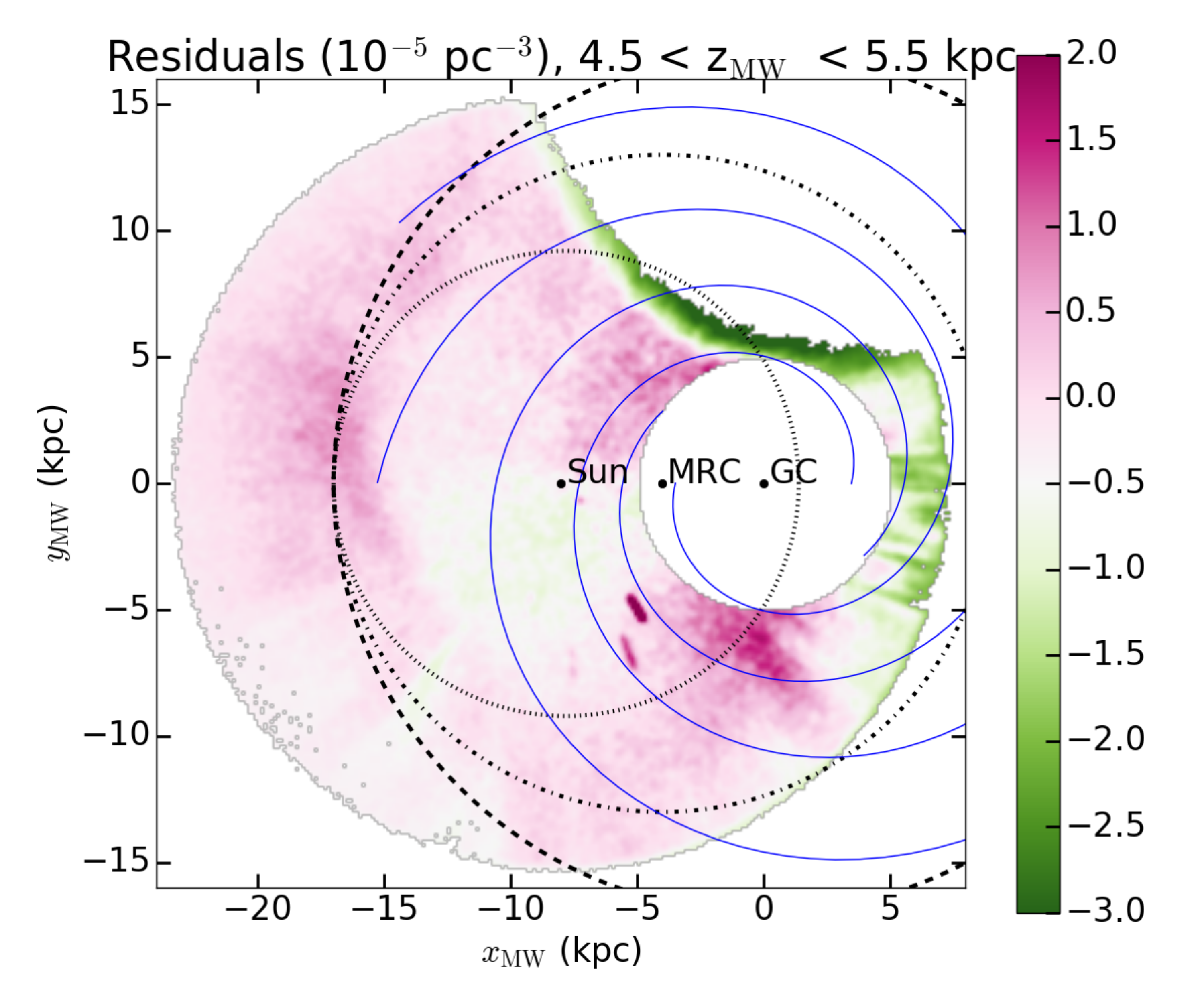}

\caption{\rm{In the top row, we have a map of residual $0.1 < g_{\rm{P1}} -r_{\rm{P1}} < 0.5$ stars after removing a Milky Way fit at Galactic height $-2.5 < z_{\rm{MW}} < -1.5$ kpc with a Galactocentric cylinder with radius 14 kpc (dashed line), an alternate cylinder 4 kpc from the Galactic center with radius 10 kpc (dot-dashed line) and a Heliocentric cylinder with radius 6 kpc (dotted line) on the left. We also have a map of all $0.1 < g_{\rm{P1}} -r_{\rm{P1}} < 0.5$ stars at Galactic height $1.5 < z_{\rm{MW}} < 2.5$ kpc with a Galactocentric cylinder with radius 17 kpc (dashed line), an alternate cylinder 4 kpc from the Galactic center with radius 13 kpc (dot-dashed line) and a Heliocentric cylinder with radius 9 kpc (dotted line) on the right. We also show model MW arms from \citet{FAUC++06} in blue. The second row contains analogous plots for Galactic height $-3.5 < z_{\rm{MW}} < -2.5$ kpc (left) and $2.5 < z_{\rm{MW}} < 3.5$ kpc (right). The third row contains analogous plots for Galactic height $-4.5 < z_{\rm{MW}} < -3.5$ kpc (left) and $3.5 < z_{\rm{MW}} < 4.5$ kpc (right). The fourth row contains analogous plots for Galactic height $-4.5 < z_{\rm{MW}} < -3.5$ kpc (left) and $3.5 < z_{\rm{MW}} < 4.5$ kpc (right). 
}}
\label{fig:multicross2}\end{figure*}



\begin{figure}[ht]
\plottwo{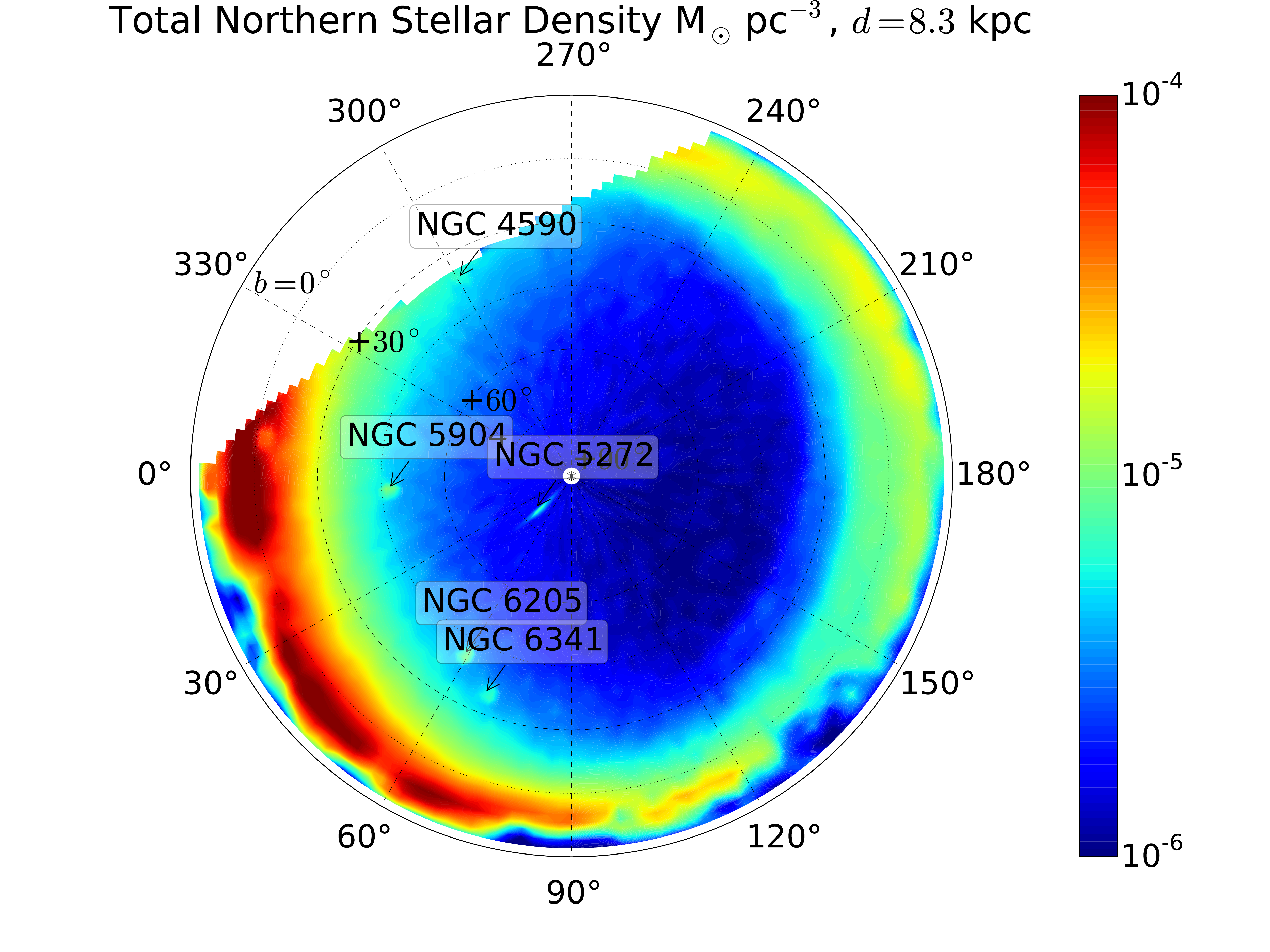}{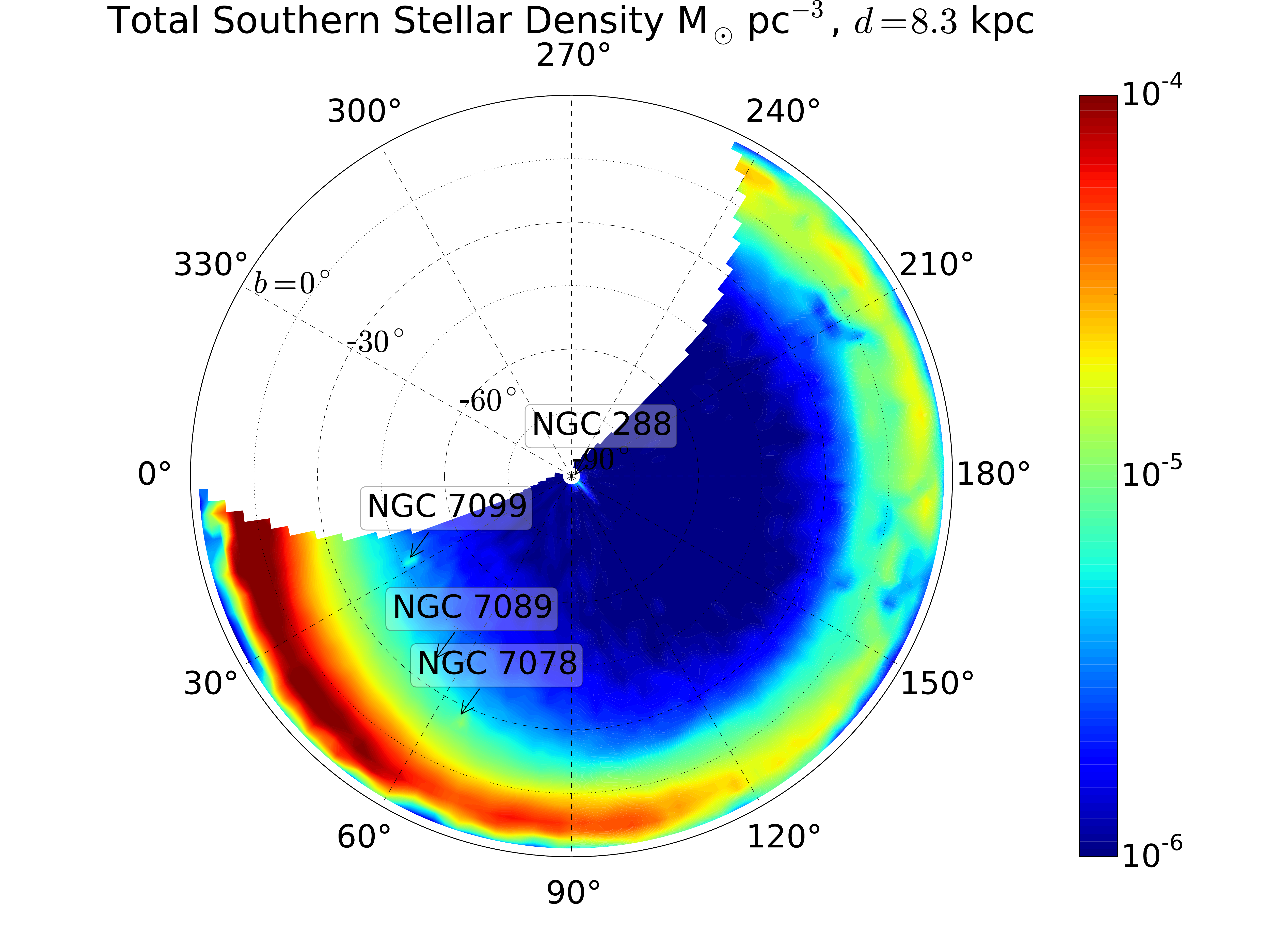}
\plottwo{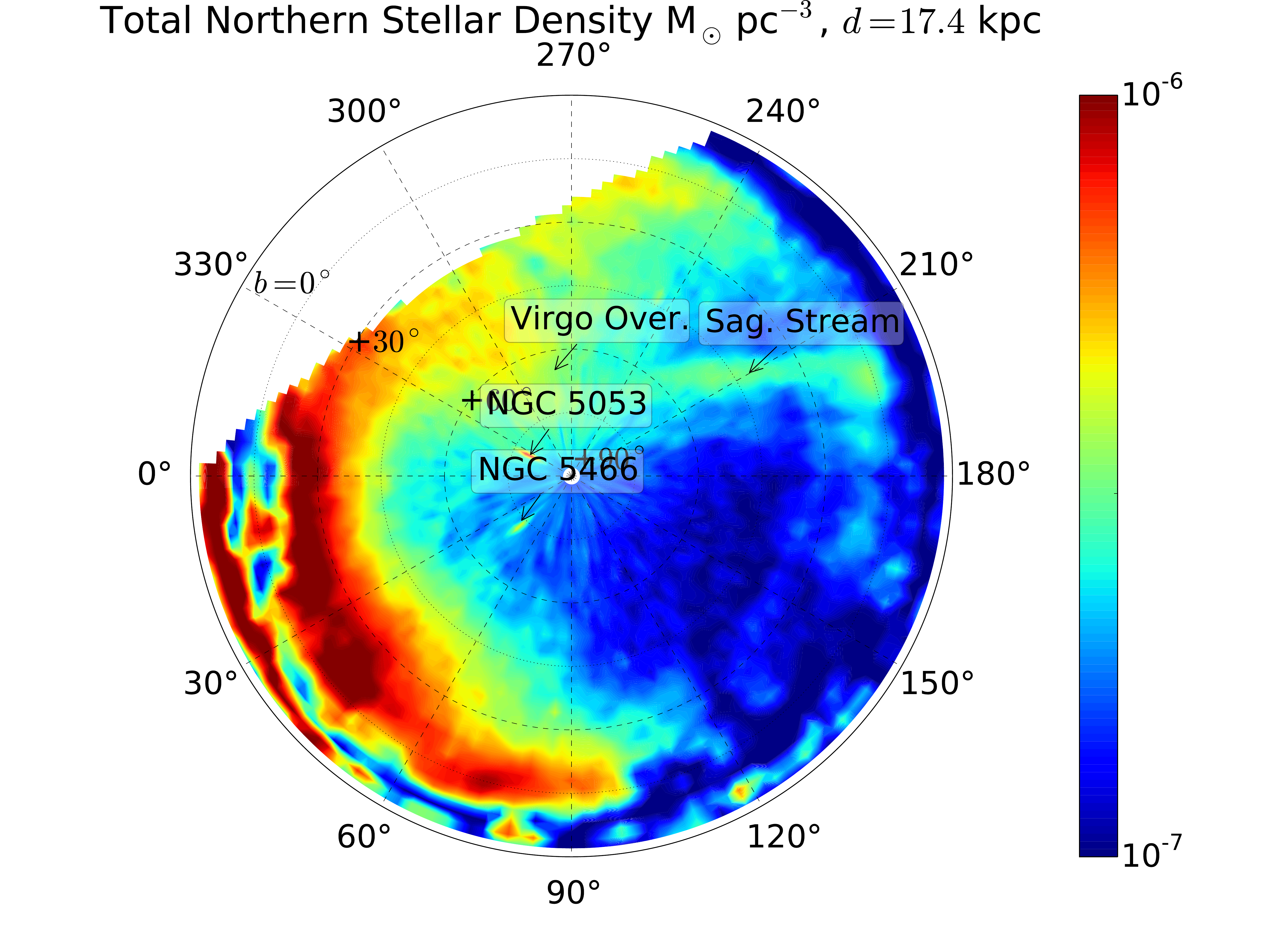}{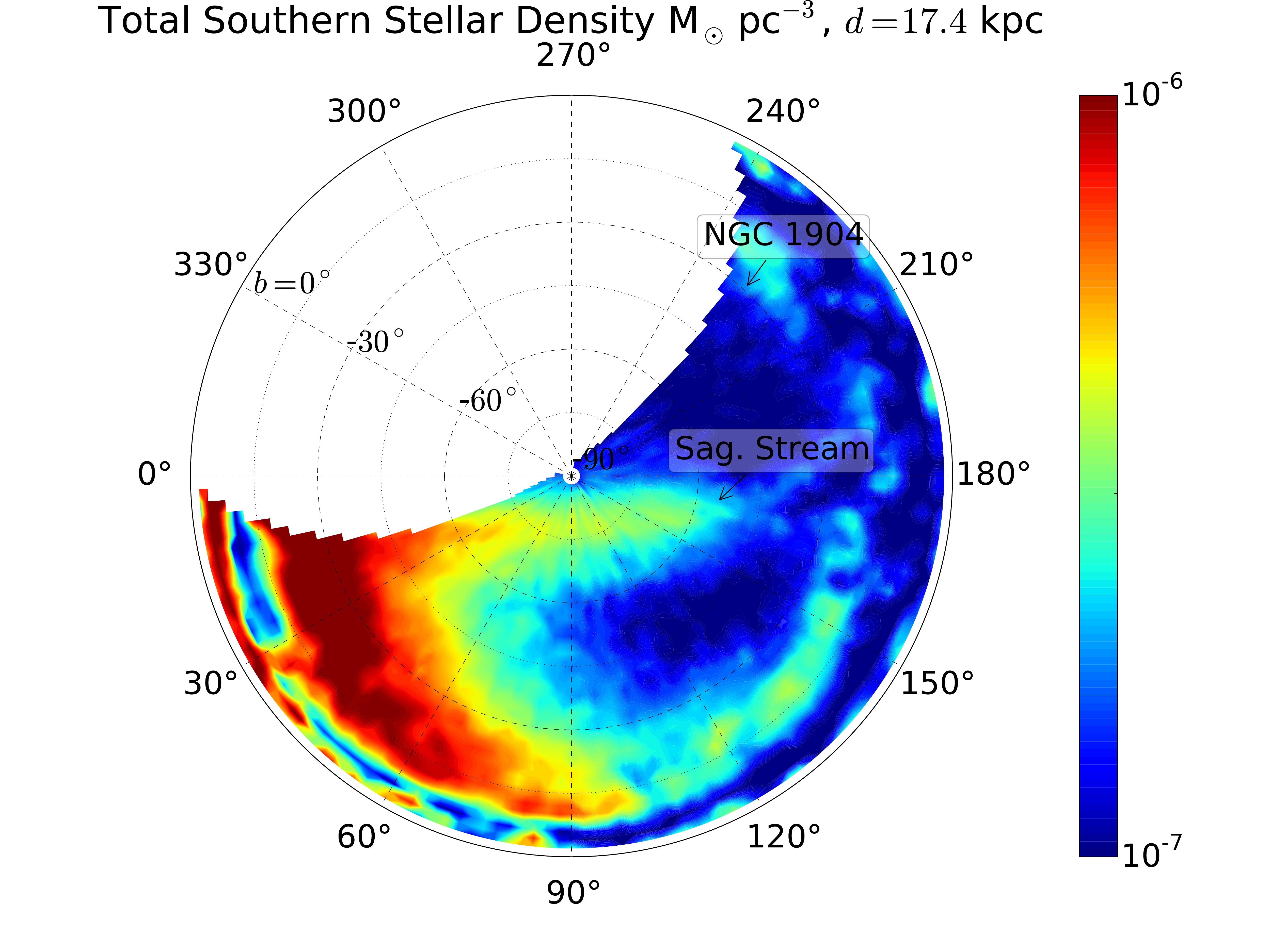}
\caption{\rm{Hemispherical density plots in at $d = 8.3$ kpc (top) and $17.4$ kpc (bottom). The density in the Northern hemisphere is shown in the left columns as while the density in the Southern hemisphere is shown in the right two columns. Prominent globular clusters and other overdensities are labeled.}}
\label{fig:hemi}\end{figure}

We can also use different projections of our data to make different Milky Way and local structures more apparent. Fig \ref{fig:hemi} shows our stellar mass density at $d = 8.3$ and $d = 17.4$ in the Northern and Southern hemispheres using a hemispherical projection in which the appropriate pole appears in the center of each plot. We label all of the clusters taken from \citet{HARR96} and \citet{KHAR++13} and noted in Table \ref{tab:cluster}. In the $d = 8.3$ kpc plots, the Monoceros Ring appears as green mass on the right side of each plot. The greater extent in the Northern hemisphere is once again obvious. In the $d = 17$ kpc plots, we can clearly see the Virgo overdensity \citep{JURI++08} and the Sagittarius Stream \citep{YANN++00,SLAT++13}. We may estimate the mass of the Sagittarius Stream with methods similar to those presented here in a future paper. 

\section{Examining Metallicity with \textsc{match}}\label{sect:addmetal}

Estimating metallicities photometrically is generally difficult, particularly when one has not properly modeled stellar age and Galactic extinction. In this paper we use a single stellar age and do not vary Galactic extinction along the line of sight. At typical MR distances of more than a few kpc and more than a few degrees away from the Galactic plane, these approximations are justified. But at lower latitudes and smaller distances, metallicity variation becomes a proxy for variations in age and extinction. With this in mind, it is worthwhile to take a cautious, qualitative look at changes in metallicity in and around the MR and to see how metallicity aliasing affects our analysis. 

\begin{figure}[ht]
\includegraphics[width=0.33\columnwidth]{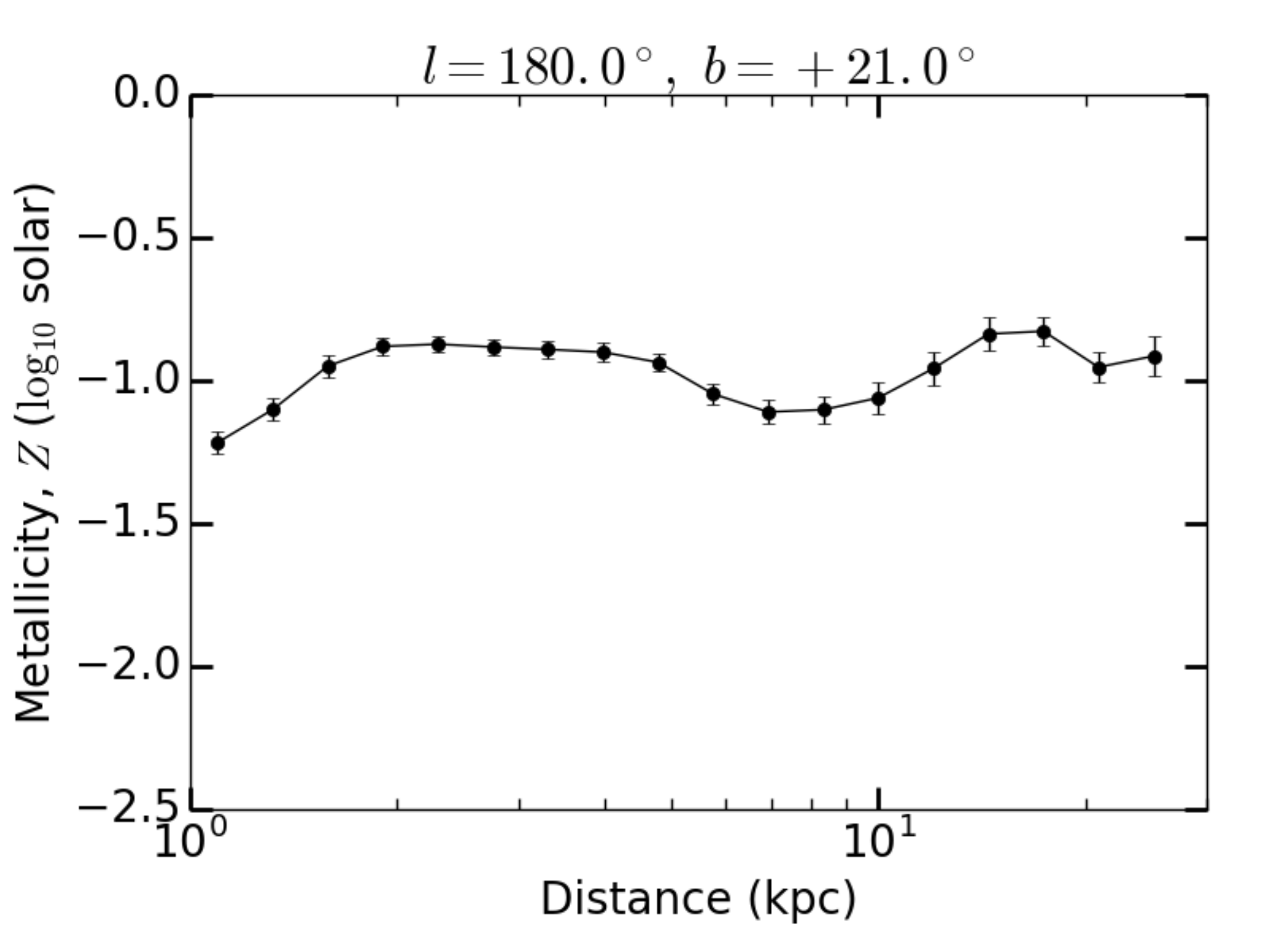}
\includegraphics[width=0.33\columnwidth]{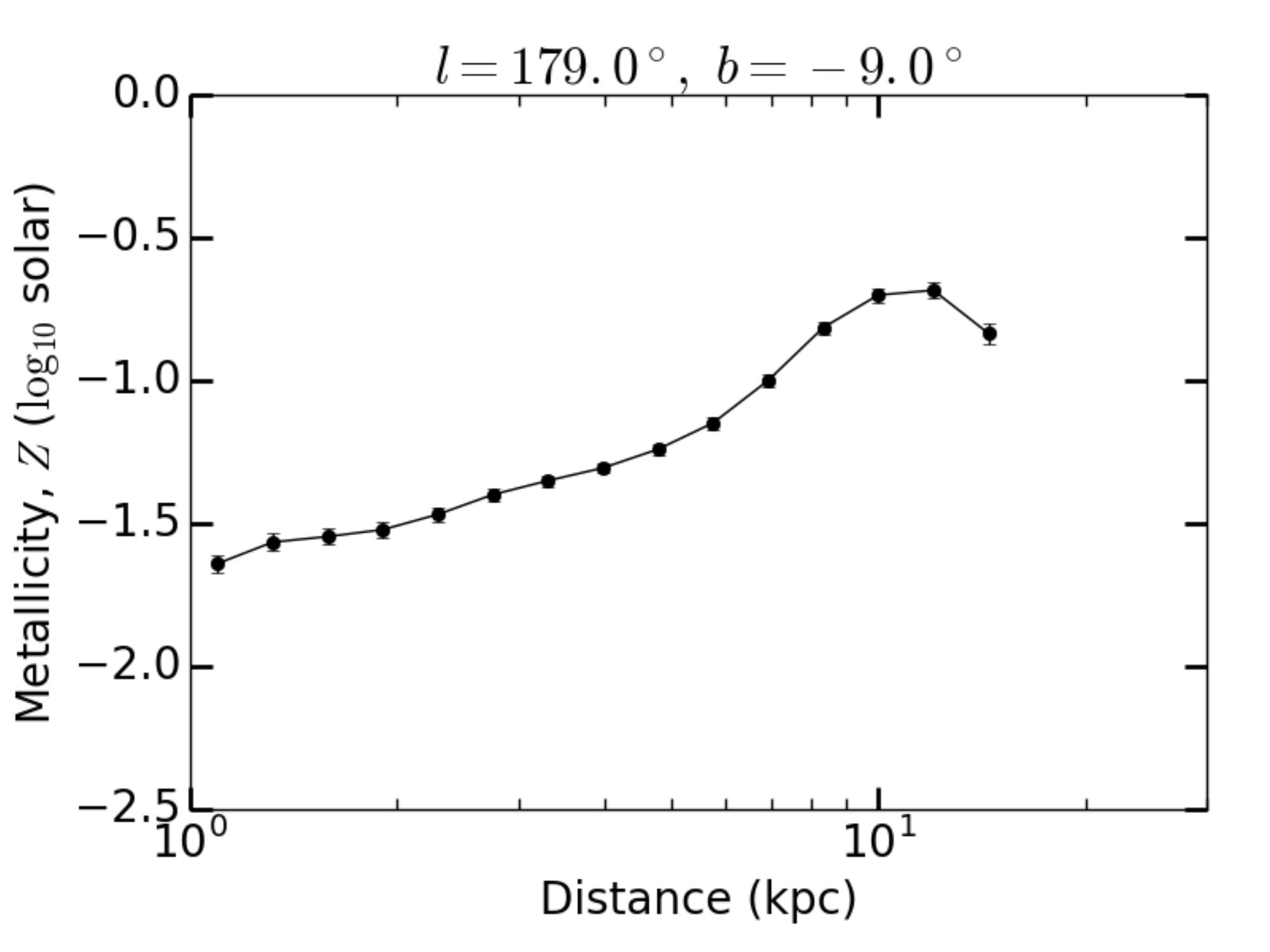}
\includegraphics[width=0.33\columnwidth]{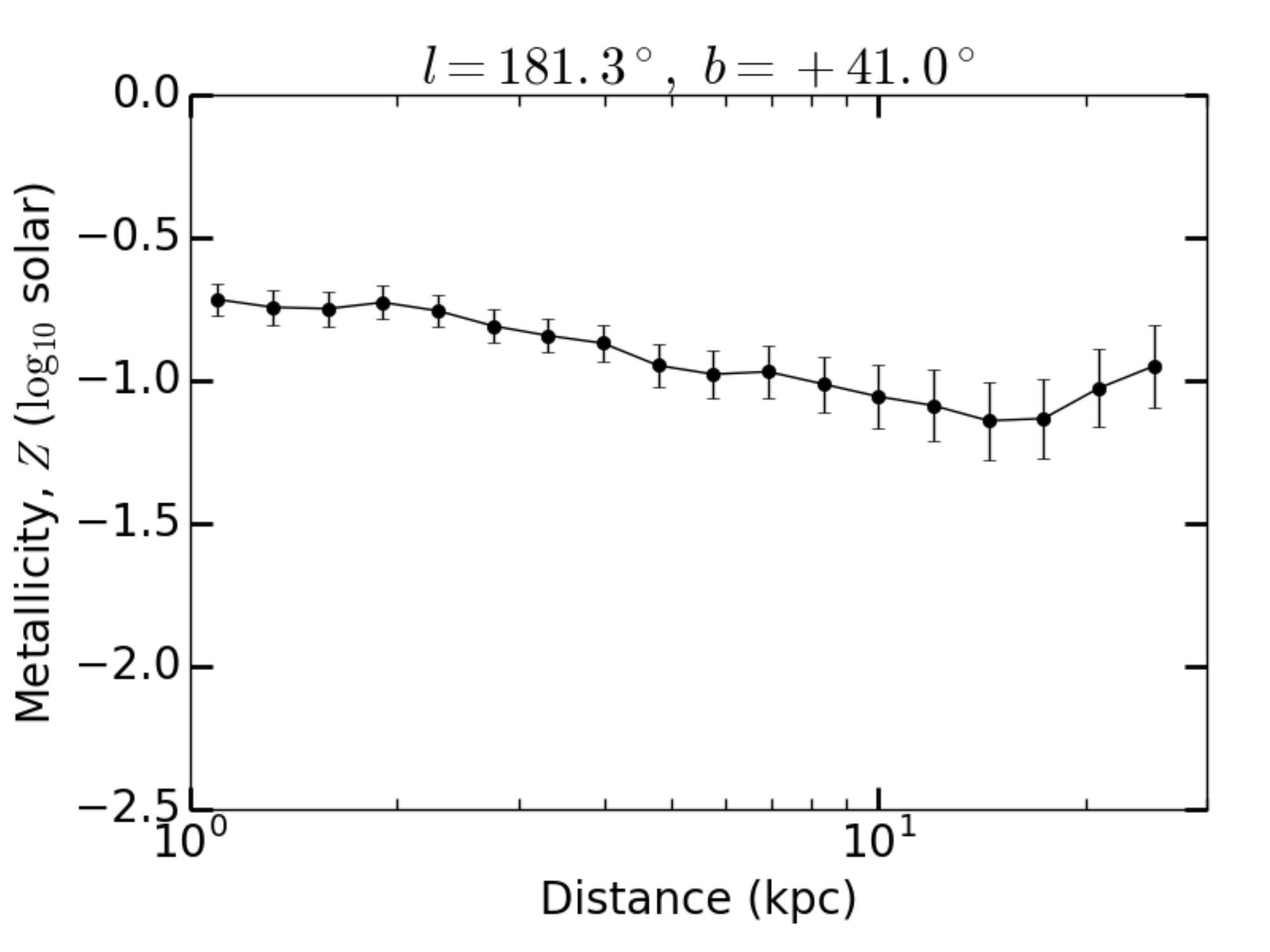}
\caption{\rm{The estimated Metallicity (Eq.\ \ref{eq:metal}) as a function of distance. The line is an interpolation as opposed to a fit. We see significant dip in the top, $b = +21^\circ$, panel corresponding to the MR. The apparent metallicity in the middle, $b = -9^\circ$ panel monotonically increasing. This is possibly due to the differential extinction along the line of sight. The metallicity slowly decreases with distance in the bottom, $b = +41^\circ$ panel.}}
\label{fig:metal}\end{figure}

Fig.\ \ref{fig:metal} shows metallicity as a function of distance, $Z(D)$, for our familiar $b = +21^\circ$, $b = -9^\circ$ and $b = +41^\circ$ ($l = 180^\circ$) stellar populations. Since we will only deal with metallicity in a semi-quantitative way in this paper, we do not fit the data, and the line is just an interpolation. To estimate uncertainty in our metallicity bins, we note that the spacing between isochrones with metallicity difference $\Delta Z$ is roughly $10(\gpo-\rpo)$. We can approximate the uncertainty in $Z$ as
\begin{equation}
\sigma_Z = 10\left(\frac{ \sum \sigma_g^2+\sigma_r^2}{N}\right)^{1/2},
\end{equation}
where the sum is over all stars in a distance bin and $N$ is the number of stars in a bin.

Along the $b = +21^\circ$ line of sight, measured metallicity starts slightly low, likely due to a younger stellar population or extinction overcorrection being aliased as a low metallicity population. Through the rest of the range, it maintains a constant value of -0.9, with a possible dip to -1.1 consistent with the MR overdensity in Fig.\ \ref{fig:distance}. Problematically, the metallicity increases in the middle ($b = -9^\circ$) panel. Here, we must remember that our dust extinction correction is a single number along the entire line of sight. At low latitudes, where there is dust to a significant distance along the line of sight, our stars apparently become redder along the line of sight, and \textsc{match} aliases this differential extinction as a change in metallicity. We estimate that this effect causes an error of less than 0.2 magnitudes in distance modulus at very short distances, and essentially disappears at $d = 8$ kpc, our region of interest. Along the $b=+41^\circ$ line of sight, metallicity falls more or less monotonically from -0.7 to -1.1 as we would expect and is roughly consistent with \citet{IVEZ++08}. \citet{AN++13} finds significantly lower metallicities ($Z \approx -2$). Including lower metallicities does not improve our density fits, and as the focus of this paper is density and not metallicity, we do not probe this further. 

\end{document}